\documentclass[twoside,final,11pt]{PAC46elsart}

\usepackage{color}
\usepackage{graphicx}
\usepackage{latexsym}
\usepackage{amssymb,amsmath,amstext}
\usepackage{multicol}
\usepackage{rotating}
\usepackage{longtable}
\usepackage{colordvi}
\usepackage{ulem}
\usepackage[numbers]{natbib}
\usepackage{epsfig}
\usepackage{graphicx}
\usepackage{subfigure}
\usepackage{marvosym}
\usepackage{subfigure}
\usepackage[dvipsnames]{xcolor}
\usepackage{siunitx}
\usepackage{pstricks}
\usepackage{fancyhdr}
\usepackage{subfigure}
\usepackage{lscape}
\usepackage{afterpage}

\newsavebox{\Imagebox}

\newcommand{\xBj}{x_{\rm B}}

\newcommand{\phigg}{\phi_{\gamma\gamma}}

\begin{document}

%

\hyphenation{brems-strah-lung}
\hyphenation{po-si-trons}

\newcommand*{\Apr}{{A^\prime}}

%
\begin{frontmatter}

{\Huge Letter-of-Intent to PAC46}

\vspace*{8pt}

{\Huge LOI12-18-004}

\vspace*{12pt}

\title{\bf\huge{Physics with Positron Beams} \\ \vspace*{8pt} \bf\huge{at Jefferson Lab 12 GeV}}
%
%
\author {\sf Andrei~Afanasev$^{29}$, }
\author {\sf Ibrahim~Albayrak$^{9}$, }
\author {\sf Salina~Ali$^{27}$, } 
\author {\sf Moskov~Amaryan$^{24}$, } 
\author {\sf Annalisa~D'Angelo$^{18,30}$, } 
\author {\sf John~Annand$^{34}$, } 
\author {\sf John~Arrington$^{6}$, } 
\author {\sf Arshak~Asaturyan$^{8}$, } 
\author {\sf Harut~Avakian$^{1}$, } 
\author {\sf Todd~Averett$^{28}$, } 
% ...

\vspace*{4pt}

\address{$^{2}$Institut de Physique Nucl\'eaire \\
Universit\'e Paris-Sud \& Paris-Saclay \\ 
15 rue Georges Cl\'emenceau, 91406 Orsay cedex, France}

\vspace*{4pt}

\address{$^{3}$ Laboratory for Nuclear Science \\
Massachusetts Institute of Technology \\
77 Massachusetts Avenue, Cambridge, MA 02139, USA}

\vspace*{4pt}

\address{$^{4}$ Istituto Nazionale di Fisica Nucleare \\ 
Sezione di Genova \\
Via Dodecaneso, 33 - 16146 Genova, Italia}

\vspace*{4pt}

\address{$^{5}$ Universit\`a di Genova \\ 
Via Balbi, 5 - 16126 Genova, Italia}

\vspace*{4pt}

\address{$^{6}$ Argonne National Laboratory \\
Physics Division \\
9700 S. Cass Avenue, Argonne, IL 60439, USA}

\vspace*{4pt}

\address{$^{7}$ University of Connecticut \\
Department of Physics \\
2152 Hillside Road, Storrs, CT U-3046, USA}

\vspace*{4pt}

\address{$^{8}$ A.~Alikhanyan National Laboratoty \\
Yerevan Physics Institute \\
Alikanian Brothers Street, 2, Yerevan 375036, Armenia}

\vspace*{4pt}

\address{$^{9}$ Akdeniz \"Universitesi \\
Pinarba\c{s}i Mahallesi, 07070 Konyaalti/Antalya, Turkey}

\vfill\eject

\address{$^{10}$ Duke University \\
134 Chapel Drive, Durham, NC 27708, USA}

\vspace*{4pt}

\address{$^{11}$ Facult\'e des Sciences de Monastir \\
Avenue de l'environnement 5019, Monastir, Tunisie}

\vspace*{4pt}

\address{$^{12}$ Fairfield University \\
1073 N Benson Road, Fairfield, CT 06824, USA}

\vspace*{4pt}

\address{$^{13}$ Florida International University \\ 
Modesto A. Maidique Campus \\
11200 SW 8th Street, CP 204, Miami, FL 33199, USA}

\vspace*{4pt}

\address{$^{14}$ Hampton University \\
Physics Department \\ 
100 E Queen Street, Hampton, VA 23668, USA}

\vspace*{4pt}

\address{$^{15}$ Institut de Recherche sur les Lois Fondamentales de l'Univers \\
Commissariat \`a l'Energie Atomique, Universit\'e Paris-Saclay \\
91191 Gif-sur-Yvette, France}

\vspace*{4pt}

\address{$^{16}$ Istituto Nazionale di Fisica Nucleare \\ 
Sezione di Catania \\ 
Via Santa Sofia, 64 - 95123 Catania, Italia}

\vspace*{4pt}

\address{$^{17}$ Istituto Nazionale di Fisica Nucleare \\ 
Sezione di Ferrara \\ 
Via Saragat, 1 - 44122 Ferrara, Italia}

\vspace*{4pt}

\address{$^{18}$ Istituto Nazionale di Fisica Nucleare \\ 
Sezione di Roma Tor Vergata \\ 
Via de la Ricerca Scientifica, 1 - 00133 Roma, Italia}

\vspace*{4pt}

\address{$^{19}$ Istituto Nazionale di Fisica Nucleare \\ 
Sezione di Torino \\ 
Via P. Giuria, 1 - 10125 Torino, Italia}

\vspace*{4pt}

\address{$^{20}$ James Madison University, \\
800 S Main Street, Harrisonburg, VA 22807, USA}

\vspace*{4pt}

\address{$^{21}$ Mississippi State University \\ 
B.S. Hood Road, Mississippi State, MS 39762, USA}

\vspace*{4pt}

\address{$^{22}$ North Carolina A\&T State University \\
1601 E Market Street, Greensboro, NC 27411, USA}

\vspace*{4pt}

\address{$^{23}$ Ohio University \\
Athens, OH 45701, USA}

\vspace*{4pt}

\address{$^{24}$ Old Dominion University \\
5115 Hampton Boulevard, Norfolk, VA 23529, USA}

\vspace*{4pt}

\address{$^{25}$ Stonybrook University \\
100 Nicolls Road, Stonybrook, NY 11794, USA}

\vspace*{4pt}

\address{$^{26}$ Temple University \\
Physics Department \\ 
1925 N 12th Street, Philadelphia, PA 19122-180, USA}

\vfill\eject

\address{$^{27}$ The Catholic University of America \\
620 Michigan Avenue NE, Washington, DC 20064, USA}

\vspace*{4pt}

\address{$^{28}$ The College of William \& Mary \\
Small Hall, 300 Ukrop Way, Williamsburg, VA 23185, USA}

\vspace*{4pt}

\address{$^{29}$ The George Washington University \\
221 I Street NW, Washington, DC 20052, USA}

\vspace*{4pt}

\address{$^{30}$ Universit\`a degli Studi di Roma Tor Vergata \\ 
Via Cracovia, 50 - 00133 Roma, Italia}

\vspace*{4pt}

\address{$^{31}$ Universit\`a di Ferrara \\ 
Via Savonarola, 9 - 44121 Ferrara, Italia}

\vspace*{4pt}

\address{$^{32}$ Universit\"at Gie\ss en \\
Luwigstra\ss e 23, 35390 Gie\ss en, Deutschland}

\vspace*{4pt}

\address{$^{33}$ University of Colorado \\
Boulder, CO 80309, USA}

\vspace*{4pt}

\address{$^{34}$ University of Glasgow \\
University Avenue, Glasgow G12 8QQ, United Kingdom}

\vspace*{4pt}

\address{$^{35}$ University of Manitoba \\
66 Chancellors Cir, Winnipeg, MB 53T 2N2, Canada}

\vspace*{4pt}

\address{$^{36}$ University of New Hampshire \\
105 Main Street, Durham, NH 03824, USA}

\vspace*{4pt}

\address{$^{37}$ University of Virginia \\
Department of Physics \\
382 McCormick Rd, Charlottesville, VA 22904, USA}

\vspace*{4pt}

\address{$^{38}$ Univerza v Ljubljani \\
Faculteta za Matematico in Fiziko \\
Jadranska ulica 19, 1000 Ljubljana, Slovenia}

\vspace*{4pt}

\address{$^{39}$ Virginia Union University \\
1500 N Lombardy Street, Richmond, VA 23220, USA}

\vspace*{20pt}

{\small
3 June 2018
\par
{\it $^{\star}$Contact persons: J.~Grames (grames@jlab.org), E.~Voutier (voutier@ipno.in2p3.fr)}
}

%
%
\newpage

\null\vfill

\begin{center}

\begin{abstract}
{\small
Positron beams, both polarized and unpolarized, are identified as essential ingredients for the experimental program at the next generation of lepton accelerators. In the context of the Hadronic Physics program at the Jefferson Laboratory (JLab), positron beams are complementary, even essential, tools for a precise understanding of the electromagnetic structure of the nucleon, in both the elastic and the deep-inelastic regimes. For instance, elastic scattering of (un)polarized electrons and positrons off the nucleon allows for a model independent determination of the electromagnetic form factors of the nucleon. Also, the deeply virtual Compton scattering of (un)polarized electrons and positrons allows us to separate unambiguously the different contributions to the cross section of the lepto-production of photons, enabling an accurate determination of the nucleon Generalized Parton Distributions (GPDs), and providing an access to its Gravitational Form Factors. Furthermore, positron beams offer the possibility of alternative tests of the Standard Model through the search of a dark photon or the precise measurement of electroweak couplings. This letter proposes to develop an experimental positron program at JLab to perform unique high impact measurements with respect to the two-photon exchange problem, the determination of the proton and the neutron GPDs, and the search for the $\Apr$ dark photon.

The ability of the CEBAF injector for the efficient production of polarized positrons was recently demonstrated. The Polarized Electrons for Polarized Positrons (PEPPo) technique offers a direct and accessible method for polarized positrons production, particularly suitable for creating such beams at JLab. The implementation of PEPPo at CEBAF is estimated to produce 100~nA polarized and 1~$\mu$A  unpolarized positron beams. Higher beam currents may be reached depending on the specific design of the positron source and the capabilities of the electron drive beam. The definition and the development of the most appropriate design, considering the many components and  constraints involved as well as beam diagnostics and transport magnets, demand a sustained R\&D effort toward a Conceptual Design Report (CDR). This letter discusses the possible options and the questions raised by the perspective of producing and accelerating positron beams at CEBAF. We are seeking the recommendation of the Jefferson Lab Program Advisory Committee to support the required resources and investment necessary to develop a CDR and sustain a commensurate level of R\&D.    
}
\end{abstract}

\end{center}
\end{frontmatter}
\vfill\eject
%
%
\tableofcontents
                                                                                
\newpage
%
%
\section{Introduction}

Quantum Electrodynamics (QED) is one of the most powerful quantum phy-sics theories. The highly accurate predictive power of this theory allows not only to investigate numerous physics phenomena at the macroscopic, atomic, nuclear, and partonic scales, but  also to test the validity of the Standard Model. Therefore, QED promotes electrons and positrons as unique physics probes, as demonstrated worldwide over decades of scientific research at different laboratories.

Both from the projectile and the target point of views, spin appears nowadays as the finest tool for the study of the inner structure of matter. Recent examples from the experimental physics program developed at the Thomas Jefferson National Accelerator Facility (JLab) include: the measurement of polarization observables in elastic electron scattering off the  nucleon~\cite{{Jon00}, {Gay02}, {Puc10}}, that established the unexpected magnitude and behaviour of the proton electric form factor at high momentum transfer (see \cite{Pun15} for a review); the experimental evidence, in the production of real photons from a polarized electron beam interacting with unpolarized protons, of a strong sensitivity to the orientation of the longitudinal polarization of the electron beam~\cite{Ste01}, that opened the investigation of the 3-dimensional partonic structure of nucleons and nuclei via the Generalized Parton Distributions (GPDs)~\cite{Mul94} measured through the Deeply Virtual Compton Scattering (DVCS)~\cite{{Ji97}, {Rad97}}; the achievement of a unique parity violation experimental program~\cite{Arm05, Ani06, And13} accessing the smallest polarized beam asymmetries ever measured  ($\sim$10$^{-7}$), which provided the first determination of the weak charge of the proton~\cite{And13} and allowed for stringent tests of the Standard Model at the TeV mass-scale~\cite{You06}; etc. Undoubtedly, polarization became an important capability and a mandatory property of the current and next generation of accelerators. 

The combination of the QED predictive power and the fineness of the spin probe led to a large but yet limited variety of  impressive physics results. Adding to this tool-kit charge symmetry properties in terms of polarized positron beams will provide a more complete and accurate picture of the physics at play, independently of the size of the scale involved. In the context of the experimental study of the structure of hadronic matter carried out at JLab, the electromagnetic interaction dominates lepton-hadron reactions and there is no intrinsic difference between the physics information obtained from the scattering of electrons or positrons off an hadronic target. However, every time a reaction process is a conspiracy of more than one elementary mechanism, the comparison between electron and positron scattering allows us to isolate the quantum interference between these mechanisms. This is of particular interest for studying limitations of the one-photon exchange Born approximation in elastic and inelastic scatterings~\cite{Gui03}. It is also essential for the experimental determination of the GPDs where the interference between the known Bethe-Heitler (BH) process and the unknown DVCS requires polarized and unpolarized electron and positron beams for a model independent extraction of the different contributions to the cross section~\cite{Vou14}. Such polarized lepton beams also provide the ability to test new physics beyond the frontiers of the Standard Model via the precise measurement of electroweak coupling parameters~\cite{Zhe09} or the search 
for new particles related to dark matter~\cite{Woj09, Mar17}.

The production of high-quality polarized positron beams to suit these many applications remains however a highly difficult  task that, until recently, was feasible only at large scale accelerator facilities. Relying on the most recent advances in high polarization and high intensity electron sources~\cite{Add10}, the PEPPo (Polarized Electrons for Polarized Positrons) technique~\cite{Abb16}, demonstrated at the injector of the Continuous Electron Beam Accelerator Facility (CEBAF), provides a novel and widely accessible approach based on the production, within a tungsten target, of polarized $e^+e^-$ pairs from the circularly polarized bremsstrahlung radiation of a low energy highly polarized electron beam. As opposed to other schemes operating at GeV lepton beam energies~\cite{Sok64, Omo06, Ale08}, the operation of the PEPPo technique requires only energies above the pair-production threshold and is therefore ideally suited for a polarized positron beam at CEBAF.

This letter presents the physics merits and technical challenges of an experimental program with high energy unpolarized and polarized positron beams at Jefferson Lab. The next  section discusses the benefit of a positron beam on the basis of the additional experimental observables available for elastic scattering and deeply virtual Compton scattering off the nucleon, and for the dark matter search. The following section discusses the possible schemes for implementing a PEPPo-based positron beam at CEBAF, particularly beam production and transport issues, and the necessary R\&D effort toward this end. The next sections are composed of specific letters describing in details the physics motivation  and experimental configuration of a given measurement using the proposed 11~GeV positron beam. The corresponding positron experimental program is summarized in the last 
section.

%
%

\section{Physics motivations}

%
%

\subsection{Elastic lepton scattering} \label{sec:elscat}

\begin{figure}[t!]
\begin{center}
\includegraphics[width=0.60\textwidth]{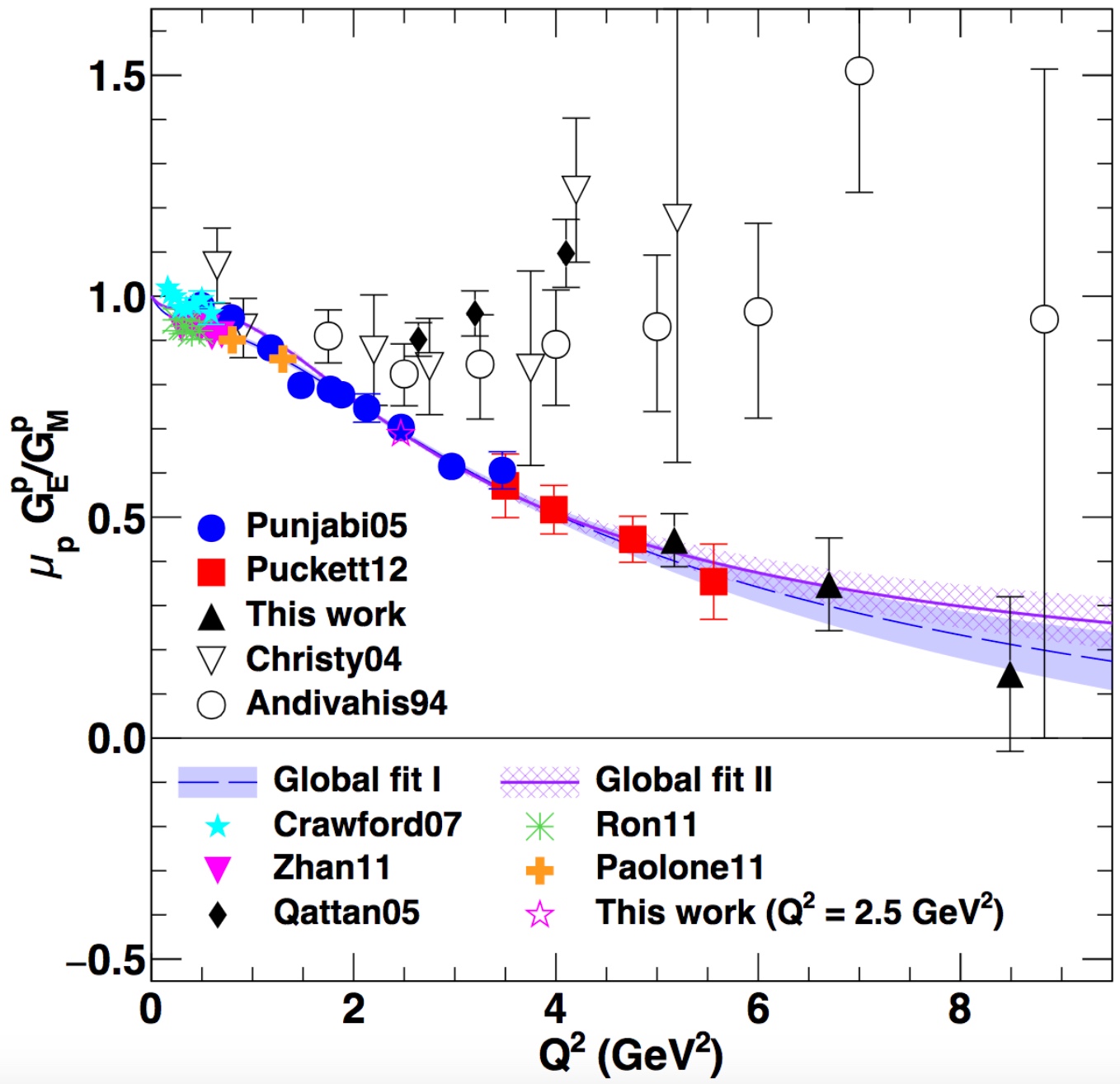}
\caption{Rosenbluth (open and diamond symbols) and polarization transfer (all other symbols) experimental data for the ratio between the electric and magnetic form factor of the proton, together with global fits of polarization data~\cite{Puc17}.}
\label{GepData}
\end{center}
\end{figure}

The measurement of the electric form factor of the nucleon ($G_{E}$) at high momentum transfer, in the perspective of the experimental assessment of perturbative Quantum Chromo-Dynamics (QCD) scaling laws~\cite{Bro81}, motivated an intense experimental effort supported by the advent of high energy continuous polarized electron beams. Indeed, the polarization observables technique~\cite{{Akh74},{Arn81}} is expected to be more sensitive to $G_E$ than the cross section method relying on a Rosenbluth separation~\cite{Ros50}. From this perspective, the strong disagreement between the results of these two experimental methods (Fig.~\ref{GepData}) came as a real surprise. Following the very first measurements of polarization transfer observables in the $^1$H$({\vec e},e{\vec p})$ reaction~\cite{Jon00}, the validity of the Born approximation for the description of the elastic scattering of electrons off protons was questioned. The eventual importance of higher orders in the $\alpha$-development of the electromagnetic interaction was suggested~\cite{Gui03} as a hypothesis to reconcile cross section and polarization transfer experimental data. This prevents a model-independent experimental determination of the nucleon electromagnetic form factors via only electron scattering.

Considering the possible existence of second-order contributions to the electromagnetic current, the so-called 2$\gamma$-exchange, the $eN$-interaction is no longer characterized by 2 real form factors but by 3 generalized complex form factors
\begin{equation}
\widetilde G_{M} = e \, G_{M} + \delta {\widetilde G}_M \, , \,\,\,\,\,\, 
\widetilde G_{E} = e \, G_{E} + \delta {\widetilde G}_E \, , \,\,\,\,\,\, 
\widetilde F_{3} = \delta {\widetilde F}_3 \, ,
\end{equation}
where $e$ represents the lepton beam charge. These expressions involve up to 8 unknown real quantities that should be recovered from  experiments~\cite{Rek04}. Considering unpolarized leptons, the non point-like structure of the nucleon can be expressed by the reduced cross section 
\begin{eqnarray}
\sigma_R^{e} = \tau G^2_{M} & + & \epsilon G^2_{E} + 2 e \, \tau G_{M} \, \Re e \left[ \delta {\widetilde G}_M \right] \\
& + & 2 e \, \epsilon \, G_{E} \, \Re e \left[ \delta {\widetilde G}_E \right] + e \, \sqrt{\tau(1-\epsilon^2)(1+\tau)} G_{M} \, \Re e \left[ \delta {\widetilde F}_3 \right] \nonumber
\end{eqnarray}
where the charge-dependent terms denote the additional contributions from the 2$\gamma$-exchange mechanisms. The variable  $\epsilon$ characterizing, in the 1$\gamma$-exchange approximation, the virtual photon polarization can be written as   
\begin{equation}
\epsilon = {\left[ 1 - 2 \frac{{\vec q} \cdot {\vec q}}{Q^2} \tan^2 \left( \frac{\theta_e}{2} \right) \right]}^{-1} \label{eq:epsi}
\end{equation}
where $\theta_e$ is the electron scattering angle, and $q \equiv ({\vec q}, \omega)$ is the virtual photon with four-mometum transfer $Q^2$=$-q \cdot q$, and $\tau$=$Q^2 / 4 M^2$ with $M$ representing the nucleon mass. In absence of lepton beams of opposite charge, the Rosenbluth method, consisting in the measurement of the reduced cross section at different $\epsilon$-values while keeping $Q^2$ constant, allows the determination of a combination of 1$\gamma$ and 2$\gamma$ electromagnetic form factors. Consequently, it requires a model-dependent input to further separate the electric and magnetic form factors. \newline
The transfer of longitudinal polarization by a lepton beam via scattering elastically off a nucleon, provides 2 additional linear combinations of the same physics quantities in the form of the transverse ($P_t^e$) and longitudinal ($P_l^e$) polarization components of the nucleon 
\begin{eqnarray}
\sigma_R^e \, P_{t}^e = - \, \lambda \, \sqrt{2\epsilon\tau(1-\epsilon)} & \bigg( & G_{E} G_{M} + e \, G_E \Re e \left[ \delta {\widetilde G}_M \right] \\
& + & e \, G_{M} \Re e { \left[ \delta {\widetilde G}_E \right] } + e \, \sqrt{\frac{1+\epsilon}{1-\epsilon}} G_E \Re e \left[ \delta {\widetilde F}_3 \right] \bigg) \nonumber \\
\sigma_R^e P_{l}^e = \lambda \, \tau \sqrt{1-\epsilon^2} \bigg( G_{M}^2 & + & e \, \left[ 2 + \sqrt{\frac{1+\tau}{\tau(1-\epsilon)}} \, \right] G_{M} 
\Re e \left[ \delta {\widetilde F}_3 \right] \bigg) \, ,
\end{eqnarray}
where $\lambda$ is the lepton-beam polarization. The combination of polarized and unpolarized beam observables for elastic electron scattering  involves up to 6 unknown real quantities, requiring at least 6 independent experimental observables. Therefore, taking into account 2$\gamma$-exchange mechanisms electron beams alone can no longer provide a pure experimental determination of the electromagnetic form factors of the nucleon. However, comparing polarized electron and positron beams, one can separate the charge-dependent and independent contributions of experimental observables, and thus separate the 1$\gamma$ and 2$\gamma$ form factors. For instance, 
\begin{eqnarray}
\frac{\sigma_R^+ + \sigma_R^-}{2} & = & \tau G^2_{M} + \epsilon G^2_{E} \\
\frac{\sigma_R^+ - \sigma_R^-}{2} & = & 2 \tau G_{M} \, \Re e \left[ \delta {\widetilde G}_M \right] \\ 
& + & 2 \, \epsilon \, G_{E} \, \Re e \left[ \delta {\widetilde G}_E \right] + \sqrt{\tau(1-\epsilon^2)(1+\tau)} G_{M} \, \Re e \left[ \delta {\widetilde F}_3 \right] \nonumber
\end{eqnarray}
and similarly for polarized observables. Consequently, the measurement of polarized and unpolarized elastic scattering of both electrons and positrons provides the necessary data for a model-independent determination of the nucleon electromagnetic form factors.

\subsection{Deep inelastic lepton scattering} \label{PhyMot-Deep}

The understanding of the partonic structure and dynamics of hadronic matter is one of the major goals of modern Nuclear Physics. The  availability of high intensity continuous polarized electron beams with high energy together with capable detector systems at different facilities is providing today an unprecedented but still limited insight into this problem. Similarly to the elastic scattering case, the combination of measurements with polarized electrons and polarized positrons in the deep inelastic regime will allow to obtain unique experimental observables enabling a more accurate and refined interpretation.

The GPD framework~\cite{Mul94} constitutes the most appealing and advanced parameterization of hadron structure. It encodes the internal structure of matter in terms of quarks and gluons and unifies within the same framework electromagnetic form factors, parton distributions, and the description of the nucleon spin (see \cite{Die03, Bel05} for a review). GPDs can be interpreted as the probability to find a parton at a given transverse position and carrying a certain fraction of the longitudinal momentum of the nucleon. The combination of longitudinal and transverse degrees of freedom is responsible for the richness of this universal framework.

\begin{figure}[t!]
\begin{center}
\includegraphics[width=0.85\textwidth]{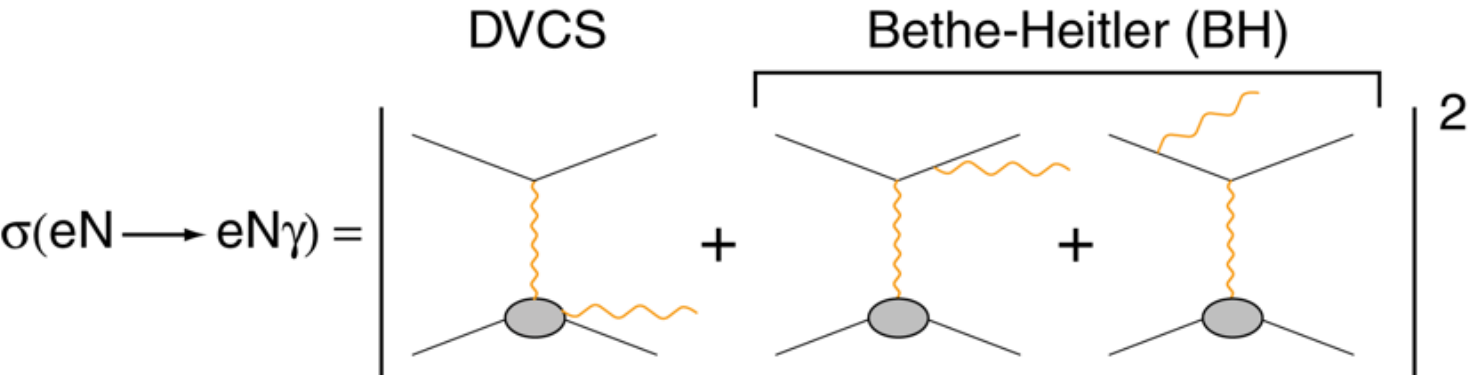}
\caption{Lowest order QED amplitude of the electroproduction of real photons off nucleons.}
\label{EAgamma}
\end{center}
\end{figure}
GPDs are involved in any deep process and are preferentially accessed in hard lepto-production of real photons (i.e. DVCS). This process competes with the known BH  reaction~\cite{Bet34} where real photons are emitted from the initial or final leptons instead of from the probed hadronic state (Fig.~\ref{EAgamma}). The lepton beam charge and polarization dependence of the $eN(A)\gamma$ cross section off nucleons(nuclei) writes~\cite{Die09}
\begin{equation}
\sigma^{e}_{\lambda 0} = \sigma_{BH} + \sigma_{DVCS} + \lambda \, \widetilde{\sigma}_{DVCS} + e \, \sigma_{INT} + e \lambda \, \widetilde{\sigma}_{INT} \label{eq:LU}
\end{equation}
where the index $INT$ denotes the interference contribution to the cross section originating from the quantum interference of the BH and DVCS processes. Polarized electron scattering provides the experimental observables 
\begin{eqnarray}
\sigma^-_{0 0} & = & \frac{\sigma^-_{+ 0} + \sigma^-_{- 0}}{2} = \sigma_{BH} + \sigma_{DVCS} - \sigma_{INT} \, , \label{eq:int00}\\
^1\Delta^-_{\lambda 0} & = & \frac{\sigma^-_{+ 0} - \sigma^-_{- 0}}{2} = \lambda \, \left[ \widetilde{\sigma}_{DVCS} -  \widetilde{\sigma}_{INT} \right] \label{eq:int10}
\end{eqnarray}
involving unseparated combinations of the unknwon $INT$ and $DVCS$ reaction amplitudes. The comparison between polarized electron and polarized positron reactions provides the additional observables
\begin{eqnarray}
\Delta \sigma_{00} & = & \frac{\sigma^+_{00} - \sigma^-_{00}}{2} = \sigma_{INT} \label{eq:int0}\\
^2\Delta_{\lambda 0} & = & \frac{^1\Delta^+_{\lambda 0} - {^1\Delta^-_{\lambda 0}} }{2} = \lambda \, \widetilde{\sigma}_{INT} \label{eq:int1}
\end{eqnarray} 
which isolate the interference amplitude. Consequently, measuring real photon lepto-production off nucleons with opposite charge polarized leptons allows to separate the four unknown contributions to the $eN\gamma$ cross section. \newline 
For a spin $s$ hadron, one can define $(2s+1)^2$ parton-helicity conserving and chiral-even elementary GPDs that can be accessed through DVCS. They appear in the reaction amplitudes in the form of unseparated linear and bi-linear expresssions. Their separation requires additional observables that can be obtained considering polarized targets ($S$)~\cite{Bel02}. The full lepton beam charge and polarizations dependence of the $eN\gamma$ cross section can be written as~\cite{Die09}
\begin{eqnarray}
\sigma^{e}_{\lambda S} & =  & \sigma^{e}_{\lambda 0} \label{eq:LL} \\
& + & S \left[ \lambda \, \Delta\sigma_{BH} + \lambda \, \Delta\sigma_{DVCS} + \Delta\widetilde{\sigma}_{DVCS} + e\lambda \, \Delta\sigma_{INT} + e \, \Delta\widetilde{\sigma}_{INT} \right] \, , \nonumber
\end{eqnarray}
where $\Delta\sigma_{BH}$ is the known sensitivity of the BH process to the target polarization and the remaining terms feature four combinations of the nucleon GPDs to be isolated. Polarized electron scattering provides the combinations
\begin{eqnarray}
^1\Delta \sigma^-_{0S} & = & \frac{\sigma^-_{0+} - \sigma^-_{0-}}{2} = S \, \left[ \Delta\widetilde{\sigma}_{DVCS} - \Delta\widetilde{\sigma}_{INT}\right] \label{eq:int20} \\
^2\Delta^-_{\lambda S} & = & \frac{^1\Delta^-_{\lambda +} - {^1\Delta^-_{\lambda -}} }{2} = S \, \lambda \, \left[ \Delta\sigma_{BH} + \Delta\sigma_{DVCS} - \Delta\sigma_{INT} \right] \label{eq:int30}
\end{eqnarray} 
and the comparison between polarized electrons and positrons yields
\begin{eqnarray}
^2\Delta \sigma_{0S} & = & \frac{^1\Delta\sigma^+_{0S} - {^1\Delta\sigma^-_{0S}} }{2} = S \, \Delta\widetilde{\sigma}_{INT} \label{eq:int2} \\
^3\Delta_{\lambda S} & = & \frac{^2\Delta^+_{\lambda S} - {^2\Delta^-_{\lambda S}} }{2} = S \, \lambda \, \Delta\sigma_{INT} \label{eq:int3} \, ,
\end{eqnarray}
which once again isolates the interference contribution and allows to separate the four reaction amplitudes of interest. \newline 
Therefore, polarized positron beams appear as a necessary complement to polarized electron beams to achieve a model-independent determination of nucleon GPDs.

\subsection{Test of the Standard Model}

The search for the evidence of Physics beyond the Standard Model (PbSM) is a long standing worldwide effort. Its existence is supported for instance by the matter-antimatter asymmetry or non-zero neutrino masses, but no direct observation has to date been reported. While most of the experimental effort for PbSM search is focussed on heavy particle candidates in the TeV mass range with high energy accelerators and/or high precision experiments, other scenarios  involving lighter gauge bosons have also been proposed~\cite{Fay80}, consistent with anomalies observed in cosmic radiations, the internal pair creation process in $^8$Be~\cite{Kra16} or the anomalous magnetic moment of the muon~\cite{Ben04}. The eventual pertinence of a U(1) symmetry, proposed many years ago as a natural extension of the Standard Model~\cite{Fay80}, is suggesting the existence of a new light {\bf U}-boson (a {\it heavy photon} or $\Apr$, also called {\it dark photon}) in the few MeV-GeV mass range, also referred to as light dark matter~\cite{Fay04} considering its ability to decay through dark matter as well as Stantard Model (SM) particles. 

\begin{figure}[t!]
\begin{center}
\includegraphics[width=0.75\textwidth]{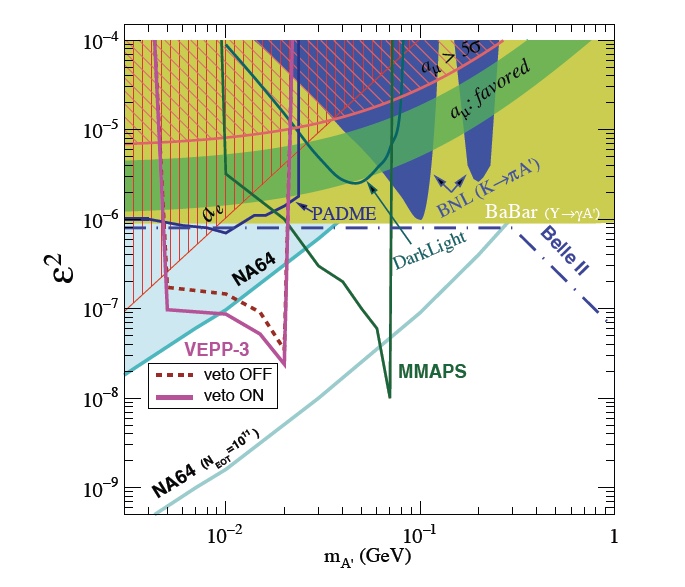}
\caption{\label{fig-limits} Parameter space covered by existing and proposed experiments with positron beams.}
\end{center}
\end{figure}
In the last few years, searches for Dark Matter (DM) extended to cover a mass region ($\sim$1~MeV/$c^2$-1~GeV/$c^2$) poorly addressed by experiments seeking for halo DM. DM with mass below 1~GeV/$c^2$ interacting with SM particles via the $\Apr$, represents a well motivated scenario that generated many theoretical and phenomenological studies, has also stimulated the re-analysis and interpretation of old data as well as promoted new experimental programs aimed to search both for the $\Apr$ and for light DM states. In this scenario the DM, charged under a new gauge symmetry $U(1)_D$, interacts with electromagnetic charged SM particles trough the $\Apr$-exchange. The coupling between SM particles and dark-photons is induced by the kinetic mixing operator. This mechanism, originally suggested by Holdom~\cite{Hol86}, can be interpreted as a {\it portal} between the SM word and a new {\it Dark Sector}~\cite{Bjo09, Iza13}. The kinetic mixing parameter $\varepsilon$ is expected to be small, in the range of $\sim 10^{-4}$-$10^{-2}$ ($\sim 10^{-6}$-$10^{-3}$) if the mixing is generated by one (two)-loops interaction. The value of the $\Apr$ mass and $\varepsilon$ should be determined  by comparison with observation. Depending on the relative mass of the $\Apr$ and the DM particles, the $\Apr$ can decay to SM particles ({\it visible } decay) or to light DM states ({\it invisible} decay).

In this context, accelerator-based experiments that make use of electron and positron beams of moderate energy ($\sim$0.5-10~GeV) show a seizable sensitivity to an uncovered area of ($\varepsilon$ vs. $m_{\Apr}$) parameter space. Several experiments proposed at Jefferson Lab (APEX, HPS, DARK LIGHT, BDX) are making use of the CEBAF electron beam to cover unexplored area of the parameter space. Positrons are a natural complementary probe to explore the Dark Sector. So far, experiments at colliders (BABAR and BELLE) used the missing mass technique ($e^++e^- \to \gamma A'$) to detect a possible signal in the mass of the missing $\Apr$ reconstructed from the photon detection. The advantage of this technique relies in the fact that no particular assumptions are requested about the $\Apr$ decay modes reducing any model dependence in the production mechanisms. Alternative approaches that exploits the larger luminosity achievable in fixed-target experiments are able to push down the exclusion limits by a significant amount for a narrow band in $m_{\Apr}$. This is related to the fact that the very efficient annihilation mechanism shows a dependence on the maximum $m_{\Apr}$ proportional to $\sqrt{E_{beam}}$. A high energy positron beam at Jefferson Lab will allow us to produce and detect $\Apr$ in an extended mass range (up to 100 MeV) with an unprecedented sensitivity.

%
%

\hyphenation{po-si-trons}

\section{Positron beams at Jefferson Lab}

The prospect of polarized or unpolarized positron beams for nuclear physics experiments
at CEBAF naturally raises many issues. Prominent among these are the generation of
positrons, their formation into beams acceptable to the 12 GeV CEBAF accelerator, and
the technical challenges associated for the magnetic transport and diagnostics of low
current positively charged beams. The following sections highlight these issues and
summarize the present thinking towards developing positron beams. Indeed, these issues
have  been explored in discussions and proceedings at previous workshops hosted by JLab
(JPos09~\cite{Jpo09} and JPos17~\cite{Jpo17}). Notably, three Ph.D. Theses were successfully
completed during this period to explore the production of polarized positrons using the
CEBAF polarized electron beam ~\cite{Dum11, Ade16} and to develop a conceptual design of
a continuous-wave positron injector compatible with CEBAF~\cite{Gol10}.

While these activities made significant strides, realizing a positron beam requires a
significant and concerted level of efforts by the Accelerator and Engineering Divisions
that is integral to Laboratory planning. The development of a complete and viable Conceptual
Design, that can be reviewed and costed, requires labor resources and advanced research
activities to address technical challenges. In this regard, a recommendation by the Program
Advisory Committee to Laboratory leadership is an essential ingredient to move in this
direction. 

\subsection{Polarized Electrons for Polarized Positrons}

\begin{figure}[t!]
\begin{center}
\includegraphics[width=0.495\textwidth,height=0.325\textheight]{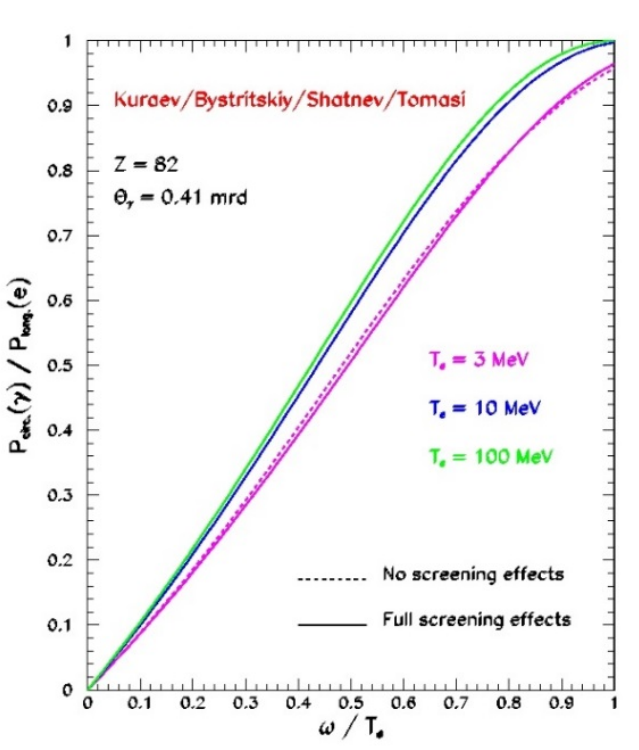}
\includegraphics[width=0.495\textwidth,height=0.325\textheight]{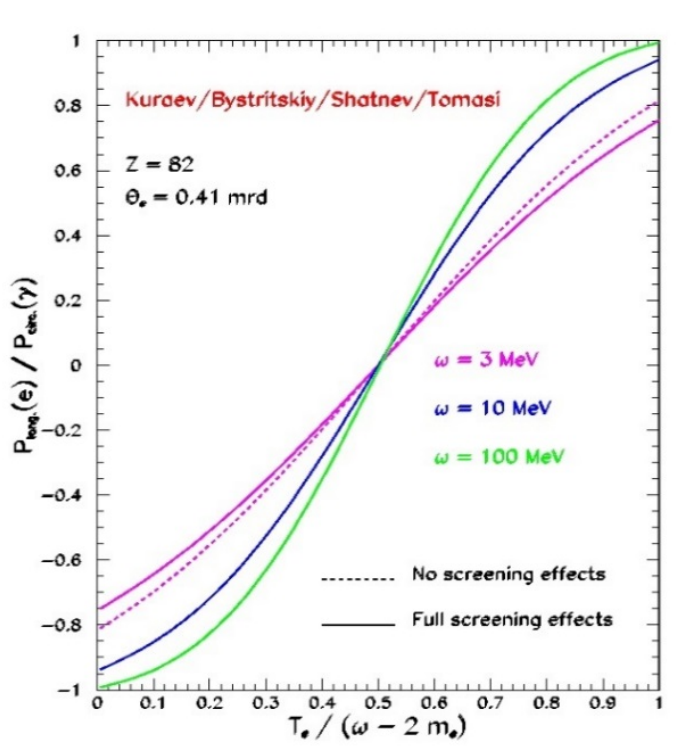}
\caption{(left) The circular polarization of the photons produced by longitudinally polarized
electrons at a fixed production angle of 0.41~mrad for several electron beam energies and
for the extreme cases of full- or no-screening as a function of the ratio of the photon
energy to the electron energy~\cite{Kur10}. (right) The electron/positron longitudinal
polarization resulting from the pair production by circularly polarized photons for the
extreme cases of full- or no- screening for several photon beam energies as a function of
the ratio of the positron energy to the photon beam energy less the sum of the electron
and positron masses~\cite{Kur10}.}
\label{Loi-Pos-Fig1}
\end{center}
\end{figure}

\begin{figure*}[h]
\begin{center}
\includegraphics[width=0.65\textwidth]{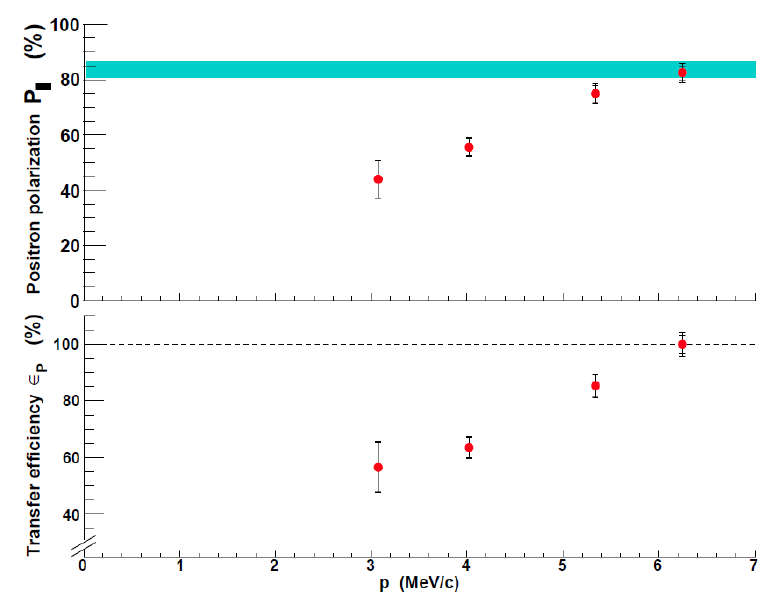}
\end{center}
\caption{PEPPo measurements of the positron polarization (top panel) and polarization
transfer efficiency (bottom panel); statistics and systematics are reported for each point,
and the shaded area indicates the electron beam polarization ~\cite{Abb16}.}
\label{Loi-Pos-Fig3}
\end{figure*}

The Polarized Electrons for Polarized Positrons (PEPPo) experiment~\cite{Abb16} demonstrated
a viable approach to the efficient production of polarized positrons for CEBAF. It  used the
highly polarized electron beam available from the CEBAF source and generated polarized positrons
through a two-step process: bremsstrahlung followed by pair production, with both reactions
taking place (in series) in the same physical target. Ultrarelativistic calculations~\cite{Ols59}
generalized to any particle energy~\cite{Kur10}, demonstrate how these two processes may
combine to produce polarized positrons. First the longitudinal polarization of the incident
electron beam is transferred to circular polarization of the bremsstrahlung photons produced
early in the target. As can be seen in Fig.~\ref{Loi-Pos-Fig1} (left), the calculations
indicate that the  transfer efficiency increases as the photon energy approaches the electron
beam energy. Fig.~\ref{Loi-Pos-Fig1} (right) shows how the polarization is then transferred
from the circularly polarized photons to the $e^+e^-$ pairs; the efficiency of the polarization
transfer increases as the positron energy approaches the energy of the $\gamma$-ray. These
calculations were tested by the PEPPo experiment. The results, for an incident polarized
electron beam of 8.2 MeV/$c$ and 85.2\% polarization (Fig.~\ref{Loi-Pos-Fig3}), showed that 
he transfer of the initial electron beam polarization to the extracted positron beam can be
very efficient, approaching 100\% as the positron beam momentum approaches the initial
electron beam momentum.


\subsubsection*{Positron polarimetry}

Positron polarimetry of GeV-scale beams can be readily accomplished using the standard
techniques of Compton and M{\o}ller polarimetry~\cite{Gas18}. In particular, the polarimeters
at JLab can potentially be used for these measurements, either with some modification or
compromise in performance. The primary challenge for the JLab Compton polarimeters is the
relatively low beam current (100 nA) projected to be feasible for polarized positron beams
at JLab. This low current leads to rather lengthy measurement times. Measurement times
could be reduced with improvements to the Compton polarimeter laser systems, although this
would require some R\&D and expense. The M{\o}ller polarimeters at JLab, on the other hand,
use magneto-optical systems designed to detect two particles of the same charge in coincidence.
M{\o}ller polarimetry with positron beams would ideally detect the scattered positron and
recoil electron. The JLab M{\o}ller polarimeters could be operated in single-arm mode,
resulting in non-trivial Mott backgrounds and potentially larger systematic uncertainties
(although the Mott backgrounds could potentially be understood by comparing single-arm
and coincidence measurements with electrons). Another option would be to replace the
quadrupole-based polarimeter optics with a dipole-based system. This would enable the
detection of the positrons and electrons in coincidence. Extra time would be needed,
though, to switch between positron and electron operating mode.

\subsection{Positron beam production at CEBAF}

PEPPo demonstrated convincingly the merits of the polarization transfer technique making
it worthwhile to explore its optimization for the production of unpolarized and polarized
positron beams in support of the physics program of the 12~GeV Upgrade. Given the rapid
increase in positron production (for positrons within a useful phase volume) with the energy
of the electron beam used to produce the positrons, one can speculate that an excellent
approach to the production of a very intense positron beam at CEBAF would be to use the
2.2~GeV beam from the first pass through CEBAF to generate positrons, transport this beam
to the injection point (adding a phase shift relative to the electron beam) and then
accelerate this beam through the full CEBAF accelerator. This would require the reversal
of the fields in all of the recirculation system and the addition of a 6th recirculation
path placed below the current system to transport the initial electron beam to the positron
production target. It would, however, be a very expensive solution. \newline

When polarized positrons are desired, it is essential to investigate the optimization of
the Figure-of-Merit ($P^2_z I$); based on the calculated behavior seen for lower energy
polarized positron production, one can speculate that it would be useful to tweak the
system so that the energy of the positrons selected for injection into CEBAF would be
roughly half of the first pass energy~\cite{Dum11}. The acceleration phase of the positron
beam in the first (North) linac could then be adjusted so that the energy of the beam
after acceleration in that linac was the nominal (1.1~GeV) energy of the usual electron
beam at that point. For the case where unpolarized positrons are desired, it is likely
that the maximum useful positron flux would be obtained by selecting positrons at energies
as low as 123~MeV (the standard energy of the injector for CEBAF at 12~GeV) and to
accelerate them through the first linac with full energy gain (i.e. at a phase 180$^\circ$
away from the standard electron phase).

\begin{figure}[t!]
\begin{center}
\includegraphics[width=0.95\textwidth]{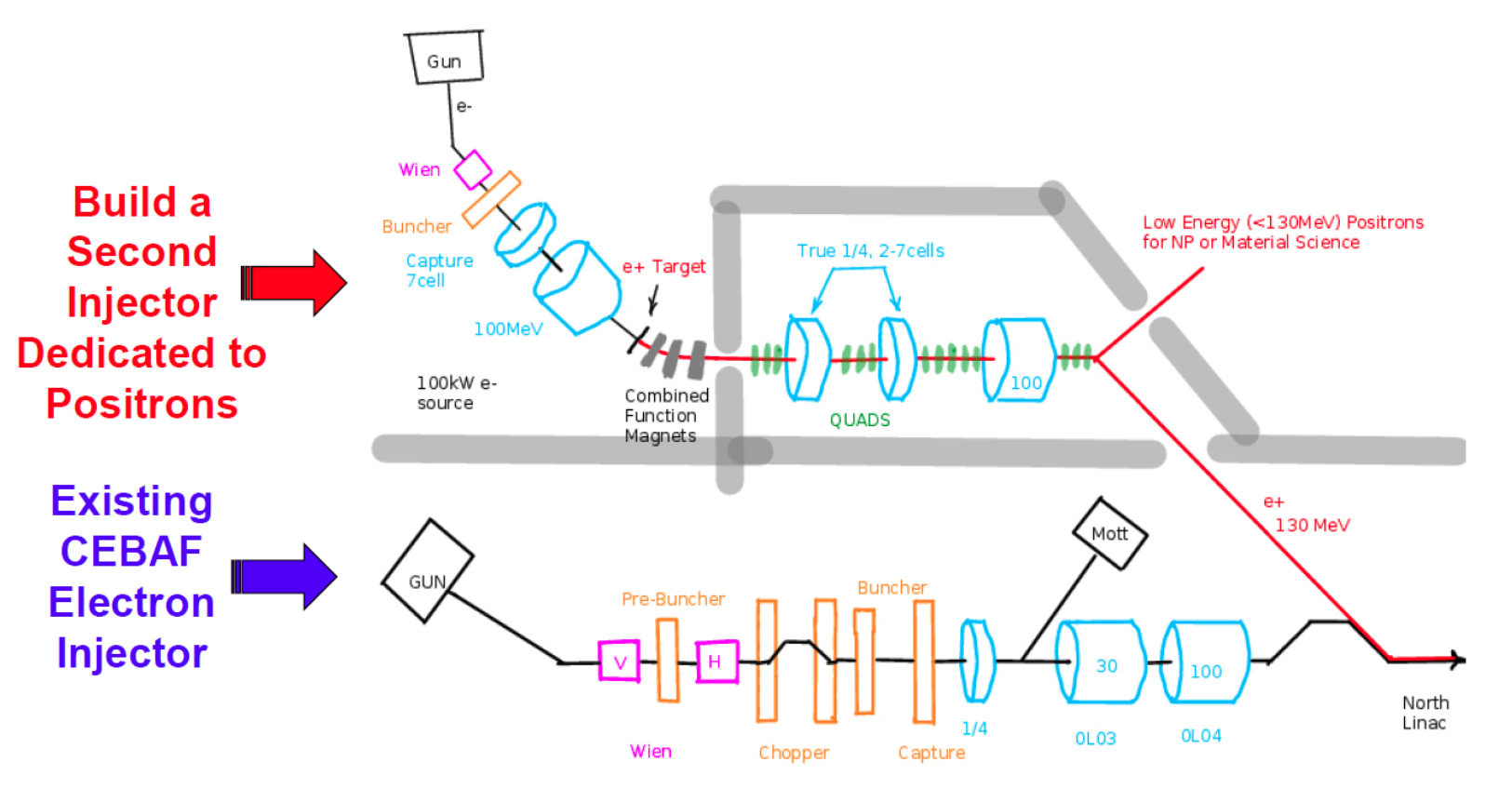}
\end{center}
\caption{An approach to adding positron capability to CEBAF~\cite{Gol10}.}
\label{Loi-Pos-Fig4}
\end{figure}
A second approach to positron beams for CEBAF is shown in Fig.~\ref{Loi-Pos-Fig4}~\cite{Gol10}.
The idea is to build a complete second injector dedicated to positron production. The new
injector would be built in a new building located adjacent to the current injector tunnel. A
123~MeV electron linac with a polarized electron source would drive a positron  production
target. A slit and magnet system that follows the production target would select the positrons
to be accelerated. They would then be passed through a second linac, accelerated to 123~MeV
for injection into CEBAF (with a 180$^\circ$ phase shift relative to the usual electron
injection phase) and accelerated in CEBAF to the desired energy (with all the magnetic fields
in the recirculation system reversed). In this approach, a 10~mA electron beam from the first
linac would comfortably produce a 3~$\mu$A polarized  positron beam. One advantage of this
design is that the positrons could also be made available at low energies appropriate for
condensed matter research without impacting CEBAF  nuclear physics operations by simply
operating the positron production system independently, and diverting the beam into a low-energy
experimental hall that could be constructed  adjacent to the positron source.

\begin{figure}[t!]
\begin{center}
\includegraphics[width=0.95\textwidth]{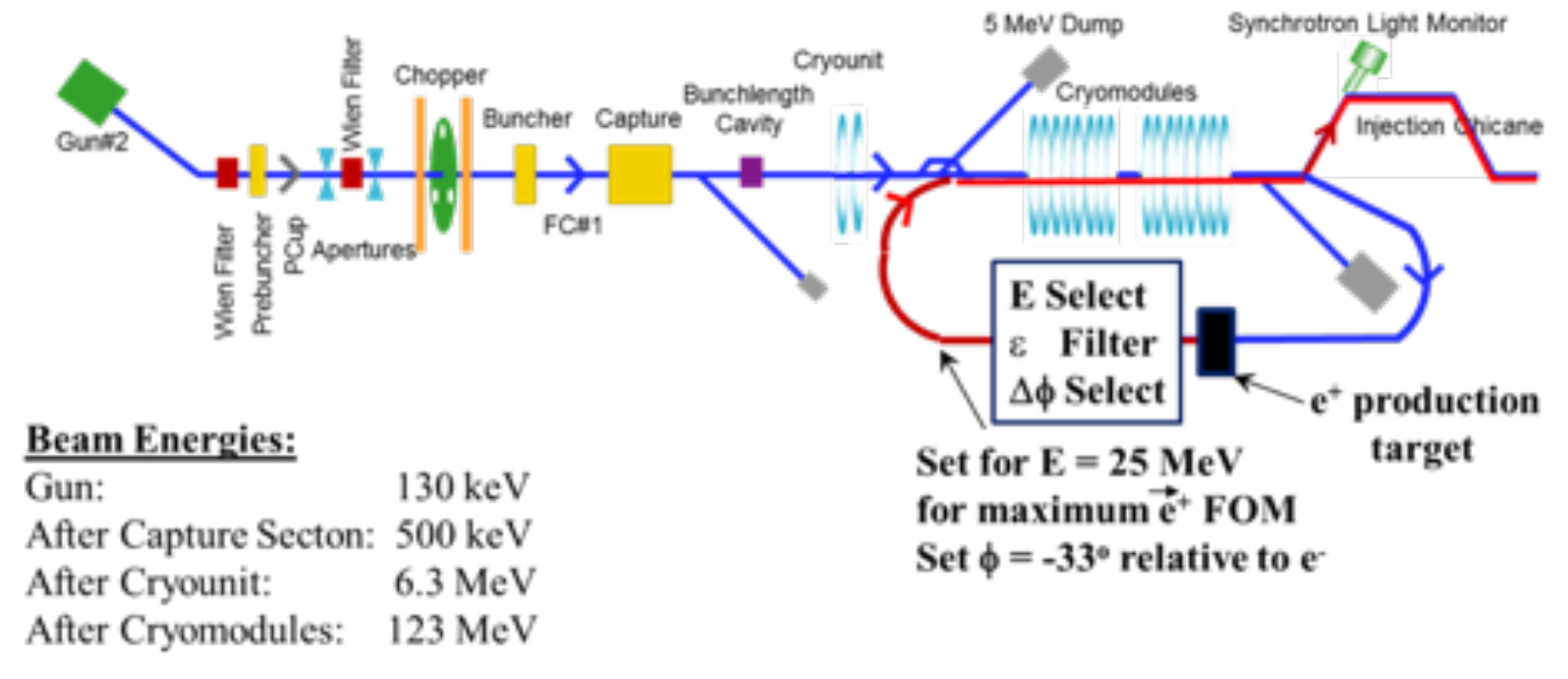}
\end{center}
\caption{An approach to the development of positron capability for CEBAF using the injector
linac both to produce the positrons and to accelerate them. The choice of an injected
positron energy of 25~MeV is based on the general behavior of the FoM for the positron
beam that peaks at about 1/4 of the production beam energy when 
beam emittance considerations are taken into account.}
\label{Loi-Pos-Fig5}
\end{figure}
\begin{figure}[t!]
\begin{center}
\includegraphics[width=0.495\textwidth,height=0.295\textheight]{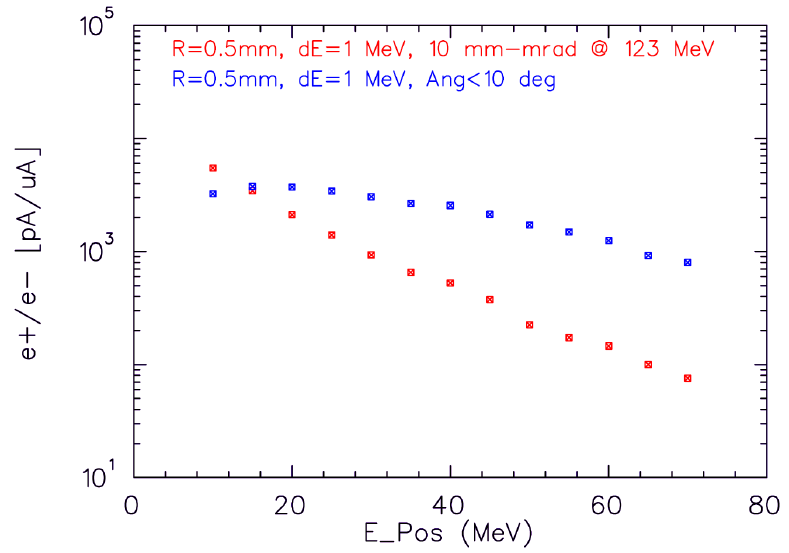}
\includegraphics[width=0.495\textwidth,height=0.295\textheight]{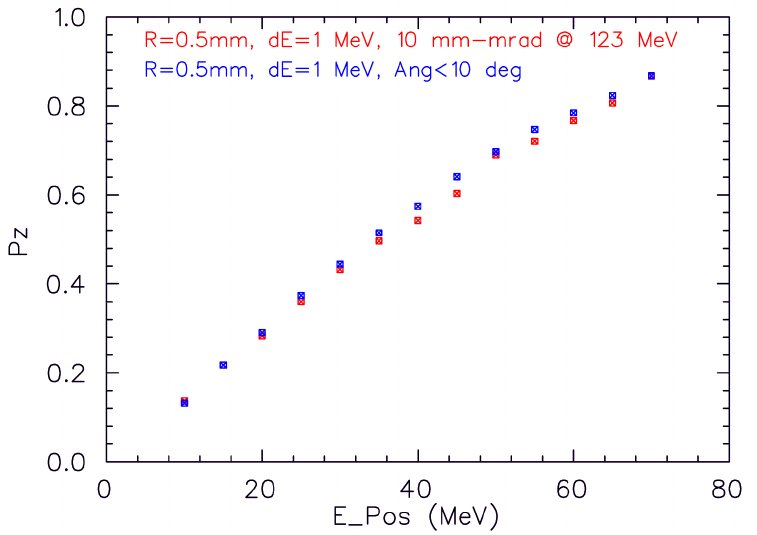}
\caption{(left) Calculations of the positron to electron ratio for positron production by a
123~MeV 85\% polarized electron beam; b (right) the longitudinal polarization of that
positron beam; both are shown as functions of the positron beam energy. The red data points 
correspond to the selection of produced positrons within an energy spread of 1~MeV and a
normalized emittance of 2400~mm$\cdot$mr, while the blue data points correspond to selection
 with the same energy spread but within a constant angular acceptance of 10$^\circ$.}
\label{Loi-Pos-Fig6}
\end{center}
\end{figure}
\begin{figure}[t!]
\begin{center}
\includegraphics[width=0.70\textwidth]{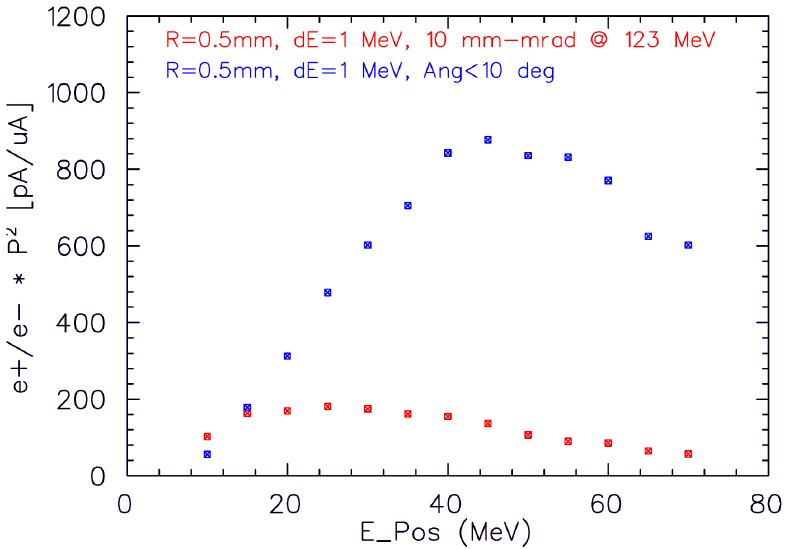}
\end{center}
\caption{The Figure-of-Merit for the positron beam produced by a 123~MeV 85\% longitudinally
polarized electron beam as a function of the positron energy. As is the case for the curves
in Fig.~\ref{Loi-Pos-Fig6}, the red data points correspond to the selection of produced
positrons within an energy spread of 1~MeV and a normalized emittance of 2400~mm$\cdot$mr,
while the blue data points correspond to selection with the same energy spread but within
a constant angular acceptance of 10$^\circ$.}
\label{Loi-Pos-Fig7}
\end{figure}

A third approach, summarized in Fig.~\ref{Loi-Pos-Fig5}, would utilize the standard CEBAF
injector, modifying it to both produce the positron beam and then accelerate it (via
recirculation at the appropriate phase) back through the injector linac to raise the positron
beam energy to the standard electron injection energy of 123~MeV. The standard operation of
the injector has a gun energy of 130~keV. In the capture section the energy is increased to
500~keV, and in the Cryounit (1/4 of a cryomodule) the energy is increased to 6.3~MeV. This
beam is then sent through a pair of standard cryomodules, increasing the energy by 116.7~MeV
to bring the beam energy to the 123~MeV used for injection into CEBAF and acceleration to
energies up to 12~GeV. As can be seen in the figure, we can modify and extend this system
to provide a polarized or  unpolarized positron beam. To begin, the 123~MeV polarized
electron beam (normally sent to CEBAF for acceleration) is instead turned 180$^\circ$ back
toward the gun and sent through a positron production target. This is followed by a (to be
designed) slit and magnet system that selects the appropriate portion of the positron beam
(setting the central energy of the beam, its energy spread, and its transverse emittance).
This would be followed by a path length adjusting system (basically similar to the path
length adjustment systems used in the recirculation of beams through CEBAF) that would
permit adjusting the phase of the positrons at the entrance of the injector cryomodules
relative to the electron beam. The positrons could then be accelerated to 123 MeV for
injection into CEBAF and then accelerated to the full energy of CEBAF by reversing the
fields in the recirculation magnets and setting the phase of the main linacs for positron 
rather than electron acceleration.  This scheme has the advantage that it probably minimizes
the cost of developing a positron beam. The low energy of the injector recirculation
system means that the cost of the magnetic elements would be modest. To help refine the
approach and estimate the positron beams it could produce, a GEANT4 model~\cite{Dum11}
consistent with the calculations of Ref.~\cite{Kur10} was used to determine the positron
flux produced with a 123~MeV electron beam 85\% polarized. For these calculations we
examined the positron beams produced by a 0.5~mm diameter polarized electron beam for
two phase space acceptances: first with a constant angular acceptance of 10$^\circ$ and
an energy spread of 1~MeV (to compare with earlier calculations); and then for an energy
spread of 1~MeV and a normalized emittance of 2400~mm$\cdot$mr (to be within the maximum
geometric acceptance of 10~mm$\cdot$mr when injecting into CEBAF from the injector linac).
The results shown in Figs.~\ref{Loi-Pos-Fig6} and ~\ref{Loi-Pos-Fig7}.

Figure~\ref{Loi-Pos-Fig6} (left) displays the calculated intensity of the positron beam
produced by a 123 MeV polarized electron beam as a function of the positron energy; the
positron yield (emerging from the emittance filter) drops by about two orders of magnitude
as the positron energy selected increases from 10 MeV to 70 MeV.  As shown in
Fig.~\ref{Loi-Pos-Fig6} (right), the polarization increases smoothly from small values
at low positron energies to the incident beam polarization as the positron energy approaches
the electron beam energy.  The combination of these two effects results in the FoM for the
positron beam shown in Fig.~\ref{Loi-Pos-Fig7} as a function of positron energy. If one
could use the full 10$^\circ$ acceptance the optimum occurs at about half the electron
beam energy as the decrease in intensity overcomes the increase in positron polarization
as the positron energy increases. This feature, of an optimum FoM at about half the electron
beam energy, is a general feature of such calculations. However, if the phase space of
positrons accepted is reduced to the maximum allowable acceptance of the first CEBAF linac
the optimum FoM occurs at a positron energy of 25 MeV, about one fourth of the electron
beam energy.

Setting the emittance filter for a positron energy of 25 MeV for a maximum FoM, we must
then adjust the phase of the positron beam relative to the cryomodule RF to about 33$^\circ$
to provide an energy gain of 98~MeV going through the two cryomodules. This would result in
a final positron energy of 123~MeV, which is correct for injection into CEBAF via the standard
injection chicane (with its fields reversed, as would be the case for all the magnetic
fields in CEBAF). It would also permit some energy compression during the acceleration,
meaning a larger energy spread could be accepted at the cryomodule entrance.  Very rough
estimates assuming the standard 100 $\mu$A, 85\% polarization electron beam used in normal
operations of CEBAF suggest that this scheme would produce positron beams with intensities
of 100-200~nA and a polarization of about 40\%. With the installed RF in the injector it
would be straightforward to increase the current available for positron production to about
200~$\mu$A, doubling the estimated positron current to 200-400~nA. The positron beam
intensity can also be increased by populating each of the sub-harmonic RF bunch and
operates at a higher bunch frequency in experimental halls. Further, one could obtain an
even higher intensity beam by increasing the intensity of the polarized electron beam.
This would require an upgrading of the RF system for the injector cryomodules. A $>$500~$\mu$A
electron beam is estimated to provide a microampere polarized positron beam. Of course
the details should be calculated precisely so one can be sure that the scheme has indeed
been optimized. 

If the positron beams needed for a particular experiment do not have to be polarized, one
can take advantage of the fact that the positron to electron ratio increases as the positron
energy selected is lowered relative to the electron beam energy (Fig.~\ref{Loi-Pos-Fig6}a).
At an extreme, one could set the filter downstream of the  positron production target for a
positron energy as low as 6.3~MeV (the nominal energy of the polarized electron beam after
the 1/4 cryomodule), and then set the phase so the positron beam is 180$^\circ$ out of
phase relative to the electron beam. It would then get the same acceleration through the
cryomodules as the electron beam does, and emerge at 123~MeV for injection into CEBAF. An
optimization might well end up with a somewhat higher positron energy so that the acceleration
would be off-crest and provide some energy compression in addition to the acceleration of
the positron beam. One anticipates that this approach would result in positron beams of
order two microamperes with a 200 $\mu$A polarized electron beam drive (and a beam of order
5 $\mu$A if we increased the drive beam to 500 $\mu$A). Currents this high have the
advantage that they are visible using the standard beam diagnostics in CEBAF, so tuning
the beam should be straightforward. One interesting possibility that may be worth considering
is the use of this beam to tune CEBAF for positron operation and then switch to polarized
positrons simply by readjusting the filter system for optimum FoM of the polarized positrons
and setting the phase shift of the recirculated positrons in the injector linac.

\subsubsection*{Proposed injector R\&D}

To begin, the anticipated yields of polarized and unpolarized positrons feasible with the
beams from the 123~MeV CEBAF injector should be calculated precisely, with the present
preliminary investigations extended to include a study of the collection system, the
bunching process for the positrons in the injector linac, and setting precise constraints
on the 123~MeV beam longitudinal and transverse emittance to ensure full acceptance in CEBAF.
This should be followed by a PEPPo-II experiment. Key apparatus  (the energy selection,
transverse emittance filter, and longitudinal emittance filter, the production target,
and the associated electron beam dump) should be designed and built. This apparatus should
be used (either at the CEBAF injector or at LERF) to measure the polarized and unpolarized
positron yields within the acceptance  specifications of CEBAF to verify the merits of
this approach and prepare for positron beams at 12 GeV. If the apparatus is built carefully
for this test, and the calculations are accurate, it should be possible to install it on
the CEBAF injector. 

On a positive note, work is already underway in many areas relevant to the production of
polarized positron beams. A key effort is work on improving the lifetime of the polarized
electron source at high beam currents~\cite{Sul18, Gra11}. Related work is improving
the photocathodes used for the production of the polarized electron beam. The distributed
Bragg reflector approach to photocathodes is of particular importance, and has demonstrated
a factor of four improvement in the quantum efficiency  relative to the familiar GaAsGaAsP
multilayer photocathodes~\cite{Liu16}. Finally, work is underway at Niowave Inc.~\cite{Bou17}
on the development of high power targets suitable for the polarized positron production;
using a liquid metal target they already demonstrated 10~kW power capability.

\subsection{Positron beam transport at CEBAF}

\begin{figure*}
\centering
\setlength\fboxsep{0pt}
   \includegraphics[width=0.9\textwidth]{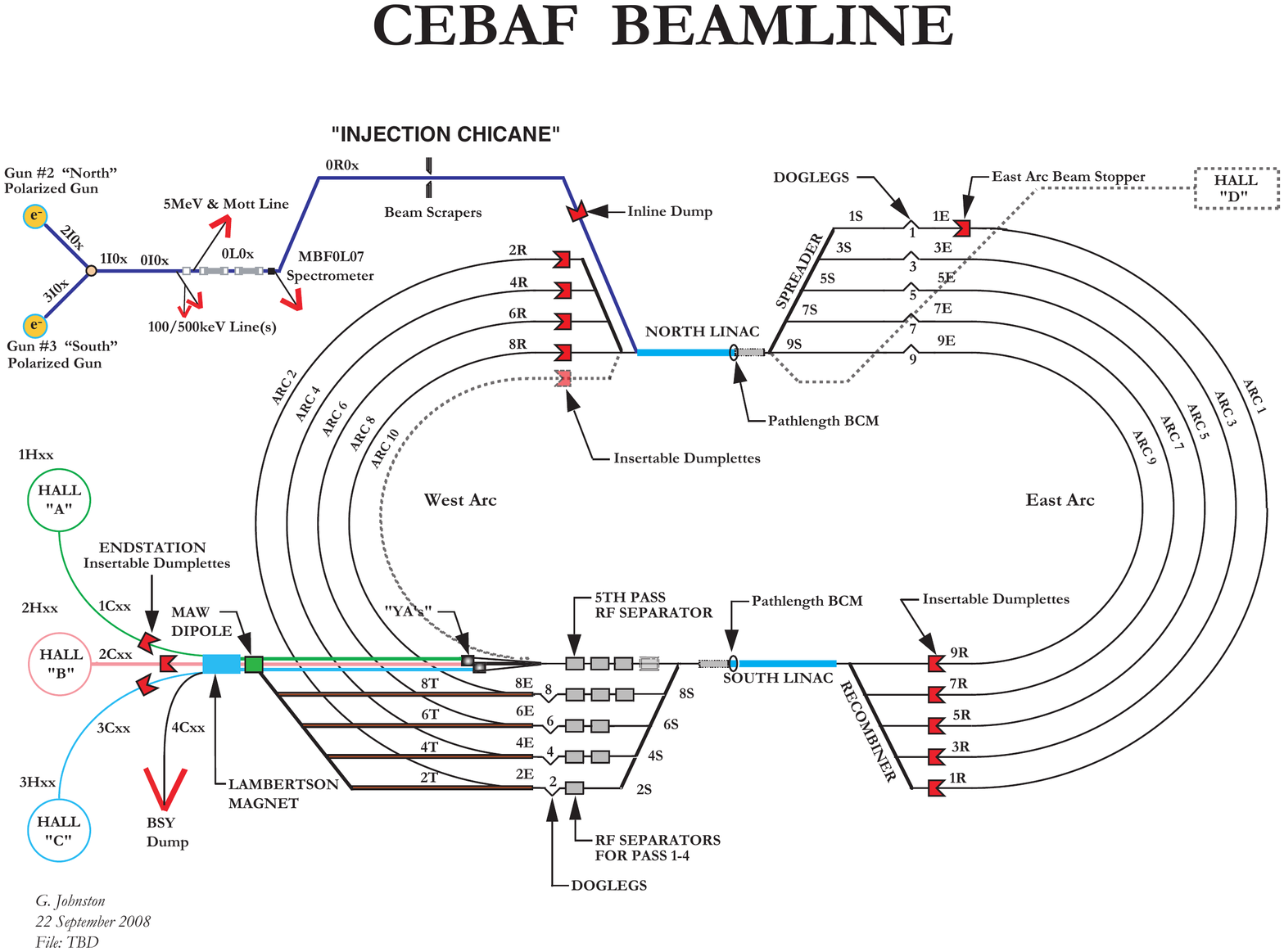}
   \caption{CEBAF accelerator schematic, showing the injection chicane merging into the North Linac.  
	Electron propagation is normally clockwise, 
        starting from the injector at the upper left corner.} 
   \label{accLayout}
\end{figure*}

The layout of the CEBAF accelerator is shown in Fig.~\ref{accLayout}.
The limitations on the beam properties at injection into CEBAF include the injector aperture and 
the injection chicane dispersion.  Both of these are normally configured for low emittance and low
momentum spread beams, but the configuration has considerable flexibility.
After injection, the beam momentum is increased by a factor of 9 in the first
linac (the North Linac).  The result of this strong adiabatic damping is that the 
momemtum acceptance of the accelerator is dominated by the injection chicane.  The transverse
emittance is similarly strongly damped, and the injection chicane again provides the principal limitation.

Operation of CEBAF requires diagnostics to configure the beam position along the accelerator, typically 
a combination of non-intercepting diagnostics (Beam Position Monitors [BPMs] and synchrotron light profile 
monitors [SLMs]) and intercepting viewscreens of various materials generating light from either 
fluorescence or transition radiation.  Finer quantitative measurement of the beam size on target
and for configuration of accelerator optics is done using either wire scanners for profile measurement
or in some cases SLMs.  Once the accelerator is configured for beam delivery, the beam position is 
normally monitored using BPMs and SLMs.  In the case of instability of accelerating RF or of steering or 
focusing magnets, the accelerator is protected from damage by a combination of photomultiplier based
beam loss monitors (BLMs) and a Beam Loss Accounting system (BLA) which compares the injected current 
to that delivered to the user(s).  Loss detected or imbalance in current {\tt IN} vs. current {\tt OUT} results in 
terminating beam delivery until the causes have been identified and corrected.

For very low current beam delivery, in particular to Hall B, it has been found adequate to configure 
the accelerator systems with low duty-factor beam with higher peak current, and then turn the beam current
down to nanoAmpere levels of continuous wave (CW) current for both final user approval and conduct of 
the experiment.  In conjunction with the BLMs, sparsely distributed SLMs in the first pass of the multi-pass 
CEBAF, one near the experiment, and a few BPMs near the experimental apparatus serve to verify that 
the beam remains appropriately on-target.  This is expected to remain practical for low-current
positron operation, although improvements in diagnostics should soon enable more extensive monitoring
of the beam in the accelerator.

\subsubsection*{Required beam parameters}

\begin{figure*}
\centering
\setlength\fboxsep{0pt}
   \includegraphics[width=0.8\textwidth]{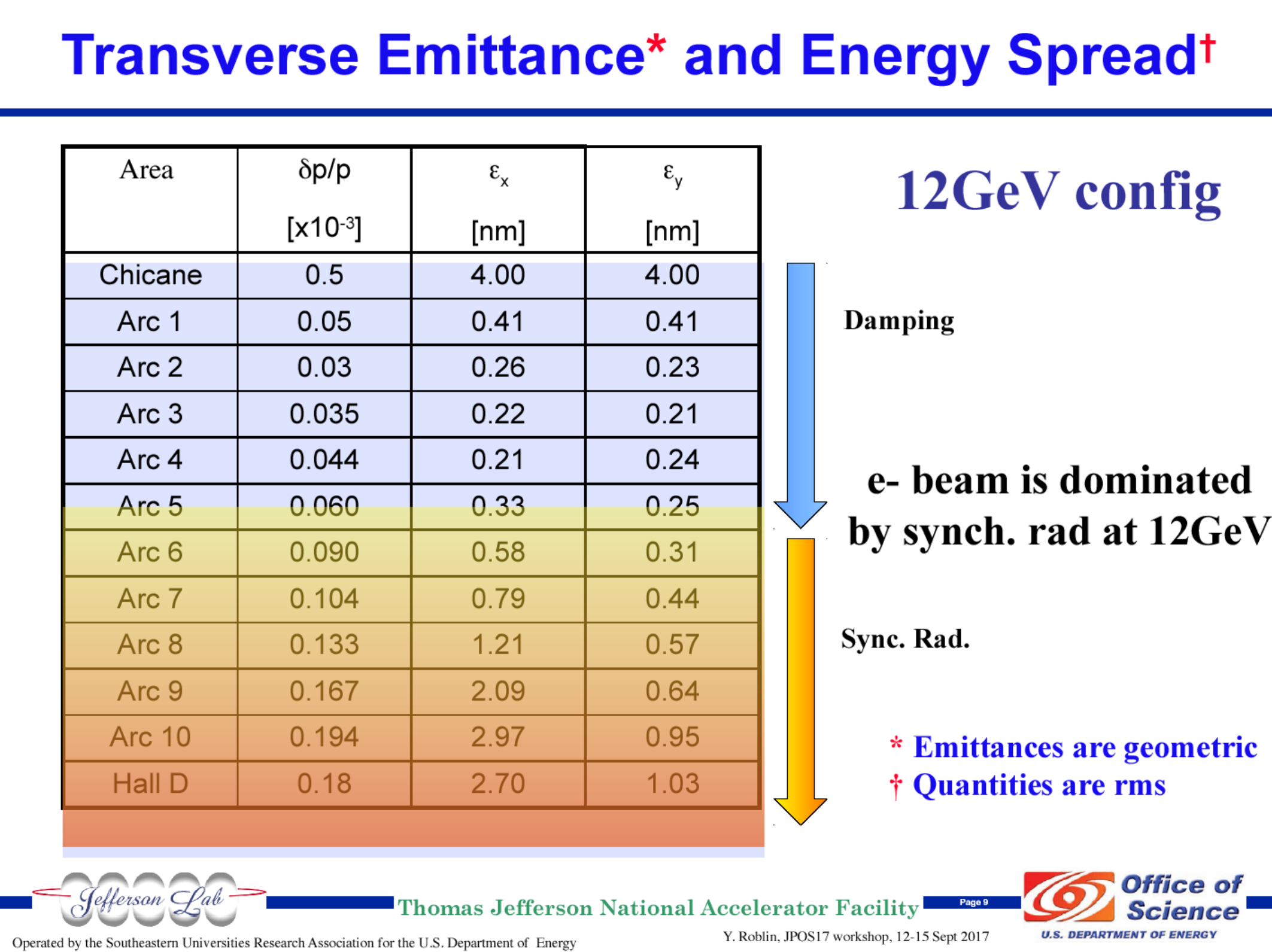}
   \includegraphics[width=0.8\textwidth]{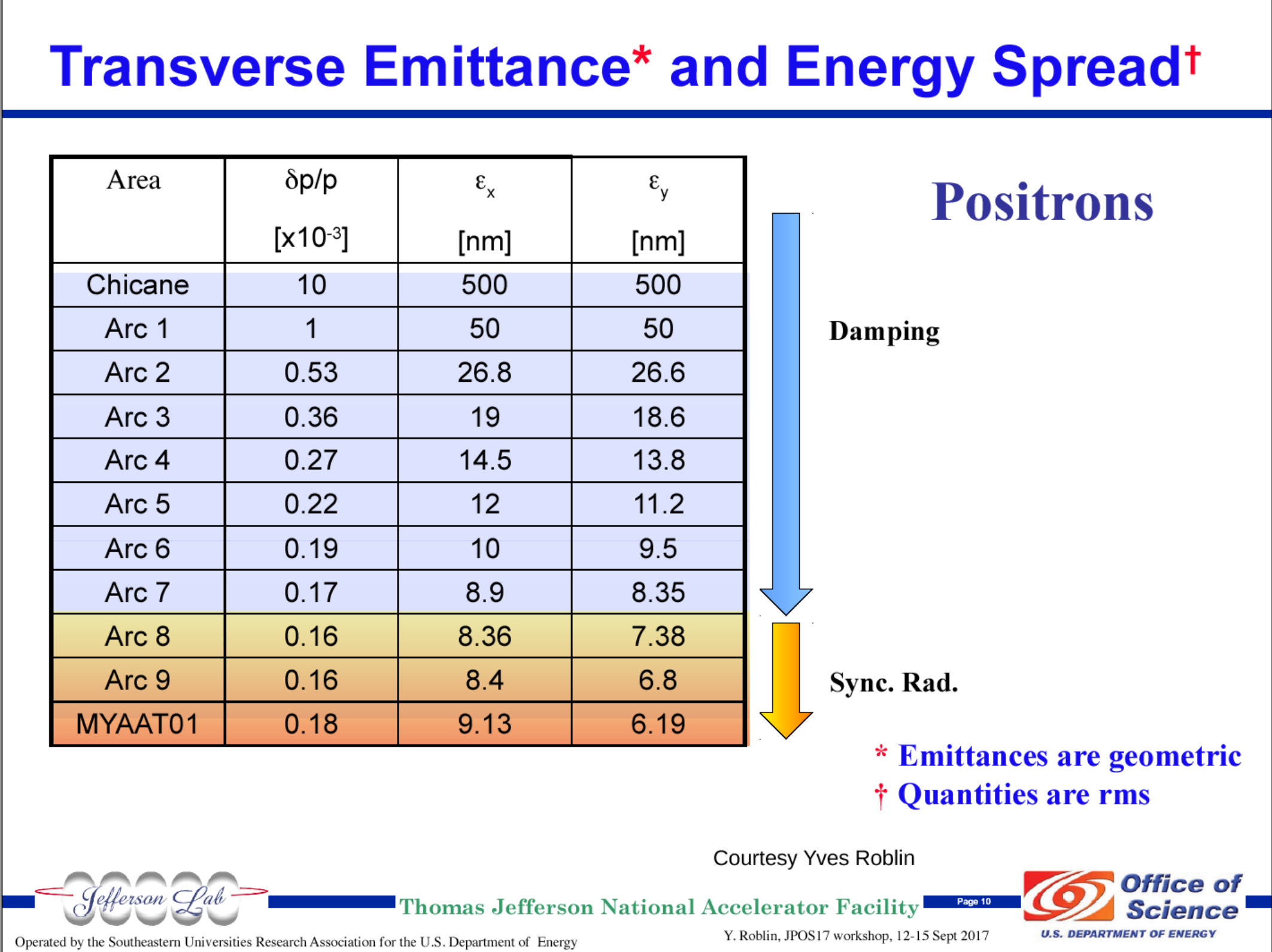}
   \caption{Beam parameter tables from a JPos17 presentation by Yves Roblin comparing anticipated positron properties
with electrons.} 
   \label{tables}
\end{figure*}

As presently constructed, the beam line aperture near injection is approximately 1~cm in radius.  In normal CEBAF operation, the peak chicane dispersion is approximately 1.6~m, but for applications with high energy spread, the dispersion can be configured to no more than 0.5~m. With this lower dispersion, the injection chicane momemtum acceptance can be as great as 2\%. The transverse acceptance (for emittance) is similarly limited by the injection aperture, but the transverse beta function in the chicane can be held to under 100~m, and typically less than 50~m at the limiting apertures. For the low currents anticiplated for position operation, the RMS beam radius may be workable at values as high as several mm. The corresponding normalized acceptance may therefore be as high as 40 mm$\cdot$mrad for beam energy near the typical injector value of $\sim$120~MeV (geometric emittance near 200~nm). Because this principal limiting aperture is very localized, it can be readily modified to increase its acceptance. Estimated beam parameters are shown in Fig.~\ref{tables} comparing electron and positron properties.
The longitudinal (bunch) structure of the beam is also very important.  The CEBAF accelerator does not rely
upon the phase stability principle underlying storage ring dynamics, but upon ``adequately well controlled
transport'' of the beam for its relatively short passage through the accelerator.  The beam occupies an
unstable fixed point in multidimensional phase space, accelerated on the crest of the RF.  The 
longitudinal dimension is the least stable, and the beam must be short enough in length that the 
momentum variation from particles drifting away from the bunch core does not result in transverse excursions
large enough to cause beam loss or to cause objectionable experimental background.  It has been the
experience in operation of CEBAF that the bunch length is very stable and does not require frequent
measurement or adjustment.  

The maximum RMS longitudinal extent typically desired for the bunch is approximately 100~$\mu$m. It is measured
using SLM imaging with enhanced dispersion optics in the first recirculation arc (1A) of CEBAF, 
combined with controlled phase shifts of the North Linac. For positron operation with maximum tolerance
to injector momentum spread, the dispersion of arc 1A should be returned to the ``standard'' optics value
for the higher arcs 3 through 10.  It can be configured transiently for metrological purposes when
positron bunch length measurement is required.

These values are much higher than typically used for CEBAF experiments.  The suitability of such beam
parameters can be examined and potentially verified in the accelerator by controlled tests using various
techniques to degrade the electron beam from the injector, as discussed below.

Finally, polarization of the beam is required for many of the proposed physics applications
for positrons at CEBAF.  Polarimetry of low-current beams is commonly done using invasive intercepting 
Moller polarimetry~\cite{Gas18}, although non-intercepting Compton polarimeters are also in 
use when the beam current is sufficiently high.  Moller (Bhabha) polarimetry is expected to serve the
purpose well.

\subsubsection*{Beam acceleration}

The CEBAF beam travels from the injector through the ``Injection Chicane'' into the North Linac (see Fig.~\ref{accLayout}). With a nominal kinetic energy of up to 123~MeV, the electrons undergo a series of 5.5$^\circ$ bends to exit the chicane along a trajectory parallel to the linac. The higher-pass beams from previous acceleration passes traverse a parallel ``reinjection chicane'' using magnets of the same $\int B\cdot dL$).  
\begin{figure}
\begin{tabular}{c}
    \includegraphics[width=0.990\linewidth]{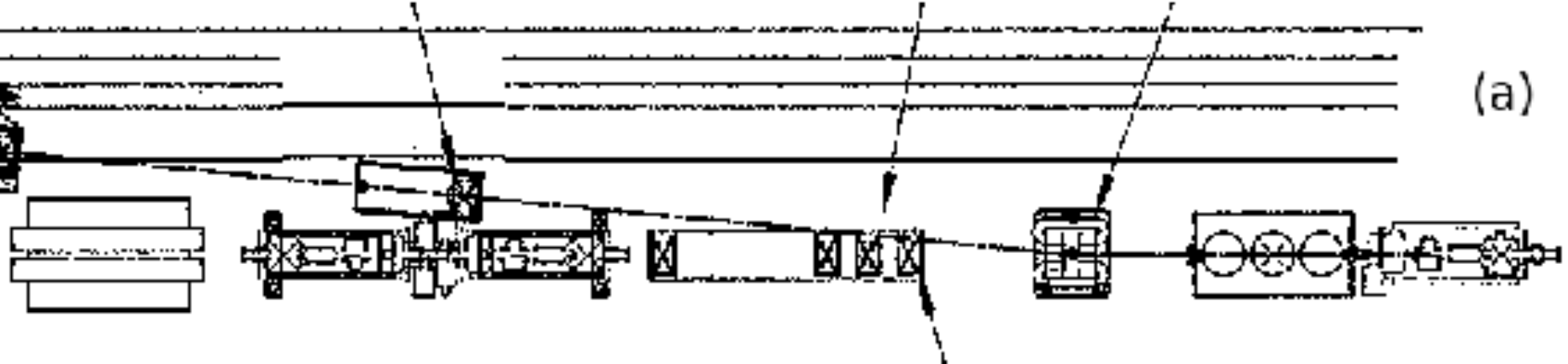}\\
    \includegraphics[width=0.990\linewidth]{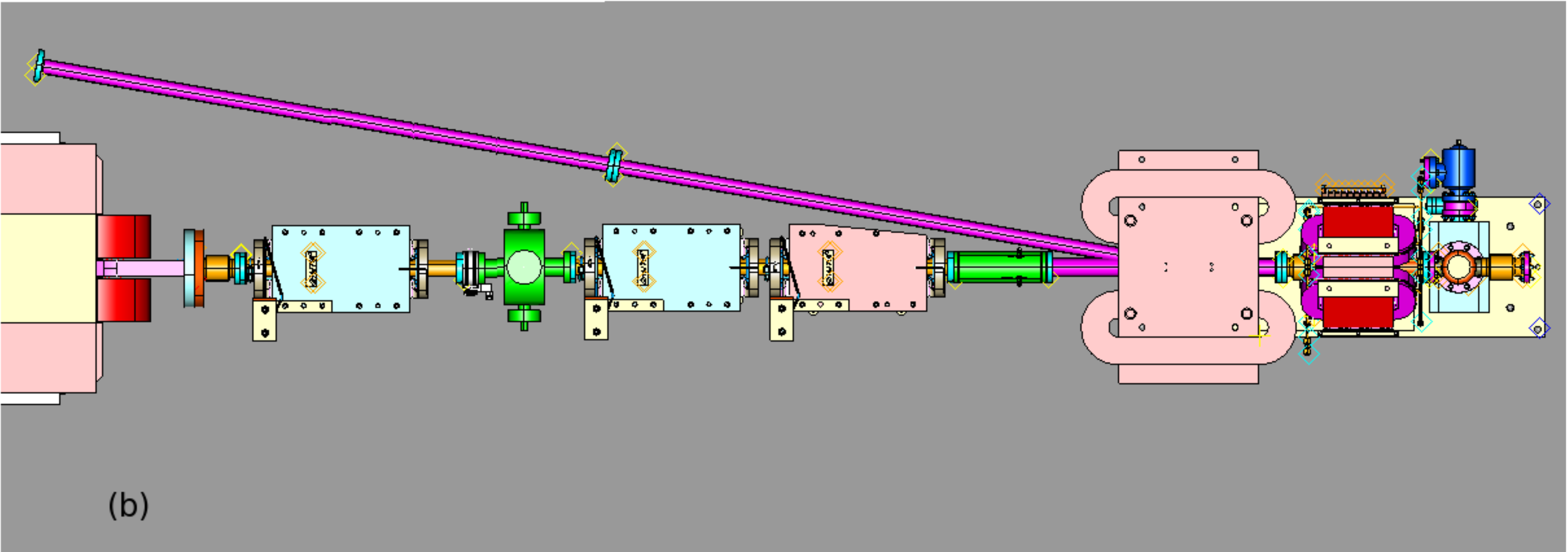}\\
\end{tabular}
   \caption{The injection chicane (a) provides for beam entrance into the North Linac, 
        propagating from left to right at the top of the diagram.  
        The chicane (b) designed to provide an exit path for injector-energy beam 
        (right to left) at the exit of the South Linac for a proposed energy recovery 
        linac demonstration enables a speculative option for antiparallel positron injection.}
   \label{fig:chicane}
\end{figure}
The final dipole of the injection chicane (Fig.~\ref{fig:chicane}) is shared by the ``re-injection chicane'', so that 
beams on each acceleration pass enter the linac with parallel trajectories.  
Separation and recombination of the various momentum beams are accomplished using vertical dipoles.  
The system is designed for equal energy gain in the two linacs. 
The tolerance to differential acceleration is useful to maintain total beam 
energy in case of hardware outages in one linac, or to improve the beam polarization
available to users on different acceleration passes.

After the injector is set up, beam is injected into the North Linac.  The RF configuration is adjusted to 
ensure that the proper energy gain is established, after which a similar process is done for the South
Linac. Beam is threaded through each linac and the following Spreader and Arc segments.  When the beam 
is re-established at the exit of the North Linac second pass, the path length through the machine is examined and
adjusted as necessary.  In each Spreader segment, the beam envelope parameters are presently measured and readjusted
to conform to design intent.

This process is repeated for each linac, and the beam is extracted from the accelerator for delivery 
on the appropriate pass to experiments. Final envelope matching is done to satisfy experimental 
requirements, and the polarization is measured and adjusted as needed.  Beam tune-up is typically done
with several microAmpere beam current in ``tune mode'' pulse structure with approximately 1.5\% duty factor.
Once the beam has been established to the user destination and configured satisfactorily, the beam current is
set to an appropriate low value, the injector is brought to CW mode, and the current is raised to the desired
level.

Along the way, BPMs report the beam position to ensure adequate clearance from walls and obstructions 
such as magnet septa.  Viewers and wire scanners report beam sizes for comparison against expectation and
to support required envelope tuning.  The RF-based path length monitor reports relative timing of the 
recirculating beams at the ends of each of North and South Linacs, as well as reporting relative current
transport so that losses can be localized and corrected.  The various BPM systems differ in sensitivity 
and stability.  All of these measurements must be possible for positron beams, as well as for electrons.

\subsection*{Beam diagnostic capabilities}

\begin{figure*}[t!]
\centering
\setlength\fboxsep{0pt}
   \includegraphics[width=0.90\textwidth]{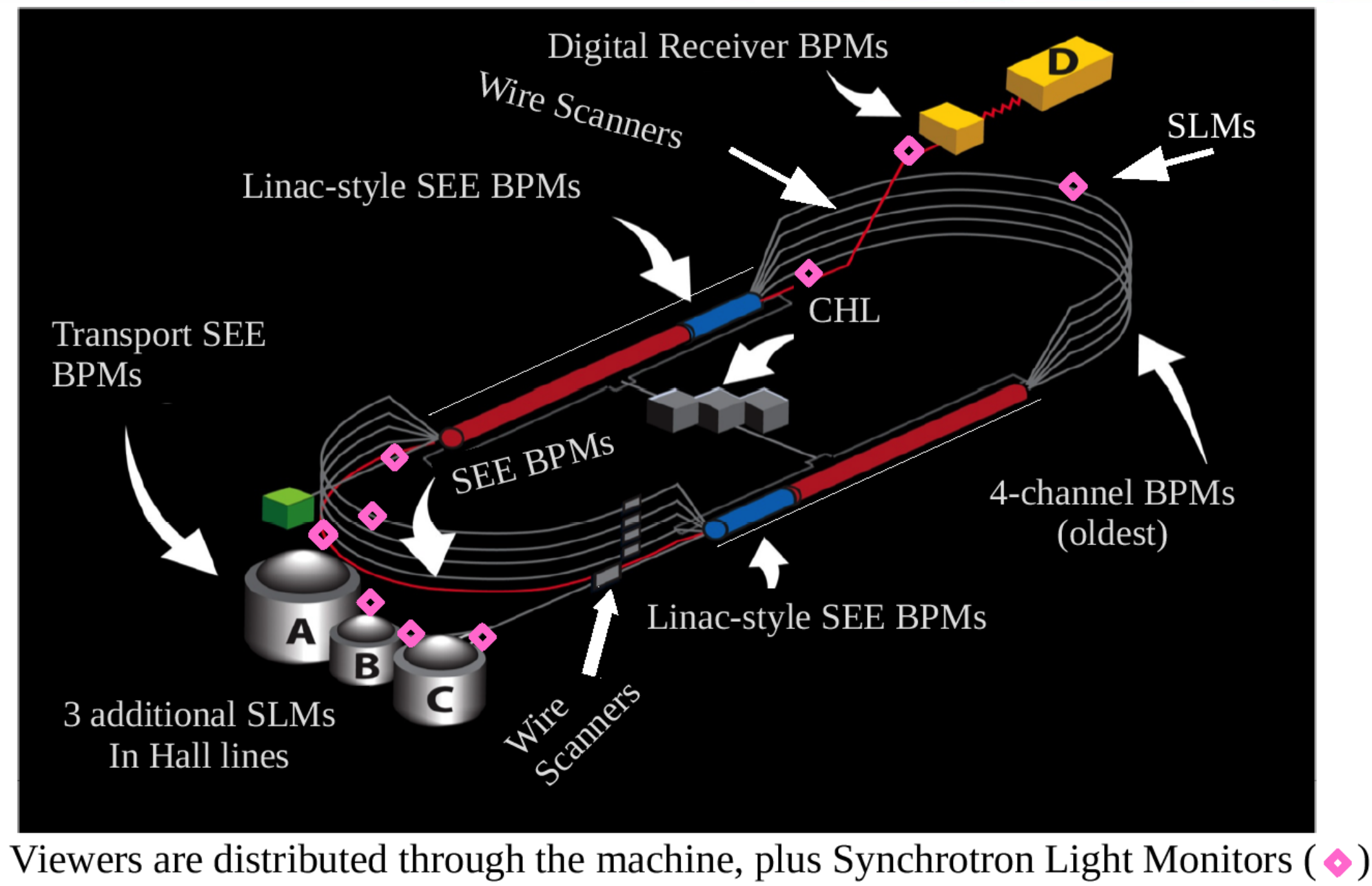}
   \caption{Beam diagnostic systems and their distribution in the CEBAF accelerator.} 
   \label{diagn}
\end{figure*}

The diagnostic systems of the CEBAF accelerator are indicated on the layout of Fig.~\ref{diagn}. It involves different devices.
\newline
Viewers: The viewer system has a usable range of average current of 1-100~nA. There are multiple materials in use, none of which are presently used for precise beam size measurements due to variable image quality, ``blooming'' of the image, and persistent (non-prompt) light emission.  Imaging for the SLMs is beginning to support precision measurement of the bunch longitudinal structure.  Viewers and SLMs function independently of whether the beam is electrons or positrons.
\newline
BPMs: The older ``4-channel'' BPMs require approximately 2~$\mu$A of beam current for reliable readings. The Switched Electrode Electronics systems (SEEs) exist in multiple configurations, some of which can operate down to approximately 200~nA CW. The ``Digital Receiver'' (DR) BPMs can report positions down to currents of approximately 30~nA. There are also a few carefully sited, cavity-coupled BPMs, which can report  positions for currents town to 1~nA.  These are called ``nanoAmp BPMs''.
\newline
SLMs: The synchrotron light monitors can provide usable images for beam currents as low as the nanoAmpere range, depending upon the camera technology used. It may be possible to extend this system capability to provide optical BPMs, opening another option for diagnostic extensions for positron beams in the event that peak currents remain small.
\newline
Cavity pickups: These are used for beam current monitoring and beam circulation (``path length'') tuning. They are useful for microAmpere average currents.
\newline
Wire scanners: These profile monitors are typically used for several microAmpere ``tune mode'' beam current, but some systems are used to measure beam profiles with as small as 5~nA of beam current in Hall B.

Extensions of the diagnostic capability of the accelerator will be required to support low-current setup if it is not possible with positron pulse compression or high intensity, low duty-factor positron source operation to raise the peak positron current to approximately 2~$\mu$A for the approximately 100~$\mu$s duration required for many of the BPMs. Envelope matching for the positron beam may be sufficiently well supported by additional SLMs which are expected to be installed as part of the electron program. It is also possible to upgrade the data acquisition for the wire scanner profile monitors already in use for envelope matching. Low current wire scanner profiles are routinely acquired for 
beams at the few nA level using scaler techniques in Hall B.

\subsubsection*{Proposed scaled beam performance R\&D}

It is possible to increase the emittance and energy spread of the CEBAF electron beam to levels commensurate with anticipated positrons, verifying the operability of various subsystems and diagnostics. For instance, Optical Transition Radiation (OTR) foils have been used in CEBAF in the past. When left in the beam path on the linacs in the hope of providing continuous beam size monitors, beam scattering was observed to increase the emittance and energy spread of the beam to levels unacceptable for users. Controlled measurements and performance comparisons such as this would provide early validation of useable positron beam parameters. 

Bunch length has been of recent concern in CEBAF operations due to parts per thousand beam loss at high beam current. Very short bunches are not required for low current operation, and it is possible to explore in CEBAF the bunch length performance required at current levels expected for positron operation.

\subsection{Magnet polarity reversal}

In order for positrons to transit CEBAF in the usual electron path, all magnetic fields must be inverted in polarity. There is no known obstacle to doing this, and no observations are recorded in CEBAF history to indicate that polarity reversal will have any detrimental effects on, for instance, calibration curves for magnetic field vs. power supply current. The dipole powering network uses common power supplies feeding strings of magnets in series. Magnets required to be independently adjustable were designed to be slightly too high in field strength, and electronic loads (called ``shunts'') are installed to shunt a controlled amount of the bus current around the magnet. All machine protection provisions for the dipole magnets remain functional with the main power supply leads inverted. The shunt hardware is unipolar, and inverting the shunt leads in coordination with the power supply leads leaves the dipole powering network fully functional and protected. 
These many power supplies and shunt connections would seem to require reversing switches to be installed with carefully designed features added to ensure that all polarity changes remain appropriately coordinated. It would be very time-consuming and likely too error-prone to rely upon manual reconnection of the exstensive set of magnet power leads.

\subsection*{Speculative: antiparallel positron operation}

It appears possible to configure CEBAF as was done for Energy Recovery Linac tests~\cite{Fre04} and may potentially be done again~\cite{Meo16, Sat16}, in a way which enables antiparallel positron injection at the exit of the South Linac. This mode of operation would not require any polarity change in the CEBAF accelerator, although it would not be able to use CEBAF's existing multi-user beam delivery mechanisms. Also, 
a beam transport line must be designed and constructed from some point along the existing West Arcs 2/4/6/8/10 to, for example, the existing experimental Hall B. Counter-propagating positron operation would require only phase changes in the linac accelerating RF, and with properly equalized acceleration in the two linacs would require no changes in quadrupole or main dipole settings. With appropriate diagnostic 
supplement (such as SLMs configured to observe antiparallel photons), the transition between electron and positron operation could be extremely simple and reversible. For further development of such a system~\cite{Tie18}.

%
%

\hyphenation{ca-lo-ri-me-ter}

\section{Executive summary}

\begin{table}[h!]
\begin{center}
\begin{tabular}{r||c|c|c||c}
                   & \multicolumn{2}{c|}{I (nA)} & Beam & Time \\  
                   & $e^-$ & $e^+$ & Polarization & (d) \\  \hline \hline
\multicolumn{5}{l}{\it Two-photon exchange}    \\ \hline
      TPE @ CLAS12 &    60 &     60 &  No & 53 \\ \hline
      TPE @ SupRos &     - &   1000 &  No & 18 \\ \hline
         TPE @ SBS & 40000 &    100 & Yes & 55 \\ \hline
\multicolumn{5}{l}{\it Generalized Parton Distributions} \\ \hline 
   p-DVCS @ CLAS12 &    75 &     15 & Yes & 83 \\ \hline
   n-DVCS @ CLAS12 &    60 &     60 & Yes & 80 \\ \hline
   p-DVCS @ Hall C &     - &   5000 &  No & 56 \\ \hline
\multicolumn{5}{l}{\it Test of the Standard Model} \\ \hline
     $\Apr$ search &     - & 10-100 & No & 180 \\ \hline
\multicolumn{4}{r||}{\bf Total Data Taking Time} & {\bf 525}\\ \hline\hline
\end{tabular}
\caption{Characteristics of a positron experimental program at Jlab.}
\label{time_req}
\end{center}
\end{table}
The prospect of polarized positron beams at JLab is attracting a lot of interest as demonstrated in previous workshops dedicated to this possibility, and by the support of the User Community to this letter. Positron beams at CEBAF will definitely enhance the scientific reach of the 12~GeV Upgrade, and could also open new windows of opportunity for the physics program at an Electron-Ion Collider~\cite{Jpo17}. \newline 
Particularly, positron beams will contribute uniquely to the 12~GeV high impact experimental programs: the physics of the two-photon exchange, the determination of Generalized Parton Distributions, and tests of the Standard Model. An example of a 12~GeV positron physics program is summarized in Tab.~\ref{time_req}. It accounts for 525~days of data taking distributed over seven experiments:
\renewcommand{\labelenumi}{\it\roman{enumi})}
\begin{enumerate}
\item{TPE @ CLAS12: this experiment proposes a measurement of the $e^+p/e^-p$ elastic cross section ratio with the {\tt CLAS12} detector, and using unpolarized positron and electron beams; it also requires an upgrade of the Central Detector, replacing the Central Neutron Detector with a new Central Electromagnetic Calorimeter based on a Tungsten Powder technology. 
}
\item{TPE @ SupRos: this experiment proposes a Super-Rosenbluth measurement of the elastic cross section off protons with an unpolarized positron beam, using standard existing spectrometers; it will provide a direct measurement of the proton electromagnetic form factors to be compared with electron data to determine the importance of two-photon effects. 
}
\item{TPE @ SBS: this experiment proposes to measure the transfer of the longitudinal polarization of electron and positron beams in the ${\overrightarrow{e^+}}p$ and ${\overrightarrow{e^-}p}$ elastic scattering with the Super Big-Bite Spectrometer for the recoil proton detection and an  electromagnetic calorimeter for detecting scattered electrons and positrons.
}
\item{p-DVCS @ CLAS12: this experiment proposes to measure the beam charge asymmetries in the DVCS reaction off protons using the {\tt CLAS12}  detector and polarized electron and positron beams; these provide unique observables, of particular interest in the determination of GPDs and in the access to the Gravitationnal Form Factors of the proton.
}
\item{n-DVCS @ CLAS12: this experiment proposes to measure the beam charge asymmetries in the DVCS reaction off neutrons using the {\tt CLAS12} detector and polarized electron and positron beams; these provide a direct access to the real part of the least known $\mathcal{E}$ Compton Form Factor, and enable the flavor separation of the corresponding GPD.
}
\item{p-DVCS @ Hall C: this experiment propose high precision measurements of the cross section for DVCS off protons with an unpolarized  positron beam, using the Hall C {\tt HMS} spectrometer and the Neutral Particle Spectrometer (NPS); measurements will be performed at selected  kinematics of the already approved NPS experimental program with electrons, for direct comparison and separation of the pure DVCS physics  signal from the interference contribution. 
}
\item{$\Apr$ search: this experiment proposes to search for the $\Apr$ dark photon, using an unpolarized positron beam and an  electromagnetic calorimeter to study $\Apr$ production in the $e^+e^-$ annihilation reaction through the invisible decay channel.
}
\end{enumerate}
The scientific context, the experimental details, and the scientific impact of each measurement is described further in the following sections gathering the corresponding specific letters. Note that the {\tt CLAS12} and Super-Rosenbluth TPE measurements are part of the same letter. Additional interest from the User Community, not described in the present letter, is already existing. For instance, the PRad/DRad  collaboration is considering using electron and positron beams to determine the deuteron radius.

The use of the CEBAF injector to make polarized positrons has been convincingly demonstrated. Initial calculations and estimates suggest options for $\sim$100~nA polarized and $\sim$1~$\mu$A unpolarized positron beams are reasonable. Higher currents can also been obtained depending on the initial production scheme, the polarized electron source capabilities, the performance of the positron collection, the emittance filter, and the capabilities of the RF system. Each of these components have limitations and upgradable possibilities. Evaluations and estimates suggest that CEBAF can accelerate and transport positron beams to experimental halls. This will require improvements of the beam diagnostics and polarity reversal capability of the transport magnets. Clearly, the thorough technical evaluation of the positron beam capabilities at CEBAF demands a sustained R\&D effort to support a Conceptual Design Report (CDR). We are seeking for the recommendation of the Jefferson Lab Program Advisory Committee to support human resources and funding investments for the R\&D and the CDR. 

%
%

\newpage

\null\vfill
	
\begin{center}

\section{\it Letter-ot-Intent: TPE @ CLAS12 \& Hall A/C}

\vspace*{15pt}

{\Large{\bf Studying two-photon exchange contributions}}

\vspace*{3pt}

{\Large{\bf in elastic e$^+$-p and e$^-$-p scattering}}

\vspace*{3pt}

{\Large{\bf at Jefferson Lab}}

\vspace*{15pt}

{\bf Abstract}

\begin{minipage}[c]{0.85\textwidth}
The proton elastic form factor ratio can be measured either via Rosenbluth separation in an unpolarized beam and target experiment, or via the use of polarization degrees of freedom. However, data produced by these two approaches show a discrepancy, increasing with $Q^2$. The proposed explanation of this discrepancy -- two-photon exchange -- has
been tested recently by three experiments. The results support the existence of a small two-photon exchange effect but cannot establish that theoretical treatment at the measured momentum transfers are valid. At larger momentum transfers, theory remains untested, and without further data, it is impossible to resolve the discrepancy. A positron beam at Jefferson Lab allows us to directly measure two-photon exchange over an extended $Q^2$ and $\varepsilon$ range with high precision. With this, we can validate whether the effect reconciles the form factor ratio measurements, and test several theoretical approaches, valid in different parts of the tested $Q^2$ range.
\end{minipage}

\vspace*{15pt} 

{\it Spokespersons: \underline{J.~Bernauer} (bernauer@mit.edu), \underline{A.~Schmidt} (schmidta@mit.edu), J.~Arrington, V.~Burkert}

\end{center}

\vfill\eject

%
%
\subsection{Introduction}

\begin{figure}[h!]
\centerline{\includegraphics[width=0.8\textwidth]{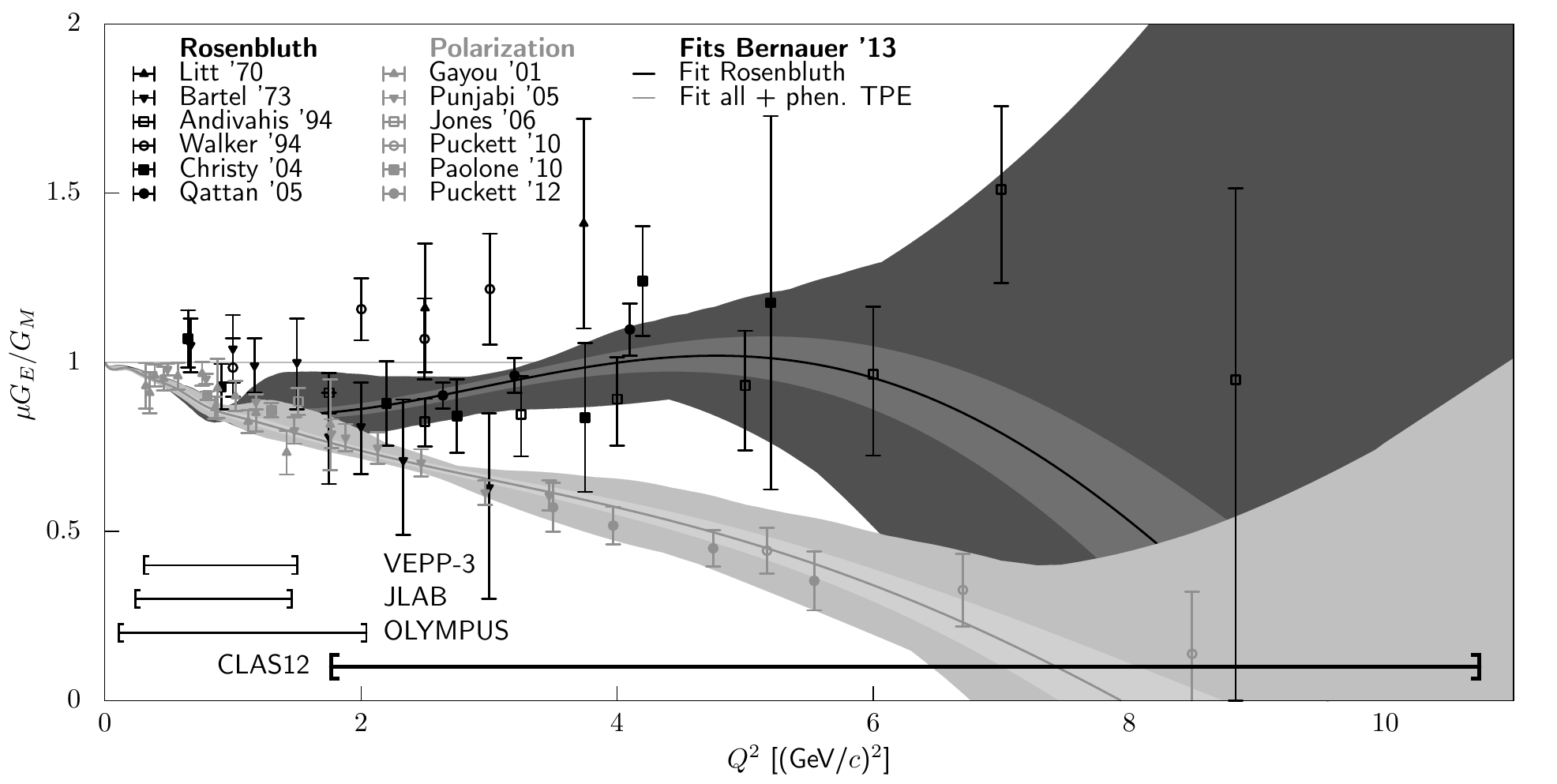}}
\caption{\label{figratio}The proton form factor ratio $\mu G_E/G_M$, as determined via Rosenbluth-type (black points, from~\cite{Lit70, Bar73, And94, Wal94, Chr04, Qat05}) and polarization-type (gray points, from~\cite{Gay01, Pun05, Jon06, Puc10, Pao10, Puc12}) experiments. While the former indicate a ratio close to 1, the latter show a distinct linear fall-off. Curves are from a phenomenological fit~\cite{Ber14}, to either the Rosenbluth-type world data set alone (dark curves) or to all data, then including a  phenomenological two-photon-exchange model. We also indicate the coverage of earlier experiments as well as of the experiment described below.}
\end{figure}

Over more than half a century, proton elastic form factors have been studied in electron-proton scattering with unpolarized beams. These experiments have yielded data over a large range of four-momentum transfer squared, $Q^2$. The form factors were extracted from the cross sections via the so-called Rosenbluth separation. Among other things, they found that the form factor ratio $\mu G_E/G_M$ is in agreement with scaling, i.e., that the ratio is constant. Somewhat more recently, the ratio of the form factors was measured using polarized beams, with different systematics and increased precision especially at large $Q^2$. However, the results indicate a roughly linearly fall-off of the ratio. The result of the different experimental methods, as well as some recent fits, are compiled in Fig.~\ref{figratio}. The two data sets are clearly inconsistent with each other, indicating that one method (or both) are failing to extract the proton's true form factors. The resolution of this "form factor ratio puzzle" is crucial to advance our knowledge of the proton form factors, and with that, of the distribution of charge and magnetization inside the proton.

The differences observed by the two methods have been attributed to two-photon exchange (TPE) effects~\cite{Gui03,Car07,Arr11,Afa17}, which are much more important in the Rosenbluth method than in the polarization transfer method, where they partially cancel out in the ratio. Two-photon exchange corresponds to a group of diagrams in the second order Born approximation of lepton scattering, namely those where two photon lines connect the lepton and proton. The so-called ``soft'' case, when one of the photons has negligible momentum, is included in the standard radiative corrections, like Ref.~\cite{Mo69,Max00}, to cancel infrared divergences from other diagrams. The ``hard'' part, where both photons can carry considerable momentum, is not. It is important to note here that the division between soft and hard part is arbitrary, and different calculations use different prescriptions. 

\begin{figure}[t!]
\centerline{\includegraphics[width=0.70\textwidth]{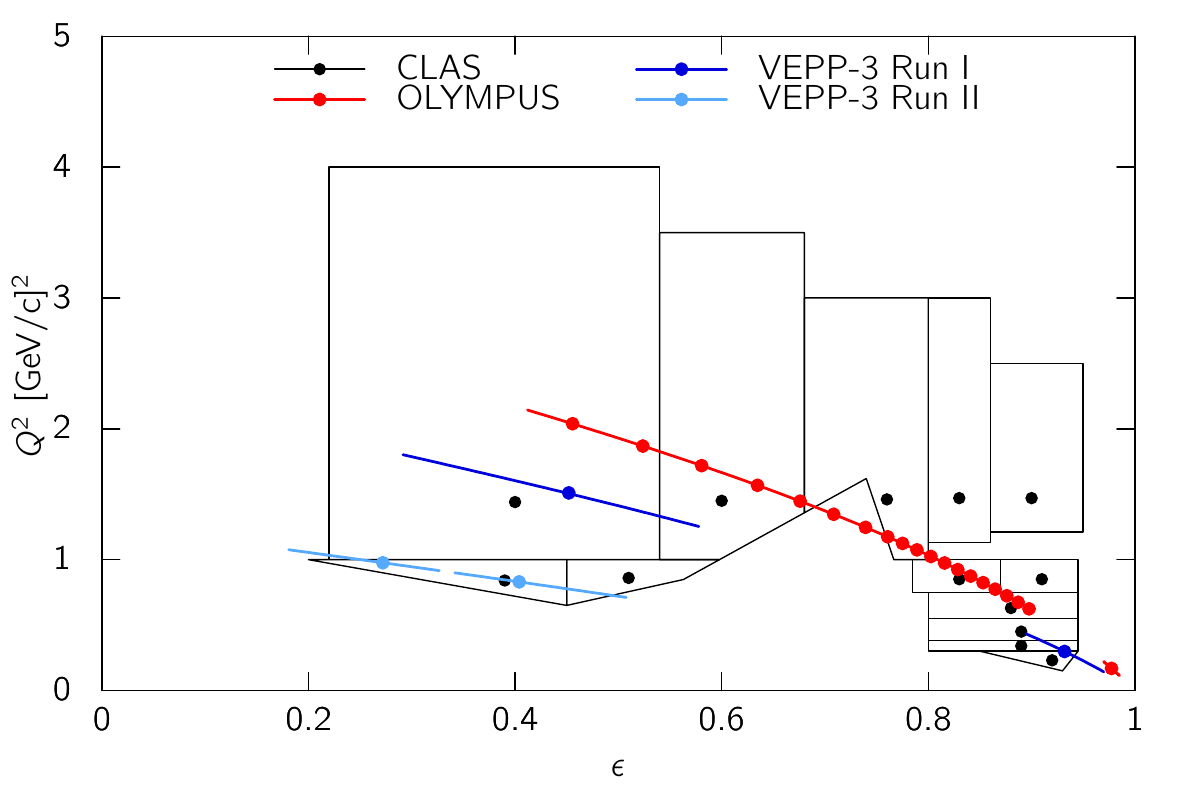}}
\caption{Kinematics covered by the three recent experiments to measure the two-photon exchange contribution to the elastic $ep$ cross section.}
\label{reach}
\end{figure}

It is obviously important to study this proposed solution to the discrepancy with experiments that have sensitivity to two-photon contributions. The most straightforward process to evaluate two-photon contribution is the measurement of the ratio of elastic $e^+p/e^-p$ scattering. Several experiments have recently been carried out to measure the 2-photon exchange contribution in elastic scattering: the VEPP-3 experiment at Novosibirsk~\cite{Rac15}, the CLAS experiment at Jefferson Lab~\cite{Mot13, Adi15, Rim17}, and the OLYMPUS experiment at DESY~\cite{Hen17}. The kinematic reach of these  experiments was limited, however, as shown in Fig.~\ref{reach}. The combined evaluation of all three experiments led the authors of the review~\cite{Afa17} to the conclusion that although the results show that the hypothesis of the absence of two-photon effects is excluded with 99.5\% confidence, "The results of these experiments are by no means definitive", and that "There is a clear need for similar experiments at larger $Q^2$ and at $\varepsilon < 0.5$".   

In this letter, we propose a new definitive measurement of the TPE effect that would be possible with a positron source at CEBAF. By alternately scattering positron and electron beams from a liquid hydrogen target and detecting the scattered lepton and recoiling proton in coincidence with the large acceptance CLAS-12 spectrometer, the magnitude of the TPE contribution between $Q^2$ values of 2 and 10~GeV$^2$ could be significantly constrained. With such a measurement, the question of whether or not TPE is at the heart of the ``proton form factor puzzle'' could be answered. 

Another option is use of the Super-Rosenbluth technique, a Rosenbluth separation using only proton detection. This approach is less sensitive to the difference between electron and positron beam runs, allowing for a precise study of TPE effects with a positron-only measurement (combined with existing electron data). The $Q^2$ range is lower, from 0.4~GeV$^2$ to 4-5~GeV$^2$, and the measurement extracts the TPE contribution to the $\varepsilon$-dependence of the cross section, rather than the cross section at a fixed value of $Q^2$ and $\varepsilon$. However, it does not require frequent changes between electron and positron beams, and is less sensitive to beam quality issues. 

\subsection{Previous work}

One significant challenge is that hard TPE cannot be calculated in a model-independent way. There are several model-dependent approaches. A full description of the available theoretical calculations are outside of the scope of this letter. Suffice it to say that they can be roughly divided into two groups: hadronic calculations, e.g.\ \cite{Blu17}, which should be valid for $Q^2$ from 0 up to a couple of GeV$^2$, and GPDs based calculations, e.g.~\cite{Afa05}, which should be valid from a couple of GeV$^2$ and up. 

Three contemporary experiments have tried to measure the size of TPE, based at VEPP-3 \cite{Rac15}, Jefferson Lab (CLAS, \cite{Mot13, Adi15, Rim17}) and DESY (OLYMPUS, \cite{Hen17}). These experiments measured the ratio of positron-proton to electron-proton elastic cross sections. The next order correction to the first order Born calculation of the elastic lepton-proton cross section contains terms corresponding to the product of the diagrams of one-photon and two-photon exchange. These terms change sign with the lepton charge sign. It is therefore possible to determine the size of TPE by measuring the ratio of positron to electron scattering:
\begin{equation} 
R_{2\gamma} = \frac{\sigma_{e^+}}{\sigma_{e^-}} \approx 1 + 2\delta_{TPE} \, .
\end{equation}
\begin{figure}[h!]
\centerline{\includegraphics[width=0.495\textwidth]{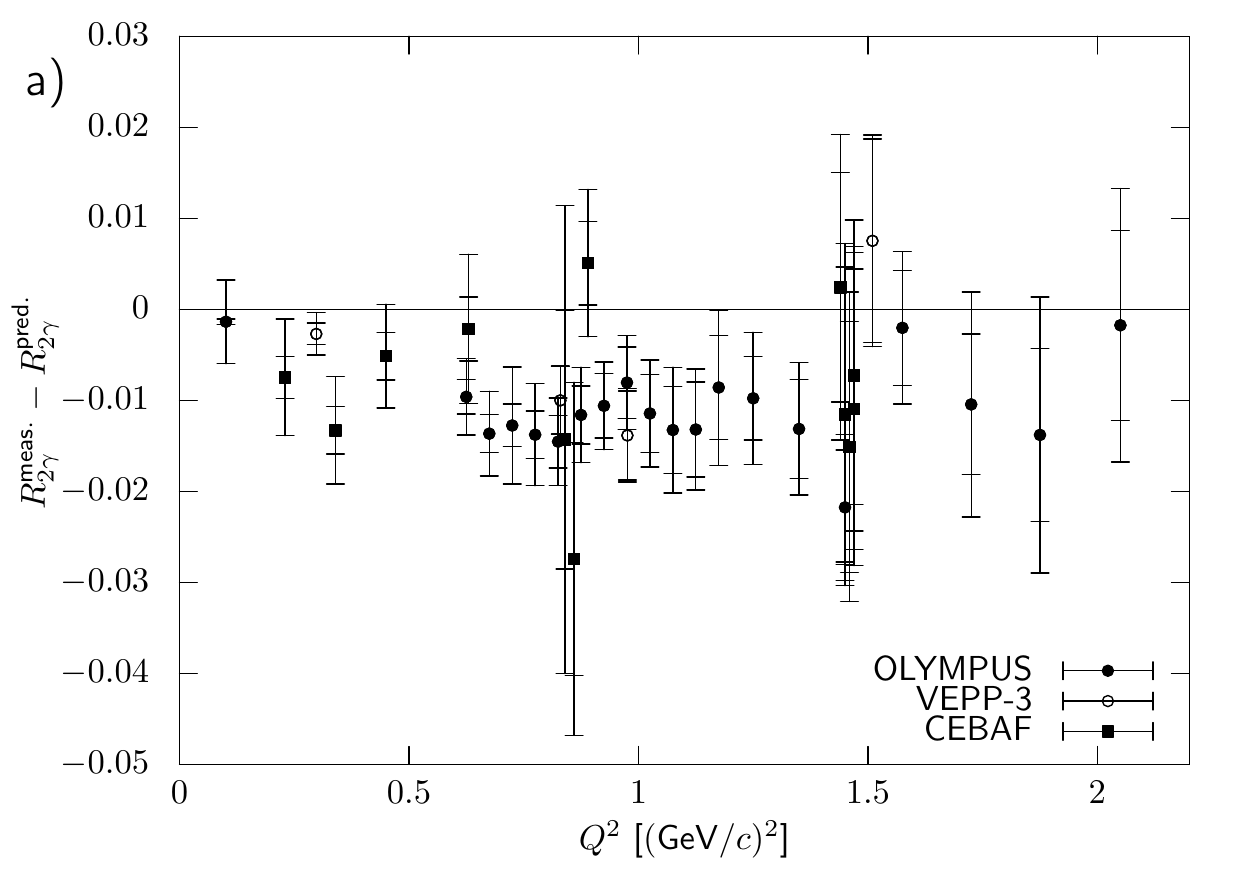}\includegraphics[width=0.495\textwidth]{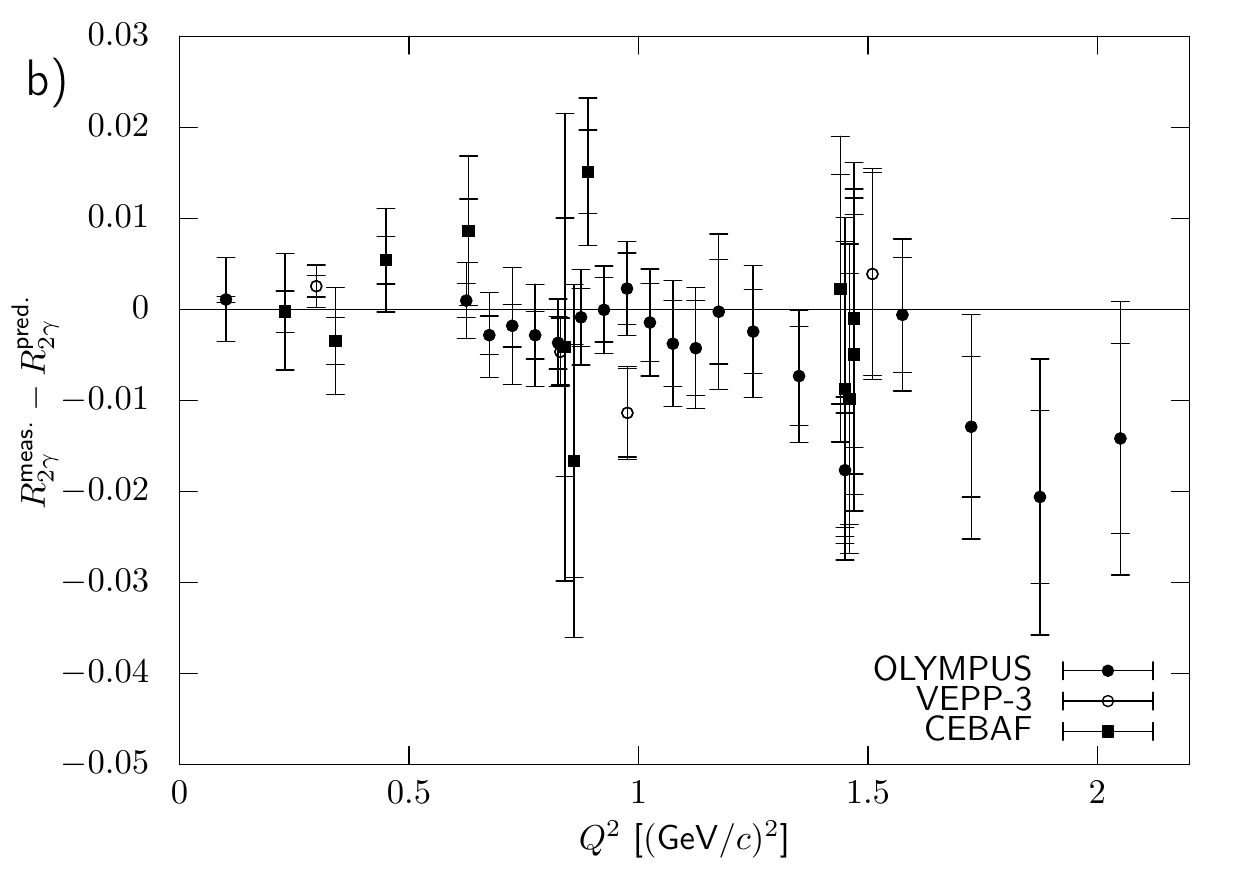}}
\caption{\label{figdiff} Difference of the data of the three recent TPE experiments~\cite{Rac15, Rim17, Hen17} to the calculation in \cite{Blu17} (a) and the phenomenological prediction from \cite{Ber14} (b).}
\end{figure}
The kinematic reach of the three experiments is shown in Fig.~\ref{reach}. The kinematic coverage in these experiments is limited to $Q^2 < 2$~GeV$^2$, and $\varepsilon > 0.5$, where the two-photon effects are expected to be small, and systematics of the measurements must be extremely well controlled. Figure~\ref{figdiff} depicts the difference of the data of the three experiments to the calculation by Blunden et al.~\cite{Blu17} and the phenomenological prediction by Bernauer et al.~\cite{Ber14}. It can be seen that the three data sets are in good agreement which each other, and appear about 1\% low compared to the calculation. The prediction appears closer for most of the $Q^2$ range, however over-predicts the effect size at large $Q^2$. This is worrisome, as this coincides with the opening of the divergence in the fits depicted in Fig.~\ref{figratio} and might point to an additional effect beyond TPE that drives the difference. The combination of the experiments prefer the phenomenological prediction with a reduced $\chi^2$ of 0.68, the theoretical calculation achieves a reduced $\chi^2$ of 1.09, but is ruled out by the normalization information of both the CLAS experiment and OLYMPUS to a 99.6\% confidence level. No hard TPE is ruled out with a significantly worse reduced $\chi^2$ of 1.53.

The current status can be summarized as such:
\begin{itemize}
\item TPE exists, but is small in the covered region;
\item Hadronic theoretical calculations, supposed to be valid in this kinematical regime, might not be good enough yet;
\item Calculations based on GPDs, valid at higher $Q^2$, are so far not tested at all by experiment;
\item A comparison with the phenomenological extraction allows for the possibility that the discrepancy might not stem from TPE alone.
\end{itemize}
We refer to \cite{Afa17} for a more in-depth review. The uncertainty in the resolution of the ratio puzzle jeopardizes the extraction of reliable form factor information, especially at high $Q^2$, as covered by the Jefferson Lab 12 GeV program. Clearly, new data are needed. 

\subsection{Experimental configuration}

Both theories and phenomenological extractions predict a roughly proportional relationship of the TPE effect with $1-\varepsilon$ and a sub-linear increase with $Q^2$. However, interaction rates drop sharply with smaller $\varepsilon$ and higher $Q^2$, corresponding to higher beam energies and larger electron scattering angles. This puts the interesting kinematic region out of reach for storage-ring experiments, and handicaps external beam experiments with classic spectrometers with comparatively small acceptance. 

\begin{figure}[t!]
\begin{center}
\includegraphics[height=230pt,width=335pt]{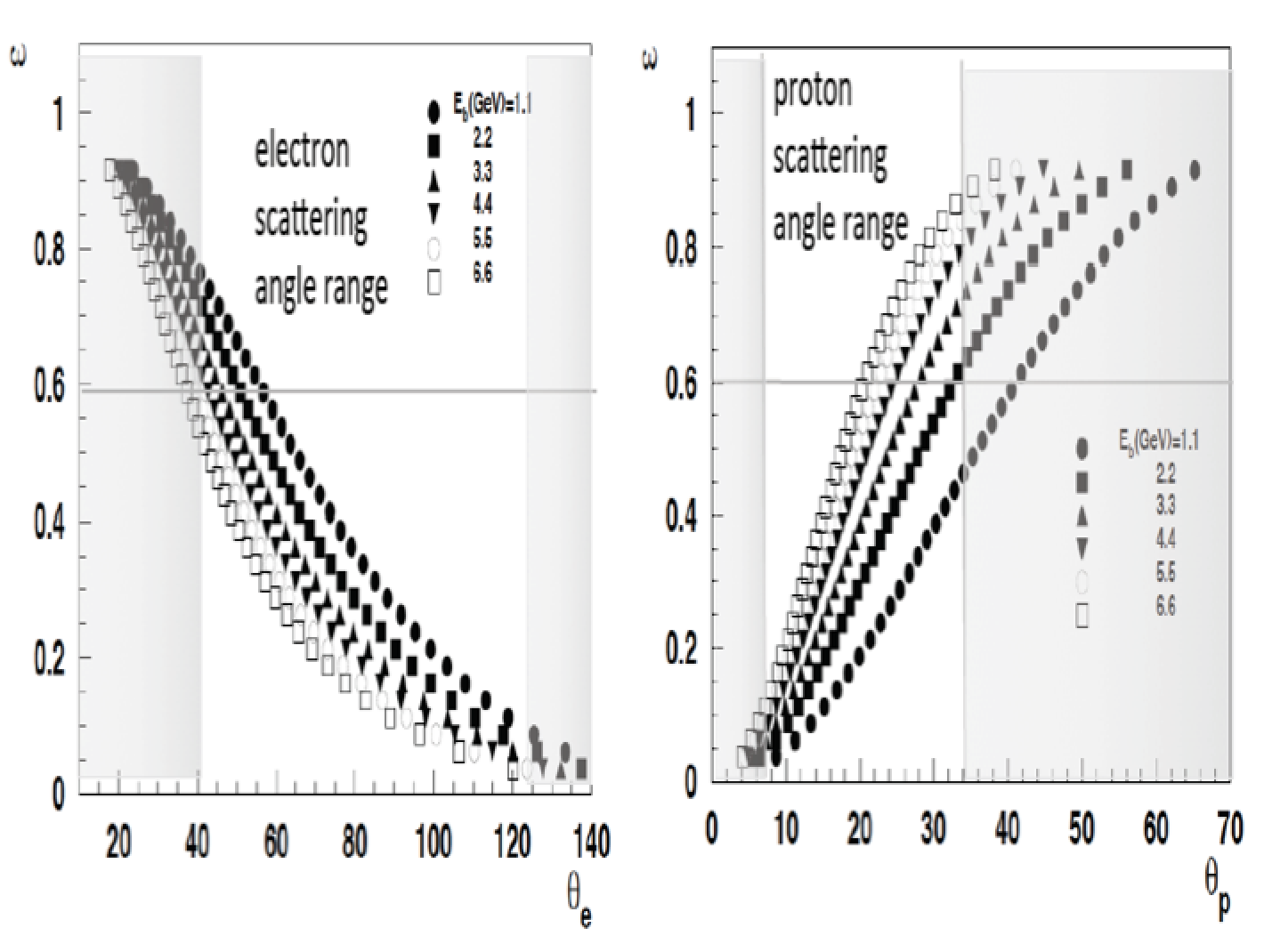}
\caption{Polar angle and $\varepsilon$ coverage for electron detection (left) and for proton detection (right). }
\label{angle_reach}
\end{center}
\end{figure}

\begin{figure}[h]
\begin{center}
\includegraphics[width=0.495\textwidth,height=0.295\textheight]{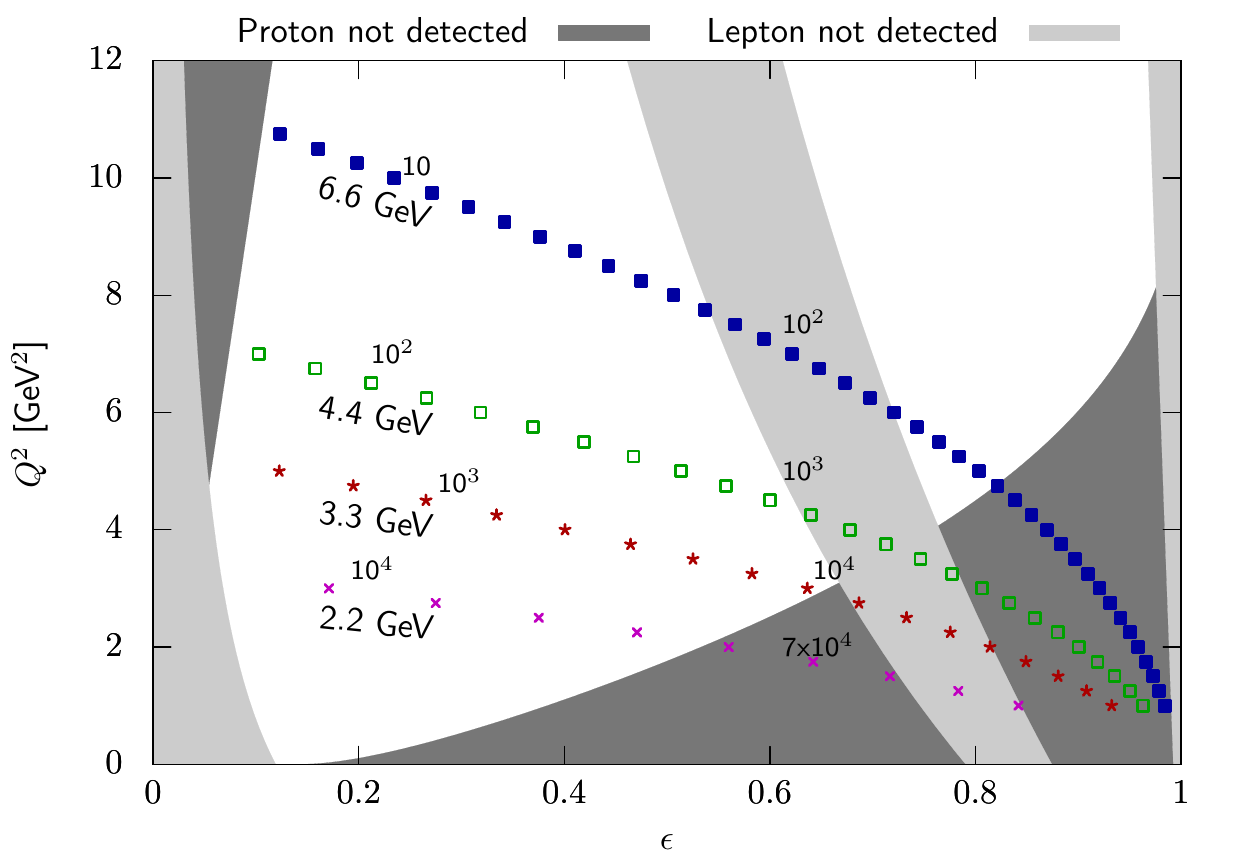}
\includegraphics[width=0.495\textwidth,height=0.295\textheight]{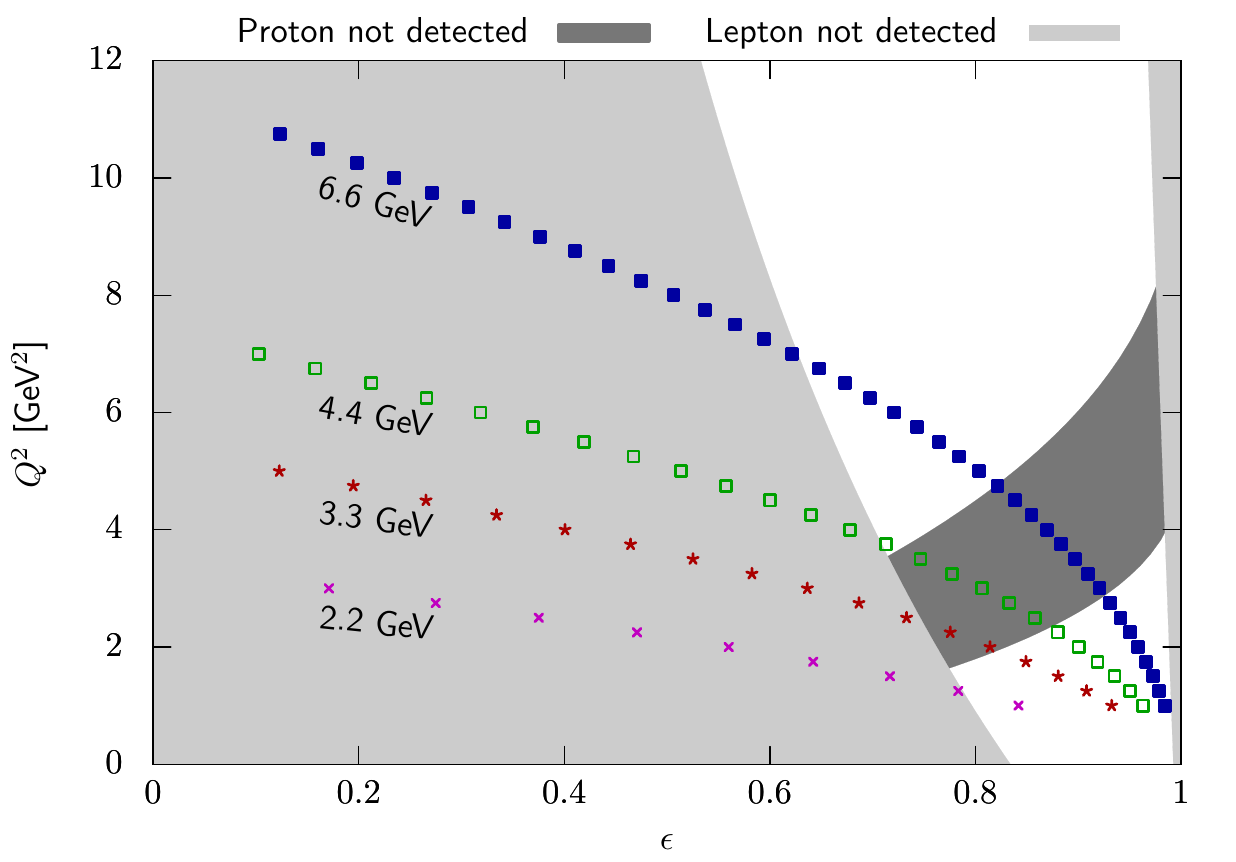}
\caption{Expected elastic event rates per hour for energies 2.2, 3.3, 4.4, 6.6 GeV in the $\varepsilon$ - $Q^2$ plane. Shaded areas are excluded by the detector acceptance. Left: proposed experiment; Right: standard setup}
\label{Q2eps}
\end{center}
\end{figure}

\begin{figure}[t!]
\begin{center}
\includegraphics[height=200pt,width=395pt]{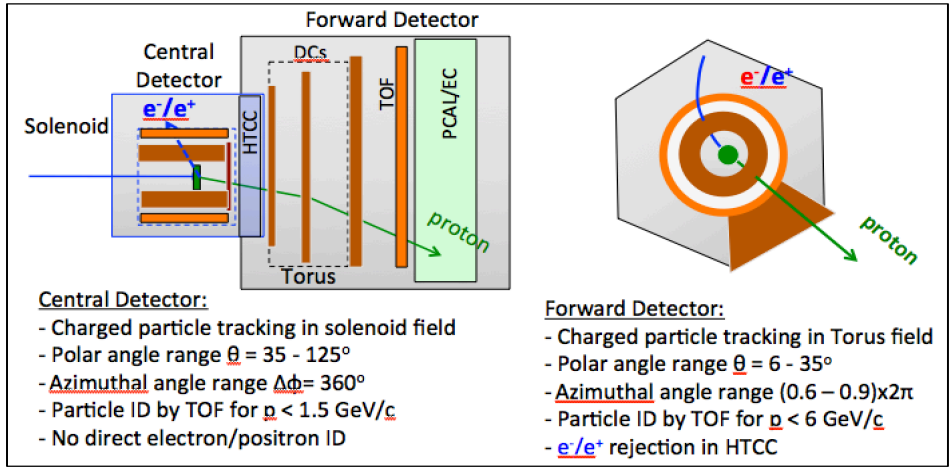}
\caption{{\tt CLAS12} configuration for the elastic $e^-p/e^+p$ scattering experiment (generic). The central detector will detect the electron/positrons, and the bending in the solenoid magnetic field will be identical for the same kinematics. The proton will be detected in the forward detector part. The torus field direction will be the same in both cases. The deflection in $\phi$ due to the solenoid fringe field will be of same in magnitude of $\Delta\phi$ but opposite in direction. The systematic of this shift can be controlled by doing the same experiment with opposite solenoid field directions that would result in the sign change of the $\Delta\phi$.}
\label{2gamma-exp}
\end{center}
\end{figure}

With the large acceptance of {\tt CLAS12}, combined with an almost ideal coverage of the kinematics, measurements of TPE across a wide kinematic range are possible, complementing the precision form factor program of Jefferson Lab, and testing both hadronic (valid at the low $Q^2$ end)  as well as GPD-based (valid at the high $Q^2$-end) theoretical approaches. Figure~\ref{angle_reach} shows the angle coverage for both the electron (left) and for the proton (right). There is a one-to-one correlation between the electron scattering angle and the proton recoil angle. For the kinematics of interest, say $\varepsilon < 0.6$ and $Q^2 > 2$~GeV$^2$ for the chosen beam energies from 2.2 to 6.6 GeV, 
nearly all of the electron scattering angles falls into a polar angle range from $40^\circ$  to $125^\circ$, and corresponding to the proton polar angle range from $8^\circ$ to $35^\circ$. These kinematics are most suitable for accessing the two-photon exchange contributions. The setup will also be able to measure the reversed kinematics with the electrons at forward angle and the protons at large polar angles. This is in fact the standard {\tt CLAS12} configuration of DVCS and most other experiments. While the two-photon exchange is expected to be small in this range, the sign change in TPE seen in the experiments, but not predicted by current theories, can be studied. \newline
Figure~\ref{Q2eps} shows the expected elastic scattering rates  covering the ranges of highest interest, with $\varepsilon < 0.6$ and $Q^2 = 2 - 10$~GeV$^2$. Sufficiently high statistics can be achieved within 10 hrs for the lowest energy and within 1000 hrs for the highest energy, to cover the full range in kinematics. Note that all kinematic bins will be measured simultaneously at a given energy, and the shown rates are for the individual bins in ($Q^2$,$\varepsilon$) phase-space.
In order to achieve the desired kinematics reach in $Q^2$ and $\varepsilon$ the {\tt CLAS12} detection system has to be used with reversed detection capabilities for electrons. The main modification will involve replacing the current Central Neutron Detector with a central electromagnetic calorimeter (CEC). The CEC will not need very good resolution, which is provided by the tracking detectors, but will only be used for trigger purposes and for electron/pion separation. The strict kinematic correlation of the scattered electron and the recoil proton should be sufficient to select the elastic events. The {\tt CLAS12} configuration suitable for this experiment is shown in Fig.~\ref{2gamma-exp}.

\subsection*{Central Electromagnetic Calorimeter}

The Central Electromagnetic Calorimeter (CEC) will be used for trigger purposes to detect electrons elastically scattered under large angles and for the separation of electrons (positrons) and charged pions. The CEC will be built based on a novel JLab Tungsten Powder/Scintillating Fiber Calorimetry technology proposed in 1999. This original proposal was to develop a compact, high-density fast calorimeter with good energy resolution at polar angles greater than 35$^\circ$ for the {\tt CLAS12} spectrometer ~\cite{Car01}, and occupy the radial space of $\approx$ 10 cm to fit inside the Central Solenoid. For the proposed elastic scattering experiment, the CEC would replace the
current Central Neutron Detector, which occupies approximately the same radial space and polar angle range. The powder calorimeter's essential features are compactness, homogeneity, simplicity, and unique readout capabilities. From the original proposal there exists a prototype calorimeter designed, built, and cosmic-ray tested. The dimensions of the active volume filled with tungsten powder are approximately length$\times$width$\times$height~=~0.1$\times$0.1$\times$0.07~m$^3$ in volume and with 5,488 fibers (Bicron BCF-12) with 0.75~mm diameter, uniformly distributed inside the tungsten powder volume. These fibers make up 35\% of the volume and the tungsten powder is filled into the remaining volume. The final density of the tungsten powder radiator is 12~g/cm$^3$, or about 5\% higher as compared with the density of bulk lead. The overall total density of the prototype active volume is $\approx$~8.0~g/cm$^3$. There is the possibility of increasing the density of the radiator to $\approx$~10.5~g/cm$^3$, which will lead to an increase of the detector absorption power. Also an additional effect can be obtained by simply decreasing sampling ratio, since having higher energy resolution is not a critical requirement. It has to be mentioned that due to the cylindrical shape of the CEC there will be no {\it side wall} effects. The estimated signal strength is  about 75 photoelectrons per MeV. The prototype can be tested and calibrated with electrons of known energy. Utilizing the unique Tungsten Powder Calorimetry expertise developed at JLab we propose to build a CEC with parameters close the the existing prototype calorimeter. The calorimeter will need to cover polar angles in a range of 40$^\circ$ to 130$^\circ$, and the full 2$\pi$ range in azimuth.

\subsection{Projected measurements at CLAS12}

\begin{figure}[t!]
\centerline{\includegraphics[width=0.75\textwidth]{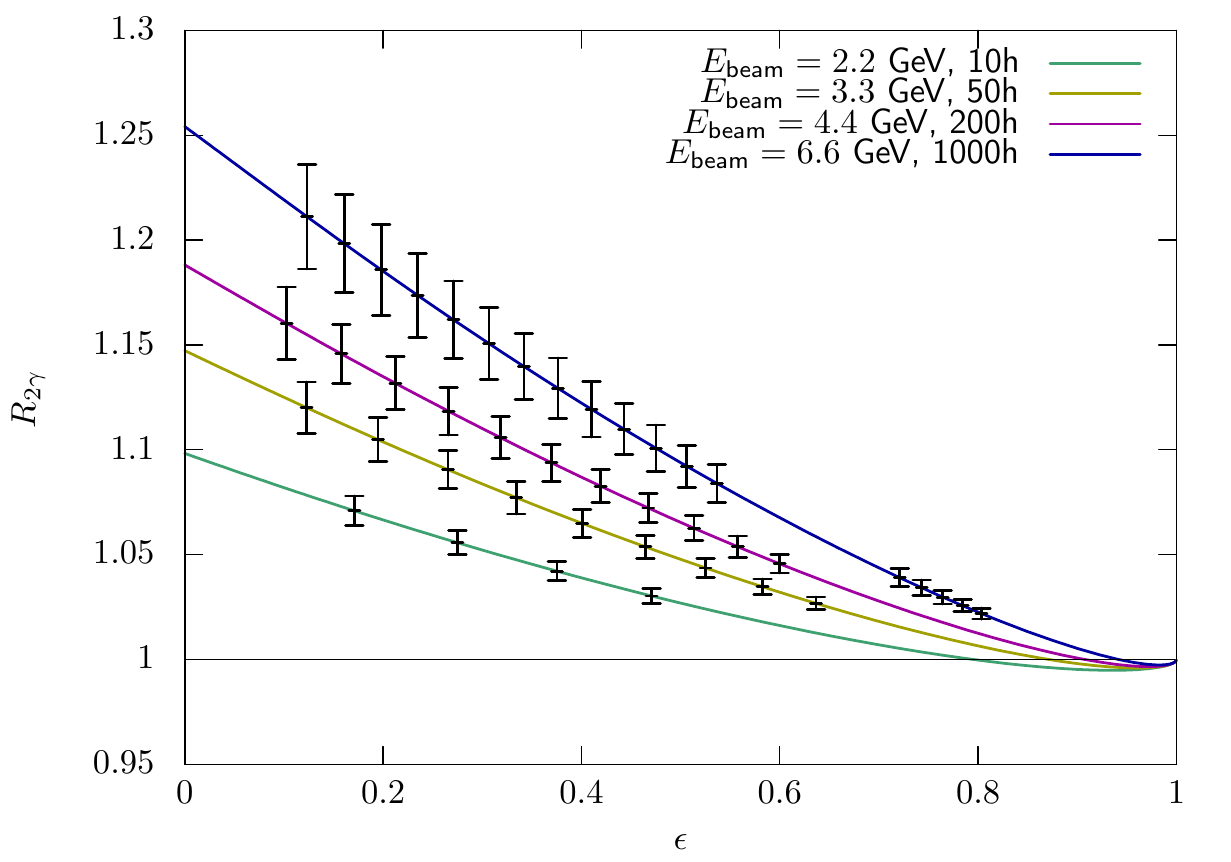}}
\caption{\label{figerrclas}Predicted effect size and estimated errors for the proposed measurement program at {\tt CLAS12}. We assume bins of constant $\Delta Q^2$=$0.25$~{GeV}$^2$.}
\end{figure}

For the rate estimates and the kinematical coverage we have made a number of assumptions that are not overly stringent: 
\renewcommand{\labelenumi}{\it\roman{enumi})}
\begin{enumerate}
\item Positron beam currents (unpolarized), $I_{e^+} \approx 60$~nA;  
\item Beam profile, $\sigma_x,~\sigma_y < 0.4$~mm;  
\item Polarization not required, so phase space at the source maybe chosen for optimized yield and beam parameters;
\item Operate experiment with 5~cm liquid H$_2$ target and luminosity of $0.8 \times 10^{35}$~cm$^{-2} \cdot$sec$^{-1}$;
\item Use the {\tt CLAS12} Central Detector for lepton ($e^+/e^-$) detection at $\Theta_l$=$40^\circ$-$125^\circ$;
\item Use {\tt CLAS12} Forward Detector for proton detection at $\Theta_p$=$7^\circ$-$35^\circ$.
\end{enumerate}   
We propose to take data at beam energies of 2.2, 3.3, 4.4 and 6.6 GeV, for 10~h, 50~h, 200~h and 1000~h respectively, split 1:1 in electron and positron running. The expected statistical errors, together with the expected effect size (phenomenological extraction from~\cite{Ber14}) are shown in Fig.~\ref{figerrclas}. The quality of the measured data will quantify hard two-photon-exchange over the whole region of precisely measured and to-be-measured cross section data, enabling a model-free extraction of the form factors from those. It will test if TPE can reconcile the form factor ratio data where the discrepancy is most significantly seen, and test for the first time GPD-based calculations. 

\subsection*{Systematics of the comparison between electron and positron measurements}

The main benefit to measure both lepton species in the same setup closely together in time is the cancellation of many systematics which would affect the result if data of a new positron scattering measurement is compared to existing electron scattering data. For example, one can put tighter limits on the change of detector efficiency and acceptance changes between the two measurements if they are close together in time, or optimally, interleaved. \newline
For the ratio, only relative effects between the species types are relevant; the absolute luminosity, detector efficiency, etc. cancel. Of special concern here is the luminosity. While an absolute luminosity is not needed, a precise determination of the species-relative luminosity is crucial. Precise relative measurement methods, for example based on M\o ller scattering, exist, but only work when the species is not changed. Switching to Bhabha scattering for the positron case and comparing with M\o ller scattering is essentially as challenging as an absolute measurement. More suitable is a measurement of the lepton-proton cross section itself at extreme forward angles, i.e., $\varepsilon\approx1$, where TPE should be negligible and the cross section is the same for both species.  To make use of these cancellations, it is paramount that the species switch-over can happen in a reasonable short time frame ($<1$~day) to keep the accelerator and detector setup stable. For the higher beam energies, where the measurement time is longer, it would be ideal if the species could be switched several times during the data taking period. To keep the beam properties as similar as possible, the electron beam should not be generated by the usual high quality source, but employ the same process as the positrons. 

The primary means of normalization for low current experiments in Hall B is the totally absorbing Faraday cup (FC) in the Hall B beam line. The absolute accuracy of the FC is better than 0.5\% for currents of 5~nA or greater. The FC can be used in $e^+$/$e^-$ beams with up to 500 W, which should not be a limitation for experiments in Hall B with {\tt CLAS12}. The relative accuracy for the ratio of electrons to positrons should be at least as good as the absolute accuracy. The only known difference between electrons and positrons is the interaction of $e^+$ and $e^-$ with the vacuum window at the entrance to the FC, which is a source of M\o ller  scattering for electrons and a source of Bhabha scattering for positrons. The FC design contains a strong permanent magnet inside the vacuum volume and just after the window. This magnet is meant to trap (most of) the low-energy M\o ller electrons to avoid over-counting the electric charge. It will also trap (most of) the Bhabha scattered electrons from the positron beam to avoid under-counting (for positrons) the electric charge. However, there may be a remaining, likely small charge asymmetry for M\o ller and Bhabha scattered electrons in the response of the FC to the different charged beams. This effect will be studied in detail with a GEANT4 simulation. 

\subsection{Direct $e^+$-$p$/$e^-$-$p$ comparisons in Hall A and C}

We also examined the possibilities for elastic measurements using the spectrometers in Halls A and C. The main kinematic considerations are the limited momentum reach of the spectrometers in Hall A and the limited angular range for the SHMS in Hall C. The SHMS in Hall C is limited to forward angles, but could be used to detect the protons instead of the leptons, providing measurements at low $\varepsilon$ with the benefit of different systematical uncertainties. BigBite in Hall A is limited in the maximum momentum. However, the large acceptance allows measurements at very low values of $\varepsilon$ with excellent precision. 

\begin{figure}[t!]
\centerline{\includegraphics[width=0.495\textwidth]{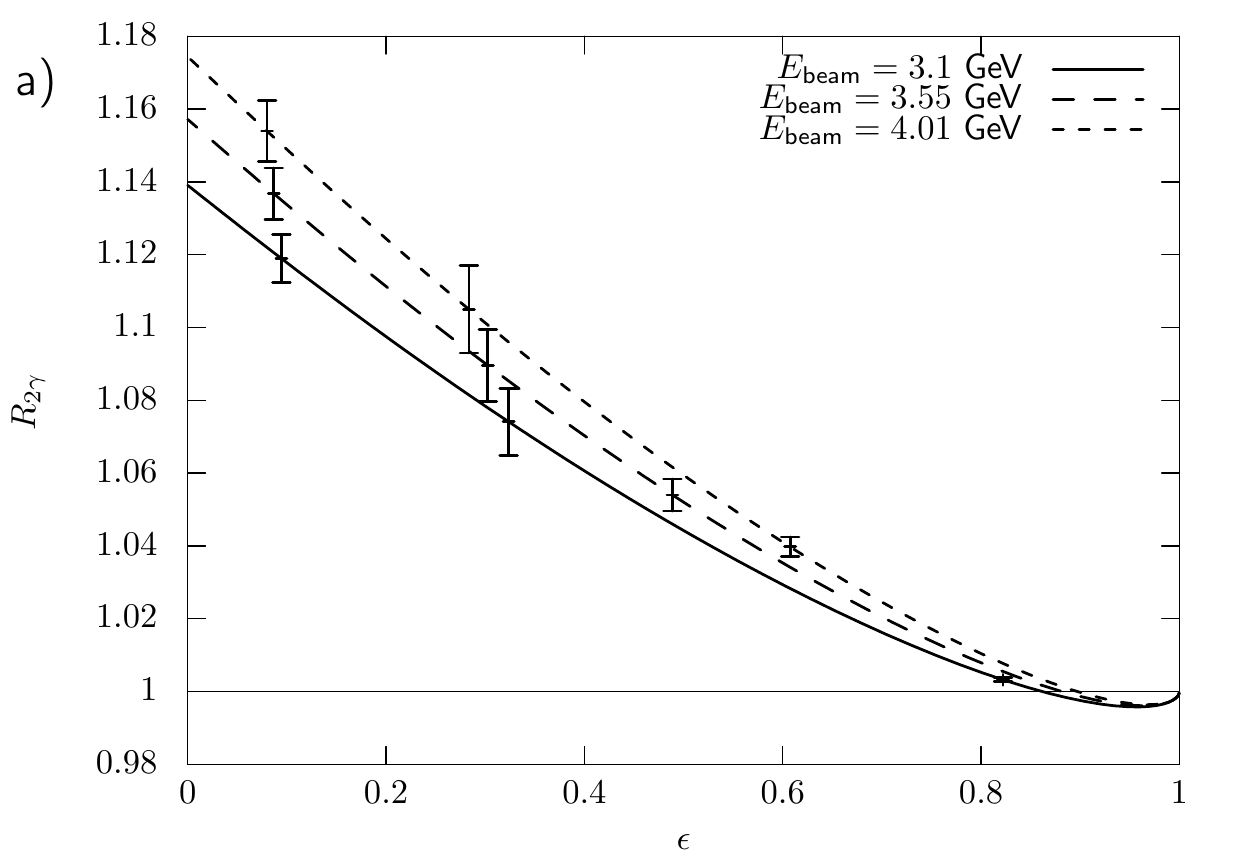}\includegraphics[width=0.495\textwidth]{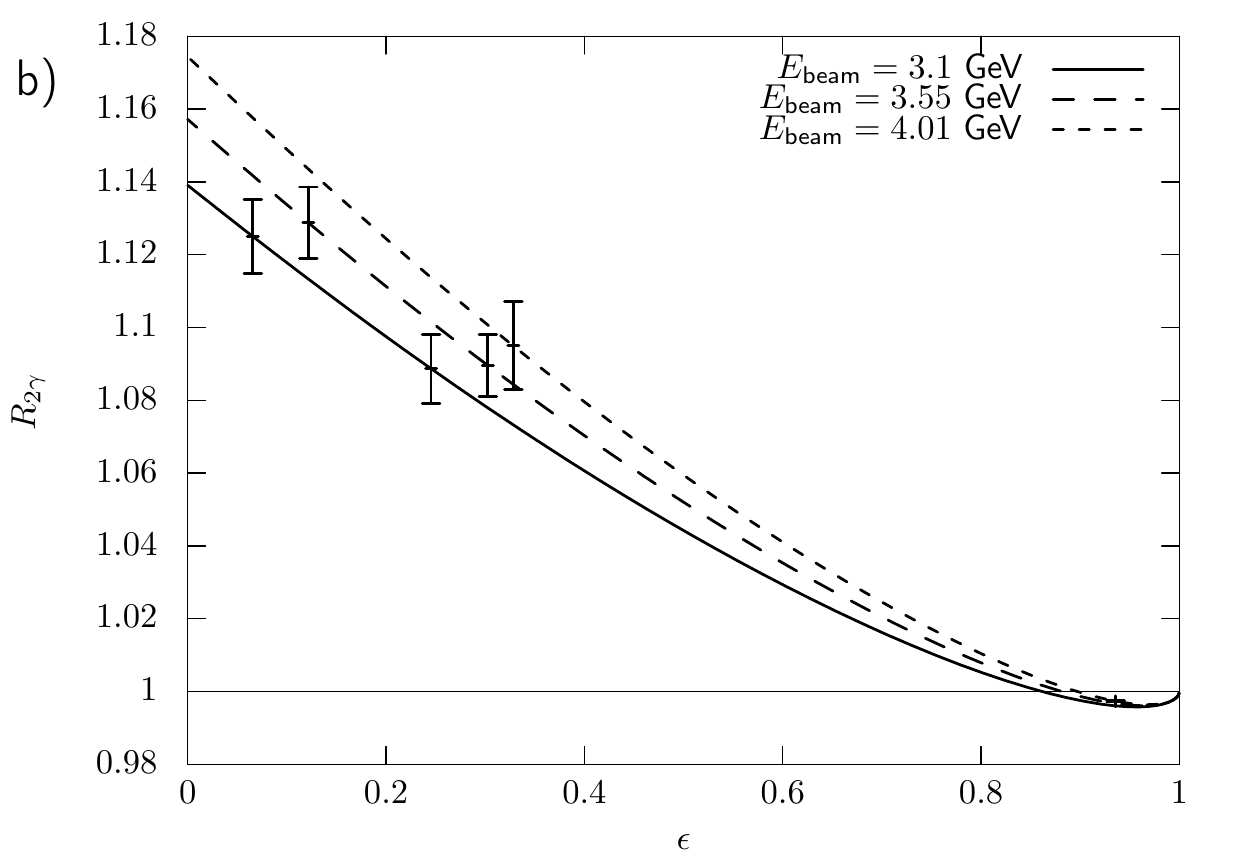}}
\caption{\label{figerrjlab}Predicted effect size and estimated errors for the proposed measurement program in Hall A (a) and Hall C (b).}
\end{figure}

\begin{table}[h!]
\begin{center}
\tabcolsep3pt
\small{
\begin{tabular}{l||ccc|ccc|ccc} \hline
$E_\mathrm{beam}$ (GeV) & \multicolumn{3}{c|}{3.10}  & \multicolumn{3}{c|}{3.55} & \multicolumn{3}{c}{4.01} \\
Spectrometer angles ($^\circ$) & \footnotesize{30} & \footnotesize{70} & \footnotesize{110} & \footnotesize{52.7} & \footnotesize{70} & \footnotesize{110} & \footnotesize{42.55} & \footnotesize{70} & \footnotesize{110} \\
$Q^2$ [$(\mathrm{GeV}/c)^2)$] & \footnotesize{1.79} & \footnotesize{3.99} & \footnotesize{4.75} & \footnotesize{3.99} & \footnotesize{4.75} & \footnotesize{5.56} & \footnotesize{3.99} & \footnotesize{5.55} & \footnotesize{6.4} \\
$\epsilon$ & \footnotesize{0.82} & \footnotesize{0.32} & \footnotesize{0.1} & \footnotesize{0.49} & \footnotesize{0.30} & \footnotesize{0.09} & \footnotesize{0.60} & \footnotesize{0.28} & \footnotesize{0.08} \\ \hline
Time [days/species] & \multicolumn{3}{c}{1}& \multicolumn{3}{c}{2}& \multicolumn{3}{c}{3}\\ \hline
\end{tabular}}
\caption{Proposed measurement program for Hall A. Angle values correspond, in order, to the central angles of the two main spectrometers and the central angle of  BigBite.} \label{tabhalla}
\end{center}
\end{table}
\begin{table}[h!]
\begin{center}
\tabcolsep3pt
\small{
\begin{tabular}{l||cc|cc|cc}
$E_\mathrm{beam}$ (GeV) & \multicolumn{2}{c|}{3.1}  & \multicolumn{2}{c|}{3.55}  & \multicolumn{2}{c|}{4.01} \\ \hline
Spectrometer angles ($^\circ$) & \footnotesize{79.7} & \footnotesize{7.64 (120)} & \footnotesize{70} & \footnotesize{9.95 (100)} & \footnotesize{18} & \footnotesize{16.57 (65)}  \\
$Q^2$ [$(\mathrm{GeV}/c)^2)$]  & \footnotesize{4.25} & \footnotesize{4.84} & \footnotesize{4.76} & \footnotesize{5.43} & \footnotesize{1.3} & \footnotesize{5.35} \\
$\varepsilon$ & \footnotesize{0.244} & \footnotesize{0.06} & \footnotesize{0.302} & \footnotesize{0.122} & \footnotesize{0.935} & \footnotesize{0.33} \\ \hline
Time [days/species] & \multicolumn{2}{c}{3} & \multicolumn{2}{c}{2} & \multicolumn{2}{c}{1} \\ \hline
\end{tabular}}
\caption{Proposed measurement program for Hall C. Central angles correspond to the HMS (leptons) and the SHMS (protons) spectrometers positions, with the equivalent lepton 
angle in parenthesis.}
\label{tabhallb}
\end{center}
\end{table}
With a beam current of \SI{1}{\micro\ampere} for unpolarized positrons on a 10 cm liquid hydrogen target, one could measure at a comparatively high luminosity of $\mathcal{L}$=2.6~pb$^{-1} \cdot$s$^{-1}$. A sketch of a possible measurement program for Hall A and Hall C is listed in Tab.~\ref{tabhalla} and Tab.~\ref{tabhallb}, respectively. While these measurements could provide precise measurements over a range of $\varepsilon$ values in a short run period, they cover a limited range of beam energies. Because they suffer from the same beam-related systematics, they would benefit from rapid change-over between positrons and electrons, as well as an independent small-angle luminosity monitor to provide checks on the luminosity of the electron and positron beams. Figure~\ref{figerrjlab} show the estimated errors and predicted effect size for Hall A (a) and Hall C (b). A high-impact measurement is possible with a comparatively small amount of beam time. Even in the case the final positron beam current is lower than assumed here, the experiment remains feasible.

\subsection{``Super-Rosenbluth'' measurements with positrons}

Both the modified {\tt CLAS12} and Hall A/C projected results shown above are direct extractions of $R_{2\gamma}$. The {\tt CLAS12} coincident detection make clean identification of elastic events at large $Q^2$ values possible even with modest kinematic resolution, and the Hall A/C measurements rely on the good resolution of the spectrometers to allow clean identification of elastic scattering from inclusive measurements. The drawback is that a direct comparison of electron and positron scattering is sensitive to differences between the electron and positron beams, as well as any time-dependent efficiency drifts if the time to change between electron and positron beams is long. These issues can be avoided by performing precise Rosenbluth separations with positrons, for direct comparison to form factors extracted using electrons. With sufficient precision, this can provide a very sensitive probe of TPE corrections, free from uncertainties associated with the different conditions of the positron and electron beams.

The so-called ``Super-Rosenbluth'' technique, involving only the detection of the struck proton, was used by JLab experiments E01-001 and E05-017 to provide a more precise Rosenbluth extraction of the ratio $G_E/G_M$ for comparison to precise polarization measurements~\cite{Qat05}. The improved precision comes from the fact that $G_E/G_M$ is independent of systematic effects that yield an overall renormalization of the measurements at a fixed $Q^2$, combined with the fact that many of the experimental conditions are unchanged when detecting $ep$ scattering at fixed $Q^2$ over a range of $\varepsilon$ values. The proton momentum is fixed, and so momentum-dependent corrections drop out in the extraction of $G_E/G_M$. In addition, the cross section dependence on $\varepsilon$ is dramatically reduced when detecting the proton, while the sensitivity to knowledge of the beam energy, spectrometer momentum, and spectrometer angle is also reduced. Finally, the large, $\varepsilon$-dependent radiative corrections also have reduced $\varepsilon$ dependence for proton detection. 

These advantages are also beneficial in making precise comparisons of electron and positron scattering. Because most of the systematic uncertainties cancel when looking at the $\varepsilon$ dependence with electrons (or positrons), the measurement does not rely on rapid change of the beam polarity, or on a precise cross normalization or comparison of conditions for electron and positron running. Because extensive data were taken using this technique with electrons during the 6 GeV era, we would propose to use only positrons and extract $G_E/G_M$ which depends only on the \textit{relative} positron cross sections as a function of $\varepsilon$. If rapid changes in the beam polarity are possible, then this approach would allow direct comparison of the cross sections with the advantage that the acceptance is unchanged, while electron detection would require a change of polarity for the Hall A/C measurements, and the overall coincidence acceptance is modified for the {\tt CLAS12} measurements. However, for this letter we assume that we would take only positron data for comparison to the existing E01-001 and E05-017 data sets. This approach can give a sensitive comparison of electron- and positron-proton scattering, with minimal systematic uncertainties and no need to cross-normalize electron and proton measurements. It does not provide direct comparisons of the $e^+p/e^-p$ cross section ratio, but does provide a direct and precise comparison of the $\varepsilon$ dependence of the elastic cross section, for which the $G_E$ contribution is identical for positrons and electrons, and the TPE contribution changes sign. 

\begin{figure}[t!]
\centerline{\includegraphics[width=0.495\textwidth]{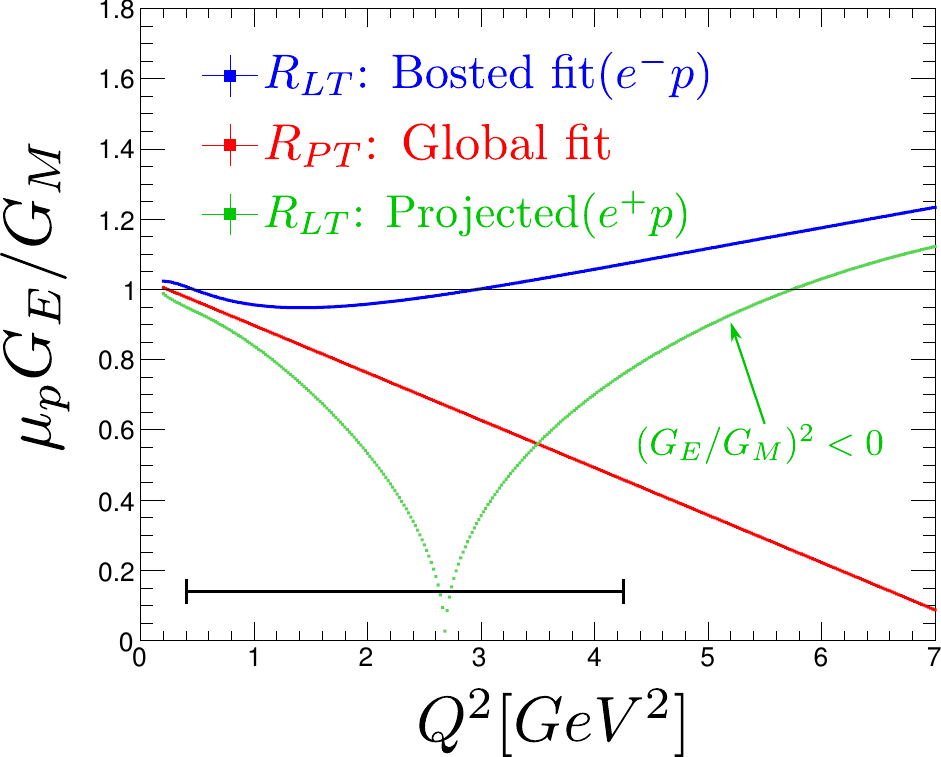}\includegraphics[width=0.495\textwidth]{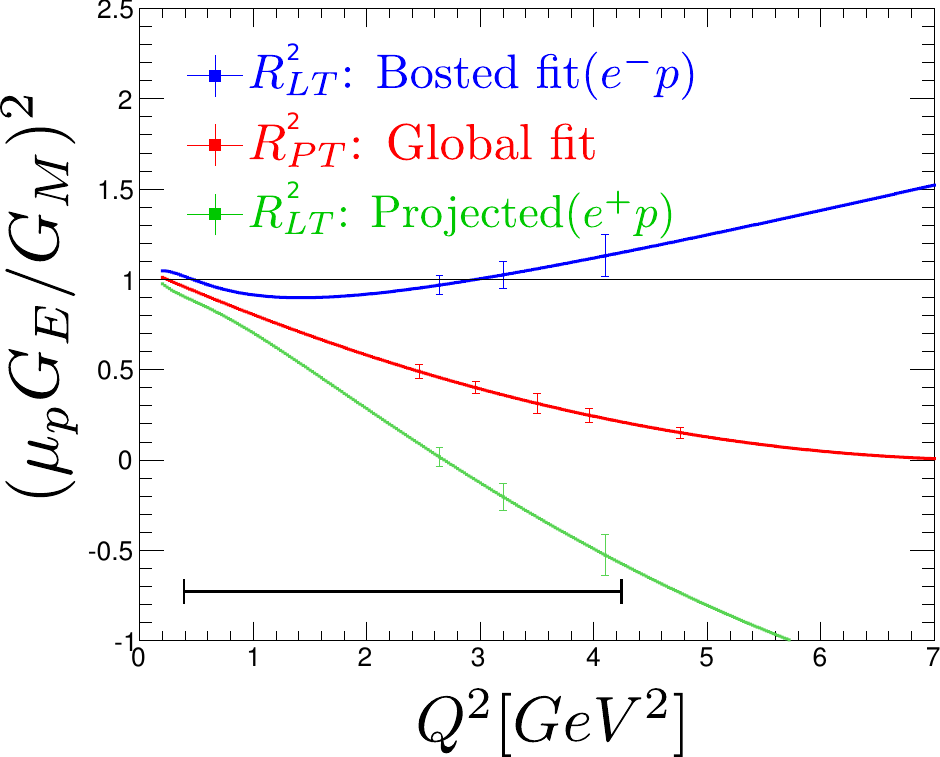}}
\caption{\label{SRproj} Parameterizations of $R=\mu_pG_E/G_M$ (left) and $R^2$ (right) from LT and polarization data, along with the results expected for positrons assuming that TPE corrections fully explain the LT-Polarization discrepancy. The right figure indicates the $Q^2$ range that could be covered under the assumptions provided in the text, and the point for the electron and positron $R^2_{LT}$ results indicate the uncertainties from the previous Hall A Super-Rosenbluth extraction~\cite{Qat05}.}
\end{figure}
The general measurements would be identical to the E05-017 experiment, with the exception of using a low intensity positron beam rather than the 30-80~$\mu$A electron beam. Assuming a 1~$\mu$A positron beam and the 4~cm LH$_2$ target used in E05-017, an 18 days run could provide measurements with sub-percent statistical uncertainties from 0.4-4.2~GeV$^2$, yielding total uncertainties comparable to the electron beam measurements. This could be extended to $>$5~GeV$^2$ with the use of a 10~cm target, or if higher  beam currents are available. Figure~\ref{SRproj} shows projections for positron Super-Rosenbluth measurements under the assumption that the discrepancy between Rosenbluth and polarization extractions is fully explained by TPE contributions. It has been shown~\cite{Arr07} that the extraction of the high-$Q^2$ form factors is not limited by our  understanding of the TPE contributions, as long as the assumption that the Rosenbluth-Polarization discrepancy is explained entirely by TPE contributions. The propose measurement would test this assumption, and also provide improved sensitivity to the overall size of the linear TPE contribution that appears as a false contribution to $G_E$ when TPE contributions are neglected. The measurement is also sensitive to non-linear contributions~\cite{Tva06} coming from TPE, and would provide improved sensitivity compared to existing electron measurements. More details are provided in Ref.~\cite{Yur17}.

\subsection{Summary}

Despite recent measurements of the $e^+p/e^-p$ cross section ratio, the proton's form factor discrepancy has not been conclusively resolved, and new measurements at higher momentum transfer are needed. With a positron source at CEBAF, the enormous capabilities of the {\tt CLAS12} spectrometer could be brought to bear on this problem and provide a wealth of new data over a widely important kinematic range. Only one major detector configuration change would be necessary to support such a measurement, the installation of the central electromagnetic calorimeter. In designing the JLab positron source, it will be crucial for this and several other experiments to keep to a minimum the time necessary to switch between electron and positron modes, in order to reduce systematic effects.

Another option, utilizing the Super-Rosenbluth technique, would allow for precise LT separations using only positron beams. This is a sensitive test of TPE contributions that does not require the rapid changeover between positrons and electrons, but it does not directly compare positron and electron scattering at fixed kinematics. Instead, it measures the impact of TPE on the Rosenbluth extraction of $\mu_p G_E/G_M$ with high precision.

The data that the proposed experiments could provide will be able to map out the transition between the regions of validity for hadronic and partonic models of hard TPE, and make definitive statements about the nature of the proton form factor discrepancy.

%
%

\newpage

\null\vfill

\begin{center}

\section{\it Letter-ot-Intent: TPE @ SBS}

\vspace*{15pt}

{\Large{\bf A measurement of polarization transfer}}

\vspace*{3pt}

{\Large{\bf in elastic polarized positron-proton scattering}}

\vspace*{15pt}

{\bf Abstract}

\begin{minipage}[c]{0.85\textwidth}
Hard two-photon exchange, the only sub-leading radiative effect that is not included in standard radiative corrections prescriptions, may be responsible for the discrepancy between polarized and unpolarized measurements of the proton's form factors. Since calculations of hard TPE are necessarily model-dependent, measurements of observables with direct sensitivity to hard TPE are needed. Much experimental attention has been focused on the unpolarized $e^+p/e^-p$ cross section ratio, but polarization transfer in polarized elastic scattering can also reveal evidence of hard TPE. Furthermore, it has a different sensitivity to the generalized TPE form factors, meaning that measurements provide new information that cannot be gleaned from unpolarized scattering alone. Both $\epsilon$-dependence of polarization transfer at fixed $Q^2$, and deviations between electron- and positron-scattering are key signatures of hard TPE. A polarized positron beam at Jefferson Lab would present a unique opportunity to make the first measurement of positron polarization transfer, and comparison with electron-scattering data would place valuable  constraints on hard TPE. In this letter, we propose a measurement program in Hall A that combines the Super BigBite Spectrometer for measuring recoil proton polarization, with a non-magnetic calorimetric detector for triggering on elastically scattered positrons. Though the reduced beam current of the positron beam will restrict the kinematic reach, this measurement will have very small systematic uncertainties, making it a clean probe of TPE.
\end{minipage}

\vspace*{15pt} 

{\it Spokespersons: \underline{A.~Schmidt} (schmidta@mit.edu), J.~Bernauer, A.~Puckett}

\end{center}

\vfill\eject

%
%

\subsection{Introduction}

The discrepancy between the ratio $\mu_p G_E / G_M$ of the the proton's electromagnetic form factor extracted from polarization asymmetry measurements, and the ratio extracted from unpolarized cross section measurements, leaves the field of form factor physics in an uncomfortable state (see \cite{Afa17} for a recent review). On the one hand, there is a consistent and viable hypothesis that the discrepancy is caused by non-negligible hard two-photon exchange (TPE) \cite{Gui03, Blu03}, the one radiative correction omitted from
the standard radiative correction prescriptions \cite{Mo69, Max00}. On the other hand, three recent measurements of hard TPE (at VEPP-3, at CLAS, and with OLYMPUS) found that the effect of TPE is small in the region of $Q^2 < 2$~GeV$^2/c^2$ \cite{Rac15, Adi15, Rim17, Hen17}. The TPE hypothesis is still viable; it is possible that hard TPE contributes more substantially at higher momentum transfers, and can fully resolve the form factor discrepancy. But the lack of a definitive conclusion from this recent set of measurements is an indication that alternative approaches are needed to illuminate the situation, and it may be prudent to concentrate experimental effort on constraining and validating model-dependent theoretical calculations of TPE. There are multiple theoretical approaches, with different assumptions and different regimes of validity~\cite{Che04, Afa05, Tom15, Blu17, Kur08}. If new experimental data could validate and solidify confidence in one or more theoretical approaches, then hard TPE could be treated in the future like any of the other standard radiative corrections, i.e., a correction that is calculated, applied, and trusted.

VEPP-3, CLAS, and OLYMPUS all looked for hard TPE through measurements of the $e^+p$ to $e^-p$ elastic scattering cross section ratio. After applying radiative corrections, any deviation in this ratio from unity indicates a contribution from hard TPE. However, this is not the only experimental signature one could use. Hard TPE can also appear in a number of polarization asymmetries. Having constraints from many orthogonal directions, i.e., from both cross section ratios and various polarization asymmetries would be valuable for testing and validating theories of hard TPE. As with unpolarized cross sections, seeing an opposite effect for electrons and positrons is a clear signature of TPE.

In this letter, we propose one such polarization measurement, that could both be feasibly accomplished with a positron beam at Jefferson Lab, and contribute new information about two photon exchange that could be used to constrain theoretical models. We propose to measure the polarization transfer (PT) from a polarized proton beam scattering elastically from a proton target, for which no data currently exist. The proposed experiments uses a combination of the future Hall A Super Big-Bite Spectrometer (SBS) to measure the polarization of recoiling protons, along with a calorimetric detector for detecting scattered positrons in coincidence. In the following sections, we review polarization transfer, sketch the proposed measurement, and discuss possible systematic uncertainties. 

\subsection{Polarization transfer}

In the Born approximation (i.e.~one-photon exchange), the polarization transferred from a polarized lepton to the recoiling proton is
\begin{align}
    P_t &= -hP_e\sqrt{\frac{2\epsilon (1-\epsilon)}{\tau}} \frac{G_E G_M}{G_M^2 + \frac{\epsilon}{\tau}G_E^2},\\
    P_l &= hP_e\sqrt{1 - \epsilon^2} \frac{G_M^2}{G_M^2 + \frac{\epsilon}{\tau}G_E^2},
\end{align}
where $P_t$ is the polarization transverse to the momentum transfer 3-vector (in the reaction plane), $P_l$ is the longitudinal polarization, $P_e$ is the initial lepton polarization, $h$ is the lepton helicity, $\tau$ is the dimensionless 4-momentum transfer squared (Sec:~\ref{sec:elscat}), $\epsilon$ is the virtual photon polarization parameter (Eq.~\ref{eq:epsi}), and $G_E$ and $G_M$ are the proton's electromagnetic form factors. The strength of the polarization transfer technique is to measure $P_t/P_l$, thereby cancelling some systematics associated with polarimetry, and isolating the ratio of the proton's form factors:
\begin{equation}
\frac{P_t}{P_l}= -\sqrt{\frac{2\epsilon}{\tau(1+\epsilon)}}\frac{G_E}{G_M}.
\end{equation}
This technique has several advantages over the traditional Rosenbluth separation technique for determining form factors. This polarization ratio can be measured at a single kinematic setting, avoiding the systematics associated with comparing data taken from different spectrometer settings. This technique allows the sign of the form factors to be determined, rather than simply their magnitudes. And furthermore, whereas the sensitivity in Rosenbluth separation to $G_E^2$ diminishes at large momentum transfer, polarization transfer retains sensitivity to $G_E$ even when $Q^2$ becomes large. When used in combination at high $Q^2$, Rosenbluth separation can determine $G_M^2$, while polarization transfer can determine $G_E/G_M$, allowing the form factors to be separately determined.

Polarization transfer using electron scattering has been used extensively to map out the proton's form factor ratio over a wide-range of $Q^2$, with experiments conducted at MIT Bates~\cite{Mil98}, Mainz~\cite{Pos01}, and Jefferson Lab~\cite{Gay01, Mac06, Ron11, Pao10, Zha11}, including three experiments, GEp-I~\cite{Jon00, Pun05}, GEp-II~\cite{Gay02}, and GEp-III~\cite{Puc10} that pushed to high momentum transfer. Another experiment, GEp-2$\gamma$, looked for hints of TPE in the $\epsilon$-dependence in polarization transfer \cite{Mez11, Puc17}. Two other experiments made equivalent measurements by polarizing the proton target instead of measuring recoil polarization \cite{Jon06,Cra07}.

While polarization transfer is less sensitive to the effects of hard TPE, it is not immune. Following the formalism of Ref. \cite{Car07}, one finds that
\begin{multline}
\frac{P_t}{P_l} = \sqrt{\frac{2\epsilon}{\tau (1+\epsilon)}}\frac{G_E}{G_M} \times \Bigg[ 1+
  \textrm{Re}\left(\frac{\delta \widetilde{G}_M}{G_M}\right) + \frac{1}{G_E}\textrm{Re}\left(\delta\widetilde{G}_E + \frac{\nu}{M^2}\widetilde{F}_3\right) \\
    - \frac{2}{G_M}\textrm{Re}\left(\delta\widetilde{G}_M + \frac{\epsilon\nu}{(1+\epsilon)M^2} \widetilde{F}_3\right) + \mathcal{O}(\alpha^4) \Bigg],
\end{multline}
with $\nu \equiv (p_e + p_{e'})_\mu (p_p + p_{p'})^\mu$, and where $\delta \widetilde{G}_E$,  $\delta \widetilde{G}_M$, and $\delta \widetilde{F}_3$ are additional form factors that become non-zero when moving beyond the one-photon exchange approximation. This particular dependence on new form factors is slightly different than one what finds when
taking a positron to electron cross section ratio:
\begin{equation}
\frac{\sigma_{e^+p}}{\sigma_{e^-p}} = 1 + 4G_M\textrm{Re}\left(\delta \widetilde{G}_M + \frac{\epsilon \nu}{M^2}\widetilde{F}_3\right)
- \frac{4\epsilon}{\tau}G_E\textrm{Re}\left(\delta \widetilde{G}_E + \frac{\nu}{M^2} \widetilde{F}_3\right) + \mathcal{O}(\alpha^4).
\end{equation}
A measurement of the difference in polarization transfer between electron and positron scattering therefore adds information about TPE in addition to what can be learned from cross section ratios alone.

The GEp-2$\gamma$ experiment looked for the effects of TPE in polarization transfer by making measurements at three kinematic points with varying values of $\epsilon$, but with $Q^2$ fixed at 2.5~GeV$^2/c^2$~\cite{Mez11}. Since in the absence of hard TPE the ratio $G_E/G_M$ has no $\epsilon$-dependence, any variation with $\epsilon$ is a sign of hard TPE. The GEp-2$\gamma$ measurement was statistically consistent with no $\epsilon$-dependence, though their measurement of purely the longitudinal component showed deviations from the one-photon exchange expectation.

A measurement with positron scattering will be useful for constraining TPE effects because deviations from the Born-approximation should have the opposite sign from those in electron scattering. This helps determine if deviations are truly caused by TPE, or if they arise from systematic effects. As the largest systematic uncertainties in polarization transfer are associated with proton polarimetry, a measurement with positrons would have largely the same systematics as an experiment with electrons. 

\subsection{Proposed measurement}

The proposed experiment copies the basic approach of earlier GEp measurements at JLab. However, since these prior experiments were able to make use of the high-current polarized electron beam, and since the proposed positron source at Jefferson Lab will be limited to currents of approximately 100~nA, several improvements have to be made relative to the GEp program for a positron experiment to be feasible.

The first major improvement will be the Super Big-bite Spectrometer (SBS) \cite{Jag10}, which is currently being designed and built for the next generation of form factor measurements in Hall A \cite{Jag10-1}. Whereas previous measurements used the current HRS and HMS spectrometers limited to less than 10~msr of acceptance, SBS is designed with 70~msr of solid-angle acceptance and much larger momentum acceptance. This will allow flexibility in choosing a momentum setting that produces an optimal bend angle for elastically recoiling protons. Furthermore, the proposed single-dipole field configuration for the SBS will greatly simplify the spin-transport properties of the spectrometer, reducing systematics. The larger angular acceptance of the SBS affords another advantage: using a longer target. Where as the GEp-III and GEp-$2\gamma$ experiments used 15~cm and 20~cm liquid hydrogen targets, the SBS can accommodate a 40~cm target at the angles relevant for a positron PT measurement. With the limited positron current, there is much reduced concern with target heating and target boiling. The third advantage is the high beam energy made possible by the 12 GeV upgrade, which will allow measurements to reach the relevant high momentum transfers at angles substantially more forward, where the cross section is comparatively higher.
\begin{figure}[t]
\begin{center}
\includegraphics[width=0.75\textwidth]{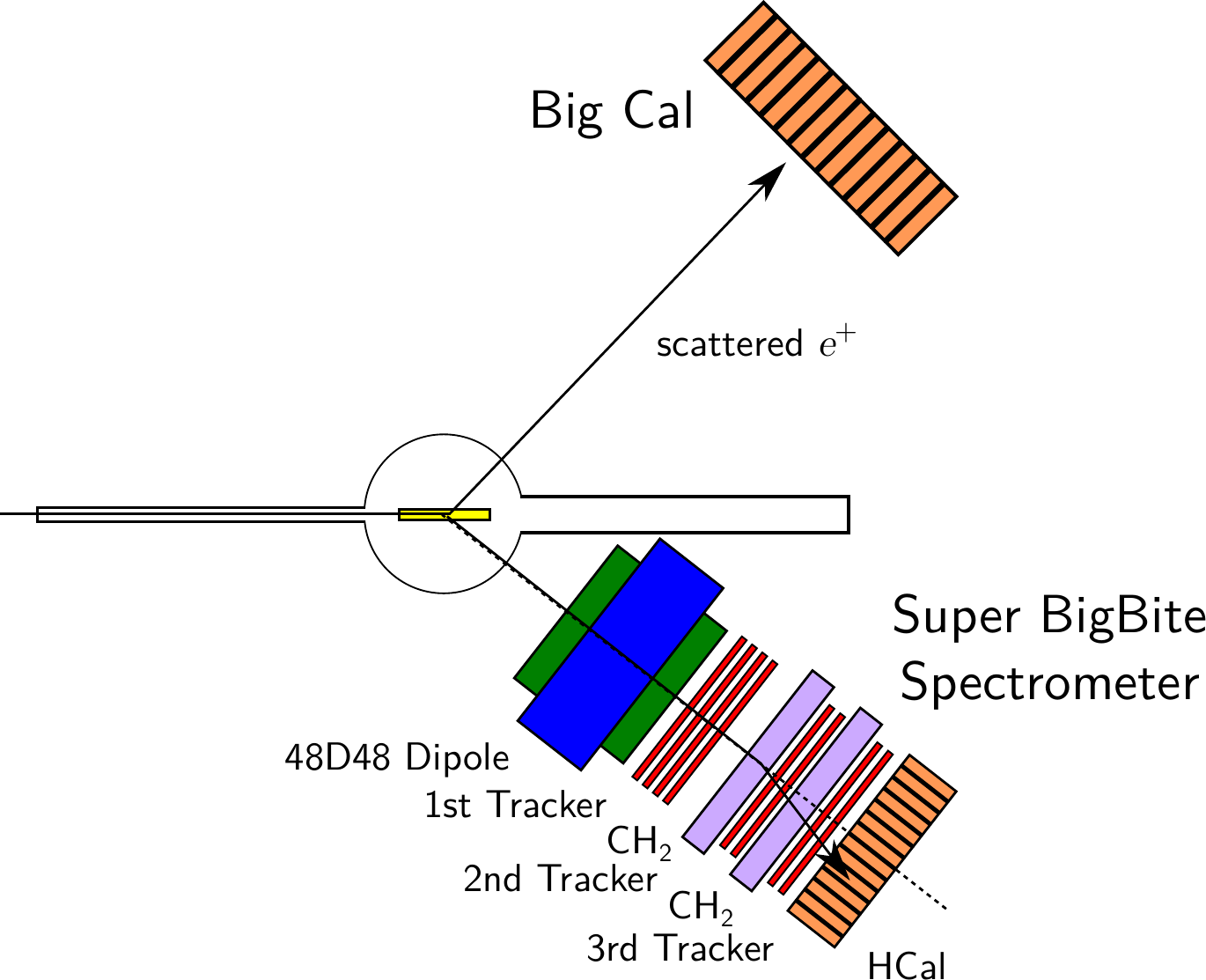}
\caption{A schematic of the proposed PT measurement.}
\label{fig:pt_schem}
\end{center}
\end{figure}

We propose a measurement set-up along the lines of the GEp-III and GEp-$2\gamma$ experiments, in which elastically scattered positrons are detected in the non-magnetic BigCal detector in coincidence with  recoiling protons being detected in the SBS, and their polarization measured by the SBS focal plane polarimeter. A schematic of the experiment is shown in Fig.~\ref{fig:pt_schem}. We have attempted to make reasonable estimates of uncertainty by scaling the achieved statistical uncertainties of the GEp-$2\gamma$ experiment. The statistical uncertainty will largely depend on the product of the magnitude of the asymmetry being measured and the achievable count-rate. That is, we assume:
\begin{equation}
\delta R \propto \left[ P_e P_p A \sqrt{ \frac{d\sigma}{d\Omega} \Omega \mathcal{L} T \varepsilon }\right]^{-1},
\end{equation}
where $P_p$ is the magnitude of the polarization transfered to the recoiling proton, $A$ is the polarimeter analyzing power, $d\sigma/d\Omega$ is the elastic cross section, $\Omega$ is the spectrometer angular acceptance, $\mathcal{L}$ is the luminosity, $T$ is the run-time, and $\varepsilon$ is the running efficiency, i.e. the live-time to wall-time ratio. Applying this assumption to the achieved uncertainties in the GEp-$2\gamma$ experiment, we find that:
\begin{equation}
    \delta R_{\text{GEp-}2\gamma}
    \approx \frac{1.2\times10^{-19} [\text{cm~sr}^{-1/2}\text{days}^{1/2}]}
    {P_p \sqrt{\frac{d\sigma}{d\Omega} T}}.
\end{equation}
Projecting to our proposed positron measurement, we assume equivalent analyzing power, and equivalent running efficiency. The luminosity will be reduced by a factor of 400 (80~$\mu$A current, 20~cm target in GEp-$2\gamma$ to 100~nA current, 40~cm target in our proposed experiment and the beam polarization reduced from $\approx80\%$ to $60\%$. However, the spectrometer acceptance will increase from 6.74~msr (Hall C HMS) to 70~msr (SBS). All of these factors combine to yield an uncertainty projection of:
\begin{equation}
    \delta R_\text{proposal} 
    \approx \frac{9.9\times10^{-19} [\text{cm~sr}^{-1/2}\text{days}^{1/2}]}
    {P_p \sqrt{\frac{d\sigma}{d\Omega} T}}.
\end{equation}

\begin{figure}[t!]
    \centering
    \includegraphics{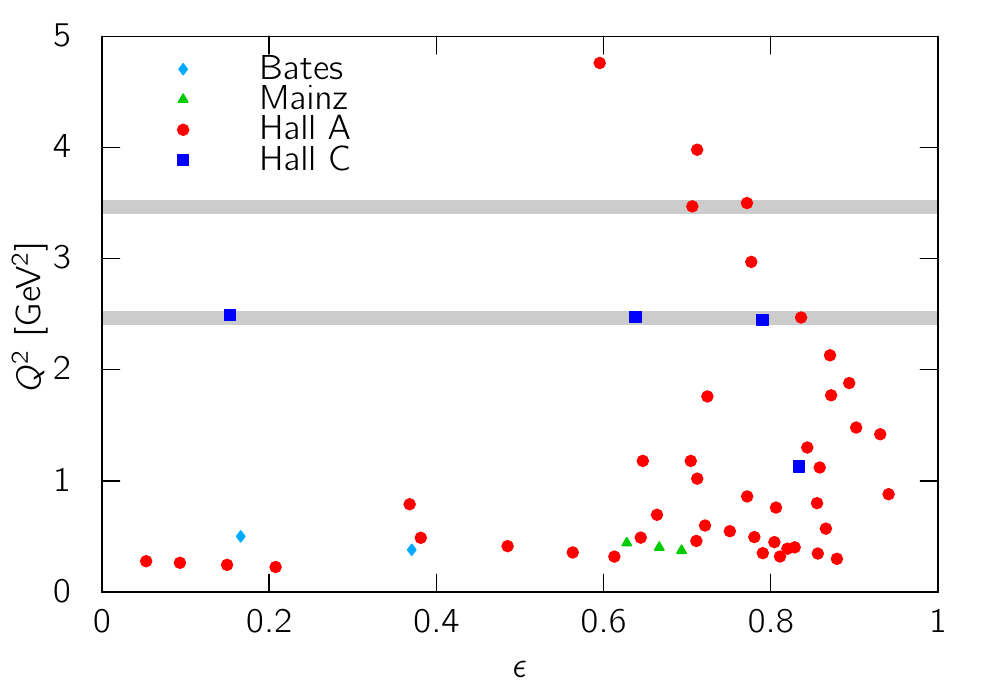}
    \caption{The kinematics of previous polarization transfer measurements with electron beams are shown. A measurement with positrons at either $Q^2=2.5$~GeV$^2$ or 3.5~GeV$^2$ would be able to compare with previous electron experiments.}
    \label{fig:target_kin}
\end{figure}
The effects of TPE will have opposite sign in electron scattering experiments relative to positron scattering experiments, and so it would be prudent for the first PT measurement with a positron beam to measure at a $Q^2$ that has already been measured with electrons. The kinematics of previous PT measurements, all with electron beams, are shown in Fig.~\ref{fig:target_kin}. We highlight $Q^2=2.5$~GeV$^2$ and $3.5$~GeV$^2$ for our proposed measurement. At these momentum transfers, the proton form factor discrepancy is significant, and both hadronic and partonic calculations of TPE are feasible. 

\begin{figure*}[t!]
\centering
\includegraphics[width=0.665\textwidth]{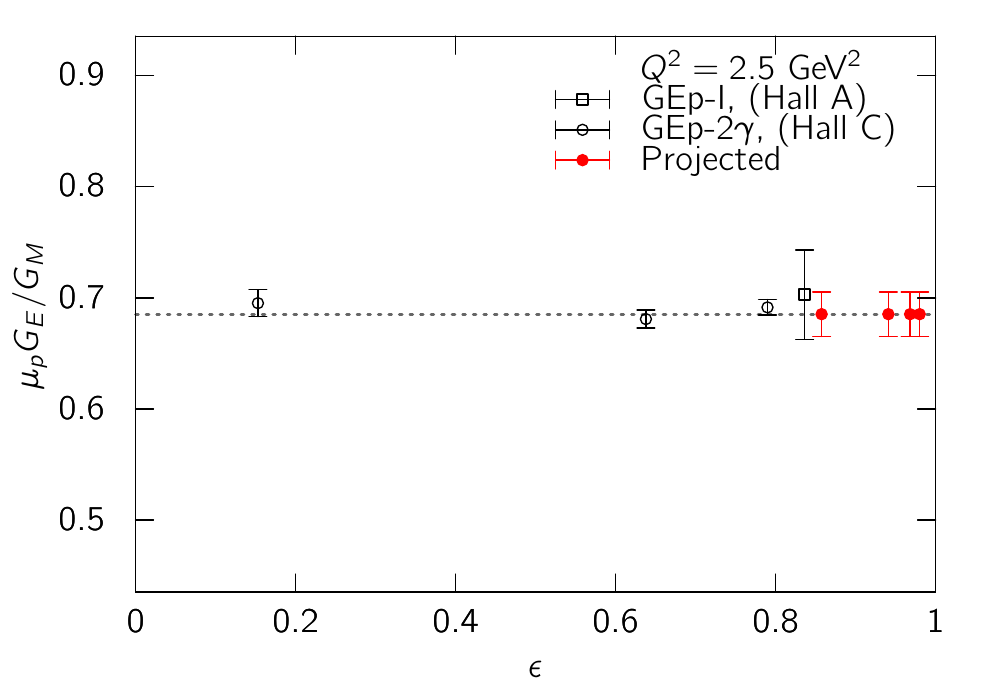} \\
\includegraphics[width=0.665\textwidth]{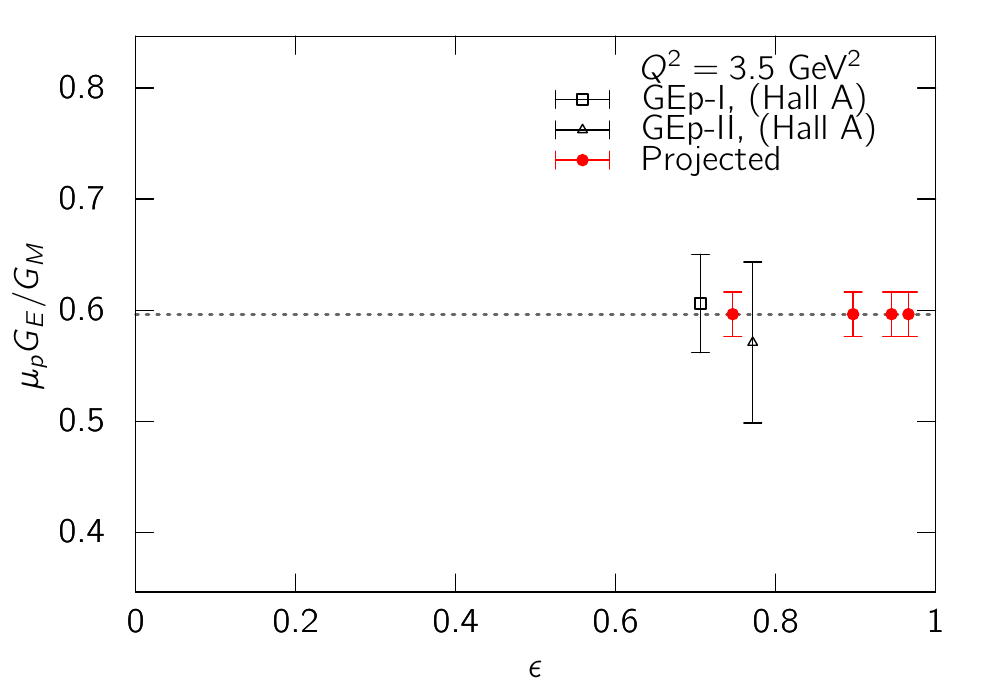}
\caption{Previous polarization transfer data taken with electrons (black) is shown as a function of $\epsilon$ in comparison to the kinematics of the projected positron measurement (red) with the target 2\% statistical uncertainties.}
\label{fig:errors}
\end{figure*}

For a competitive first measurement, we believe 2\% statistical uncertainty is a reasonable goal. Tables \ref{tab:kin250} and \ref{tab:kin350} show the kinematics for these values of momentum transfer as well as the number of measurement days that would be necessary to achieve the 2\% statistical uncertainty goal. For example, a 55~day measurement period could cover $Q^2=2.5$~GeV$^2$ at 2$^\text{nd}$ pass, 3$^\text{rd}$ pass and 5$^\text{th}$ pass, as well as $Q^2=3.5$~GeV$^2$ at 5$^\text{th}$ pass, with 48 hours available for pass and configuration changes. The accessible kinematic data points at $Q^2 = 2.5$~GeV$^2$ and 3.5~GeV$^2$ are shown in Fig.~\ref{fig:errors} along with previous PT data taken with electrons. 

{
\begin{table}[h!]
    \centering
    \begin{tabular}{c c c c c c c c c}
    \hline
    Pass & $E_{e^+}$ & $\epsilon$ & $\theta_{e^+}$ [$^\circ$] & $p_{e^+}$ & $\theta_p$ [$^\circ$] & $p_p$ & Days to 2\% \\
    \hline\hline
2$^\text{nd}$ & 4.4 & 0.858 & 24.9 & 3.07 & 38.6 & 2.07 & 12.3 \\
3$^\text{rd}$ & 6.6 & 0.941 & 15.4 & 5.27 & 42.6 & 2.07 & 9.0 \\
4$^\text{th}$ & 8.8 & 0.968 & 11.2 & 7.47 & 44.5 & 2.07 & 7.9 \\
5$^\text{th}$ & 11.0 & 0.980 & 8.8 & 9.67 & 45.6 & 2.07 & 7.3 \\
    \hline
    \end{tabular}
    \caption{Kinematics for measurements at $Q^2=2.5$~GeV$^2$, all energies and momenta in units of GeV ($c=1$).}
    \label{tab:kin250}
\end{table}
\begin{table}[h!]
    \centering
    \begin{tabular}{c c c c c c c c c}
    \hline
    Pass & $E_{e^+}$ & $\epsilon$ & $\theta_{e^+}$ [$^\circ$] & $p_{e^+}$ & $\theta_p$ [$^\circ$] & $p_p$ & Days to 2\% \\
    \hline\hline
2$^\text{nd}$ & 4.4 & 0.747 & 32.5 & 2.53 & 31.1 & 2.64 & 56.7 \\
3$^\text{rd}$ & 6.6 & 0.897 & 19.3 & 4.73 & 36.3 & 2.64 & 33.8 \\
4$^\text{th}$ & 8.8 & 0.945 & 13.8 & 6.93 & 38.6 & 2.64 & 27.3 \\
5$^\text{th}$ & 11.0 & 0.966 & 10.7 & 9.13 & 40.0 & 2.64 & 24.4 \\
    \hline
    \end{tabular}
    \caption{Kinematics for measurements at $Q^2=3.5$~GeV$^2$, all energies and momenta in units of GeV ($c=1$).}
    \label{tab:kin350}
\end{table}
}

The necessary run times from Tab.~\ref{tab:kin250}-\ref{tab:kin350} can be used to design a measurement  program. We propose, as an example a 55-day measurement program, with configurations listed in Tab.~\ref{tab:run}. This program would be able to cover $Q^2=2.5$~GeV$^2$ at 2nd pass, 3rd pass, and 5th pass, as well as $Q^2=3.5$~GeV$^2$ at 5th pass. This time includes the machine duty factor, but does not include time for configuration and pass changes. This program includes a small amount of time for electron running, 
which would require much less time if the traditional polarized electron source were used. Compared to a 100~nA positron beam, a conservative 40~$\mu$A electron beam would provide a factor of 20 increase in rate. This conservative factor was used in generating Tab.~\ref{tab:run}. The ordering of the runs might be further optimized to reduce the necessary time for configuration changes. For example, if the time to switch between the $e^-$ and $e^+$ modes were significantly longer than the time to switch beam energies, the positron runs at different energies could all be taken consecutively. 

\subsection{Systematic uncertainties}

The dominant sources of systematic uncertainty in the GEp campaign are associated with the proton polarimetry, meaning that measurements with electrons and positrons will have systematic offsets in the same direction. Combining electron and positron measurements therefore will not lead to more accurate determination of the proton's Born-level form factors. However, these systematic effects will cancel in determinations of TPE, making polarization transfer an extremely clean technique.

It would therefore be sensible to include, for any positron scattering measurement, a complementary electron scattering measurement. Such an electron measurement could be performed at higher beam currents (so long as target boiling are kept under control) to reduce run times. The SBS magnetic field setting could be kept at the exact same value. If fast-switching between electron and positron modes were possible, the electron and positron running should be inter-leaved further reducing time-dependent systematic effects.

Even without accompanying measurements with electrons, polarization transfer is already a systematically clean technique, and the design of the SBS may lead to further reduction in systematics. One of the leading systematic effects in the GEp-III and GEp-$2\gamma$ experiments was the knowledge of the spectrometer magnetic field, which must be known to fully understand the proton spin precession through the spectrometer. The SBS's single-dipole design will greatly simplify the proton spin-precession. Furthermore, tracking in the HMS polarimeter was complicated due to left-right ambiguities in the design of the drift chambers. The SBS polarimeter, which uses large-area GEM detectors for tracking are being designed to avoid such ambiguities. 

Because of the small systematic uncertainties involved, the uncertainties in a polarization-transfer measurement with the proposed positron source at Jefferson Lab will almost certainly be statistically dominated. 

\subsection{Summary}

In this letter, we lay out a feasible approach to measuring polarization transfer in elastic positron-proton scattering with the proposed positron source at Jefferson Lab. Such a measurement would add valuable information that could constrain calculations of two-photon exchange and would be complementary to that from measurements of the unpolarized $e^+p/e^-p$ cross section ratio. Our proposed experiment would take advantage of the upcoming Super Big-Bite Spectrometer to overcome the limitations in luminosity that would be inevitable with a positron beam. 

Several important steps must still be taken, most crucially, the successful completion and commissioning of the SBS spectrometer. The estimates laid out in this letter are based on scaling the uncertainties from previous measurements. Sophisticated simulations of a fully-realized detector will make this estimates much more concrete and trustworthy. Lastly, the proposed measurement focuses on the high-$\epsilon$ region, accessible in realistic experiment time-frames. The current best theoretical calculations of hard TPE are needed for PT at these kinematics to understand how much value such an experiment will add.
\begin{table}[h!]
\centering
\begin{tabular}{c c c c c} \hline
Species & Energy & $Q^2$ & $\epsilon$ & Days\\ \hline\hline
$e^-$ &  4.4 & 2.5 &  0.858 & 0.6  \\
$e^+$ &  4.4 & 2.5 &  0.858 & 12.2 \\
$e^-$ &  6.6 & 2.5 &  0.941 & 0.4  \\
$e^+$ &  6.6 & 2.5 &  0.941 & 9.0  \\
$e^-$ & 11.0 & 2.5 &  0.980 & 0.4  \\
$e^+$ & 11.0 & 2.5 &  0.980 & 7.3  \\
$e^-$ & 11.0 & 3.5 &  0.966 & 1.1  \\
$e^+$ & 11.0 & 3.5 &  0.966 & 24.0 \\ \hline
      &      &     & Total: & 55   \\ \hline
\end{tabular}
\caption{Beam time allotment for a 55-day experiment that includes measurements with both positrons and electrons, all energies and momenta in units of GeV ($c=1$).}
\label{tab:run}
\end{table}

%
%

\newpage

\null\vfill

\begin{center}

\section{\it Letter-ot-Intent: p-DVCS @ CLAS12}

\vspace*{15pt}

{\Large{\bf A polarized positron beam}}

\vspace*{3pt}

{\Large{\bf for DVCS on the proton with CLAS12}}

\vspace*{3pt}

{\Large{\bf at Jefferson Lab}}

\vspace*{15pt}

{\bf Abstract}

\begin{minipage}[c]{0.85\textwidth}
The measurement of Deeply Virtual Compton Scattering on the proton with a polarized positron beam in {\tt CLAS12} can give access to a complete set of observables for the extraction of Generalized Parton Distributions with the upgraded 11 GeV CEBAF. This provides a clean separation of the real and imaginary parts of the amplitudes, greatly simplifies the analysis, and provides a crucial handle on the model dependences and associated systematic uncertainties. The real part of the amplitude is in particular sensitive to the $D$-term which parameterizes the Gravitational Form Factors of the nucleon. Azimuthal dependences and $t$-dependences of the azimuthal moments for Beam Charge Asymmetries on unpolarized Hydrogen are estimated using a 1000 hours run with a luminosity of $2\times10^{34}$ cm$^{-2}\cdot$s$^{-1}$ and 80\% beam polarization.
\end{minipage}

\vspace*{15pt} 

{\it Spokespersons: \underline{V.~Burkert} (burkert@jlab.org), L.~Elouadrhiri, F.-X.~Girod}

\end{center}

\vfill\eject

%
%

\subsection{Introduction}
\label{intro}

The challenge of understanding nucleon electromagnetic structure still continues after six decades of experimental scrutiny. From the initial measurements of elastic form factors to the accurate determination of parton distributions through deep inelastic scattering, the experiments have increased in statistical and systematic precision. 
During the past two decades it was realized that the parton distribution functions represent special cases of a more general, much more powerful, way to characterize the structure of the nucleon, the generalized parton distributions (GPDs) (see~\cite{Mul94} for the original work and~\cite{Die03, Bel05} for reviews). The GPDs are the Wigner quantum phase space distribution of quarks in the nucleon describing the simultaneous distribution of particles with respect to both position and momentum in a quantum-mechanical system. In addition to the information about the spatial density and momentum density, these functions reveal the correlation of the spatial and momentum distributions, {\it i.e.} how the spatial shape of the nucleon changes when probing quarks of different momentum fraction of he nucleon. 

The concept of GPDs has led to completely new methods of ``spatial imaging'' of the nucleon in the form of (2+1)-dimensional tomographic images, with 2 spatial dimensions and 1 dimension in momentum~\cite{Bur02, Ji03, Bel04}. The second moments of GPDs are related to form factors that allow us to quantify how the orbital motion of quarks in the nucleon contributes to the nucleon spin, and how the quark masses and the forces on quarks are distributed in transverse space, a question of crucial importance for our understanding of the dynamics underlying nucleon structure and the forces leading to color confinement.   

\begin{figure}[t!]
\centerline{\includegraphics[height=180pt,width=350pt]{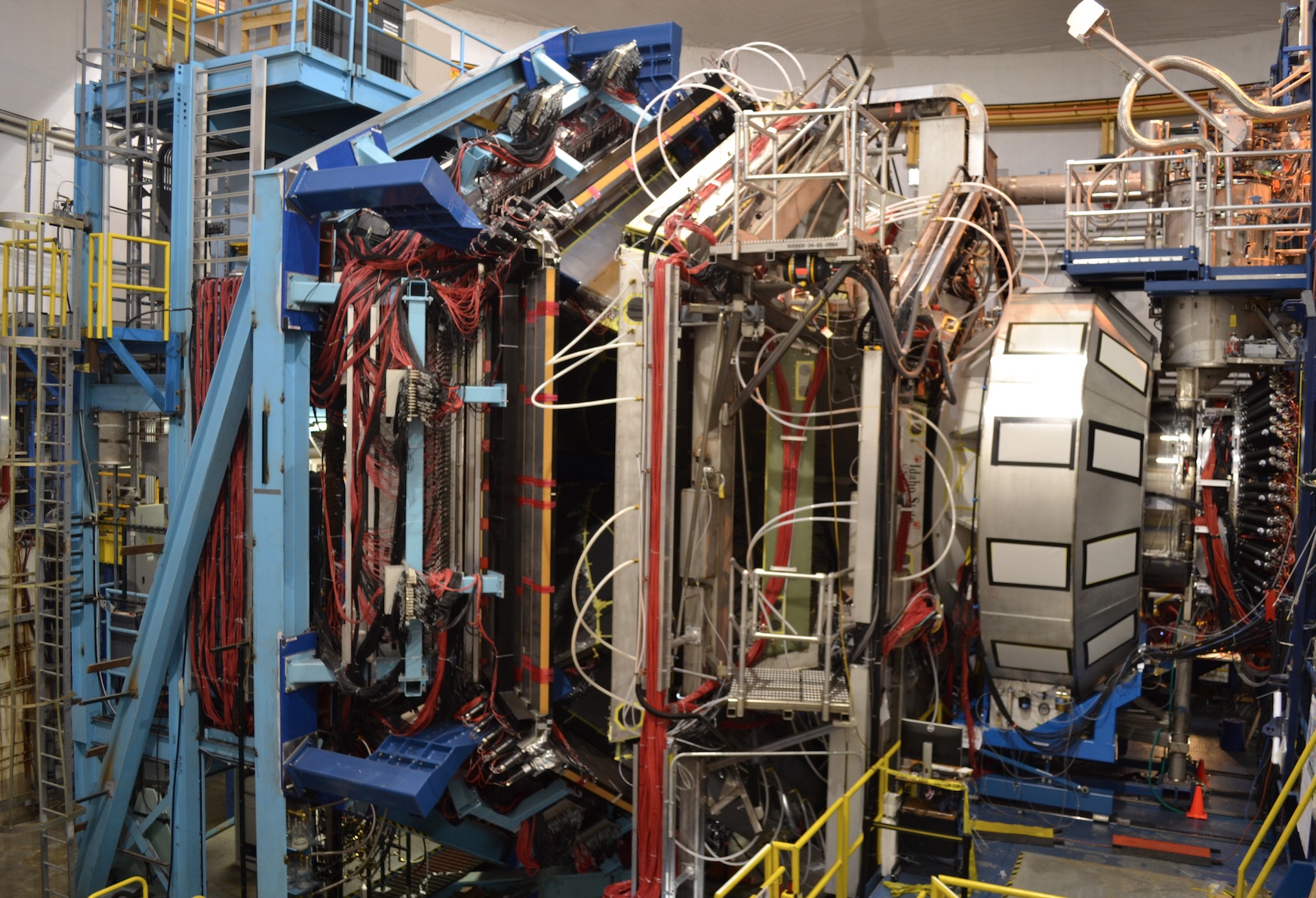}}
\caption{The {\tt CLAS12} detector in Hall B. The beam line is running from the right to the left. The liquid hydrogen target is centered in the solenoid magnet with 5~T central magnetic field, and is surrounded by tracking and particle identification detectors covering the polar angle range from $40^\circ$ to $125^\circ$. The forward detector consists of the $2\pi$ gas \v{C}erenkov counter (large silvery box to the right), the tracking chambers around the superconducting torus magnet, 2 layers of time-of-flight systems and two layers of electromagnetic calorimeters for electron triggering and photon detection to the far left.} \label{fig:clas12}
\end{figure}

The four leading twist GPDs $H$, $\widetilde{H}$, $E$, and $\widetilde{E}$, depend on the 3 variable $x$, $\xi$, and $t$, where $x$ is the longitudinal momentum fraction of the struck quark, $\xi$ is the longitudinal momentum transfer to the quark ($\xi \approx x_B/(2-x_B)$), and $t$ is the invariant 4-momentum transfer to the proton. The mapping of the nucleon GPDs, and a detailed understanding of the spatial quark and gluon structure of the nucleon, have been widely recognized as key objectives of nuclear physics of the next decades. This requires a comprehensive program, combining results of measurements of a variety of processes in $eN$ scattering with structural information obtained from theoretical studies, as well as with expected results from future lattice QCD simulations. The {\tt CLAS12} detector (Fig.~\ref{fig:clas12}) has recently been completed and has begun the experimental science program in the 12 GeV era Jefferson Lab.  

\subsection{Accessing GPDs in DVCS}
\label{gpds} 

\begin{figure}[t!]
\begin{center}
\includegraphics[width=0.60\textwidth]{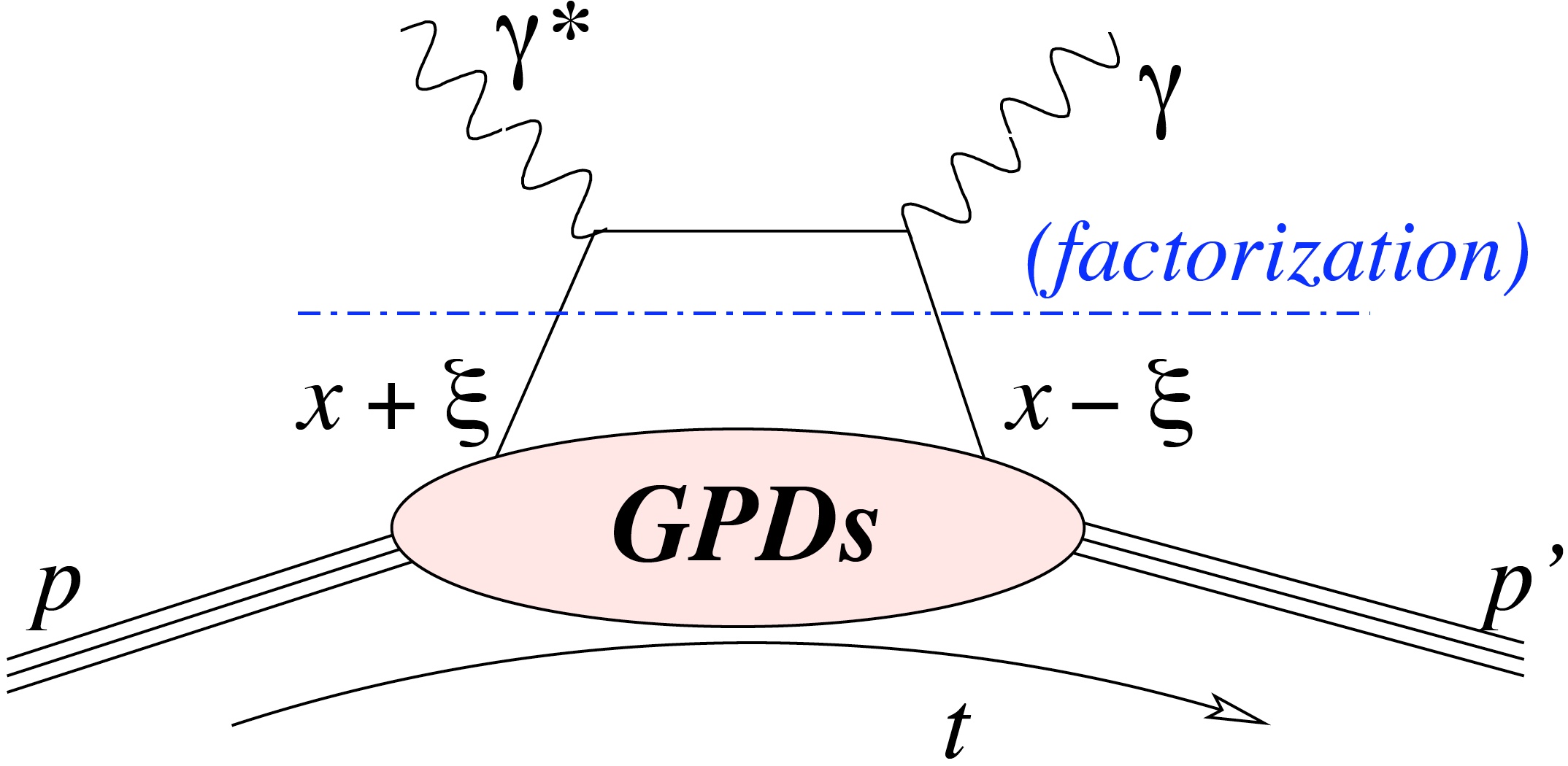}
\caption{Leading order contributions to the production of high energy single photons from protons. The DVCS handbag diagram contains the information on the unknown GPDs.}
\label{fig:handbag}
\end{center}
\end{figure}

The most direct way of accessing GPDs at lower energies is through the measurement of DVCS in a kinematical domain where the so-called handbag diagram (Fig.~\ref{fig:handbag})  makes the dominant contributions. However, in DVCS as in other deeply virtual reactions, the GPDs do not appear directly in the cross section, but in convolution integrals called Compton Form Factors (CFFs), which are complex quantities defined as, e.g. for the GPD $H^q$:  
\begin{equation}
{\cal{H^q(\xi,\it{t})}} \equiv \int_{-1}^{+1}{\frac{H^q(x,\xi,t)dx}{x - \xi + i\epsilon}} = \int_{-1}^{+1}{\frac{H^q(x,\xi,t)dx}{x - \xi }} + i\pi H^q(\xi,\xi,t)~,
\end{equation}
where the first term on the r.h.s. corresponds to the real part and the second term to the imaginary part of the scattering amplitude. The superscript $q$ indicates that GPDs depend on the quark flavor. From the above expression it is obvious that GPDs, in general, can not be accessed directly in measurements. However, in some kinematical regions the BH process where high energy photons are emitted from the incoming and scattered electrons, can be important. Since the BH amplitude is purely real, the interference with the DVCS amplitude isolates the imaginary part of the DVCS amplitude. The interference of the two processes offers the unique possibility to determine GPDs directly at the singular kinematics $x=\xi$. At other kinematical regions a deconvolution of the cross section is required to determine the kinematic dependencies of the GPDs. It is therefore important to obtain all possible independent information that will aid in extracting information on GPDs. The interference terms for polarized beam $I_{LU}$, longitudinally polarized target $I_{UL}$, transversely (in scattering plane) polarized target $I_{UT}$, and perpendicularly (to scattering plane) polarized target $I_{UP}$ are given by the expressions: 
\begin{eqnarray}
I_{LU} & \sim & \sqrt{\tau^\prime} [F_1 H + \xi (F_1+F_2) \widetilde{H} + \tau F_2 E] \\
I_{UL} & \sim & \sqrt{\tau^\prime} [F_1 \widetilde{H} + \xi (F_1 + F_2) H + (\tau F_2 - \xi F_1)\xi \widetilde{E}] \\
I_{UP} & \sim & {\tau}[F_2 H - F_1 E + \xi (F_1 + F_2)\xi \widetilde{E} \label{perptar} \\
I_{UT} & \sim & {\tau}[F_2 \widetilde{H} + \xi (F_1 + F_2) E - (F_1+ \xi F_2) \xi \widetilde{E}] \\
\end{eqnarray}
\noindent
where $\tau = -t/4M^2$ and $\tau^\prime = (t_0 - t)/4M^2$. By measuring all 4 combinations of interference terms one can separate all 4 leading twist GPDs at the specific kinematics $x=\xi$. Experiments at JLab using 4 to 6 GeV electron beams have been carried out with polarized beams~\cite{Ste01, Mun06, Gir07, Gav09,Jo15} and with longitudinal
target~\cite{Che06, Sed14, Pis15}, showing the feasibility of such measurements at relatively low beam energies, and their sensitivity to the GPDs. Techniques of how to extract GPDs from existing DVCS data and what has been learned about GPDs can be found in ~\cite{Kum12, Gui13}. 

The structure of the differential cross section for polarized beam with unpolarized target, and polarized beam with polarized target is reported in Eq.~\ref{eq:LU} and  Eq.~\ref{eq:LL}. In these expressions, $\sigma_i$ and $\Delta\widetilde{\sigma}_i$ are even in the azimuthal angle $\phi$ and beam-polarization independent, while $\widetilde{\sigma}_i$ and $\Delta\sigma_i$ are odd in $\phi$ and beam-polarization dependent. The interference terms 
\begin{eqnarray}
\sigma_{INT} & \sim & {\cal R}{\mathrm e} \left[ {\it A}({\gamma^*N\rightarrow \gamma N}) \right] \\  
\widetilde{\sigma}_{INT} & \sim & {\cal I}{\mathrm m}  \left[ {\it A}({\gamma^*N\rightarrow \gamma N}) \right] \\
\Delta\sigma_{INT} & \sim & {\cal R}{\mathrm e} \left[ {\it A}({\gamma^* {\vec N} \rightarrow \gamma N}) \right] \\  
\Delta\widetilde{\sigma}_{INT} & \sim & {\cal I}{\mathrm m}  \left[ {\it A}({\gamma^* {\vec N} \rightarrow \gamma N}) \right] 
\end{eqnarray}
are the real and imaginary parts of the Compton amplitude. The unpolarized and polarized beam $e^+ - e^-$ charge difference for unpolarized and polarized targets  
determines uniquely the interference contributions (Eq.~\ref{eq:int0}-\ref{eq:int1}-\ref{eq:int2}-\ref{eq:int3}). If only a polarized electron beam is available, the beam helicity asymmetry and average determine a combination of the inteference and pure DVCS amplitudes (Eq.~\ref{eq:int00}-\ref{eq:int10}-\ref{eq:int20}-\ref{eq:int30}). One can separate these contributions using the Rosenbluth technique~\cite{Ros50}. This requires measurements at two significantly different beam energies, which reduces the kinematical coverage that can be achieved with this method. The combination of polarized electron and polarized positron beams does not suffer this limitation, and it offers a separation 
over the full kinematic range available at the maximum beam energy.   

\subsection{Estimates of experimental uncertainties} 

\subsubsection*{The CLAS12 Detector}

The experimental program will use the {\tt CLAS12} detector (Fig.~\ref{fig:clas12}) for the detection of the hadronic final states. {\tt CLAS12} consists of a Forward Detector (FD) and a Central Detector (CD). The Forward Detector is comprised of six symmetrically arranged sectors defined by the six coils of the superconducting torus magnet. Charged particle tracking is provided by a set of 18 drift chambers with a total of  36 layers in each sector. Additional tracking at $5^\circ$-$35^\circ$ is achieved by a set of 6 layers of micromesh gas detectors (micromegas) immediately downstream of the target area and in front of the High Threshold \v{C}erenkov Counter (HTCC). Particle identification is provided by time-of-flight information from two layers of scintillation counter detectors (FTOF). Electron, photon, and neutron detection are provided by the triple layer  electromagnetic calorimeter, PCAL, EC(inner), and EC(outer). The heavy-gas \v{C}erenkov Counter (LTCC) provides separation of high momentum pions from kaons and protons. The Central Detector consists of 6 to 8 layers (depending on the configuration) of silicon strip detectors with stereo readout and 6 layers of micromegas arranged as a barrel around the target, a barrel of scintillation counters to measure the particle flight time from the target (CTOF), and a scintillation-counter based Central Neutron Detector (CND). 

\subsubsection*{Beam charge asymmetries on protons}

\begin{figure}[h!]
\begin{center}
\includegraphics[height=.22\textheight,width=0.99\textwidth]{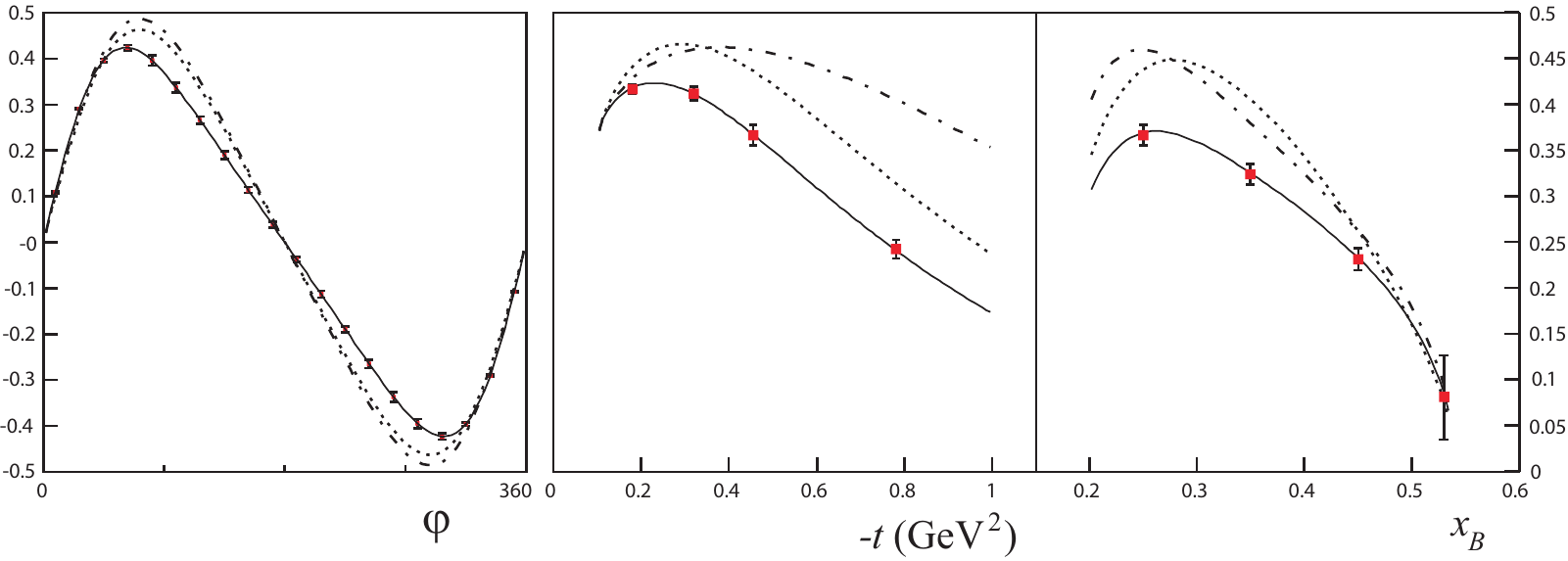}
\caption{The beam spin asymmetry showing the DVCS-BH interference for 11 GeV beam energy~\cite{Sab11}: (left panel) $x=0.2$, $Q^2=3.3$~GeV$^2$, $-t=0.45$~GeV$^2$; (middle and right panels) $\phi=90^{\circ}$, other parameters same as in left panel. Many other bins will be measured simultaneously. The curves represent various parameterizations within the VGG model~\cite{Van99}. Projected uncertainties are statistical.}
\label{fig:dvcs_alu_12gev}
\end{center}
\end{figure}
Beam spin asymmetries of polarized electrons for the DVCS process have been measured at lower energies and are known to be large, up to 0.3-0.4. Figure~\ref{fig:dvcs_alu_12gev} shows projections of the Beam Spin Asymmetry (BSA) for some specific kinematics at an electron beam energy of 11 GeV. The uncertainties are estimated assuming an experiment of 
1000 hours at an instantaneous luminosity of ${\cal L} = 10^{35}$cm$^{-2}\cdot$s$^{-1}$. The asymmetry is the results of the interference term $\tilde{\sigma}_{INT}$ in Eq.~\ref{eq:LU}). Note that the magnitude of the interference amplitude is independent of the electric charge, but the BSA sign is opposite for electrons and positrons.      

\begin{figure}[t!]
\begin{center}
\includegraphics[height=.49\textheight,width=0.925\textwidth]{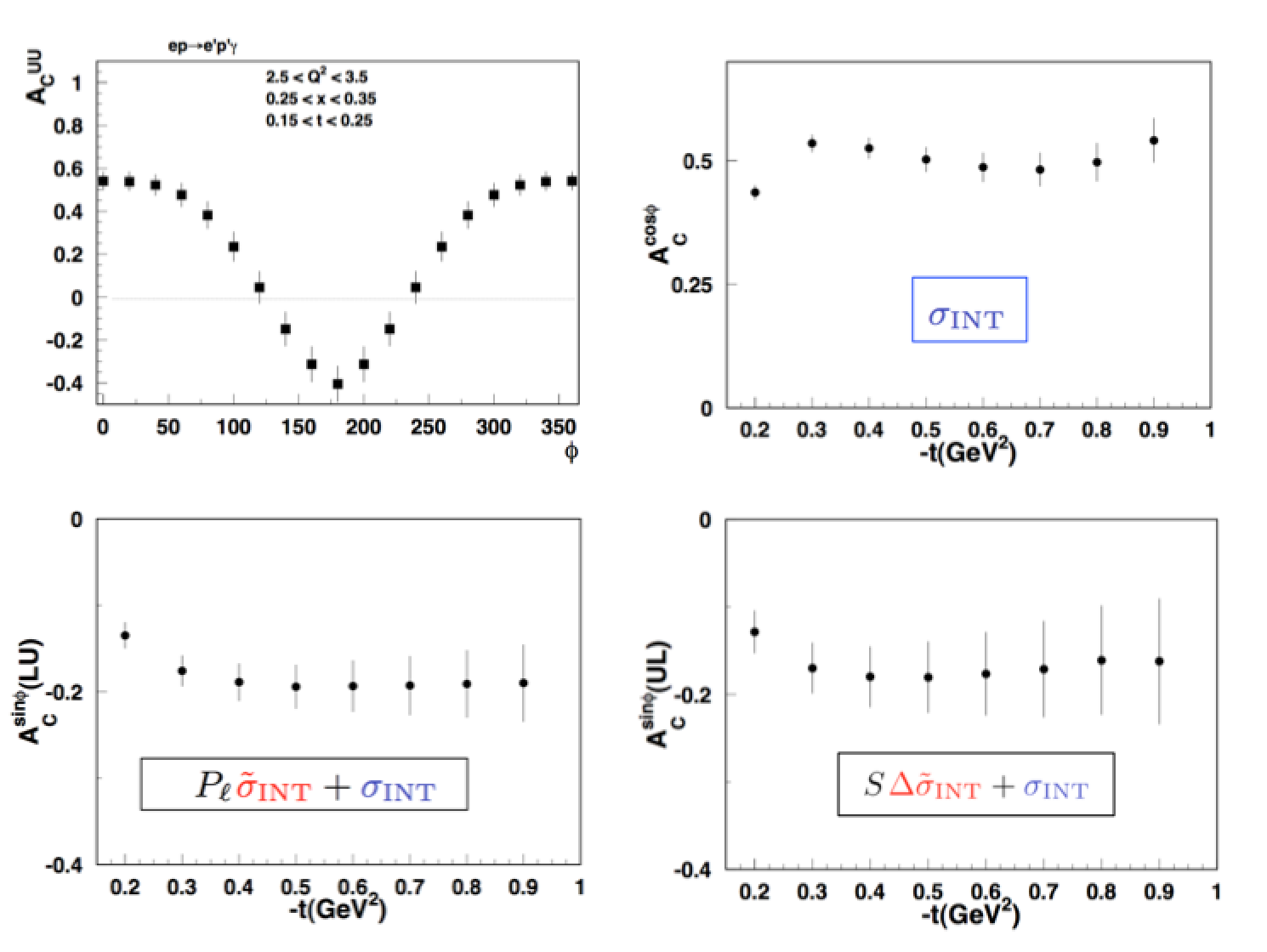}
\caption{Electron-positron DVCS charge asymmetries: (top-left) azimuthal dependence of the charge asymmetry for positron and electron beam at 11 GeV beam; (top-right) moment in $\cos(\phi)$ of the charge asymmetry versus momentum transfer $t$ to the proton; (bottom-left) charge asymmetries for polarized electron and positron beams at fixed polarization (LU); (bottom right) charge asymmetry for longitudinally polarized protons at fixed polarization (UL). The error bars are estimated for a 1000 hours run with positron beam and luminosity ${\cal L} = 2 \times 10^{34}$~cm$^{-2}\cdot$sec$^{-1}$ at a beam polarization $P=0.6$, and a 1000 hours electron beam run with luminosity ${\cal L} = 10 \times 10^{34}$~cm$^{-2}\cdot$sec$^{-1}$ and beam polarization $P = 0.8$. The error bars are statistical for a single bin in $Q^2$, $x$, and $t$ as shown in the top-left panel. Other bins are measured simultaneously.}
\label{fig:cross_section}
\end{center}
\end{figure}
Eq.~\ref{eq:int0} also shows that the term $\sigma_{INT}$ can be isolated in the difference of unpolarized electron and positron cross sections. Examples of the charge difference and the charge asymmetry are shown in Fig.~\ref{fig:cross_section}. The unpolarized charge asymmetry $A_c^{UU}$ and its $\cos{\phi}$ moment $A^{\cos{\phi}}$ can both be large for the dual model assumed in our estimate. For quantitative estimates of the charge differences in the cross sections we use the acceptance and luminosity achievable with {\tt CLAS12} as basis for measuring the process $ep \rightarrow ep\gamma $ at different beam and target conditions. A 5~cm long liquid hydrogen is assumed with an electron current of 75~nA, corresponding to an operating luminosity of $10^{35}$cm$^{-2}\cdot$sec$^{-1}$. For the positron beam a 5 times lower beam current of 15~nA is assumed. In either case 1000 hours of beam time is used for the rate projections. For quantitative estimates of the cross sections the dual model~\cite{Guz06,Guz09} is used. It incorporates parameterizations of the GPDs $H$ and $E$. As shown in Fig.~\ref{fig:cross_section}, effects coming from the charge asymmetry can be large. In case of unpolarized beam and unpolarized target the cross section for electron scattering has only a small dependence on azimuthal angle $\phi$, while the corresponding positron cross section has a large $\phi$ modulation. The difference is directly  related to the term $\sigma_{INT}$ in Eq.~\ref{eq:int0}.

\subsection{The science case for DVCS with polarized positrons}

The science program for DVCS with electrons beams has been well established, and several approved experiments for 12~GeV operation have already been carried out or are currently in the process and planned for the next few years. What do polarized positron beams add which makes a most compelling case for experiments with {\tt CLAS12}? In this section we discuss one example of the impact of DVCS measurements with polarized positron beams, and corresponding to the unraveling of the force distribution on quarks in the proton. Here We refer to the recent publication in the journal Nature of the results of an analysis on the pressure distribution in the proton~\cite{Bur18}. 

This analysis is based on the results of BSA and unpolarized DVCS cross section DVCS measured with CLAS in Hall B. The determination of the pressure distribution proceeds in several steps:
\renewcommand{\labelenumi}{\it\roman{enumi})}
\begin{enumerate}
\item We begin with the sum rules that relate the second Mellin moments of the GPDs to the Gravitational Form Factors (GFFs)~\cite{Ji97}; 
\item We then define the complex CFF~$\cal{H}$ directly related to the experimental observables describing the DVCS process, i.e., the BSA and the differential cross section;  
\item The real and imaginary parts of $\cal{H}$ can be related through a dispersion relation~\cite{Die07, Ani08, Pas14} at fixed $t$, where the $D(t)$-term appears as a  subtraction constant~\cite{Pol99}; 
\item We recover $d_1(t)$ from the expansion of the $D(t)$-term in the Gegenbauer polynomials of $\xi$, the momentum transfer to the struck quark; 
\item We finally proceed with the fits to the data and extract $D(t)$ and determine $d_1(t)$;
\item The pressure distribution is then determined from the relation of $d_1(t)$ and $p(r)$ through a Bessel integral. 
\end{enumerate}
The sum rules that relate the second Mellin moments of the chiral-even GPDs to the GFFs are~\cite{Ji97}:
\begin{eqnarray}
\int \mathrm{d}x \, x \left[ H(x, \xi, t) + E(x, \xi, t)\right] & = & 2 J(t) \\
\int \mathrm{d}x \, x H(x, \xi, t) & = & M_2(t) + \frac{4}{5} \xi^2 d_1(t) \, , 
\end{eqnarray}
where $M_2(t)$ and $J(t)$ respectively correspond to the time-time and time-space components of the Energy Momentum Tensor (EMT), and give access to the mass and total angular momentum distributions carried by the quarks in the proton. The quantity $d_1(t)$ corresponds to the space-space components of the EMT, and encodes the shear forces and pressure acting on the quarks. We have some constraints on $M_2(t)$ and $J(t)$, notably at $t=0$ they are fixed to the proton's mass and spin. By contrast, almost nothing is known on the equally fundamental quantity $d_1(t)$. For instance, considering the physics content of $d_1(t)$, we can expect the existence of a zero sum rule ensuring the total pressure and forces to vanish, thus preserving the stability of the dynamics of the proton. The observables are parameterized by the CFFs, which for the GPD $H$ are the real quantities ${\cal R}{\mathrm e} \left[ {\mathcal H} \right]$ and ${\cal I}{\mathrm m} \left[ {\mathcal H} \right]$ defined by:
\begin{eqnarray}
{\cal R}{\mathrm e} \left[ {\mathcal H}(\xi,t) \right] & + & i \, {\cal I}{\mathrm m} \left[ {\mathcal H}(\xi,t) \right] \\
& = & {\int_{-1}^{1} dx \left[ \frac{1}{\xi-x-i\epsilon } -  \frac{1}{\xi+x-i\epsilon} \right] \it H(x,\xi,t)} \, . \nonumber 
\end{eqnarray}
The average quark momentum fraction $x$ is not observable in the process; it is integrated over with the quark propagators. Analytical properties of the amplitude in the Leading Order (LO) approximation lead to the dispersion relation:
\begin{equation}
{\cal R}{\mathrm e} \left[ {\mathcal H}(\xi,t) \right] \stackrel{\rm LO}{=} D(t) + \frac{1}{\pi}{\mathcal P}\int_{0}^{1} dx  \left
(\frac{1}{\xi-x}-\frac{1}{\xi+x}\right) \rm{Im}{\mathcal H}(\it x,\it t)
\label{disp}
\end{equation}
where the subtraction constant is the so-called $D$-term. The dispersion relation allows us trading-off the two CFFs as unknowns with one CFF and the $D$-term~\cite{Rad13,Rad13-1}. For our purpose we recover the $d_1(t)$ as the first coefficient in the Gegenbauer expansion of the $D$-term. Here, we truncate this expansion to $d_1(t)$ only:
\begin{equation} 
D(t)  = \frac{1}{2} \, \int_{-1}^1 \rm{d}z \, \frac{\it D(z,t)}{1-z}
\end{equation} 
with 
\begin{equation}
D(z,t) = (1-z^2)\left[d_1(t) C^{3/2}_1(z) +  \cdots\right]
\end{equation} 
and 
\begin{equation}
{-1<z=\frac{x}{\xi}<1} \, .
\end{equation}

Our starting points in the analysis are the global fits presented in~\cite{Kum10,Mul13}, referred to as KM parameterization. The imaginary part of the amplitude is calculated from a parameterization of the GPDs along the diagonal $x=\xi$. The real part of the amplitude is then reconstructed assuming LO dominance and applying the dispersion relation. The $\xi$-dependence of the $D$-term is completely generated by the Gegenbauer expansion, restricted to the $d_1(t)$ term only. Finally, the momentum transfer dependence of the $d_1(t)$ term is given as a functional form, with three parameters $d_1(0)$, $M$, and $\alpha$:
\begin{equation}
d_1(t) = d_1(0) \left( 1-\frac{t}{M^2} \right)^{-\alpha} 
\end{equation}
where the chosen form of $d_1(t)$ with $\alpha$=$3$ is consistent with the asymptotic behavior required by the dimensional counting rules in QCD~\cite{Lep80}. We adjust and fix the central values of the model parameters to the data at 6 GeV~\cite{Gir07, Jo15}. They include unpolarized and polarized beam cross-sections over a wide phase-space in the valence region, and support the model indicating that the GPD $H$ largely dominates these observables. An illustration of a fit to the $d_1(t)$ dependence is provided in  Fig.~\ref{d1}. The data points correspond to the values extracted from the fit to the unpolarized cross section data. The experimental analysis shows that $d_1(0)$ has a negative sign. This is consistent with several theoretical studies~\cite{Goe07, Kim12, Pas14}. The fit results in a $d_1(0)$ value of
\begin{equation}
d_1(0) = -2.04 \pm 0.14 ({\rm stat.}) \pm 0.33 ({\rm syst.}) \, .
\end{equation}
The negative sign of $d_1(0)$ found in this analysis seems deeply rooted in the spontaneous breakdown of chiral symmetry~\cite{Kiv01}, which is a consequence of the transition of the microsecond old universe from its state of de-confined quarks and gluons to the state of confined quarks in stable protons. It is thus intimately connected to the  stability of the proton~\cite{Goe07} and of the visible universe. We finally can relate the GFF $d_1(t)$ to the pressure distribution {\it via} the spherical Bessel integral:
\begin{equation}
d_1(t) \propto \int\rm{d}^3{\bf r}\;\frac{j_0(r\sqrt{-t})}{2\it{t}}\; p(r) \, .
\label{d1}
\end{equation}
\begin{figure}[t!]
\begin{center}
\includegraphics[height=.30\textheight]{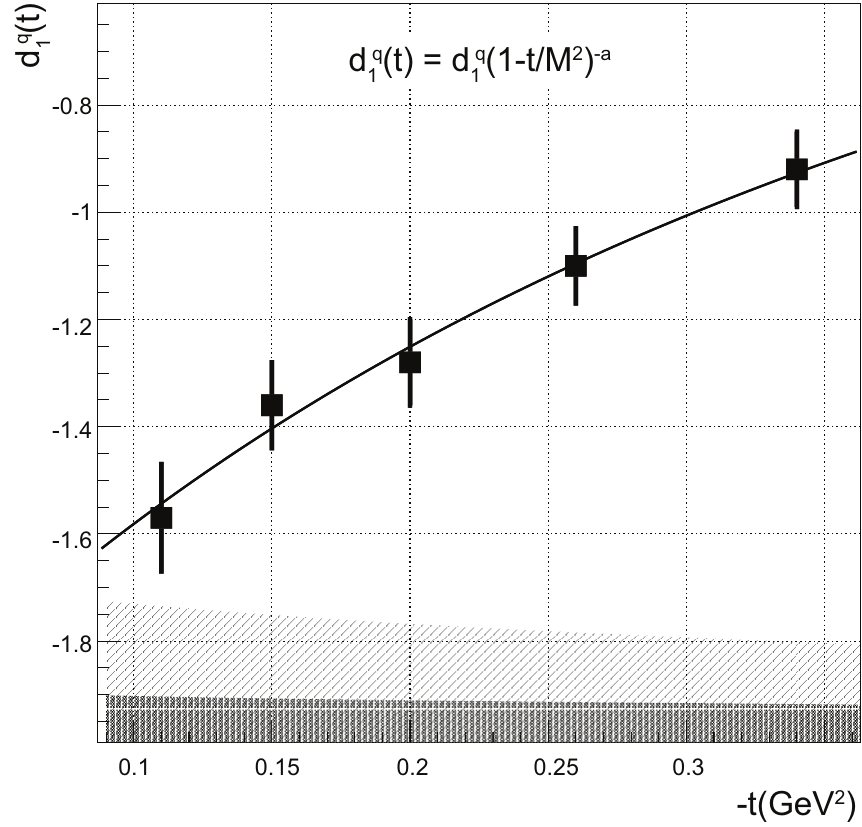}
\caption{Example of a fit to $d_1(t)$. The error bars are from the fit to the cross sections at fixed value of $-t$. The single-shaded area at the bottom corresponds to the uncertainties from the extension of the fit into regions without data and is reflected in the green shaded are in Fig.~\ref{pressure}. The double-shaded area corresponds with the projected uncertainties from future experiment~\cite{Elo16}, as shown in Fig.~\ref{pressure} with the red shaded area. Uncertainties represent 1 standard gaussian deviation.}
\end{center}
\end{figure}

\begin{figure}[t!]
\centerline{\includegraphics[height=190pt,width=265pt]{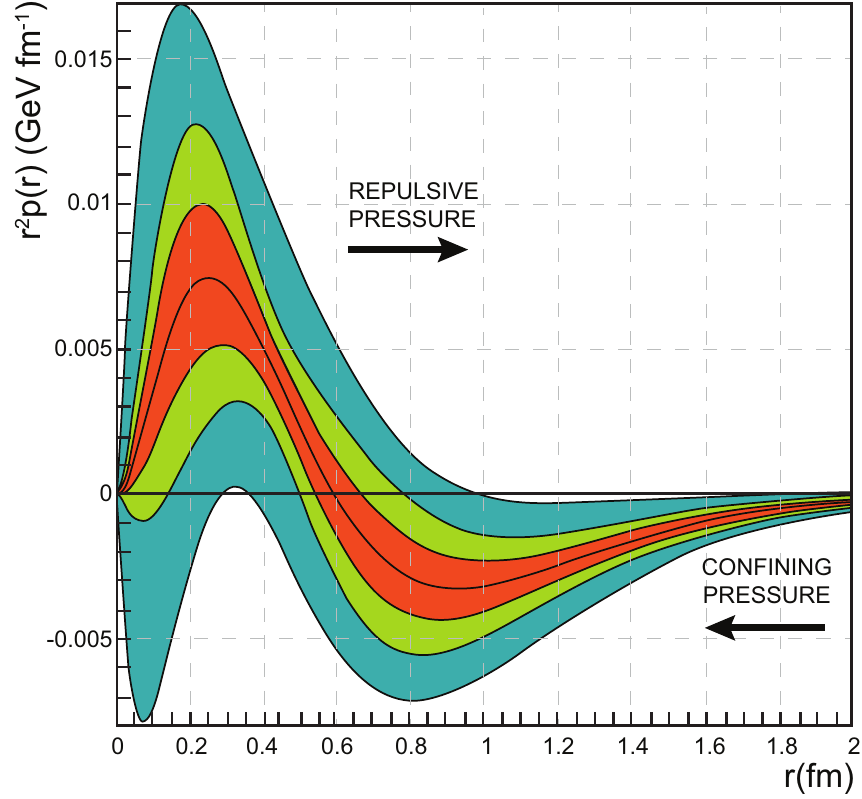}}
\caption{The radial pressure distribution in the proton. The graph shows the pressure distribution $r^2 p(r)$ resulting from the interactions of the quarks in the proton versus the radial distance from the center in femtometer. The black central line corresponds to the pressure extracted from the $D$-term parameters fitted to the published data at 6 GeV~\cite{Jo15}. The corresponding estimated uncertainties are displayed as the shaded area shown in light green. Uncertainties represent 1 standard deviation.}
\label{pressure}
\end{figure}

Our results on the quark pressure distribution in the proton are illustrated in Fig.~\ref{pressure}. The black central line corresponds to pressure distribution $r^2p(r)$ extracted from the $D$-term parameters fitted to the published 6 GeV data~\cite{Jo15}. The corresponding estimated uncertainties are displayed as the shaded area shown in light green. There is a positive core and a negative tail of the $r^2p(r)$ distribution as a function of the radial distance from the proton's center with a zero-crossing near 0.6~fm from that center. We also note that the regions where repulsive and binding pressures dominate are separated in radial space, with the repulsive distribution peaking near 
$r$=$0.25$~fm, and the maximum of the negative pressure responsible for the binding occurring near $r$=$0.8$~fm. The outer shaded area shown in dark green in Fig.~\ref{pressure} corresponds with the $D$-term uncertainties obtained in the global fit results from previous research~\cite{Kum10, Mul13}. They exhibit a shape similar to the light green area and confirm the robustness of the analysis procedure to extract the $D$-term. Here we remark that the pressure $p(r)$ must satisfy the stability condition 
\begin{equation}
\int_0^\infty \it r^2 p(r) dr  = 0 \, ,
\end{equation}
which is realized within the uncertainties of our analysis. The shape of the radial pressure distribution mimics closely the results obtained within the chiral quark soliton model~\cite{Goe07}. In this model, the proton is modeled as a chiral soliton in which constituent quarks are bound by a self-consistent pion field. The comparison with our  results suggests that the pion field is significantly relevant for the description of the proton as a bound state of quarks. 
 
\subsection*{What positrons will add to this program?} 
 
There are a couple of limiting factors in the analyses presented above. These are related to the limited experimental information that can be obtained from having just polarized electron beam available:  
\renewcommand{\labelenumi}{\it\roman{enumi})}
\begin{enumerate} 
\item The use of the dispersion relation in Eq.~\ref{disp} to determine ${\cal R}{\mathrm e} \left[ \cal{H(\xi, \it{t)}} \right]$;
\item The need to extrapolate the $t$-dependence of the formula in Eq.~\ref{d1}. 
\end{enumerate}
While the extrapolation is unavoidable when extracting the pressure distribution over the entire radial distance, applying the dispersion relation in Eq.~\ref{disp} at large -$t$ values, where issues with convergence may occur, is problematic. It is therefore highly desirable to determine the subtraction term $D(t)$ directly from the DVCS data without the need for applying the dispersion relation. Such a procedure requires to determine both the real and imaginary parts of the CFF $\cal{H}(\xi, \it{t})$ in Eq.~(\ref{disp})  directly from experiment. The term $D(t)$ can then be directly extracted. By isolating the terms $\sigma_{INT}$ and $\widetilde{\sigma}_{INT}$, the real and imaginary parts of the Compton amplitude can be separated. This is achieved by measuring the difference in the unpolarized cross sections and the helicity-dependent cross sections for (polarized) electrons and (polarized) positrons. From Fig.~\ref{fig:cross_section}, we can infer that both of these observables can result in large cross section differences and polarization asymmetries, and can be well measured already with modest positron currents, by making use of the large acceptance capabilities of {\tt CLAS12}.    

While our focus for this letter is the determination of the pressure distribution and the shear forces in the proton, using a transversely spin polarized target and polarized electron and positron beams, the term $\Delta\sigma_{INT}$ in Eq.~\ref{eq:LL} can be isolated. It is related to the GPD $E$ thorugh the CFF $\cal{E}(\xi,\it t)$, and thus to the angular momentum distribution in the proton. Measurement of $\cal{E}(\xi,\it t)$ will allow for the extraction of the radial dependence of the angular momentum density in  protons and can be determined in a fashion similar to the one described for the pressure distribution. 

\subsection{Experimental setup for DVCS experiments} \label{dvcs}

\begin{figure}[t!]
\centerline{\includegraphics[height=195pt,width=390pt]{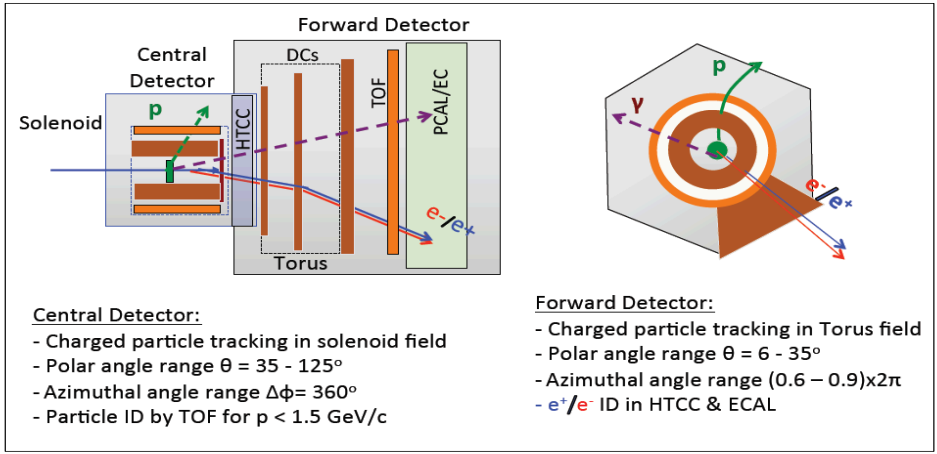}}
\caption{Generic {\tt CLAS12} configuration for the electron-proton and the positron-proton experiments. The central detector will detect the protons, and the bending in the magnetic solenoid field will be identical for the same kinematics. The electron and the positron, as well as the high-energy DVCS photon will be detected in the forward detector part.  The electron and positron will be deflected in the torus magnetic field in the same way as the torus field direction will be opposite in the two experiments. The deflection in $\phi$ due to the solenoid fringe field will be of same magnitude $\Delta\phi$ but opposite in direction. The systematic of this shift can be controlled by doing the same  experiment with opposite solenoid field directions that would result in the sign change of the $\Delta\phi$.}
\label{dvcs_experiment}
\end{figure}

Figure~\ref{dvcs_experiment} shows generically how the electron-proton and the positron-proton DVCS experiments would be configured. Electrons and positrons will be detected in the forward detection system of {\tt CLAS12}. However, for the positron run the torus magnet would have the reversed polarity so that positron trajectories would look identical to the electron trajectories in the electron-proton experiment, and limit systematic effects in acceptances. The recoil proton in both cases would  be detected in the Central Detector at the same solenoid magnet polarity, also eliminating most systematic effects in the acceptances. However, there is a remaining systematic difference in the two configuration, as the forward scattered electron/positron would experience different transverse field components in the solenoid, which will cause the opposite azimuthal motion in $\phi$ 
in the forward detector. A good understanding of the acceptances in both cases is therefore important. The high-energy photon is, of course, not affected by the magnetic field configuration.

\subsection{Summary}

In this letter, we described the use of a new polarized positron beam in conjunction with the already available polarized electron beam to significantly enhance the program to study the generalized parton distribution and to extract physical quantities that are related to the mechanical properties of the proton, such as the distribution of shear forces, the pressure distribution, the mechanical radius of the proton, and the angular momentum distribution. These quantities have never been measured before as they couple directly only to the gravitational field. The development of the generalized parton distributions and their relationship to the gravitational form factors through the second 
Mellin moments made this feasible in an indirect way. First results have been obtained recently~\cite{Bur18}. An experiment has been approved by PAC44 using a polarized electron beam to improved the precision of the pressure distribution. The use of the {\tt CLAS12} detector to broaden this program is natural as the expected polarized positron current is much lower than what can be achieved with polarized electron beams, and fits naturally with the capabilities of the {\tt CLAS12}. Simulations have been made with realistic beam currents and beam polarization that show that the relevant observables can be measured with good accuracy and will have a very significant scientific impact.    

%
%

\newpage

\null\vfill

\begin{center}

\section{\it Letter-ot-Intent: n-DVCS @ CLAS12}

\vspace*{15pt}

{\Large{\bf Beam Charge Asymmetries for}}

\vspace*{3pt}

{\Large{\bf Deeply Virtual Compton Scattering}}

\vspace*{3pt}

{\Large{\bf on the Neutron}}

\vspace*{3pt}

{\Large{\bf with CLAS12 at 11 GeV}}

\vspace*{15pt}

{\bf Abstract}

\begin{minipage}[c]{0.85\textwidth}
Measuring DVCS on a neutron target is a necessary step to deepen our understanding of the structure of the nucleon in terms of GPDs. The combination of neutron and proton targets allows to perform a flavor decomposition of the GPDs. Moreover, DVCS on a neutron target plays a complementary role to DVCS on a transversely polarized proton target in the determination of the GPD $E$, the least known and  constrained GPD that enters Ji's angular momentum sum rule. We propose to measure, for the first time, the beam charge asymmetry (BCA) in the $e^{\pm} d\to e^{\pm}n\gamma(p)$ reactions, with the upgraded 11 GeV CEBAF positron/electron beams and the CLAS12 detector. The exclusivity of the final state will be ensured by detecting in CLAS12 the scattered lepton, the photon (including the Forward Tagger at low polar angles), and the neutron. Running 80 days on a deuterium target at the maximum CLAS12 luminosity ($10^{35}$~cm$^{-2}\cdot$s$^{-1}$) will yield a rich BCA data set in the 4-dimensional ($Q^2$, $x_B$, $-t$, $\phi$) phase space. This observable will significantly impact the experimental determination of the real parts of the ${\cal E}_n$ and, to a lesser extent, $\widetilde{{\cal H}_n}$ Compton form factors.
\end{minipage} 

\vspace*{15pt} 

{\it Spokespersons: \underline{S.~Niccolai} (niccolai@ipno.in2p3.fr), E.~Voutier}

\end{center}

\vfill\eject

%
%

\subsection{Introduction}

It is well known that the fundamental particles which form hadronic matter are the quarks and the gluons, whose interactions are described by the QCD Lagrangian. However, exact QCD-based calculations cannot yet be performed to explain all the properties of hadrons in terms of their constituents. Phenomenological functions need to be used to connect experimental observables with the inner dynamics of the constituents of the nucleon, the partons. Typical examples of such functions include form factors, parton densities, and distribution amplitudes. The GPDs are nowadays the object of intense research effort in the perspective of unraveling nucleon structure. They describe the correlations between the longitudinal momentum and transverse spatial position of the partons inside the nucleon, they give access to the contribution of the orbital momentum of the quarks to the nucleon, and they are sensitive to the correlated $q\bar{q}$ components of the nucleon wave function~\cite{Mul94, Die03, Bel05}.

\begin{figure}[h!]
\begin{center}
\includegraphics[scale=0.60]{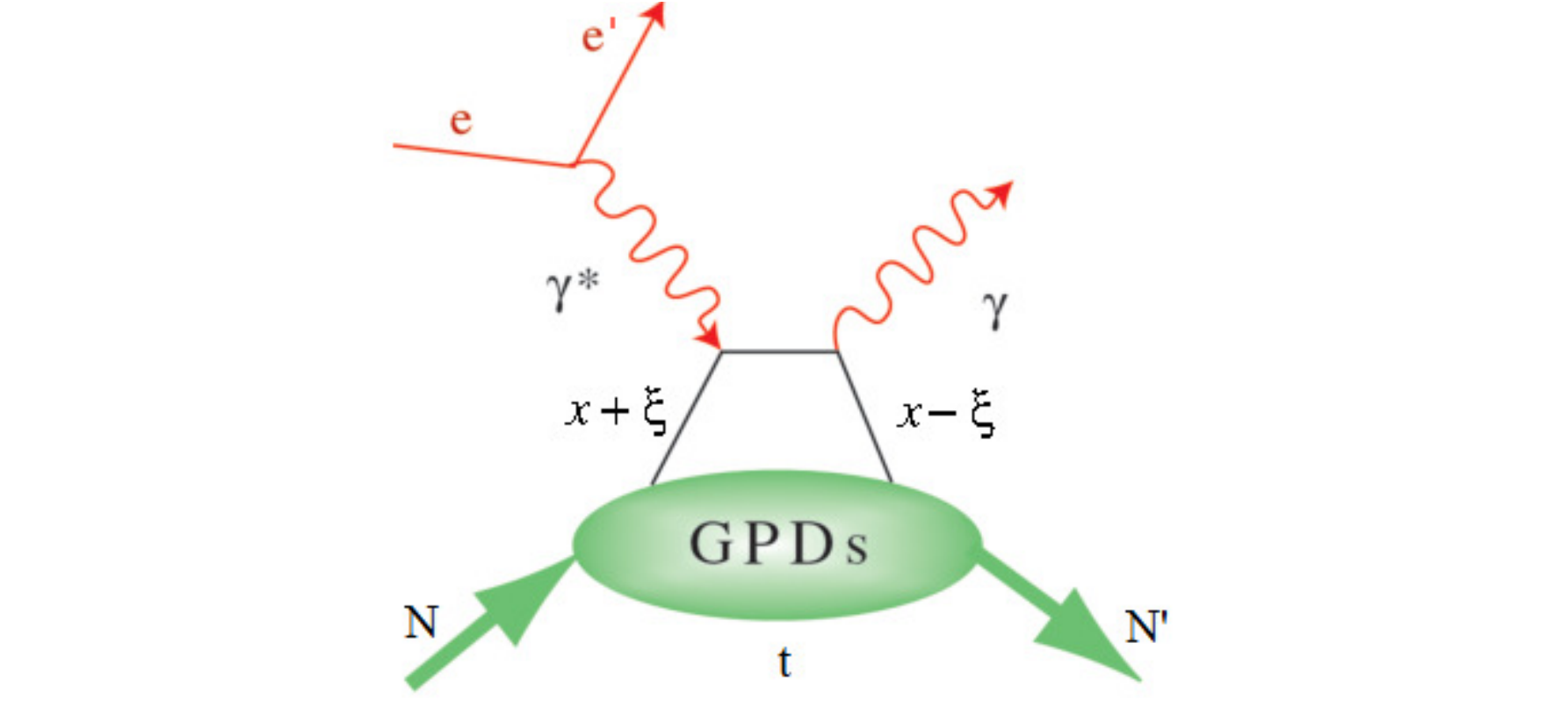}
\caption[Handbag diagram for the DVCS process] {The handbag diagram for the DVCS process on the nucleon $eN \to e'N'\gamma'$; here $x+\xi$ and $x-\xi$ are the longitudinal momentum fractions of the struck quark before and after scattering, respectively, and $t=(N-N')^2$ is the squared four-momentum transfer between the initial and final nucleons. $\xi\simeq x_B/(2-x_B)$ is proportional to the Bjorken scaling variable $x_B=Q^2/2M\nu$, where $M$ is the nucleon mass and $\nu$ is the energy transferred to the quark.}
\label{fig:dvcs}
\end{center}
\end{figure}
The nucleon GPDs are the structure functions which are accessed in the measurement of the exclusive leptoproduction of a photon (i.e. DVCS) or of a meson on the nucleon, at sufficiently large photon virtuality ($Q^2$) for the reaction to happen at the parton level.  Figure~\ref{fig:dvcs} illustrates the leading process for DVCS, also called ``handbag diagram''. At leading-order QCD and at leading twist, considering only quark-helicity conserving quantities and the quark sector, the process is described by four GPDs: $H^q, \widetilde{H^q} , E^q, \widetilde{E^q}$, one for each quark flavor $q$, that account for the possible combinations of relative orientations of the nucleon spin and quark helicities between the initial and final states. $H^q$ and $E^q$ do not depend on the quark helicity and are therefore called unpolarized GPDs while $\widetilde{H^q}$ and $\widetilde{E^q}$ depend on the quark helicity and are called polarized GPDs. $H^q$ and $\widetilde{H^q}$ conserve the spin of the nucleon, whereas $E^q$ and $\widetilde{E^q}$ correspond to a nucleon-spin flip. \newline
The GPDs depend upon three variables, $x$, $\xi$ and $t$: $x+\xi$ and $x-\xi$ are the longitudinal momentum fractions of the struck quark before and after scattering, respectively, and $t$ is the squared four-momentum transfer between the initial and final nucleon (see caption of Fig.~\ref{fig:dvcs} for definitions). The transverse component of $t$ is the Fourier-conjugate of the transverse position of the struck parton in the nucleon. Among these three variables, only $\xi$ and $t$ are experimentally accessible with DVCS. 

The DVCS amplitude is proportional to combinations of integrals over $x$ of the form 
\begin{equation}\label{dvcs-ampl}
\int_{-1}^{1} d x \, F(\mp x,\xi,t)\left[\frac{1}{x - \xi + i \epsilon}\pm\frac{1}{x + \xi - i \epsilon}\right],
\end{equation}
where $F$ represents one of the four GPDs. The top combination of the plus and minus signs applies to unpolarized GPDs ($H^q, E^q$), and the bottom combination of signs applies to the polarized GPDs ($\widetilde{H^q}, \widetilde{E^q}$). Each of these 4 integrals or Compton Form Factors (CFFs) can be decomposed into their real and imaginary parts, as following:
\begin{eqnarray}\label{def_cffs1}
\Re{\rm e} \left[{\cal F} (\xi,t)\right] & = & {\cal P}  \int_{-1}^{1}dx\left[\frac{1}{x-\xi}\mp\frac{1}{x+\xi}\right]F(x,\xi,t) \\
\Im{\rm m} \left[{\cal F} (\xi,t)\right] & = & -\pi [F(\xi,\xi,t)\mp F(-\xi,\xi,t)], \label{def_cffs2}
\end{eqnarray}
where ${\cal P}$ is Cauchy's principal value integral and the sign convention is the same as in Eq.~\ref{dvcs-ampl}. The information that can be extracted from the experimental data at a given ($\xi,t$) point depends on the measured observable. $\Re{\rm e} [{\cal F}]$ is accessed primarily measuring observables which are sensitive to the real part of the DVCS amplitude, such as double-spin asymmetries, beam charge asymmetries or unpolarized cross sections. $\Im{\rm m} [{\cal F}]$ can be obtained measuring observables sensitive to the imaginary part of the DVCS amplitude, such as single-spin asymmetries or the difference of polarized cross-sections. However, knowing the CFFs does not define the GPDs uniquely. A model input is necessary to deconvolute their $x$-dependence. \newline
The DVCS process is accompanied by the BH process (Fig.~\ref{EAgamma}), in which the final-state real photon is radiated by the incoming or scattered electron and not by the nucleon itself. The BH process, which is not sensitive to GPDs, is experimentally indistinguishable from DVCS and interferes with it at the amplitude level (Sec.~\ref{PhyMot-Deep}). However, considering that the nucleon form factors are well known at small $t$, the BH process is precisely calculable.

\subsection{Neutron GPDs and flavor separation}

The importance of neutron targets in the DVCS phenomenology was clearly established in the pioneering Hall A experiment, where the polarized-beam cross section difference off a neutron, from a deuterium target, was measured~\cite{Maz07}. Measuring neutron GPDs in complement to proton GPDs allows for their quark-flavor separation. For instance, the ${\cal E}$-CFF of the proton and of the neutron can be expressed as 
\begin{eqnarray}
{\cal E}_p(\xi, t) & = & \frac{4}{9}{\cal E}^u(\xi, t)+\frac{1}{9}{\cal E}^d(\xi, t) \\
{\cal E}_n(\xi, t) & = & \frac{1}{9}{\cal E}^u(\xi, t)+\frac{4}{9}{\cal E}^d(\xi, t)  
\end{eqnarray}
(and similarly for ${\cal H}$, ${\widetilde {\cal H}}$ and ${\widetilde {\cal E}}$). The $u$- and $d$-quark CFFs can be determined as:
\begin{eqnarray}
{\cal E}^u(\xi, t)=\frac{9}{15} \left[4 {\cal E}_p(\xi, t)-{\cal E}_n(\xi, t)\right] \\
{\cal E}^d(\xi, t)=\frac{9}{15} \left[4 {\cal E}_n(\xi, t)-{\cal E}_p(\xi, t)\right] \, .  
\end{eqnarray}
An extensive experimental program dedicated to the measurement of the DVCS reaction on a proton target has been approved at  Jefferson Lab, in particular with CLAS12. Single-spin asymmetries with polarized beam and/or linearly or transversely polarized proton targets, as well as unpolarized and polarized cross sections, will be measured with high precision over a vast kinematic coverage. A similar experimental program on the neutron will allow the quark flavor separation of the various GPDs. The beam spin asymmetry for n-DVCS, particularly sensitive to the GPD $E_n$ will be soon measured at CLAS12, involving direct detection of the active neutron~\cite{Nic11}, unlike the pioneer Hall A measurement~\cite{Maz07}. Additionally, the measurement of single- and double-spin asymmetries with a longitudinally polarized neutron target is also foreseen for the nearby future at CLAS12~\cite{Nic15}. The present letter focuses on the extraction of one more observable, the beam charge asymmetry. The next sections outline the benefits of this observable for the determination of the CFFs. 

\subsection{Beam charge asymmetry}

Considering unpolarized electron and positron beams, the sensitivity of the $eN \to e N \gamma$ cross section to the lepton-beam charge (Sec.~\ref{PhyMot-Deep}) can be expressed with the beam-charge asymmetry observable~\cite{Hos16} 
\begin{equation}
A_{\rm C}(\phi) = \frac{d^4\sigma^+ - d^4\sigma^-} {d^4\sigma^+ + d^4\sigma^-} =  
                  \frac{d^4\sigma_{UU}^{\rm I}} {d^4\sigma_{UU}^{\rm BH} + d^4\sigma_{UU}^{\rm DVCS}} \, , \label{eq:AC}
\end{equation}
which isolates the BH-DVCS interference contribution at the numerator and the DVCS amplitude at the denominator. Following the harmonic decomposition of observables proposed in Ref.~\cite{Bel02},
\begin{eqnarray}
d^4 \sigma_{UU}^{\rm BH} & = & \frac{K_1}{\mathcal{P}_1(\phi) \, \mathcal{P}_2(\phi)} \sum_{n=1}^2 c_{n, \rm {unp}}^{\rm BH} \cos(n\phi) \\
d^4 \sigma_{UU}^{\rm DVCS} & = & \frac{K_3}{Q^2} \sum_{n=0}^2 c_{n, \rm {unp}}^{\rm DVCS}\cos(n\phi) \, , 
\end{eqnarray}
and
\begin{equation}
d^4 \sigma_{UU}^{\rm I} = \frac{K_2}{\mathcal{P}_1(\phi) \, \mathcal{P}_2(\phi)} \sum_{n=0}^3 c_{n, \rm {unp}}^{\rm I}\cos(n\phi) \label{eq:int}
\end{equation}
where $K_i$'s are kinematical factors, and $P_i(\phi)$'s are the BH propagators. Because of the $1/Q^2$ kinematical suppression of the DVCS amplitude, the dominant contribution to the denominator of Eq.~\ref{eq:AC} originates from the BH amplitude. At leading twist, the dominant coefficients to the numerator are $c_{0,{\rm{unp}}}^{\rm I}$ and $c_{1,{\rm{unp}}}^{\rm I}$  
\begin{eqnarray}
c_{0,\rm{unp}}^{\rm I} & \propto & - \frac{\sqrt{-t}}{Q} \, c_{1,\rm{unp}}^{\rm I} \label{eq:r} \\
\label{eq:cosphi_term_bca}
c_{1,{\rm{unp}}}^{\rm I} & \propto & \Re{\rm e} \left[F_1 { \cal H}+\xi (F_1 + F_2)\widetilde{\cal H}-\frac{t}{4M^2}F_2{\cal E} \right] \, .  \label{c1cof}
\end{eqnarray}
Given the relative strength of $F_1$ and $F_2$ at small $t$ for a neutron target, the beam charge asymmetry becomes
\begin{equation}
A_{\rm C}(\phi) \propto \frac{1}{F_2} \, \Re{\rm e} \left[ \xi \widetilde{{\cal H}_n} - \frac{t}{4M^2} {\cal E}_n \right] \, .
\end{equation}
Therefore, the BCA is mainly sensitive to the real part of the GPD $E_n$ and, for selected kinematics, to the real part of the GPD $\widetilde{H_n}$.

Considering polarized electron and positron beams, two additional observables can be constructed: the charge difference ($\Delta_{\rm C}^{LU}$) and the charge average ($\Sigma_{\rm C}^{LU}$) beam helicity asymmetries~\cite{Hos16}: 
\begin{eqnarray}
\Delta_{\rm C}^{LU}(\phi) & = & \frac{(d^4\sigma^+_+ - d^4\sigma^+_-)-(d^4\sigma^-_+ - d^4\sigma^-_-)}{d^4\sigma^+_+ + d^4\sigma^+_- + d^4\sigma^-_+ + d^4\sigma^-_-} = \frac{d^4\sigma_{LU}^{\rm I}} {d^4\sigma_{UU}^{\rm BH} + d^4\sigma_{UU}^{\rm DVCS}} \label{eq:BSA_diff}
\\
\Sigma_{\rm C}^{LU} (\phi) & = & \frac{(d^4\sigma^+_+ - d^4\sigma^+_-)+(d^4\sigma^-_+ - d^4\sigma^-_-)}{d^4\sigma^+_+ + d^4\sigma^+_- + d^4\sigma^-_+ + d^4\sigma^-_-} = \frac{d^4\sigma_{LU}^{\rm DVCS}} {d^4\sigma_{UU}^{\rm BH} + d^4\sigma_{UU}^{\rm DVCS}} \label{eq:BSA_sum} 
\end{eqnarray}
which single out the sensitivity to the beam polarization of the interference and DVCS amplitudes. Following Ref.~\cite{Bel02}, these can be written as:
\begin{eqnarray}
d^4 \sigma_{LU}^{\rm I} & = & \frac{K_2}{\mathcal{P}_1(\phi) \, \mathcal{P}_2(\phi)} \sum_{n=1}^2 s_{n, \rm {unp}}^{\rm I}\sin(n\phi) \\
d^4 \sigma_{LU}^{\rm DVCS} & = & \frac{K_3}{Q^2} \, s_{1, \rm {unp}}^{\rm DVCS} \sin(\phi) \, ,
\end{eqnarray}
where $s_{1, \rm {unp}}^{\rm I}$ is the dominant twist-2 contribution, proportional to the imaginary part of $c_{1,{\rm{unp}}}^{\rm I}$ (Eq.~\ref{c1cof}), and $s_{2, \rm {unp}}^{\rm I}$ and $s_{1, \rm {unp}}^{\rm DVCS}$ are twist-3 contributions with distinct GPD dependence and different harmonic behaviour. Eq.~\ref{eq:BSA_diff} should be compared to the beam-spin asymmetry (BSA) observable 
\begin{equation}
A_{LU}(\phi) = \frac{d^4\sigma^-_+ - d^4\sigma^-_-} {d^4\sigma^-_+ + d^4\sigma^-_-} =  
               \frac{d^4\sigma_{LU}^{\rm DVCS}-d^4\sigma_{LU}^{\rm I}} {d^4\sigma_{UU}^{\rm BH} - d^4\sigma_{UU}^{\rm I}  + d^4\sigma_{UU}^{\rm DVCS}} \, , \label{eq:BSA}
\end{equation}
to be measured at CLAS12 with a polarized electron beam. Eq.~\ref{eq:BSA} can be expressed as
\begin{equation}
A_{LU} = \frac{\Sigma_{\rm C}^{LU}} {1 - A_{\rm C}} - \frac{\Delta_{\rm C}^{LU}}{1 - A_{\rm C}} \, .
\end{equation}
At leading twist $\Sigma_{\rm C}^{LU}$=$0$, and $A_{LU}$ differs from $\Delta_{\rm C}^{LU}$ due to the contribution of the polarization-independent part of the interference amplitude in the denominator. In that sense, $\Delta_{\rm C}^{LU}$ offers a complementary observable for the extraction of the CFFs. On the other hand, $\Sigma_{\rm C}^{LU}$ represents a new observable measuring the effects of higher twist in the $eN \gamma$ reaction. However, corresponding asymmetries are assumed to be small and then difficult to assess with precision. While the present letter concerns the BCA measurement, it should be stressed that polarization observables will come from free if, as expected, the proposed positron beam at JLab operates similarly to the actual electron beam. 

\subsection{Experimental set-up}

We are proposing to measure the beam charge asymmetry for the electroproduction of photons on the neutron using a liquid deuterium target, the 11 GeV CEBAF electron beam, and the proposed 11 GeV positron beam. The scattered electrons/positrons and photons  will be detected with the CLAS12 detector in its baseline configuration, completed at small angles with the Forward Tagger (FT)~\cite{Bat11}.  The detection of the active neutrons will be accomplished with the CND (Central Neutron Detector) and the CTOF (Central Time-of-Flight) at backwards angles, and the FEC (Forward Electromagnetic Calorimeter), the PCAL (Preshower Calorimeter), and the FTOF (Forward Time-of-Flight) at forward angles. In order to match the detector acceptance for the different lepton beam charges, the positron data taking will be performed with opposite polarities for the CLAS12 torus and solenoid, with respect to the electron data taking. 

\begin{figure}[t!]  
\begin{center}
\begin{minipage}[l]{0.495\textwidth}
\begin{center}
\includegraphics[width=0.95\textwidth]{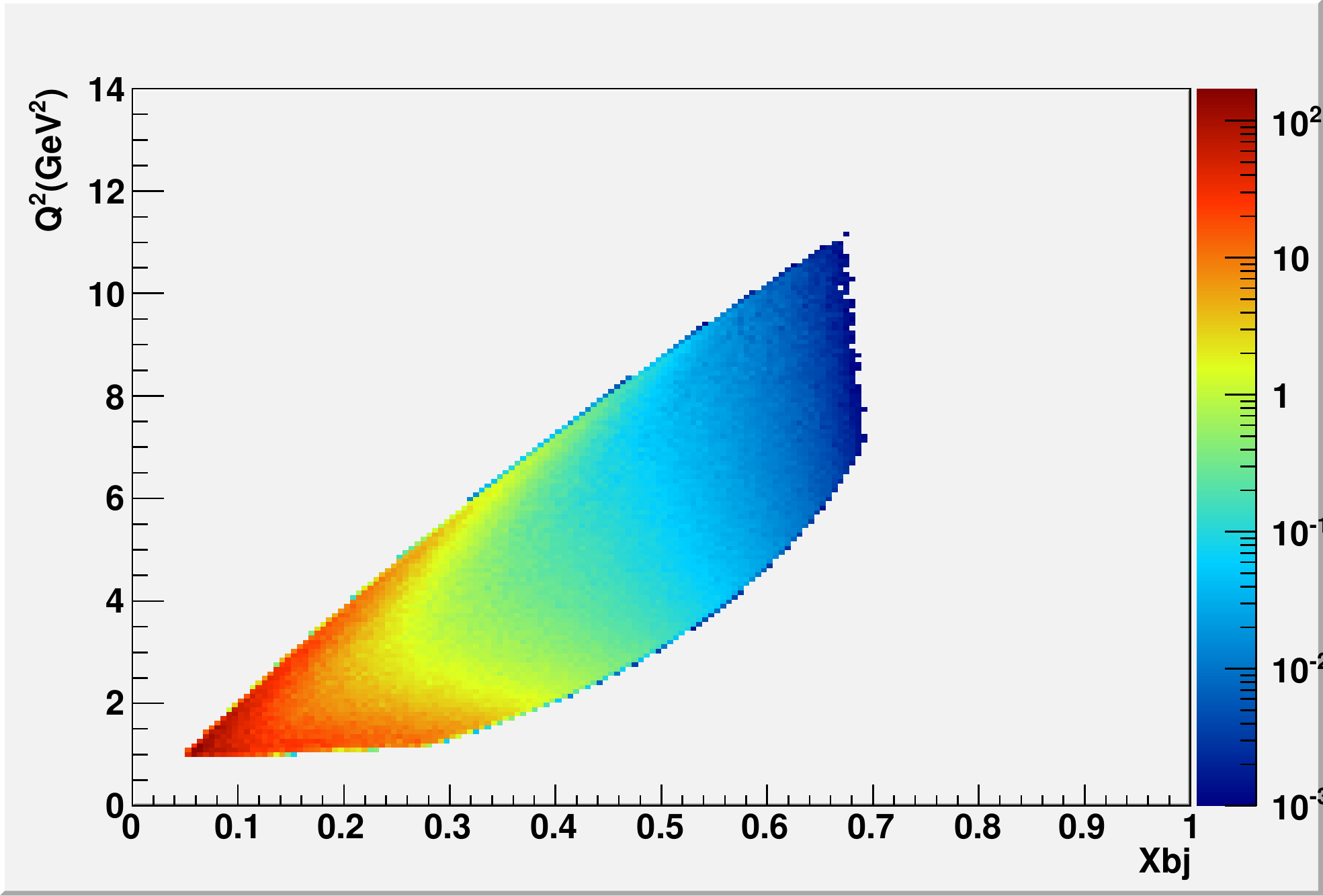}
\includegraphics[width=0.95\textwidth]{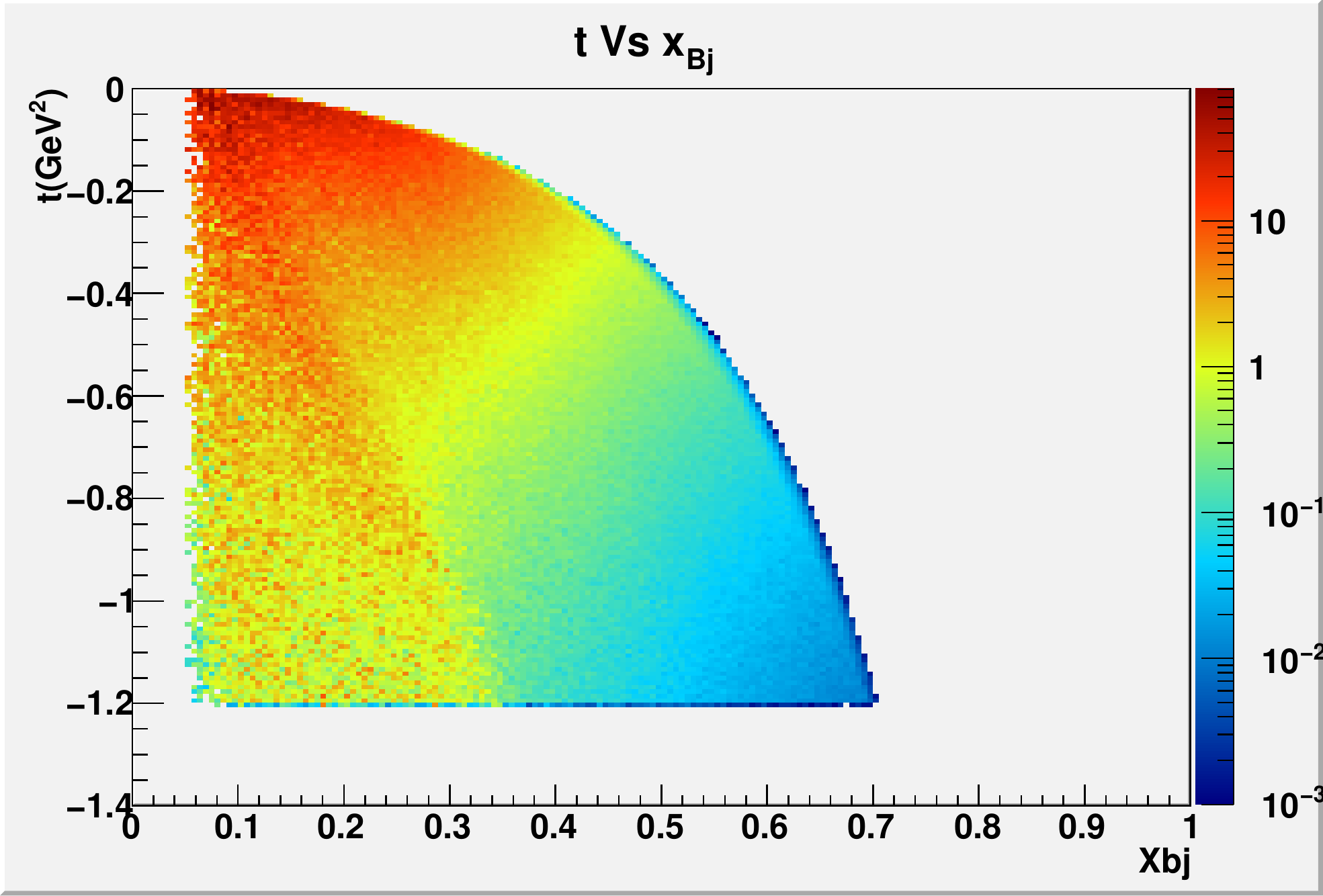}
\includegraphics[width=0.95\textwidth]{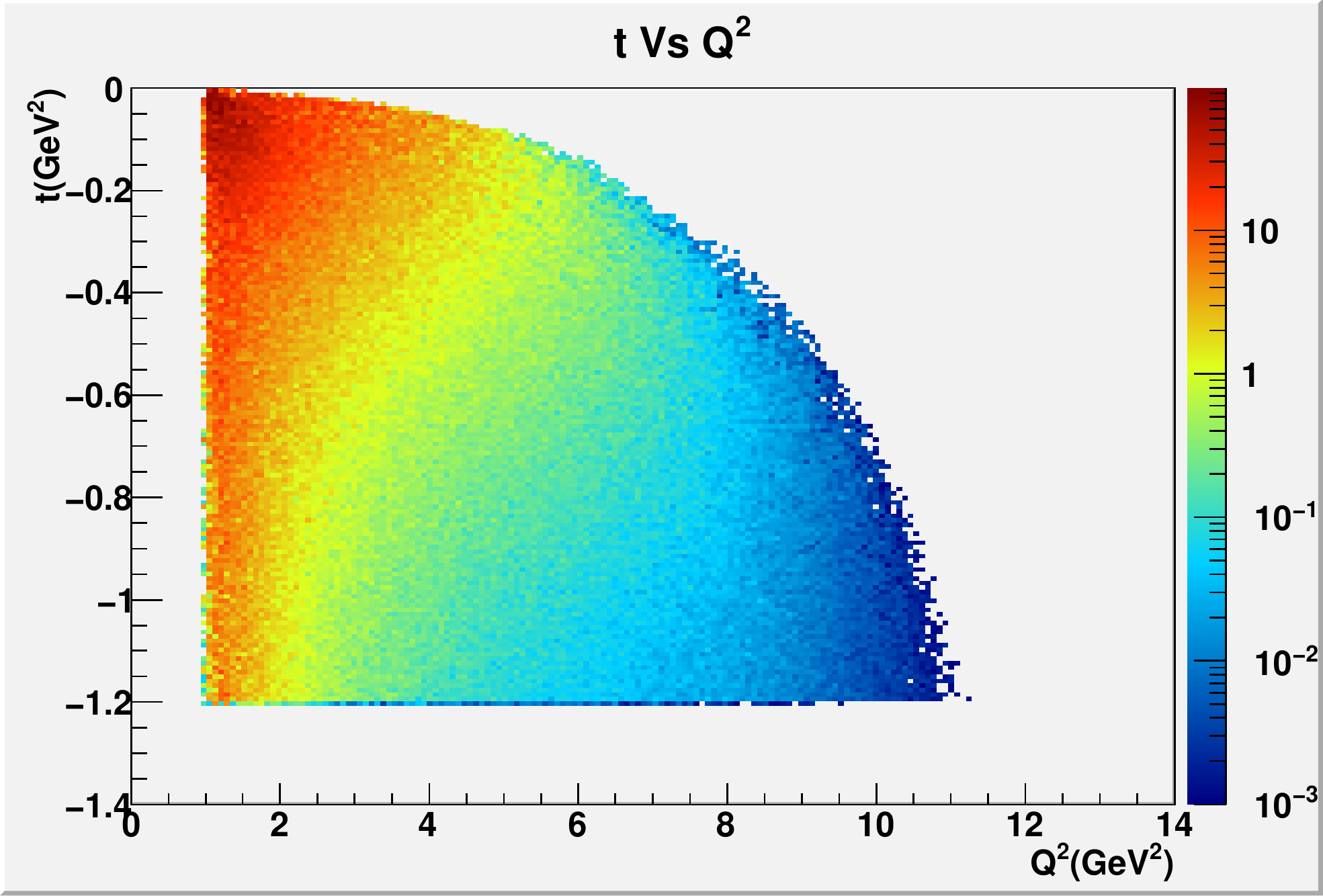}
\end{center}
\end{minipage}
\begin{minipage}[r]{0.495\textwidth}
\begin{center}
\includegraphics[width=0.95\textwidth,height=0.195\textheight]{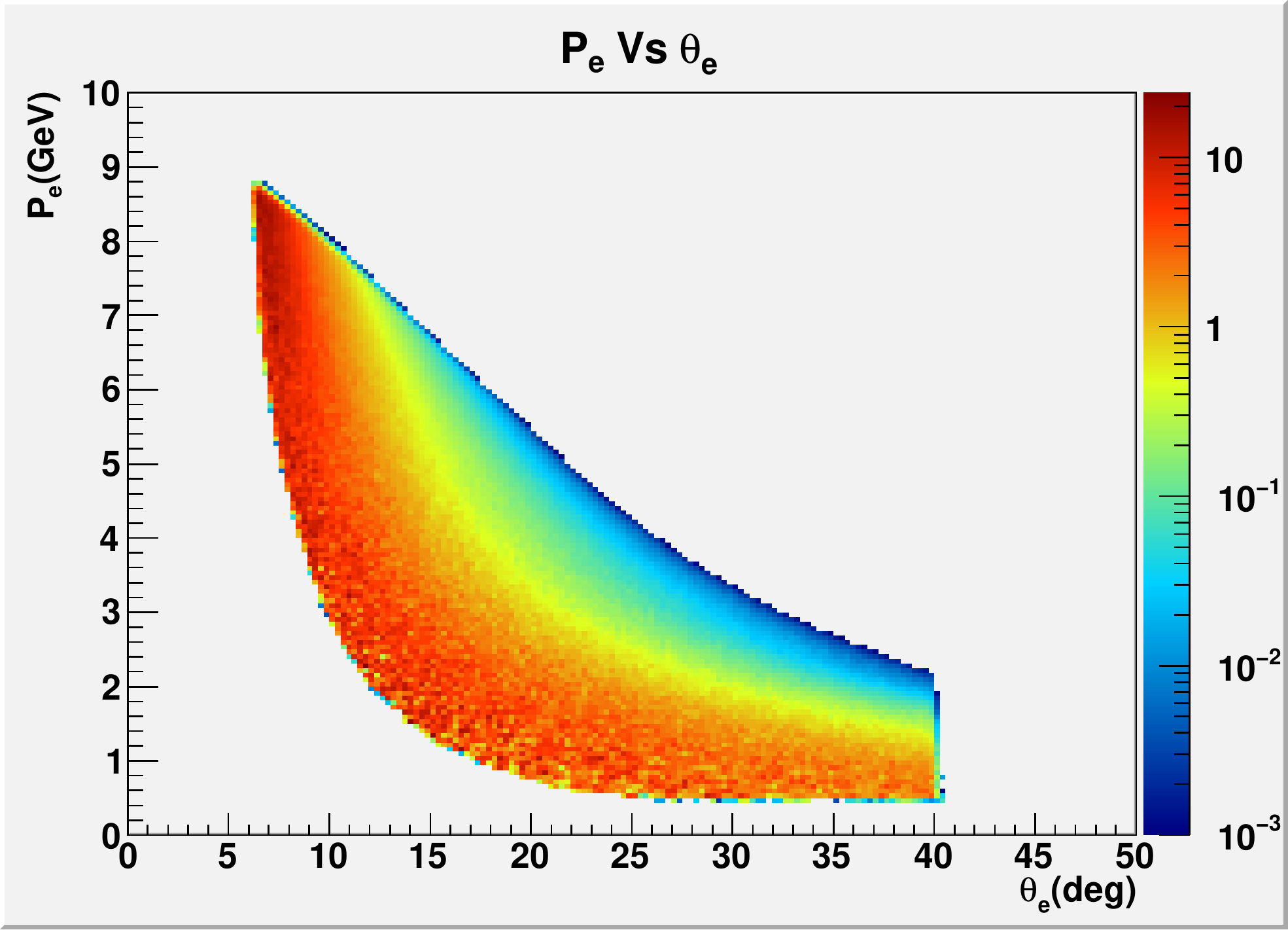}
\includegraphics[width=0.95\textwidth]{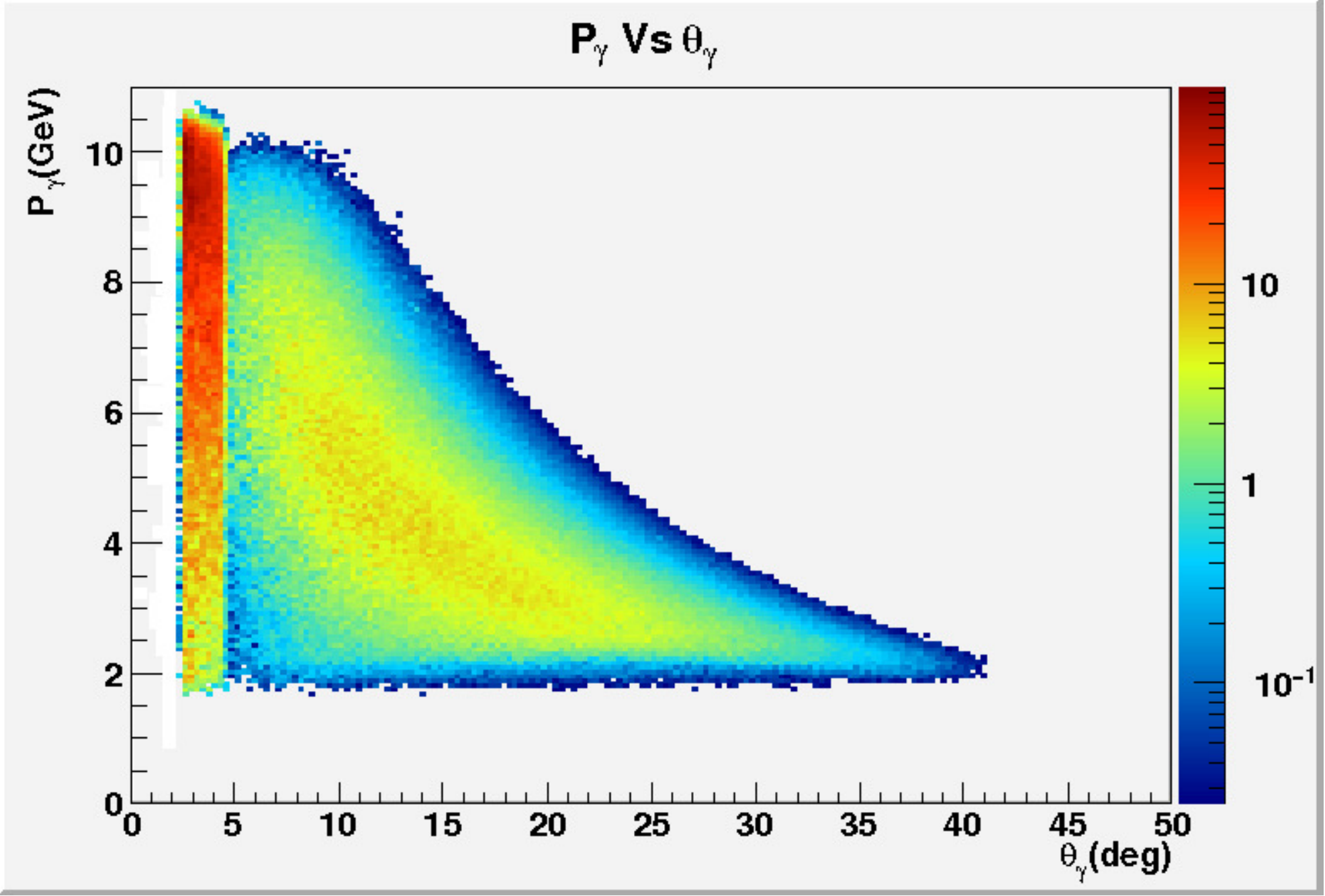}
\includegraphics[width=0.95\textwidth]{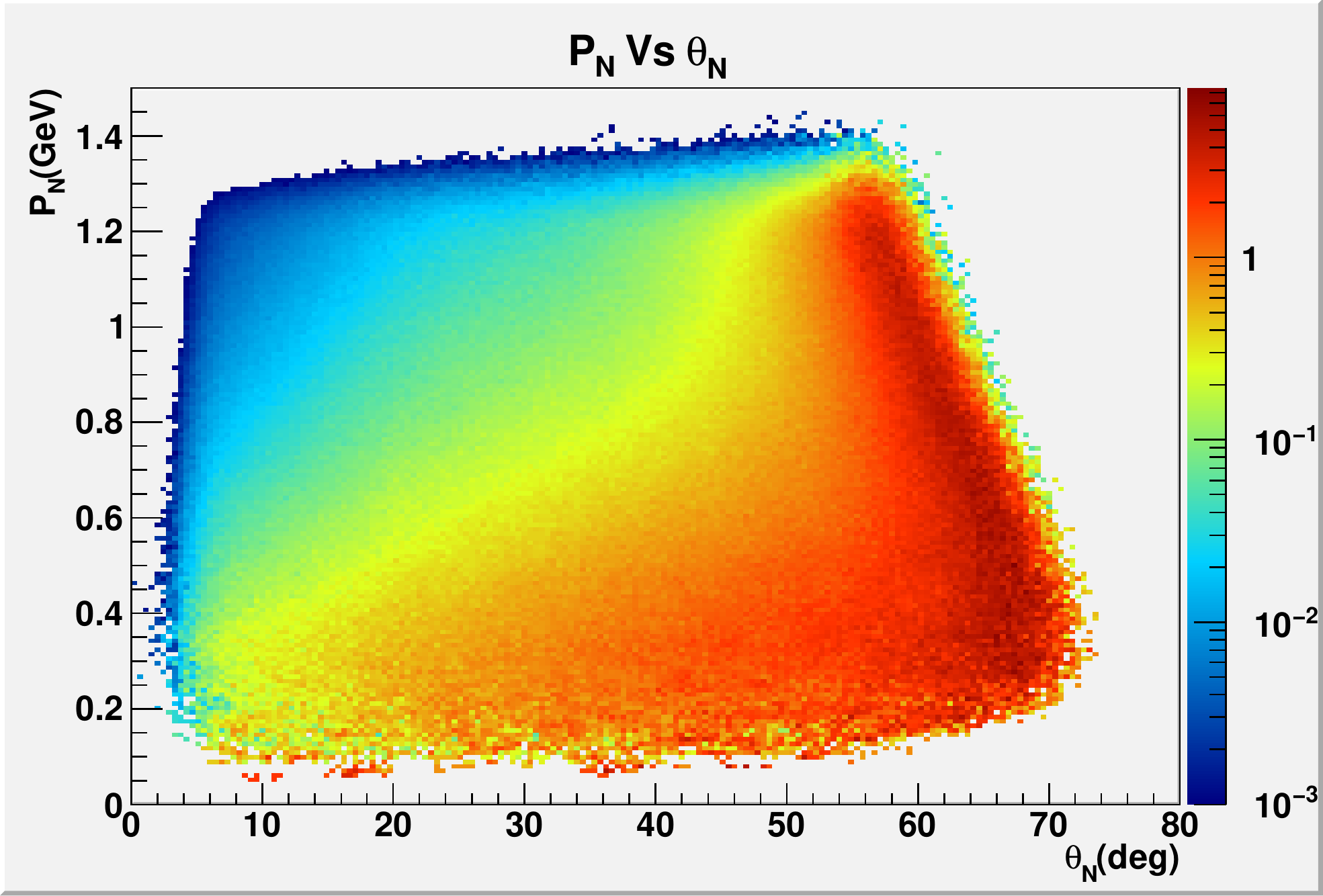}
\end{center}
\end{minipage}
\caption{(left) Distributions of the kinematic variables for n-DVCS events, including acceptance and physics cuts: $(Q^2,x_B)$ phase space (top), $(t,x_B)$ phase space (middle), and $(t,Q^2)$ phase space (bottom). (right) Momentum distribution as function of polar angle for the $e n \gamma(p)$ final state: electron/positron (top), photon (middle), and neutron momentum (bottom).}
\label{kine_vars}
\end{center}
\end{figure}
An event generator (GENEPI) for the DVCS, the BH and exclusive $\pi^0$ electroproduction processes on the neutron inside a deuterium target was developed~\cite{Ala09}. The DVCS amplitude is calculated according to the BKM formalism~\cite{Bel02}, while the GPDs are taken from the standard CLAS DVCS generator~\cite{Van99,Goe01}. The initial Fermi-motion distribution of the neutron is determined from the Paris potential~\cite{Lac80}. The output of the event generator was fed through CLAS12 FASTMC, to simulate acceptance and resolution effects in CLAS12. Kinematic cuts to ensure the applicability of the GPD formalism  ($Q^2>1$~GeV$^2$/$c^2$, $t>-1.2$~GeV$^2$/$c^2$, $W>2$ GeV) have been applied. Figure~\ref{kine_vars} (left) shows the coverage in $Q^2$, $x_B$ and $t$ obtained for the D$(e,e n \gamma)p$ reaction with an electron or positron beam energy of 11 GeV and the appropriate magnet polarities. The three plots in Fig.~\ref{kine_vars} (right) show the energy/momentum distributions of the final state particles: as expected,the scattered leptons and the photons are mostly emitted at forward angles, while the recoil neutrons populate dominantly the backward angles region.

\subsection{Projections for the beam-charge asymmetry} 

The expected number of reconstructed $e n \gamma(p)$ events was determined as a function of the kinematics. An overall 10\% neutron-detection efficiency for neutrons with $\theta>40^o$ was assumed (CND+CTOF). The detection efficiencies for electrons/positrons and photons are assumed to be 100\%, within the fiducial cuts. Considering the always-improving performance of the CLAS12 data-acquisition system, the operation of CLAS12 at its design luminosity ${\cal L}=10^{35}$~cm$^{-2} \cdot$s$^{-1}$ per nucleon, corresponding to 60~nA electron and positron beam currents, is assumed for the present data projections. An overall data 
\begin{figure}[t!] 
\begin{center}
\includegraphics[width=0.860\textwidth]{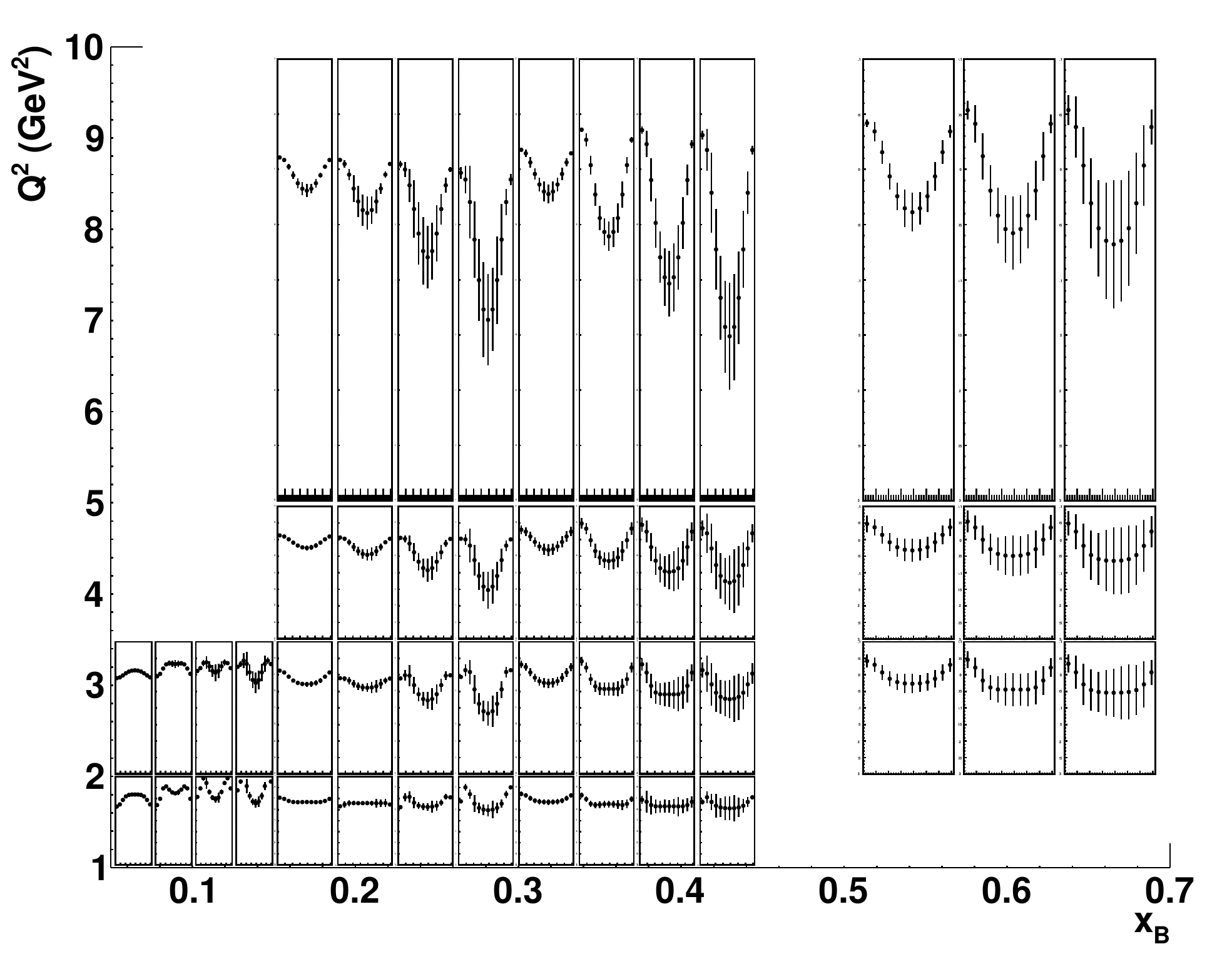}
\includegraphics[width=0.860\textwidth]{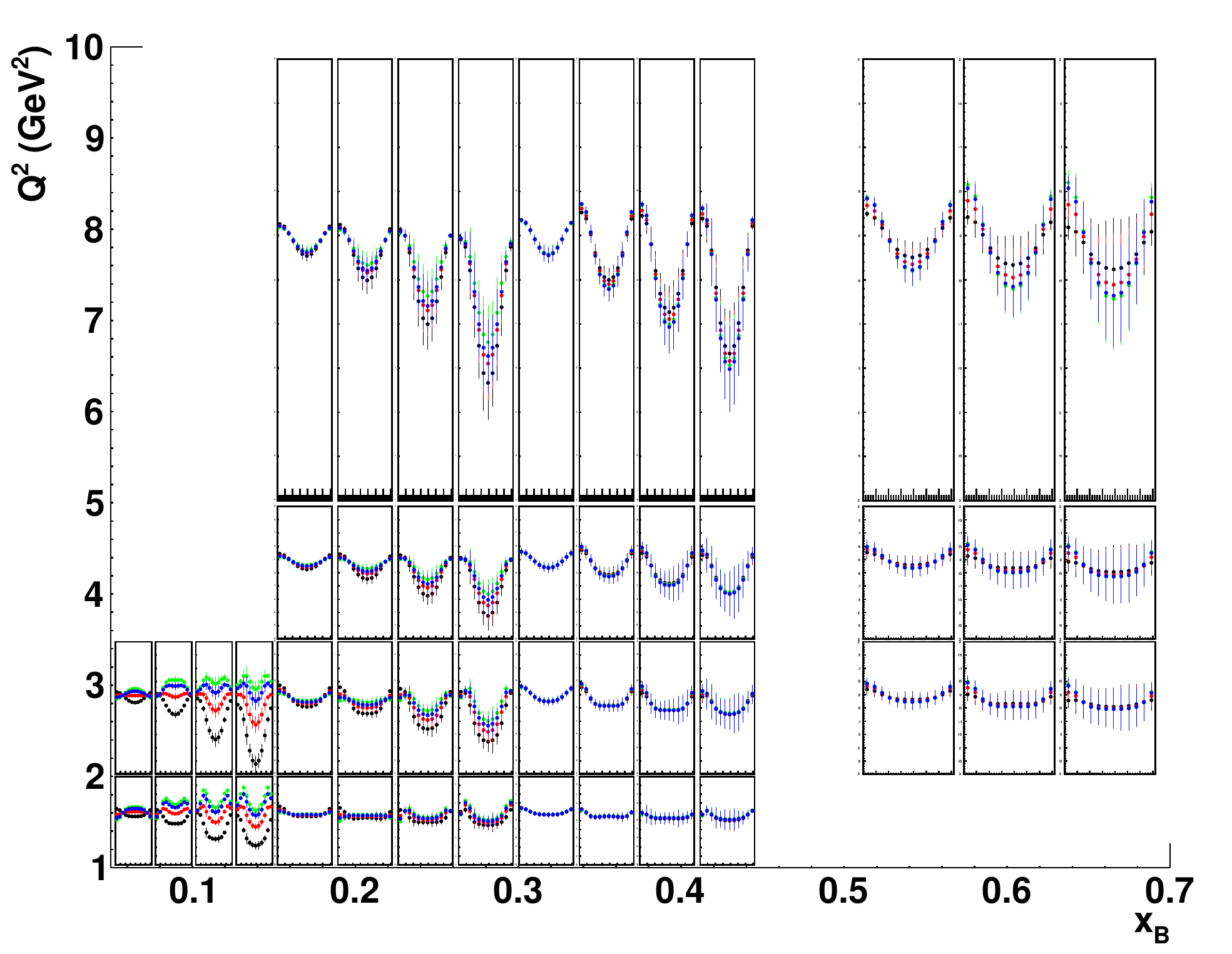}
\caption {Projected BCA data for the D$(e,e n \gamma)p$ reaction as predicted by the VGG model for $(J_u,J_d)=(0.3,0.1)$ (top) and alternative combinations (bottom). The bottom plot compares $(J_u,J_d)$: $(0.3,0.1)$ (black), $(0.2,0.0)$ (red), $(0.1,$-$0.1)$ (green), and $(0.3,$-$0.1)$ (blue). The vertical axis scale ranges from -0.3 to 0.1 for the top plot and from -0.3 to 0.2 for the bottom plot. The error bars reflect the expected statistical uncertainties for 80 days of beam time at a luminosity of $10^{35}$~cm$^{-2} \cdot$s$^{-1}$ per nucleon.}
\label{asym_ndvcs}
\end{center}
\end{figure}
taking time of 80 days, equally shared between electrons and positrons, is also considered. The following 4-dimensional grid of bins has been adopted:
\begin{itemize}
\item{4 bins in $Q^2~[1, 2, 3.5, 5, 10$ GeV$^2$/$c^2]$;}
\item{4 bins in $-t~ [0 ,0.2 ,0.5 ,0.8 , 1.2$ GeV$^2$/$c^2]$;}
\item{4 bins in $x_B~ [0.05,0.15,0.3,0.45,0.7]$;} 
\item{12 bins in $\phi$, each $30^o$ wide.}
\end{itemize}
For each bin, the beam charge asymmetry (BCA) is experimentally reconstructed as 
\begin{equation}
A_{\rm C} = \frac{(N^+/Q^+)-(N^-/Q^-)}{(N^+/Q^+)-(N^-/Q^-)}
\end{equation}
where $Q^{\pm}$ is the integrated charge for lepton beam of each polarity ($Q^+$=$Q^-$ in the present evaluation), and $N^{\pm}$ is the corresponding number of $e n \gamma(p)$ events. For each bin $N^{\pm}$ is computed as: 
\begin{equation}
N^{\pm} = {\cal L}^{\pm} \cdot T \cdot \frac{d\sigma}{dQ^2dx_Bdtd\phi} \cdot \Delta t\cdot \Delta Q^2 \cdot\Delta x_B\cdot \Delta\phi \cdot {\cal A} \cdot \epsilon_n \, ,
\end{equation}
where ${\cal L}^{\pm}$ is the beam luminosity, $T$ is the running time, $d^4\sigma/dQ^2dx_Bdtd\phi$ is the 4-fold differential cross section, $\Delta Q^2 \Delta x_B \Delta t \Delta \phi$ is the full bin width, ${\cal A}$ is the bin by bin acceptance, and $\epsilon_n$ is the neutron-detection efficiency. The statistical errors on the BCA depend on the BCA magnitude via the formula: 
\begin{equation}
\sigma \left( A_{\rm C} \right) = \sqrt{\frac{1-A_{\rm C}^2}{N}}
\end{equation}
where $N$=$N^++N^-$ is the total number of events in each bin. Figure~\ref{asym_ndvcs}(top) shows the expected statistical accuracy of the proposed BCA measurement. The magnitude of the BCA is obtained for each bin with the VGG model assuming $J_u=0.3$ and $J_d=0.1$. 
Figure~\ref{asym_ndvcs}(bottom) shows the BCA for different $(J_u,J_d)$ values. It should be noted that the BCA is particularly  sensitive to $(J_u,J_d)$ at small $x_B$, in comparison to the beam-spin asymmetry which depends linearly on $(J_u,J_d)$. This is most likely an effect of the $x$-dependence of GPDs. \newline
Summing $N^{\pm}$ over for the full grid of bins, about 25$\times 10^6$ $e n \gamma (p)$ events are expected to be collected over the full kinematic range for 80 days of running.

\subsection{Extraction of Compton form factors} \label{sec_cff}

In order to establish the impact of proposed experiment on the CLAS12 n-DVCS program, the four sets of projected asymmetries BSA~\cite{Nic11}), TSA and DSA~\cite{Nic15}, and BCA (Fig.~\ref{asym_ndvcs}(top)), for all kinematic bins, were processed using a fitting procedure~\cite{Gui08,Gui13} to extract the neutron CFFs. This approach is based on a local-fitting method at each given experimental $(Q^2, x_B,-t)$ kinematic point.  In this framework, there are eight real CFF-related quantities 
\begin{eqnarray}
F_{Re}(\xi,t) & = & \Re{\rm e} \left[ {\cal F}(\xi,t) \right] \\
F_{Im}(\xi,t) & = & -\frac{1}{\pi} \Im{\rm m} \left[ {\cal F}(\xi,t) \right] = \left[ F(\xi,\xi,t)\mp F(-\xi,\xi,t) \right], 
\end{eqnarray}
\begin{figure}[t!]  
\begin{center}
\includegraphics[width=0.870\textwidth]{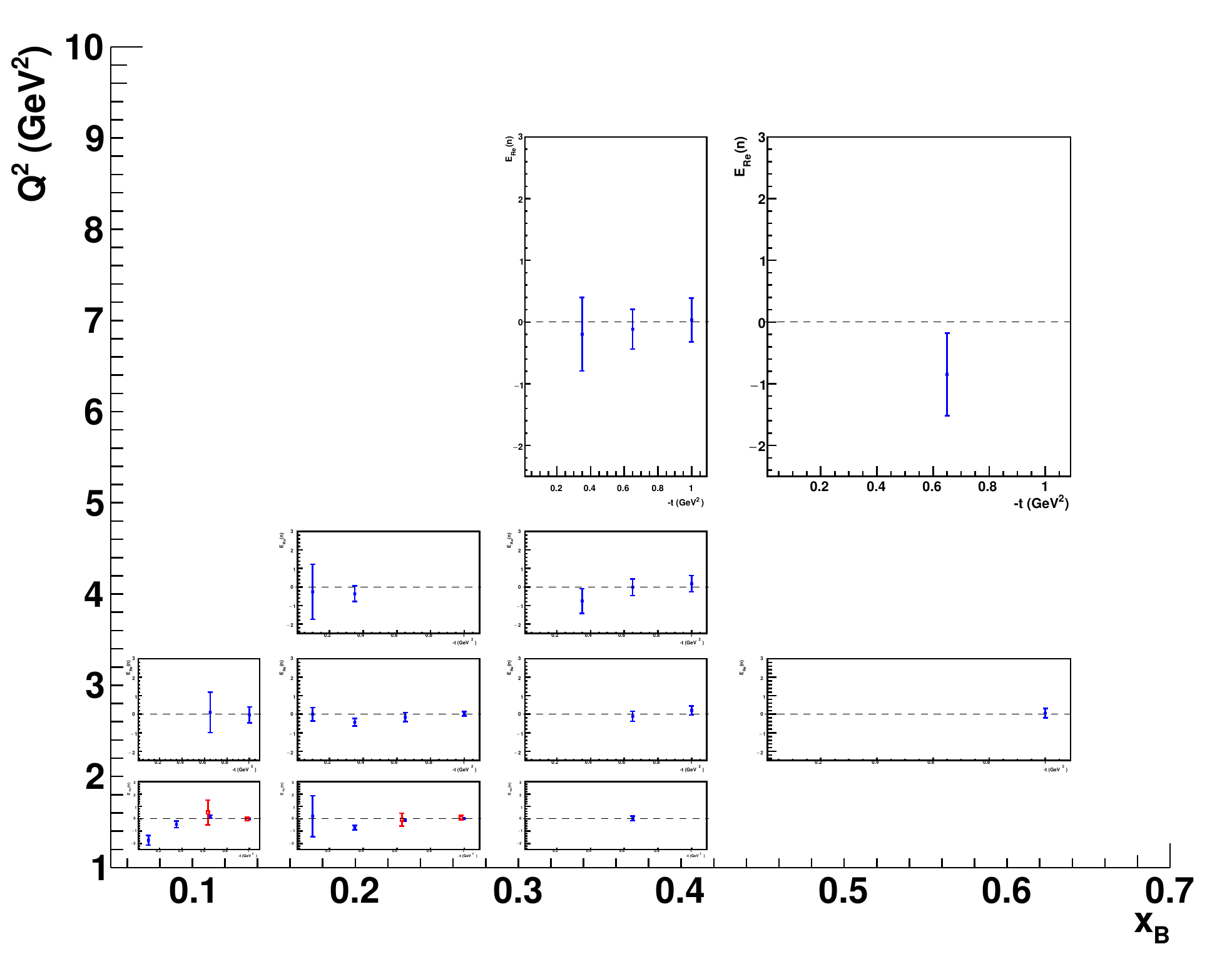}
\caption{$E_{Re}(n)$ as a function of $-t$, for all bins in $Q^2$ and $x_B$. The blue points are the results of the fits including the proposed BCA while the red ones include only already approved experiments.}
\label{cff_ere}
\end{center}
\end{figure}
\begin{figure}[h!]  
\begin{center}
\includegraphics[width=0.870\textwidth]{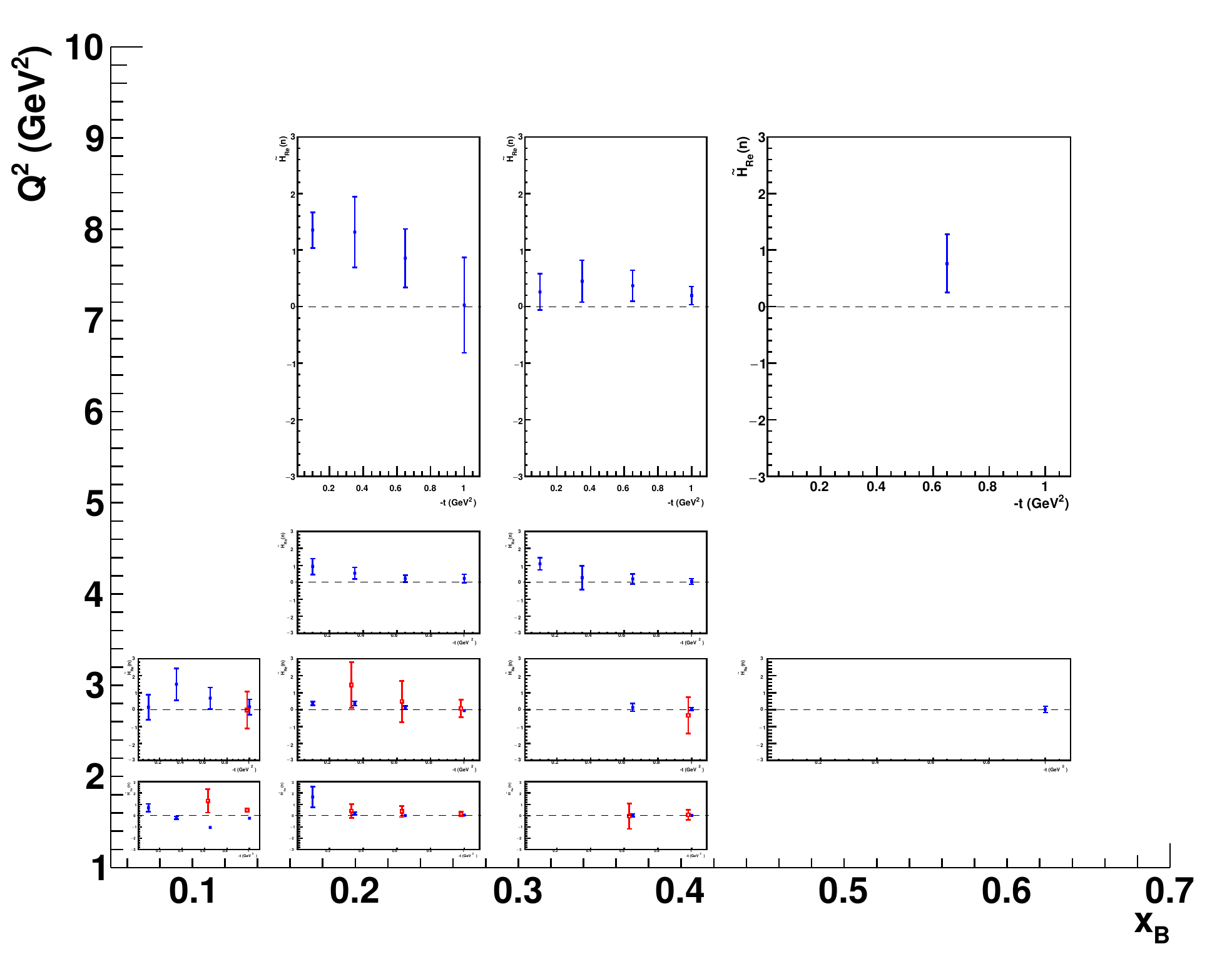}
\caption{$\widetilde{{H}_{Re}}(n)$ as a function of $-t$, for all bins in $Q^2$ and $x_B$. The blue points are the results of the fits including the proposed BCA while the red ones include only already approved experiments.}
\label{cff_htre}
\end{center}
\end{figure}

\begin{figure}[t]  
\begin{center}
\includegraphics[width=0.870\textwidth]{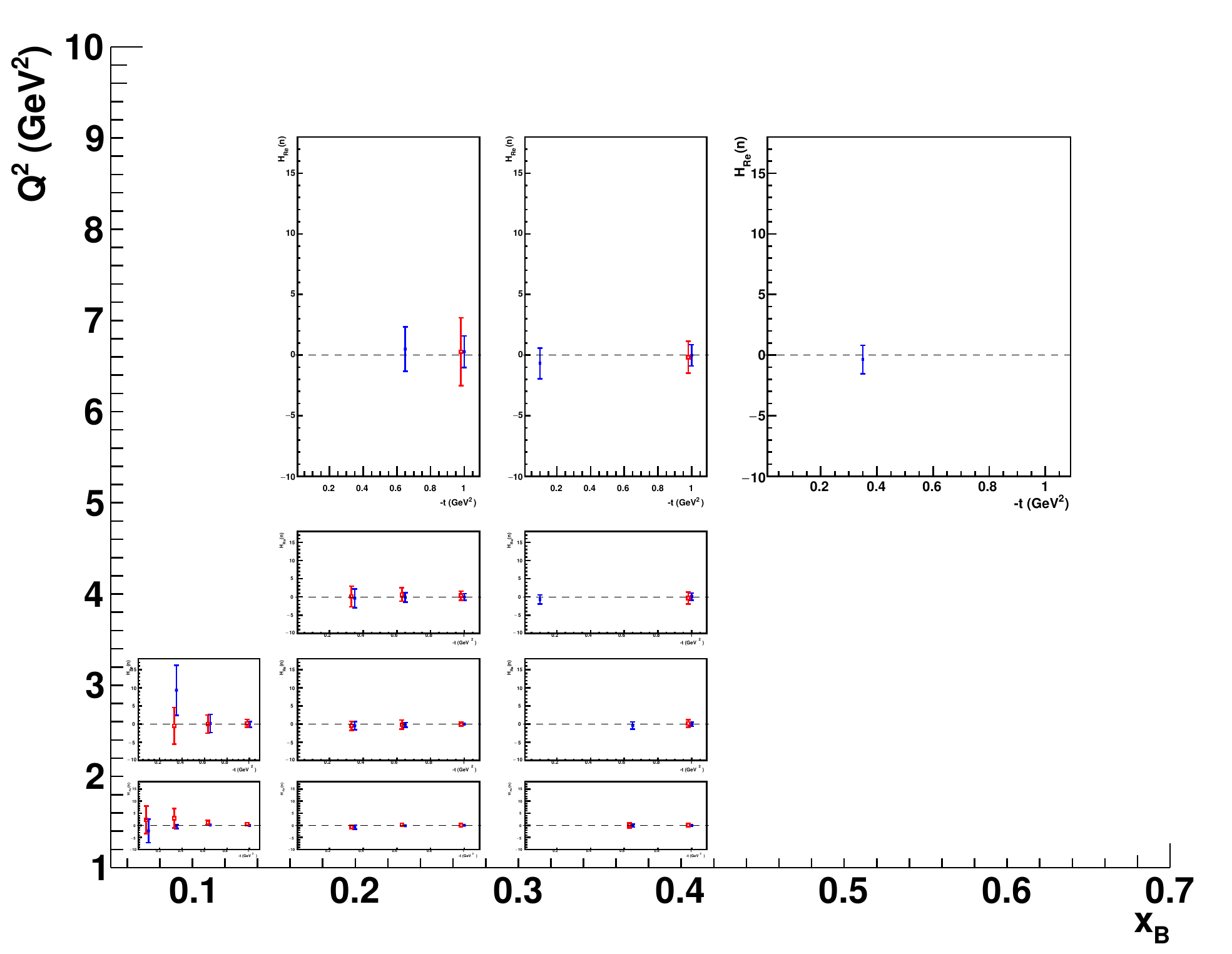}
\caption{$H_{Re}(n)$ as a function of $-t$, for all bins in $Q^2$ and $x_B$. The blue points are the results of the fits including the proposed BCA while the red ones include only already approved experiments.}
\label{cff_hre}
\end{center}
\end{figure}
\begin{figure}[h!]  
\begin{center}
\includegraphics[width=0.870\textwidth]{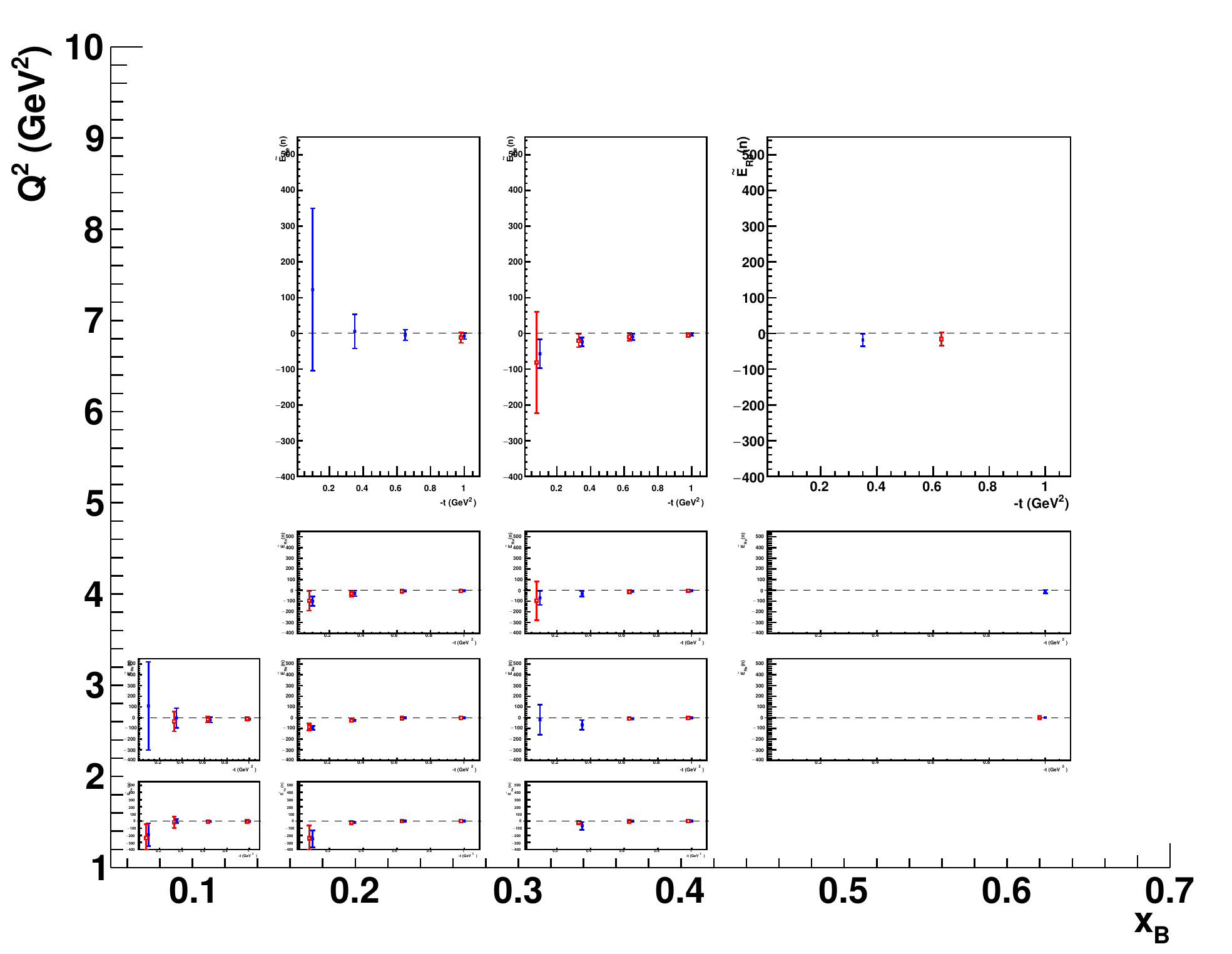}
\caption{$\widetilde{{E}_{Re}}(n)$ as a function of $-t$, for all bins in $Q^2$ and $x_B$. The blue points are the results of the fits including the proposed BCA while the red ones include only already approved experiments.}
\label{cff_etre}
\end{center}
\end{figure}

\begin{figure}[t!]  
\begin{center}
\includegraphics[width=0.870\textwidth]{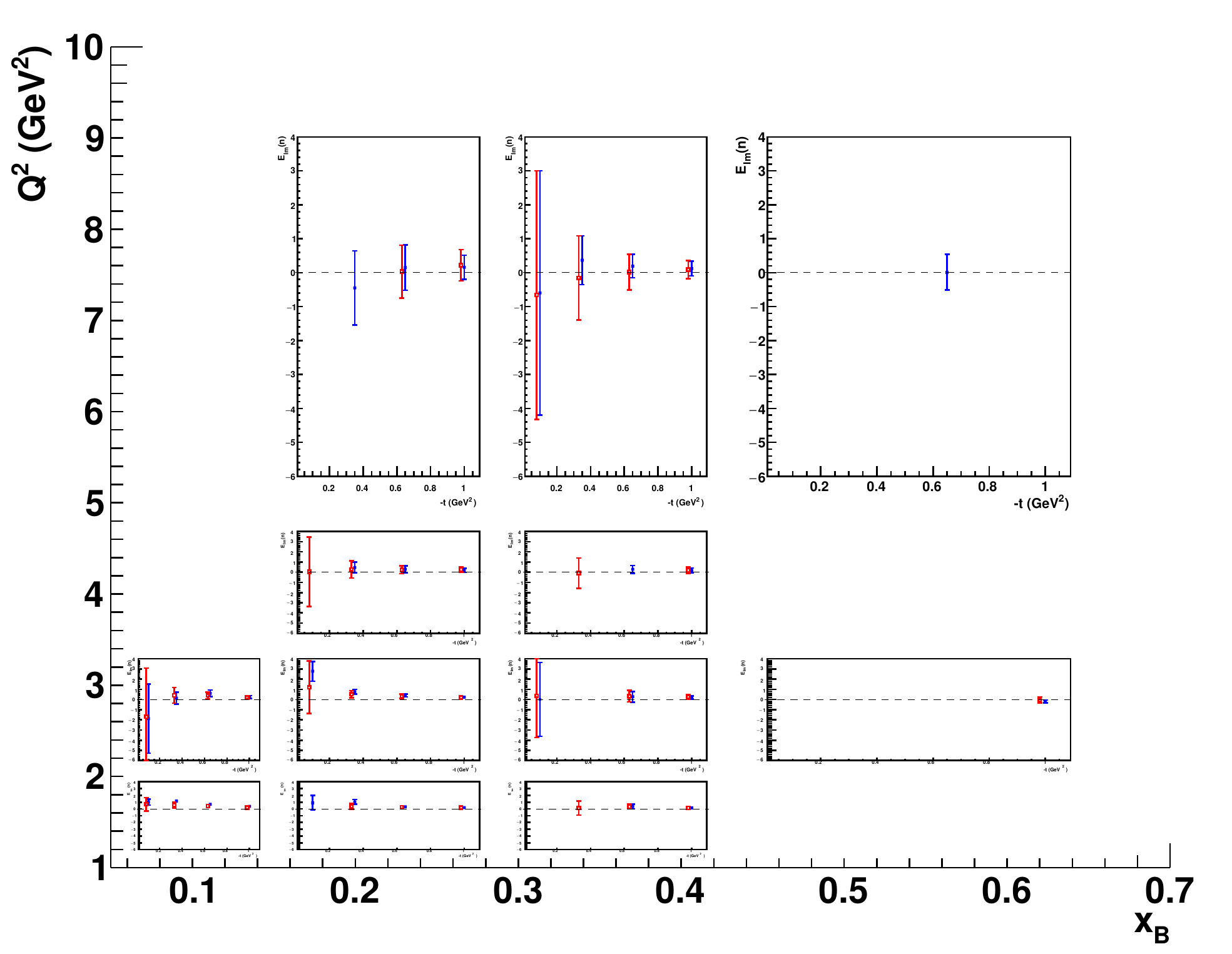}
\caption{$E_{Im}(n)$ as a function of $-t$, for all bins in $Q^2$ and $x_B$. The blue points are the results of the fits including the proposed BCA while the red ones include only already approved experiments.}
\label{cff_eim}
\end{center}
\end{figure}
\begin{figure}[h!]  
\begin{center}
\includegraphics[width=0.870\textwidth]{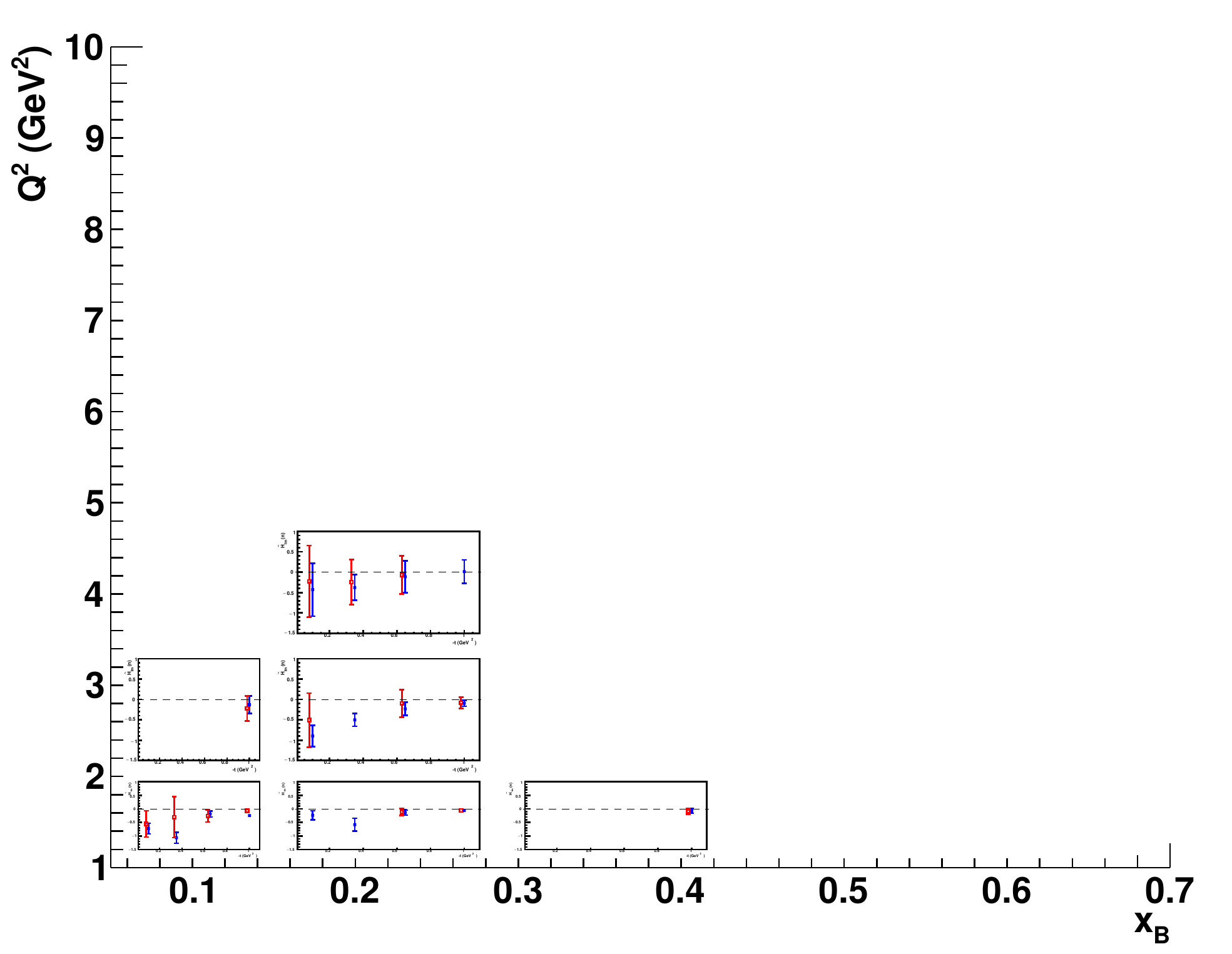}
\caption{$\widetilde{{H}_{Im}}(n)$ as a function of $-t$, for all bins in $Q^2$ and $x_B$. The blue points are the results of the fits including the proposed BCA while the red ones include only already approved experiments.}
\label{cff_htim}
\end{center}
\end{figure}

\begin{figure}[t!]  
\begin{center}
\includegraphics[width=0.870\textwidth]{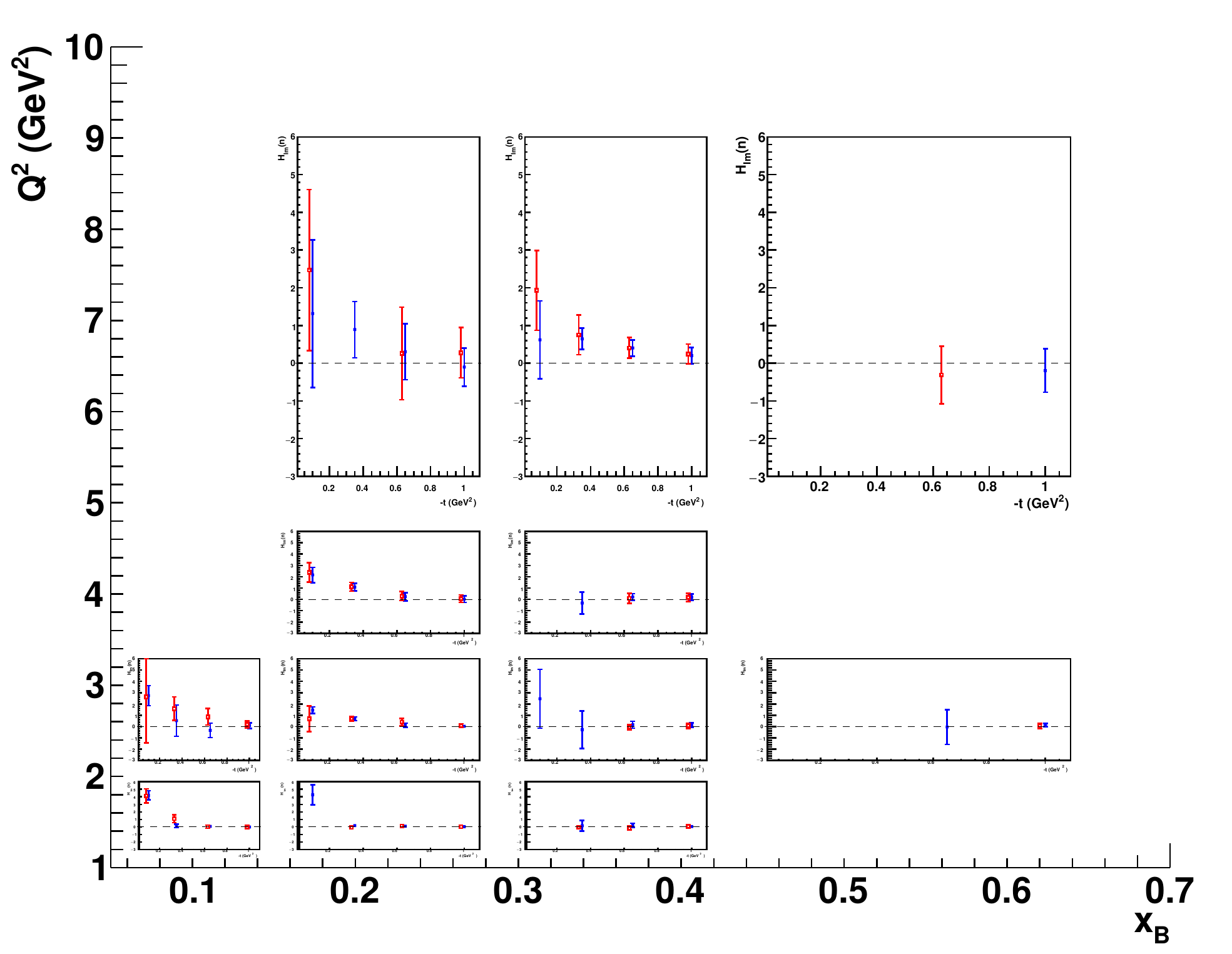}
\caption{$H_{Im}(n)$ as a function of $-t$, for all bins in $Q^2$ and $x_B$. The blue points are the results of the fits including the proposed BCA while the red ones include only already approved experiments.}
\label{cff_him}
\end{center}
\end{figure}
where the sign convention is the same as for Eq.~\ref{dvcs-ampl}. These CFFs are the almost-free\footnote{The values of the CFFs are allowed to vary within $\pm 5$ times the values predicted by the VGG model~\cite{Van99,Gui05}.} parameters to be extracted from DVCS observables using the well-established theoretical description of the process based on the DVCS and BH mechanisms. The BH amplitude is calculated exactly while the DVCS one is determined at the QCD leading twist~\cite{Van99}. 

As there are eight CFF-related free parameters, including more observables measured at the same kinematic points  will result in tighter constraints on the fit and will increase the number of extracted CFFs and their accuracy. In the adopted version of the fitter code, $\widetilde{E_{Im}}(n)$ is set to zero, as $\widetilde{E_n}$ is assumed to be purely real. Thus, seven out of the eight real and imaginary parts of the CFFs are left as free parameters in the fit. The results for the 7 neutron CFFs are shown in Figs.~\ref{cff_ere}-\ref{cff_htim}, as a function of $-t$, and for each bin in $Q^2$ and $x_B$. The blue points are the CFFs resulting  from the fits of the four observables, while the red ones are the CFFs obtained fitting only the projections of the currently approved n-DVCS experiments. The error bars reflect both the statistical precision of the fitted observables and their sensitivity to that particular CFF. Only results for which the error bars are non zero, and therefore the fits properly converged, are included in the figures. \newline
The major impact of the proposed experiment is, as expected, on $E_{Re}(n)$, for which the already approved projections have hardly any sensitivity. Thanks to the proposed BCA measurement, $E_{Re}(n)$ will be extracted over the whole phase space. A considerable extension in the coverage will be obtained also for $\widetilde{{H}_{Re}}(n)$. An overall improvement to the precision on the other CFFs, as well as an extension in their kinematic coverage will also be induced by the proposed n-DVCS BCA dataset. 

\subsection{Systematic uncertainties}

The goal of this experiment is to measure beam charge asymmetries which are ratios of absolute cross sections. In this ratio, 
several charge-independent

\begin{table}[h]
\begin{center}
\begin{tabular}{|c|c|}
\hline
             Source of error & $\sigma (A_{\rm C})^{Sys.}$ \\  \hline \hline
     Beam charge measurement &                        3\%  \\ \hline
       $\pi^0$ contamination &                        5\%  \\ \hline
                  Acceptance &                        3\%  \\ \hline
       Radiative corrections &                        3\%  \\ \hline
n-$\gamma$ misidentification &                        5\%  \\ \hline \hline
Total & 9\%  \\ \hline
\end{tabular}
\caption{Expected systematic uncertainties of the proposed measurement.}
\label{table_syst}
\end{center}
\end{table}
terms, such as acceptances, efficiencies, and radiative corrections, cancel out at first order. The BCA systematics comprises several contributions (Tab.~\ref{table_syst}) of comparable magnitude. The $\pi^0$-background evaluation, which depends on the accuracy of the description of the detector acceptance and efficiency, will contribute 5\% to the overall systematic uncertainties. A similar contribution is expected from n-$\gamma$ misidentification. Due to its strong variation as a function of $\phi$, the acceptance will bring an additional 3\% systematic error. A summary of the uncertainties induced by the various sources is reported in Tab.~\ref{table_syst}. The total systematic uncertainty is expected to be of the order of 9\%.

\subsection{Summary}

The strong sensitivity to the real part of the GPD $E^q$ of the beam charge asymmetry for DVCS on a neutron target makes the measurement of this observable particularly important for the experimental GPD program of Jefferson Lab. \newline
GEANT4-based simulations show that a total of 80 days of beam time at full luminosity with {\tt CLAS12} will allow to collect good statistics for the n-DVCS BCA over a large phase space. The addition of this observable to already planned measurements with {\tt CLAS12}, will permit the model-independent extraction of the real parts of the ${\cal E}_n$ and $\widetilde{{\cal H}_n}$ CFF of the neutron over the whole available phase space. Combining all the neutron and the proton CFFs obtained from the fit of n-DVCS and p-DVCS observables to be measured at {\tt CLAS12}, will ultimately allow the quark-flavor separation of all GPDs. 

%
%

\newpage

\null\vfill

\begin{center}

\section{\it Letter-ot-Intent: p-DVCS @ Hall C}

\vspace*{15pt}

{\Large{\bf Deeply Virtual Compton Scattering}}

\vspace*{3pt}

{\Large{\bf using a positron beam in Hall C}}

\vspace*{15pt}

{\bf Abstract}

\begin{minipage}[c]{0.85\textwidth}
We propose to use the High Momentum Spectrometer of Hall C combined with the Neutral Particle Spectrometer (NPS) to perform high precision measurements of the Deeply Virtual Compton Scattering (DVCS) cross section using a beam of positrons. The combination of measurements with opposite charge incident beams provide the only unambiguous way to disentangle the contribution of the DVCS$^2$ term in the photon electroproduction cross section from its interference with the Bethe-Heitler amplitude. A wide range of kinematics accessible with an 11 GeV beam off an unpolarized proton target will be covered. The $Q^2$-dependence of each contribution will be measured independently.
\end{minipage}

\vspace*{15pt} 

{\it Spokesperson: \underline{C.~Mu\~noz Camacho} (munoz@ipno.in2p3.fr)}

\vspace*{10pt} 

{and the}

\vspace*{7pt} 

{\large Neutral Particle Spectrometer (NPS) Collaboration}

\end{center}

\vfill\eject

%
%

\subsection{Introduction}

Deeply Virtual Compton Scattering refers to  the reaction $\gamma^*p\rightarrow p\gamma$ in the Bjorken limit of Deep Inelastic Scattering (DIS). Experimentally, we can access DVCS through electroproduction of real photons $ep\to ep\gamma$, where the DVCS amplitude interferes with the so-called Bethe-Heitler process. The BH contribution is calculable in QED since it corresponds to the emission of the photon by the incoming or the outgoing electron. 

DVCS is the simplest probe of a new class of light-cone (quark) matrix elements, called Generalized Parton Distributions. The GPDs offer the exciting possibility of the first ever spatial images of the quark waves inside the proton, as a function of their wavelength~\cite{Mul94, Ji97, Ji97-1, Ji97-2, Rad96, Rad97}. The correlation of transverse spatial and longitudinal momentum information contained in the GPDs provides a new tool to evaluate the contribution of quark orbital angular momentum to the proton spin.

GPDs enter the DVCS cross section through integrals, called Compton Form Factors. CFFs are defined in terms of the vector GPDs $H$ and $E$, and the axial vector GPDs $\widetilde{H}$ and $\widetilde{E}$. For example ($q\in\{u,d,s\}$) \cite{Bel02}:
\begin{eqnarray}
{\mathcal H}(\xi,t) = \sum_{q}  \left[\frac{e_q}{e}\right]^2 &\Biggl\{& i \pi \left[H^q(\xi,\xi,t) - H^q(-\xi,\xi,t)\right] \nonumber \\
& + & {\mathcal P} \int_{-1}^{+1} dx \left[ \frac{1}{\xi-x} - \frac{1}{\xi+x} \right]   H^q(x,\xi,t)\Biggr\}.
\label{eq:CFF}
\end{eqnarray}
Thus, the imaginary part accesses GPDs along the line $x=\pm\xi$, whereas the real part probes GPD integrals over $x$. The {\it diagonal} GPD, $H(\xi,\xi,t=\Delta^2)$ is not a positive-definite probability density, however it is a transition density with the momentum transfer $\Delta_\perp$ Fourier-conjugate to the transverse distance $r$
between the active parton and the center-of-momentum of the spectator partons in the target~\cite{Bur07}. Furthermore, the real part of the Compton form factor is determined by a dispersion integral over the diagonal $x=\pm \xi$ plus the $D$-term \cite{Ter05, Die07, Ani07, Ani08}:
\begin{eqnarray}
\Re\text{e}\left[\mathcal H (\xi,t) \right] & = & \\
\int_{-1}^1 & dx & \left\{\left[H(x,x,t)+H(-x,x,t)\right]\left[\frac{1}{\xi-x} - \frac{1}{\xi+x}\right] + 2 \, \frac{D(x,t)}{1-x} \right\} \nonumber
\end{eqnarray}
The $D$-term~\cite{Pol99} only has support in the ERBL region $|x|<\xi$ in which the GPD is determined by $q\overline{q}$ exchange in the $t$-channel.

\subsection{Physics goals}

In this letter, we propose to exploit the charge dependence provided by the use of a positron beam in order to cleanly separate the DVCS$^2$ term from the DVCS-BH interference in the photon electroproduction cross section.

\begin{figure}[h!]
\begin{center}
\includegraphics[width=0.95\textwidth]{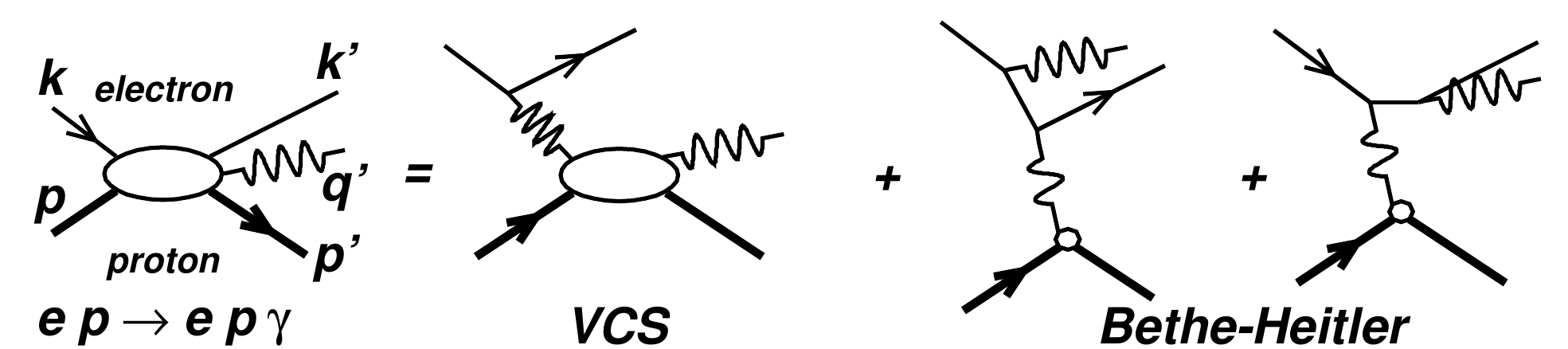}
\caption{\baselineskip 13 pt Lowest order QED amplitude for the $ep\rightarrow ep\gamma$ reaction. The momentum four-vectors of all external particles are labeled at left.
The net four-momentum transfer to the proton is $\Delta_\mu=(q-q')_\mu=(p'-p)_\mu$. In the virtual Compton scattering (VCS) amplitude, the (spacelike) virtuality of the  incident photon is $Q^2=-q^2=-(k-k')^2$. In the Bethe-Heitler (BH) amplitude, the virtuality of the incident photon is $-\Delta^2=-t$. Standard $(e,e')$ invariants are  $s_e=(k+p)^2$, $\xBj=Q^2/(2q\cdot p)$ and $W^2=(q+p)^2$.}
\label{fig:epepg}
\end{center}
\end{figure}
The photon electroproduction cross section of a polarized lepton beam of energy $k$ off an unpolarized target of mass $M$ is sensitive to the coherent interference of the DVCS amplitude with the Bethe-Heitler amplitude (see Fig.~\ref{fig:epepg}). It can be written as:
\begin{eqnarray}
\frac{d^5\sigma(\lambda,\pm e)}{d^5\Phi} & = & \frac{d\sigma_0}{dQ^2 dx_B} \left| \mathcal T^{BH}(\lambda) \pm \mathcal T^{DVCS}(\lambda) \right|^2/|e|^6 \nonumber \\
 & = &  \frac{d\sigma_0}{dQ^2 dx_B}   \left[ \left|\mathcal T^{BH}(\lambda) \right|^2 + \left| \mathcal T^{DVCS}(\lambda)\right|^2  \mp \mathcal I(\lambda)   \right]\frac{1}{e^6} \label{eq:dsigDVCS}
\end{eqnarray} 
with
\begin{equation} 
\frac{d\sigma_0}{dQ^2 dx_B} = \frac{\alpha_{\rm QED}^3}{16\pi^2} \, \frac{1}{(s_e-M^2)^2 x_B} \, \frac{1}{\sqrt{1+\epsilon^2}}  \, , \label{eq:dsig0}
\end{equation}
and 
\begin{eqnarray}
\epsilon^2 & = & 4 M^2 x_B^2/Q^2 \, ,\\
s_e & = & 2 M k + M^2 \, , \\
d^5\Phi & = & dQ^2 dx_B d\phi_e dt d\phi_{\gamma\gamma} \, .
\end{eqnarray}
Here, $\lambda$ is the electron helicity and the $+$($-$) stands for the sign of the charge of the lepton beam. The BH contribution is calculable in QED, given our 
$\approx 1\%$  knowledge of the proton elastic form factors at small momentum transfer. The other two contributions to the cross section, the interference and the DVCS$^2$ terms, provide complementary information on GPDs. It is possible to exploit the structure of the cross section as a function of the angle $\phigg$ between the leptonic and hadronic plane to separate up to a certain degree the different contributions to the total cross section~\cite{Die97}. The angular separation can be supplemented with a beam energy separation. The energy separation has been successfully used in previous experiments~\cite{Def17} at 6~GeV and is the goal of already approved experiments at 
12~GeV~\cite{Mun13}. The $|\mathcal T^{BH}|^2$ term is given in~\cite{Bel02}, and only its general form is reproduced here:
\begin{equation}
|\mathcal T^{BH}|^2 = \frac{e^6}{x_B^2 t y^2 (1+\epsilon^2)^2  \mathcal P_1(\phigg) \mathcal P_2(\phigg)} \sum_{n=0}^2 c_n^{BH} \cos(n\phi_{\gamma\gamma}) \, .
\label{eq:BHPhi}
\end{equation}
The harmonic coefficients $c_n^{BH}$ depend upon bilinear combinations of the ordinary elastic form factors $F_1(t)$ and $F_2(t)$ of the proton. The factors $\mathcal P_i$ are the electron propagators in the BH amplitude~\cite{Bel02}. The interference term in Eq.~\ref{eq:dsigDVCS} is a linear combination of GPDs, whereas the DVCS$^2$ term is 
a bilinear combination of GPDs. These terms have the following harmonic structure:
\begin{eqnarray}
\mathcal I & = & \frac{e^6}{x_B y^3 \mathcal P_1(\phigg) \mathcal P_2(\phigg) t } \bigg\{ c_0^{\mathcal I}  \label{eq:IntPhi} \\
& & \,\,\,\,\,\,\,\,\,\,\,\,\,\,\,\,\,\,\,\,\,\,\,\,\,\,\,\,\,\,\,\,\,\,\,\,\,\,\,\,\,\,\,\,\,\,\,\,\,\,\,\,\,\,\,\,\,\,\,\,\,\, + \sum_{n=1}^3 \left[ c_n^{\mathcal I}(\lambda)\cos(n\phigg) + \lambda s_n^{\mathcal I}\sin(n\phigg) \right] \bigg\} \, ,\nonumber
\end{eqnarray}
and 
\begin{eqnarray}
\left| \mathcal T^{DVCS}(\lambda) \right|^2 & = & \frac{e^6}{y^2 Q^2} \bigg\{ c_0^{DVCS} \label{eq:DVCSPhi} \\
& & \,\,\,\,\,\,\,\,\,\,\,\,\,\,\,\,\,\,\,\,\,\, +\sum_{n=1}^2 \left[ c_n^{DVCS}  \cos(n\phigg) + \lambda s_n^{DVCS} \sin(n\phigg) \right] \bigg\} \, .\nonumber
\end{eqnarray}
The $c_0^{DVCS, \mathcal I}$, and $(c,s)_1^{\mathcal I}$ harmonics are dominated by twist-two  GPD terms, although they do have twist-three admixtures that must be quantified by the $Q^2$-dependence of each harmonic.  The $(c,s)_1^{DVCS}$ and $(c,s)_2^{\mathcal I}$ harmonics are dominated by twist-three matrix elements, although the same twist-two GPD terms also contribute (but with smaller kinematic coefficients than in the lower Fourier terms).  The $(c,s)_2^{DVCS}$ and $(c,s)_3^{\mathcal I}$ harmonics stem from twist-two double helicity-flip gluonic GPDs alone. They are formally suppressed by $\alpha_s$ and will be neglected here. They do not mix, however, with the  twist-two quark amplitudes. The exact expressions of these harmonics in terms of the quark CFFs of the nucleon are given in~\cite{Bel10}.

Eq.~\ref{eq:dsigDVCS} shows how a positron beam, together with measurements with electrons, provides a way to separate without any assumptions the DVCS$^2$ and BH-DVCS interference contributions to the cross section. With electrons alone, the only approach to this separation is to use the different beam energy dependence of the DVCS$^2$ and BH-DVCS interference. This is the strategy that will be used in approved experiment E12-13-010. However, as recent results have shown~\cite{Def17} this technique has limitations due to higher order contributions (next-to-leading order in $\alpha_s$ or higher twists) and some assumptions needed. A positron beam, on the other hand, will be able to pin down each individual term. The $Q^2$ dependence of each of them can later be used to study the nature of the higher order contributions by comparing it to the predictions of the leading twist diagram.

\subsection{Experimental setup}

We propose to make a precision coincidence setup measuring charged particles (scattered positrons) with the existing HMS and photons using the Neutral Particle Spectrometer (NPS), currently under construction. The NPS facility consists of a PbWO$_4$ crystal calorimeter and a sweeping magnet in order to reduce electromagnetic backgrounds. A high luminosity spectrometer+calorimeter (HMS+PbWO$_4$) combination is ideally suited for such measurements. The sweeping magnet will allow to achieve low-angle photon detection. Detailed background simulations show that this setup allows for $\ge 10 \mu A$ beam current on a 10~cm long cryogenic LH$_2$ target at the very smallest NPS angles, and much higher luminosities at larger $\gamma,\pi^0$ angles~\cite{Mun13}.

\subsubsection*{High Momentum Spectrometer}

The magnetic spectrometers benefit from relatively small point-to-point uncertainties, which are crucial for absolute cross section measurements. In particular, the optics properties and the acceptance of the HMS have been studied extensively and are well understood in the kinematic range between 0.5 and 5~GeV, as evidenced by more than 200 L/T separations ($\sim$1000 kinematics)~\cite{Lia04}. The position of the elastic peak has been shown to be stable to better than 1~MeV, and the precision rail system and rigid pivot connection have provided reproducible spectrometer pointing for more than a decade.

\subsubsection*{Neutral Particle Spectrometer}

\begin{figure}[t!]
\centering
\subfigure[\label{fig:exp-setup} \it ]
{\includegraphics[width=0.490\textwidth]{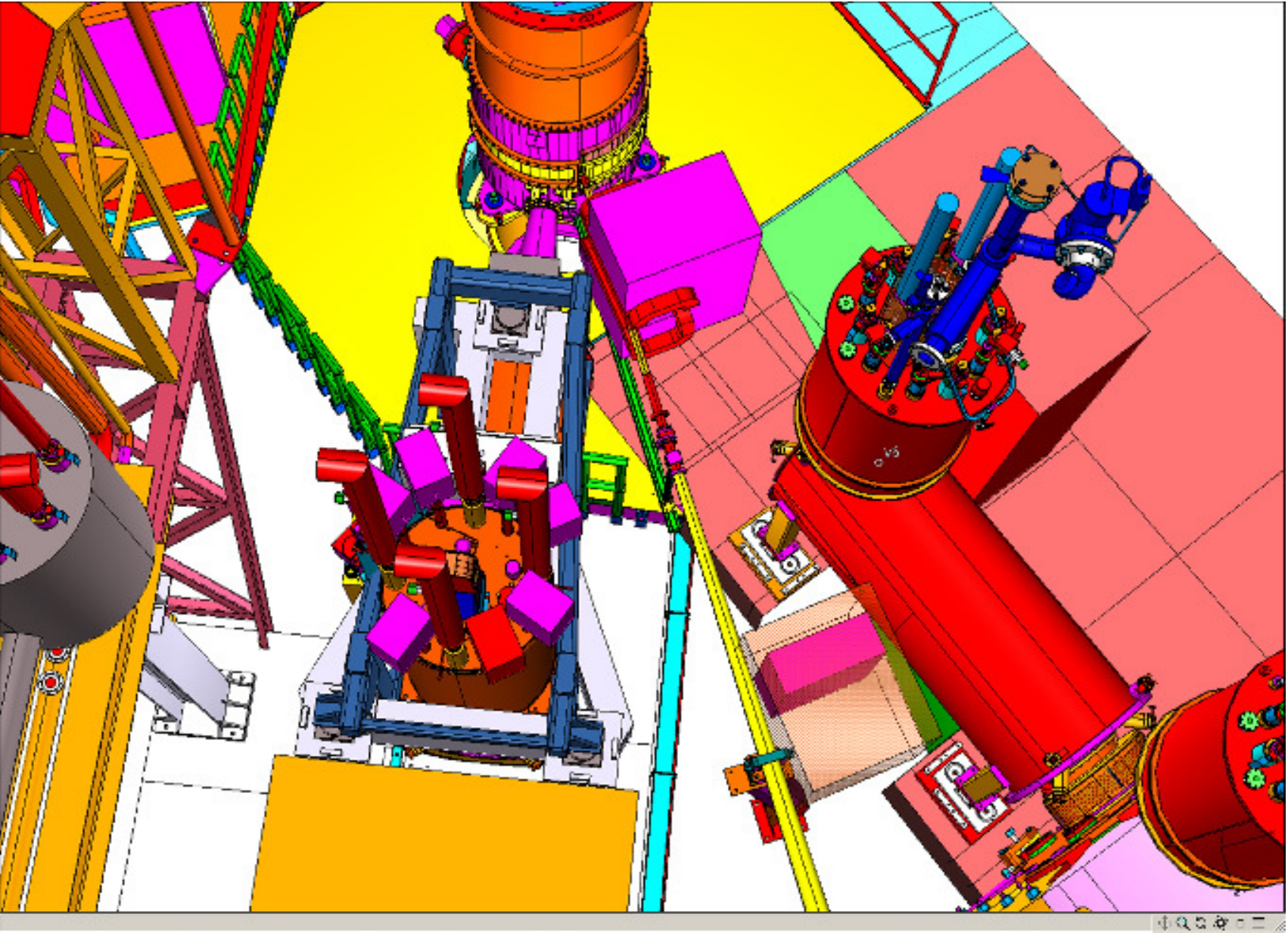}}
\subfigure[\label{fig:PbWO4_block} \it ] 
{\includegraphics[width=0.481\textwidth]{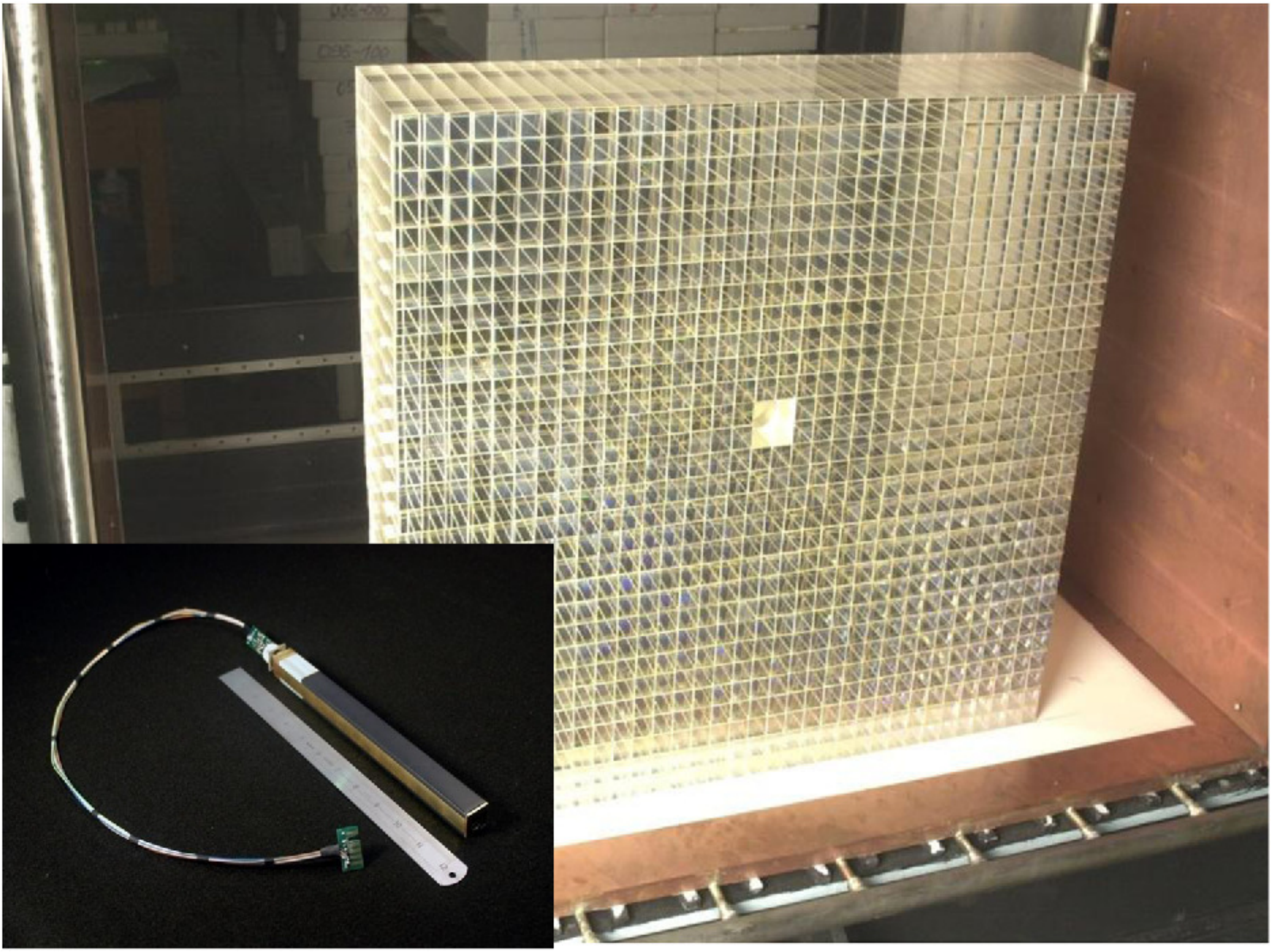}}
\caption{\label{fig:pi0_detector_hallc} \baselineskip 13 pt (a)~The DVCS/$\pi^0$ detector in Hall C. The cylinder at the top center is the (1~m diameter) vacuum chamber containing the 10~cm long liquid-hydrogen target. The long yellow tube emanating from the scattering chamber on the lower right is the downstream beam pipe. 
To the left of the beam pipe is the HMS. Only the liquid He and liquid N$_2$ lines for the large superconducting quadrupoles at the entrance of the spectrometer are clearly visible. To the right of the beam line, the first quadrupole of the SHMS and its cryogenic feed lines are shown. This spectrometer will be used as a carriage to support the PbWO$_4$ calorimeter (shown in its light-tight and temperature control box next to the beam line) and the associated sweep magnet. (b) The high resolution PbWO$_4$ part of the HYCAL~\cite{Kub06} on which the present NPS design is based.}
\end{figure}

We will use the general-purpose and remotely rotatable NPS system for Hall C. A floor layout of the HMS and NPS is shown in Fig.~\ref{fig:exp-setup}. The NPS system consists of the following elements:
\begin{itemize}
\item{A sweeping magnet providing 0.3~Tm field strength;}
\item{A neutral particle detector consisting of 1080 PbWO$_4$ blocks (similar to the PRIMEX~\cite{Gas02} experimental setup, see Fig.~\ref{fig:PbWO4_block}) in a temperature controlled frame, comprising a 25~msr device at a distance of 4~m;}
\item{Essentially deadtime-less digitizing electronics to independently sample the entire pulse form for each crystal, allowing for background subtraction and 
identification of pile-up in each signal;}
\item{A new set of high-voltage distribution bases with built-in amplifiers for operation in high-rate environments.}
\item{Cantilevered platforms on the SHMS carriage, to allow for precise and remote rotation around the Hall C pivot of the full neutral-pion detection system, over an angle range between 6$^\circ$ and 30$^\circ$;}
\item{A dedicated beam pipe with as large critical angle as possible to reduce backgrounds beyond the sweeping magnet.}
\end{itemize}

\subsubsection*{The PbWO$_4$ electromagnetic calorimeter}

The energy resolution of the photon detection is the limiting factor of the experiment. To ensure exclusivity of the reaction by the missing mass technique, we plan to use a PbWO$_4$ calorimeter 56 cm wide and 68 cm high. This corresponds to 30 by 36 PbWO$_4$ crystals of 2.05 by 2.05 cm$^2$ (each 20.0 cm long). We managed one crystal on each side to properly capture showers, and thus designed the PbWO$_4$ calorimeter to consist of 30 by 36 PbWO$_4$ crystals, or 60 by 72 cm$^2$. This amounts to a requirement of 1080 PbWO$_4$ crystals. To reject very low-energy background, a thin absorber could be installed in front of the PbWO$_4$ detector. The space between the sweeper magnet and the proximity of the PbWO$_4$ detector will be enclosed within a vacuum channel (with a thin exit window, further reducing low-energy background) to minimize the decay photon conversion in air. Given the temperature sensitivity of the scintillation light output of the PbWO$_4$ crystals, the entire calorimeter must be kept at a constant temperature, to within $0.1^\circ$ to guarantee 0.5\% energy stability for absolute calibration and resolution. The high-voltage dividers on the PMTs may dissipate up to several hundred Watts, and this power similarly must not create temperature gradients or instabilities in the calorimeter. The calorimeter will thus be thermally isolated and be surrounded on all four sides by water-cooled copper plates.

At the anticipated background rates, pile-up and the associated baseline shifts can adversely affect the calorimeter resolution, thereby constituting the limiting factor for the beam current. The solution is to read out a sampled signal, and perform offline shape analysis using a flash ADC (fADC) system. New HV distribution bases with built-in pre-amplifiers will allow for operating the PMTs at lower voltage and lower anode currents, and thus protect the photocathodes or dynodes from damage. To take full advantage of the high-resolution crystals while operating in a high-background environment, modern flash ADCs will be used to digitize the signal. They continuously sample the signal every 4\,ns, storing the information in an internal FPGA memory. When a trigger is received, the samples in a programmable window around the threshold crossing are read out for each crystal that fired. Since the readout of the FPGA does not interfere with the digitizations, the process is essentially deadtime free.

The PbWO$_4$ crystals are 2.05 x 2.05 cm$^2$. The typical position resolution is 2-3 mm. Each crystal covers 5~mrad, and the expected angular resolution is 
0.5-0.75~ mrad, which is comparable with the resolutions of the HMS and SOS, routinely used for Rosenbluth separations in Hall C.

\subsection{Exclusivity of the DVCS reaction} \label{sec:mm2}

\begin{figure}[t!]
\begin{center}
\includegraphics[width=0.59\linewidth]{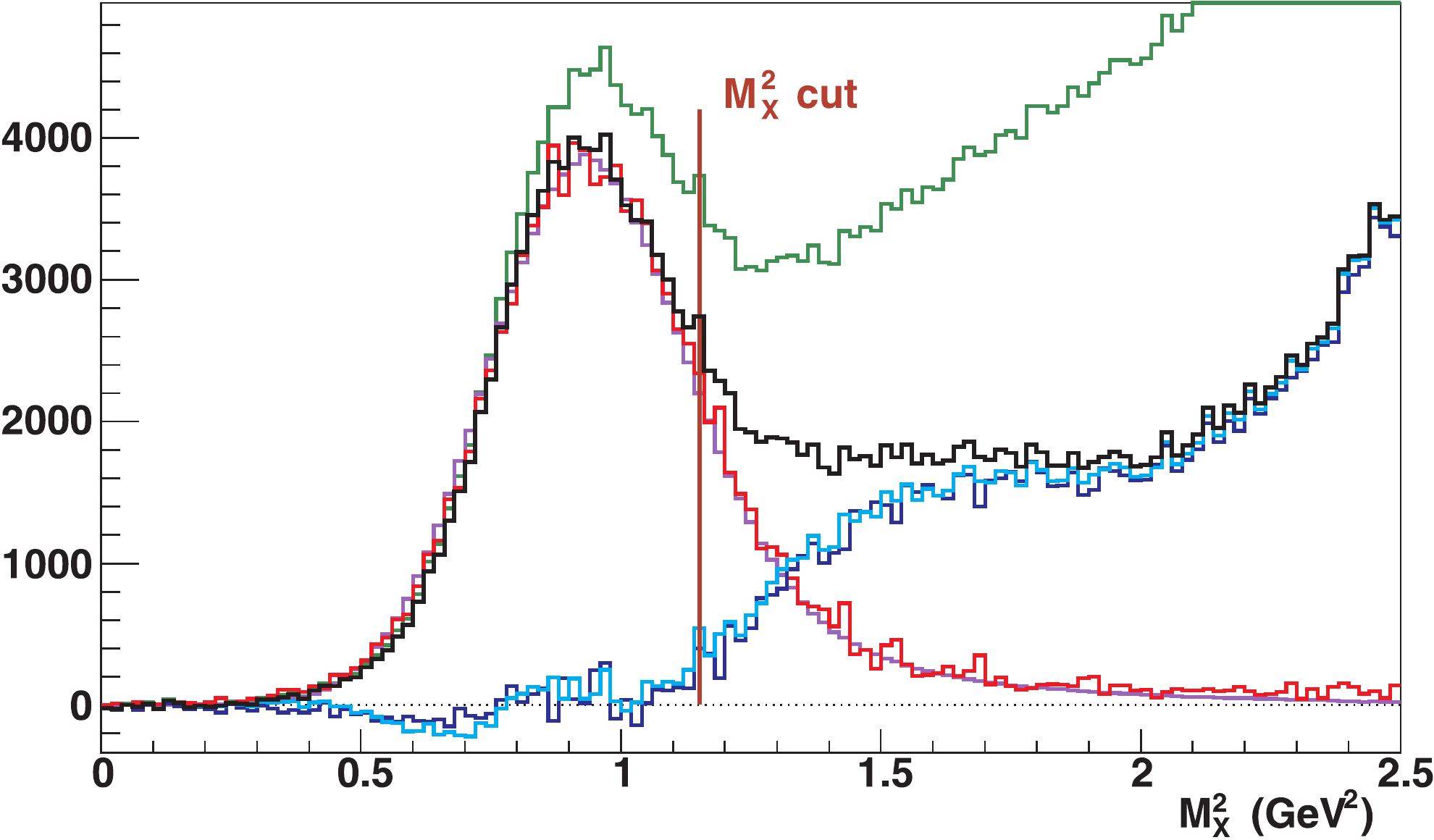}
\includegraphics[width=0.39\linewidth]{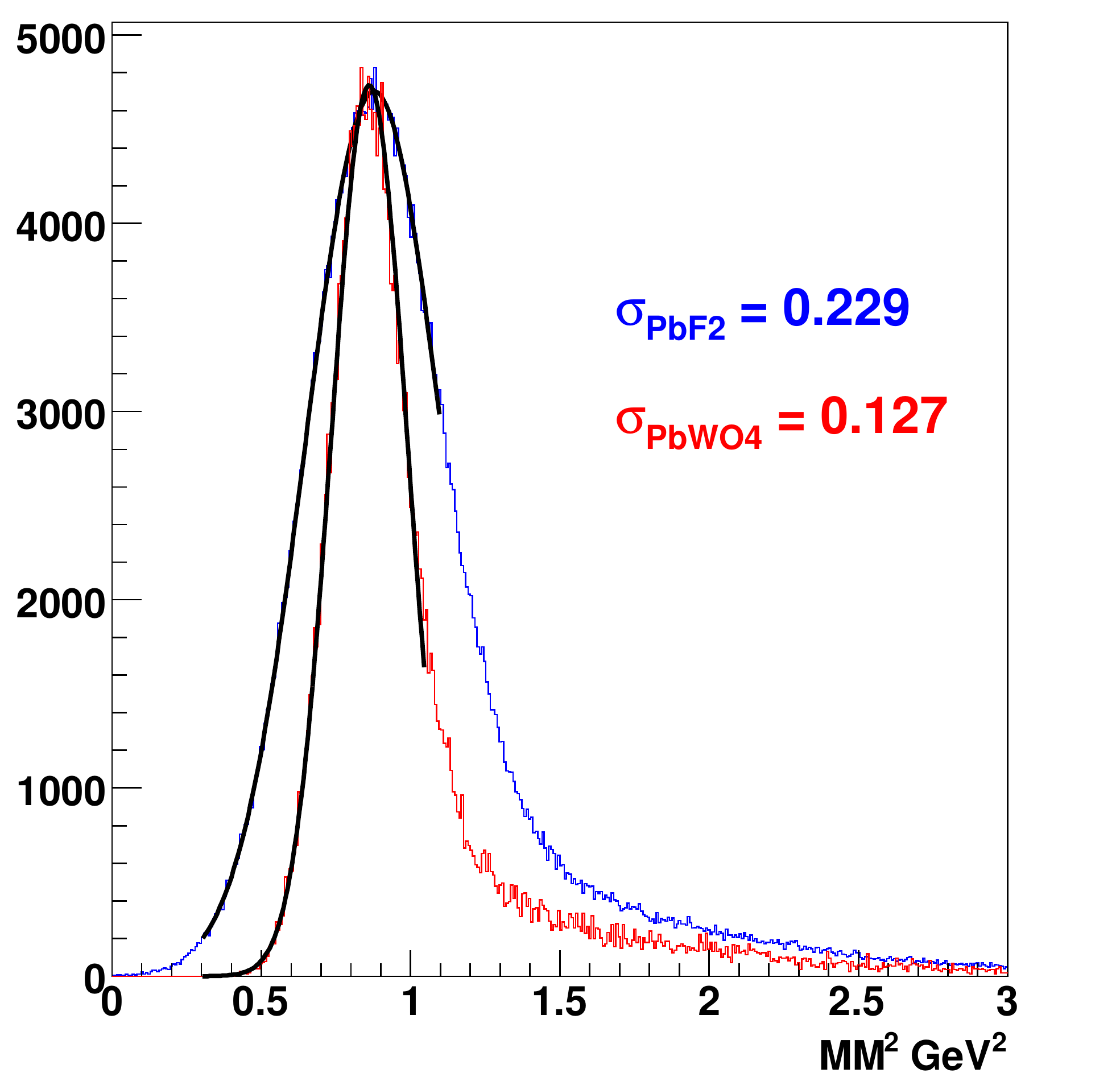}
\caption{\baselineskip 13 pt
(left) Missing mass squared in E00-110 for H$(e,e'\gamma)X$ events (green curve) at $Q^2=2.3$ GeV$^2$ and $-t\in[0.12,0.4]$ GeV$^2$, integrated over the azimuthal angle of the photon $\phi_{\gamma\gamma}$. The black curve shows the data once the H$(e,e'\gamma)\gamma X'$ events have been subtracted. The other curves are described in the text. (right) Projected missing mass resolution for a similar kinematic setting ($E_b=6.6$~GeV, $Q^2=3$~GeV$^2$, $x_B=0.36$). By using PbWO$_4$ instead of PbF$_2$, the missing mass resolution will be considerably improved. Values are given in Tab.~\ref{tab:DVCS-Kin} and are to be compared to the value $\sigma(M_X^2)=0.2$~GeV$^2$ obtained in previous experiments in Hall A.}
\label{fig:mm2}
\end{center}
\end{figure}

The exclusivity of the DVCS reaction will be based on the missing mass technique, successfully used during Hall A experiments E00-110 and E07-007. Fig.~\ref{fig:mm2} (left) presents the missing mass squared obtained in E00-110 for H$(e,e'\gamma)X$ events, with coincident electron-photon detection. After subtraction of an accidental coincidence sample, our data is essentially background free: there is negligible contamination of non-electromagnetic events in the HRS and PbF$_2$ spectra. In addition to H$(e,e'\gamma) p$, however, we do have the following competing channels: H$(e,e'\gamma) p\gamma$ from $e p \rightarrow e \pi^0 p$, $e p \rightarrow e \pi^0 N\pi$, $e p \rightarrow e \gamma N\pi$, $e p \rightarrow e \gamma N\pi\pi\ldots$. From symmetric (lab-frame) $\pi^0$-decay, we obtain a high statistics sample of H$(e,e'\pi^0)X'$ events, with two photon clusters in the PbF$_2$ calorimeter. From these events, we determine the statistical sample of [asymmetric] H$(e,e'\gamma)\gamma X'$ events that must be present in our H$(e,e'\gamma)X$ data. The $M_X^2$ spectrum displayed in black in Fig.~\ref{fig:mm2} (left) was obtained after subtracting this $\pi^0$ yield from the total (green) distribution. This is a $14\%$ average subtraction in the exclusive window defined by {\it $M_X^2$ cut} in Fig.~\ref{fig:mm2} (left). Depending on the bin in $\phi_{\gamma\gamma}$ and $t$, this subtraction varies from 6\% to 29\%. After $\pi^0$ subtraction, the only remaining channels, of type H$(e,e'\gamma)N\pi$, $N\pi\pi$, {\it etc.\/} are kinematically constrained to $M_X^2 >(M+m_\pi)^2$. This is the chosen value for truncating the missing mass signal integration. \newline
Resolution effects can cause the inclusive channels to contribute below this cut. To evaluate this possible contamination, during E00-110 we used an additional proton array (PA) of 100 plastic scintillators. The PA subtended a solid angle (relative to the nominal direction of the {\bf q}-vector) of $18^\circ<\theta_{\gamma p}<38^\circ$ and $45^\circ < \phi_{\gamma p} = 180^\circ-\phi_{\gamma\gamma} < 315^\circ$, arranged in 5 rings of 20 detectors. For H$(e,e'\gamma)X$ events near the exclusive region, we can predict which block in the PA should have a signal from a proton from an exclusive H$(e,e'\gamma p)$ event. The red histogram is the $X=(p+y)$ missing mass squared distribution for H$(e,e'\gamma p)y$ events in the predicted PA block, with a signal above an effective threshold $30$ MeV (electron equivalent). The blue curve shows our inclusive yield, obtained by subtracting the normalized triple coincidence yield from the H$(e,e'\gamma)X$ yield. The (smooth) violet curve shows our simulated H$(e,e'\gamma)p$ spectrum, including radiative and resolution effects, normalized to fit the data for $M_X^2\le M^2$. The cyan curve is the  estimated inclusive yield obtained by subtracting the simulation from the data. The blue and cyan curves are in good agreement, and show that our exclusive yield has less than $2\%$ contamination from inclusive processes.

In the presently proposed experiment we plan to use a PbWO$_4$ calorimeter with a resolution more than twice better than the PbF$_2$ calorimeter (Fig.~\ref{fig:mm2} (right) and Tab.~\ref{tab:DVCS-Kin}) used in E00-110. While the missing mass resolution will be slightly worse at some high beam energy, low $x_B$ kinematics, the better energy resolution of the crystals will largely compensate for it, and the missing mass resolution in this experiment will be significantly better than ever before.  

\subsection{Proposed kinematics and projections} \label{sec:Projections}

Tab.~\ref{tab:DVCS-Kin} details the kinematics and beam time requested. $Q^2$ scans at 4 different values of $x_B$ were chosen in kinematics with already approved electron data~\cite{Mun13}. The positron beam current assumed is 5~$\mu$A (unpolarized beam). Beam time is calculated in order to match the statistical precision of the electron data, which in turn corresponds to the typical values of the expected systematic uncertainties.

\begin{figure}[t!]\begin{center}
\includegraphics[width=0.75\linewidth]{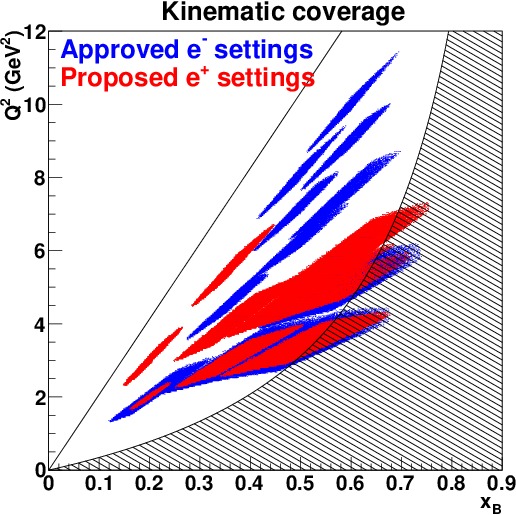}
\caption{\baselineskip 13 pt Display of the different kinematic settings proposed. Blue corresponds to the settings already approved in Hall A and Hall C using an electron beam. Red shows the proposed kinematics with positrons. Shaded areas show the resonance region $W<2$~GeV and the line $Q^2=(2M_pE_b)x_B$ limits the physical region for a maximum beam energy $E_b$=$11$~GeV.}
\end{center} \label{fig:kin}
\end{figure}

The different kinematics settings are represented in Fig.~\ref{fig:kin} in the $Q^2$--$x_B$ plane. The area below the straight line $Q^2=(2M_pE_b)x_B$ corresponds to the physical region for a maximum beam energy $E_b$=$11$~GeV. Also plotted is the resonance region $W<2$~GeV. We have performed detailed Monte Carlo simulation of the experimental setup and evaluated counting rates for each of the settings. For this purpose, we have used a recent global fit of world data with LO sea evolution~\cite{Mul16}. This fit reproduces the magnitude of the DVCS cross section measured in Hall A at $x_B=0.36$ and is available up to values of $x_B\le 0.5$. For the high $x_B$ settings we used the GPD parametrization of Ref.~\cite{Kro13} fitted to deeply virtual meson production data, together with a code to compute DVCS cross sections~\cite{Mou13, Gui08-1}. Notice that for DVCS, counting rates and statistical uncertainties will be driven {\em at first order} by the well-known BH cross section.
\begin{table}[t!]
\begin{center}
\begin{tabular}{|c|c|c|c|c|c|c|c|c|c|}  
\hline\hline
$x_\text{Bj}$ &\multicolumn{2}{c|}{0.2}&\multicolumn{3}{c|}{0.36}&\multicolumn{2}{c|}{0.5}&\multicolumn{2}{c|}{0.6}\\ \hline
                \hline
                $Q^2\,\text{(GeV)}^2$ &2.0&3.0&3.0&4.0&5.5&3.4&4.8&5.1&6.0\\\hline
                $k\ \text{(GeV)}$ & \multicolumn{2}{c|}{11}&\multicolumn{2}{c|}{8.8}&11&\multicolumn{2}{c|}{11}&8.8&11\\\hline
                $k'\ \text{(GeV)}$ &5.7&3.0&4.4&2.9&2.9&7.4&5.9&4.3&5.7\\\hline
                $\theta_\text{Calo}\,\text{(deg)}$ &10.6&6.3&14.7&10.3&7.9&21.7&16.6&17.8&17.2\\\hline
                $D_\text{Calo}$ (m) & 4&6&3&4&4&3&3&3&3\\\hline
                $\sigma_{M_X^2}$(GeV$^2$) &0.17&0.22&0.13&0.15&0.19&0.09&0.11&0.09&0.09\\\hline
                Days & 2&2&4&2&10&4&10&2&20\\\hline
\hline
\end{tabular}
\end{center}
\caption{\baselineskip 13pt
DVCS kinematics with positrons in Hall C. The incident and scattered beam energies are $k$ and $k'$, respectively. The calorimeter is centered at the angle $\theta_\text{Calo}$, which is set equal to the nominal virtual-photon direction.  The front face of the calorimeter is at a distance $D_\text{Calo}$ from the center of the target, and is adjusted to optimize multiple parameters: first to maximize acceptance, second to ensure sufficient separation of the two clusters from symmetric $\pi^0\rightarrow \gamma\gamma$ decays, and third to ensure that the edge of the calorimeter is never at an angle less than $3.2^\circ$ from the beam line.}
\label{tab:DVCS-Kin}
\end{table}

\begin{table}[b!]
\begin{center}
\begin{tabular}{||l|c|c||} \hline
\hspace*{1.5cm}Source                   &  pt-to-pt &  scale   \\
                                        &  (\%)     &  (\%)    \\ \hline 
  Acceptance                            &   0.4     &   1.0    \\
  Electron/positron PID                 &   $<$0.1  &   $<$0.1 \\
  Efficiency                            &   0.5     &   1.0    \\
  Electron/positron tracking efficiency &   0.1     &   0.5    \\
  Charge                                &   0.5     &   2.0    \\
  Target thickness                      &   0.2     &   0.5    \\
  Kinematics                            &   0.4     &  $<$0.1  \\
  Exclusivity                           &   1.0     &   2.0    \\
  $\pi^0$ subtraction                   &   0.5     &   1.0    \\
  Radiative corrections                 &   1.2     &   2.0    \\ \hline 
  Total                                 & 1.8--1.9  & 3.8--3.9 \\ \hline
\end{tabular}
\end{center}
\caption{\baselineskip 13pt Estimated systematic uncertainties for the proposed experiment based on previous Hall C experiments.} \label{tab:sys} 
\end{table}
The HMS is a very well understood magnetic spectrometer which will be used here with modest requirements (beyond the momentum), defining well the ($x_B,Q^2$) kinematics. Tab.~\ref{tab:sys} shows the estimated systematic uncertainties for the proposed experiment based on previous experience from Hall C equipment and Hall A experiments.

\begin{sidewaysfigure}[h!]
\centering\includegraphics[width=0.9\linewidth]{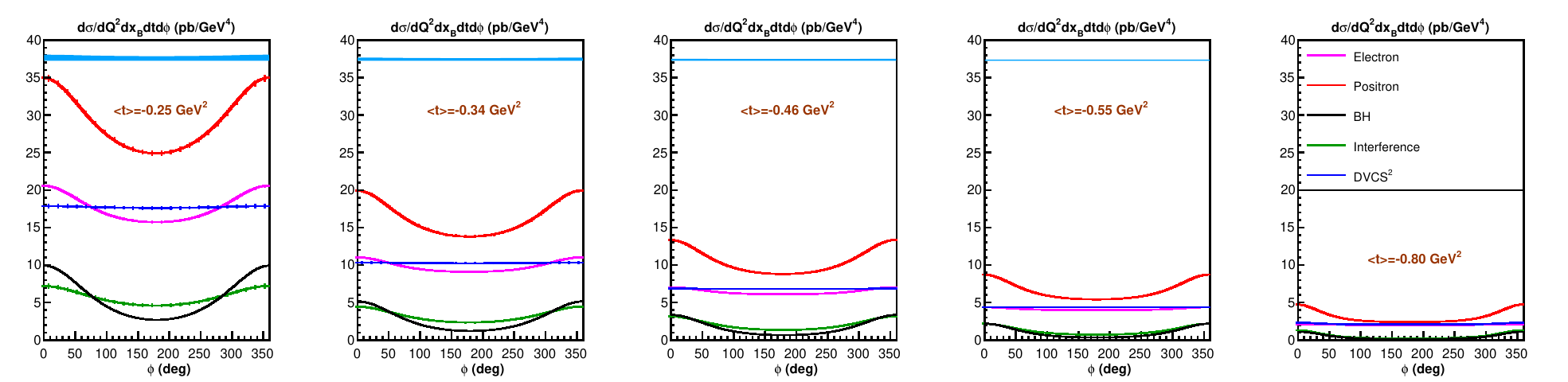}\\
\centering\includegraphics[width=0.9\linewidth]{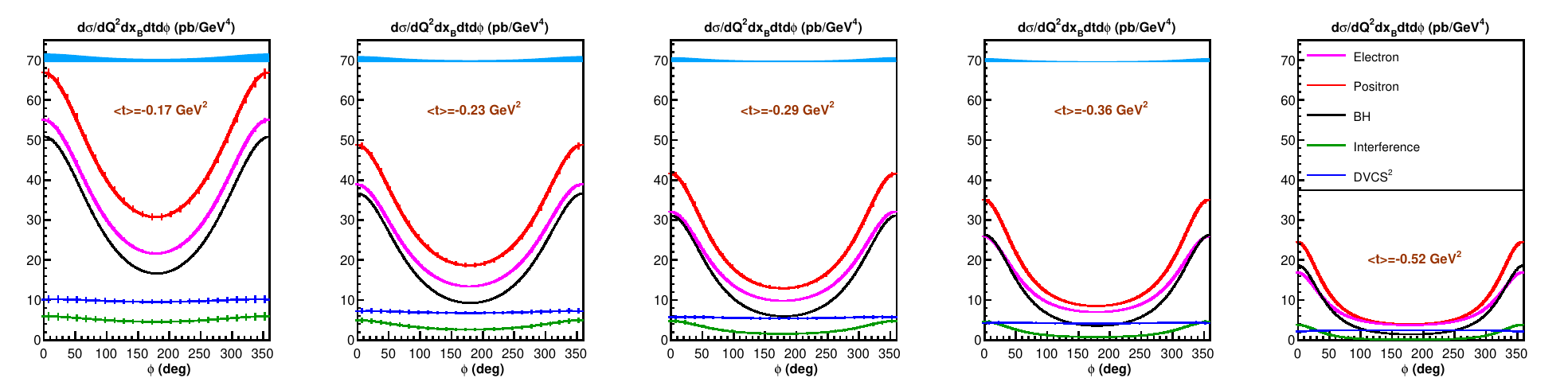}\\
\centering\includegraphics[width=0.9\linewidth]{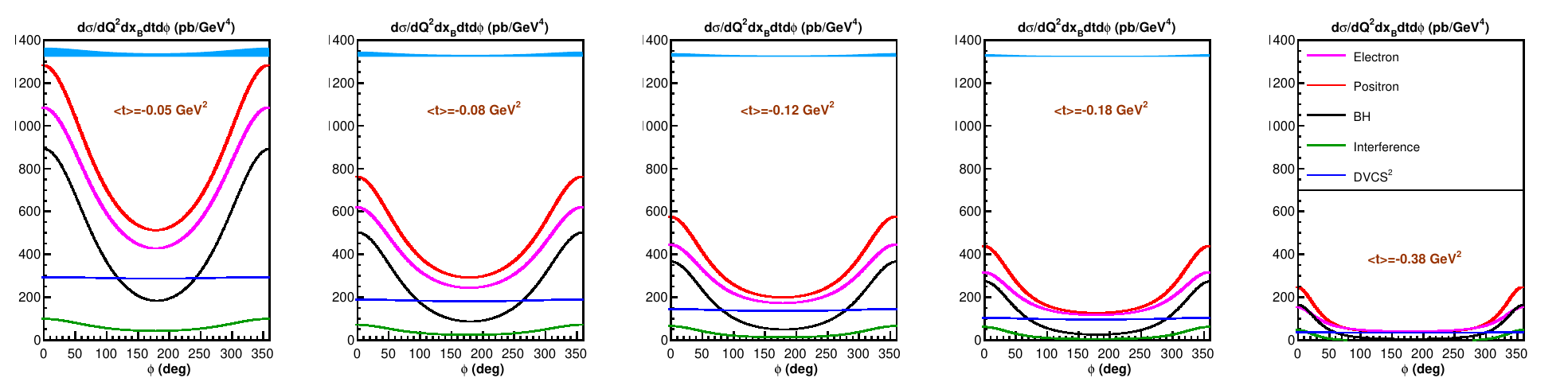}
\caption{\baselineskip 13 pt
Experimental projections for 3 of the proposed settings: $x_B=0.2$, $Q^2=2.0$~GeV$^2$ (top),  $x_B=0.36$, $Q^2=4.0$~GeV$^2$ (middle) and $x_B=0.5$, $Q^2=3.4$~GeV$^2$ (bottom). Red points show the projected positron cross sections with statistical uncertainties. Electron cross sections that will be measured in experiment E12-13-010 are shown in magenta. The combination of $e^-$ and $e^+$ cross sections allow the separation of the DVCS$^2$ contribution (blue) and the DVCS-BH interference (green). For reference, the BH cross section is displayed in black. Systematic uncertainties are shown by the cyan band.}
\label{fig:ros}
\end{sidewaysfigure}

Fig.~\ref{fig:ros} shows the projected results for 3 selected settings at different values of $x_B$=$0.2, 0.36, 0.5$. Statistical uncertainties are shown by 
error bars and systematic uncertainties are represented by the cyan bands. The DVCS$^2$ term (which is $\phi$ independent at leading twist) can be very cleanly separated from the BH-DVCS interference contribution, without any assumption regarding the leading-twist dominance. The $Q^2$ dependence of each term will be measured (cf. Tab.~\ref{tab:DVCS-Kin}) and compared to the asymptotic prediction of QCD. The extremely high statistical and systematic precision of the results illustrated in Fig.~\ref{fig:ros} will be crucial to disentangle higher order effects (higher twist or next-to-leading order contributions) as shown by recent results~\cite{Def17}.

\afterpage{\clearpage}

\subsection{Summary}

We propose to measure the cross section of the DVCS reaction accurately using positrons in the wide range of kinematics allowed by a set of beam energies up to 11 GeV. We will exploit the beam charge dependence of the cross section to separate the contribution of the BH-DVCS interference and the DVCS$^2$ terms. The $Q^2$ dependence of each individual term will be measured and compared to the predictions of the handbag mechanism. This will provide a quantitative estimate of higher-twist effects to the GPD formalism at JLab kinematics.

We plan to use Hall C High-Momentum Spectrometer, combined with a high resolution PbWO$_4$ electromagnetic calorimeter. In order to complete this full mapping of the DVCS cross section with positrons over a wide range of kinematics, we request 56 days of (unpolarized) positron beam ($I>5\mu$A).

%
%

\newpage

\null\vfill

\begin{center}

\section{\it Letter-ot-Intent: $\Apr$ search}

\vspace*{15pt}

{\Large{\bf Searching for Dark Photon}}

\vspace*{3pt}

{\Large{\bf with Positrons at Jefferson Lab}}

\vspace*{15pt}

{\bf Abstract}

\begin{minipage}[c]{0.85\textwidth}
The interest in the Dark Photon ($\Apr$) has recently grown, since it could act as a light mediator to a new sector of Dark Matter particles. In this paradigm, the electron-positron annihilation can rarely produce a $\gamma  \Apr$  pair. Various experiments have been proposed to detect this process using positron beams impinging on fixed targets. In such experiments, the energy of the photon from the $e^+ e^- \rightarrow \gamma \Apr$ process is measured with an electromagnetic calorimeter and the missing mass is computed. However, the $\Apr$ mass range that can be explored is limited by the accessible energy in the center of mass frame, scaling as the square root of the beam energy. The realization of a high energy positron beam at Jefferson Lab would allow to search for $\Apr$ masses up to $\sim 100$~MeV, reaching unexplored regions of the $\Apr$ parameter space. We propose in this letter a PADME-like experiment at Jefferson Lab, assuming a 11~GeV positron beam with a $\sim$100~nA current. The achievable sensitivity of this experiment was estimated, studying the main sources of background using CALCHEP and GEANT4 simulations.

\end{minipage} 

\vspace*{15pt} 

{\it Spokespersons: M.~Battaglieri, A.~Celentano, \underline{L.~Marsicano} (lmarsicano@ge.infn.it)}

\end{center}

\vfill\eject

%
%

\subsection{Theoretical background}

The Standard Model (SM) of elementary particles and interactions is able to describe with an extraordinary precision ordinary matter in a variety of different environments and energy scales. However, some phenomena such as Dark Matter (DM), neutrino masses and matter-antimatter asymmetry do not fit in the scheme, calling for new physics beyond the SM. DM existence is highly motivated by various astrophisical observations but its fundamental properties remain to date unknown. Experimental efforts have been mainly focused, until today, in the WIMPs search (Weakly Interacting Massive Particles): in this paradigm, DM is made of particles with mass of order of $\sim 100-1000$ GeV interacting with the Standard Model via Weak force. Despite attaining the highest energy ever reached at accelerators, LHC has not yet been able to provide evidence for WIMPs-like particles. The same null results in direct detection of halo DM strongly constrains this class of models. 

\begin{figure}[h!]
\begin{center}
\includegraphics[width=0.65\textwidth]{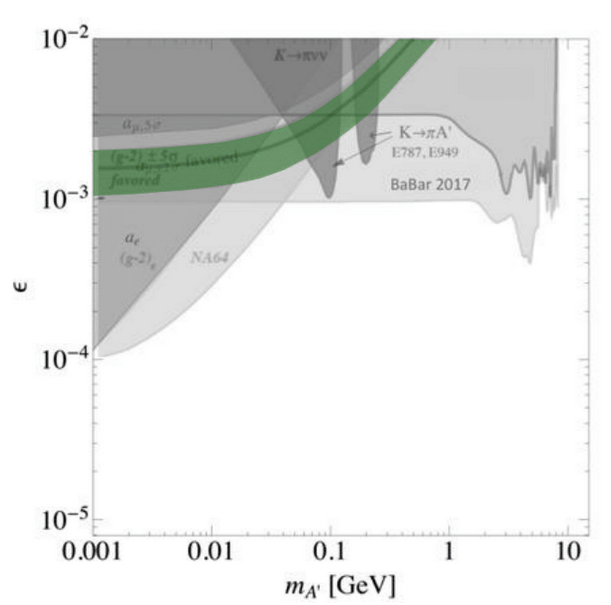}
\caption{Current exclusion limits for $\Apr$ invisible decay.}
\label{fig:inv}
\end{center}
\end{figure}
Recently, the interest in new scenarios predicting DM candidates with lower masses has grown. Various models postulate the existence of a hidden sector interacting with the visible world through new {\it portal} interactions that are constrained by the symmetries of the SM. In particular, DM with mass below 1~GeV/$c^2$ interacting with the Standard model particles via a light boson (a heavy photon or $\Apr$, also called dark photon) represents a well motivated scenario that generated many theoretical and phenomenological studies. In this specific scenario the DM, charged under a new gauge symmetry $U(1)_D$~\cite{Hol86}, interacts with electromagnetic charged SM particles through the exchange of a dark photon. The interaction between the $\Apr$ and SM particles is generated effectively by a {\it kinetic mixing} operator. The low energy effective Lagrangian extending the SM to include dark photons can thus be written as
\begin{equation}
\mathcal{L}_{eff} = \mathcal{L}_{SM} - \frac{\epsilon}{2} F^{\mu\nu} F'_{\mu\nu} -\frac{1}{4} F'^{\mu\nu} F'_{\mu\nu} +\frac{1}{2}m_{\Apr}^2\Apr_\mu \Apr^\mu 
\end{equation}
where $F_{\mu\nu}$ is the usual electromagnetic tensor, $F'_{\mu\nu}$ is the $\Apr$ field strength, $m_{\Apr}$ is the mass of the heavy photon, and $\epsilon$ is the mixing coupling constant. In this scenario, SM particles acquire a dark {\it millicharge} proportional to $\epsilon^2$. The value of $\epsilon$ can be so small as to preclude the discovery of the $\Apr$ in the experiments carried out so far.

The decay of the $\Apr$ depends on the ratio between its mass and the mass of the dark sector particles: if the dark photon mass is  smaller than twice the muon mass and no dark sector particle lighter than the $\Apr$ exists, it can only decay to $e^+e^-$ pairs ({\it Visible Decay}). Instead, if new $\chi$ particles with $2m_\chi < m_{\Apr}$  exist in the dark sector, the dominant dark photon decay mode is $\Apr\rightarrow \chi \bar{\chi}$ ({\it Invisible Decay}). In this letter, we only address the second scenario (see Fig.~\ref{fig:inv} for the current state of the $\Apr$ research in the {\it Invisible Decay} scenario).

\subsection{Annihilation induced $\Apr$ production}

\begin{figure}[h!]
\centerline{\includegraphics[width=0.80\textwidth]{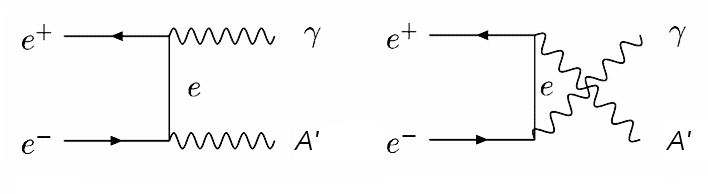}}
\caption{$\Apr$ production via $e^+e^-$ annihilation.}
\label{fig:fay}
\end{figure}
The $\Apr$ can be produced in $e^+e^-$ annihilation, via the $e^+e^- \to \gamma \Apr$ reaction (Fig.~\ref{fig:fay}). Several experiments have been proposed to search for the production of $\Apr$ in this process (e.g. PADME@LNF~\cite{Rag14}, and VEPP-3~\cite{Woj17}). The first $e^+$ on target experiment searching for $\Apr$ is PADME (Positron Annihilation into Dark Matter Experiment) which uses the 550~MeV positron beam provided by the DA$\Phi$NE linac at LNF (Laboratori Nazionali di Frascati) impinging on a thin diamond target.

The experiment involves the detection of the photons from the annihilation process with a BGO electromagnetic calorimeter placed $\sim 2$~m  downstream of the interaction target. The $\Apr$ leaves the detector area without interacting. A magnetic field of $\sim$1~T bends away  from the calorimenter the positron beam and all the charged particles produced in the target. A single kinematic variable, the missing mass, is computed for each event
\begin{equation}
M^2_{miss} = (P_{e^-} + P_{beam} -P_\gamma)^2 \, .
\end{equation}
The corresponding distribution peaks at $M^2_{\Apr}$ in case of production of the $\Apr$. All processes resulting in a single $\gamma$ hitting the calorimeter constitute the experimental background. Among these, the most relevant are: bremsstrahlung, annihilation into $2\gamma$ ($e^+e^- \to \gamma \gamma$), annihilation into $3\gamma$ ($e^+e^- \to \gamma \gamma \gamma$). In order to reduce the bremsstrahlung background, the PADME detector features an active veto system composed of plastic scintillators: positrons losing energy via bremmstrahlung in the target are detected in the vetos, allowing to reject the event. However, the pile-up of the bremsstrahlung events is an issue for this class of experiments, limiting the maximum viable beam current of the beam. For this reason, a beam with a continuous structure would be the best option for a PADME-like experiment. 

The sensitivity of PADME-like experiments in the $\Apr$ parameter space is constrained by the available energy in the center of mass frame: with a beam energy of $\sim$500~MeV, PADME can search for masses up to 22.5~MeV. Higher energy positron beams are required to exceed these limits. In this letter, the achievable sensitivity of a Dark Photon experiment using the proposed 11~GeV continuous positron beam at JLab is discussed.

\subsection{Searching for $\Apr$ with positrons at Jefferson Lab}

\begin{figure}[h]
\begin{center}
\includegraphics[width=0.80\textwidth]{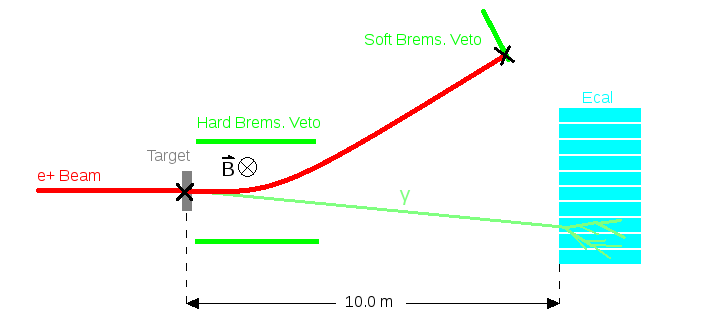}
\caption{Schematic of the proposed experiment at Jefferson Lab.}
\label{fig:jlab}
\end{center}
\end{figure}
The perspective of a high energy continuous positron beam at Jlab is particularly attractive to enlarge the reach of the $\Apr$ search in the annihilation channel. For a 11~GeV positron beam, the mass region up to $\sim$106~MeV can be investigated. The experimental setup foreseen for such an experiment at JLab is presented in Fig.~\ref{fig:jlab}. It features:
\renewcommand{\labelenumi}{\it\roman{enumi})}
\begin{enumerate}
\item{A 100 $\mu$m thick carbon target, as a good compromise between density and a low $Z/A$ ratio to minimize bremsstrahlung production;}
\item{A 50~cm radius highly segmented (1$\times$1$\times$20~cm$^3$ crystals) electromagnetic calorimeter placed 10~m downstream of 
the target, and with the energy resolution $\sigma(E)/E$=$0.02/\sqrt{E(GeV)}$;}
\item{An active veto system with a detection efficiency higher than $99.5$\% for charged particles;}
\item{A magnet supporting a field of 1~T over a 2~m region downstream of the target, and bend the positron beam.} 
\end{enumerate}
Experimental projections are evaluated assuming an adjustable beam current betwen 10-100~nA, a momentum dispersion beeter than 1\%, and an angular dispersion better than 0.1~mrad. It should be noticed that momentum and angular dispersion are critical parameters for such an experiment, since a good knowledge of the beam particles initial state is fundamental for the missing mass calculation. Given the low current involved, a natural location for this experimental setup could be JLab Hall B; in this case 
the electromagnetic calorimeter could be placed in the downstream alcove.

\subsection{Experimental projections}

The study of the reconstructed missing mass distribution for the background events serves as a basic criteria to evaluate the sensitivity of the proposed experimental setup. As discussed previously, the main background processes of this experiment are bremsstrahlung and electron-positron annihilation into 2 or 3 photons, which can result in a single hit in the calorimeter. Different strategies were adopted to study the impact of these backgrounds.

\begin{figure}[h!]
\begin{center}
\includegraphics[width=0.75\textwidth]{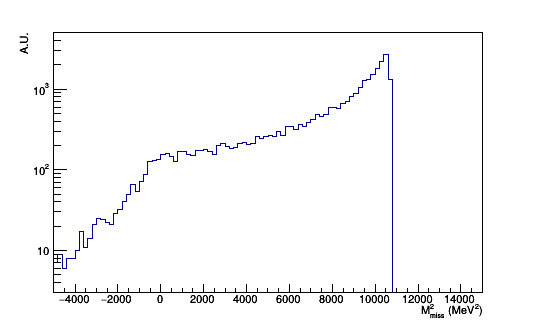}
\caption{Calculated missing mass spectrum of bremsstrahlung events.}
\label{fig:bre}
\end{center}
\end{figure}
\begin{figure}[h!]
\begin{center}
\includegraphics[width=0.75\textwidth]{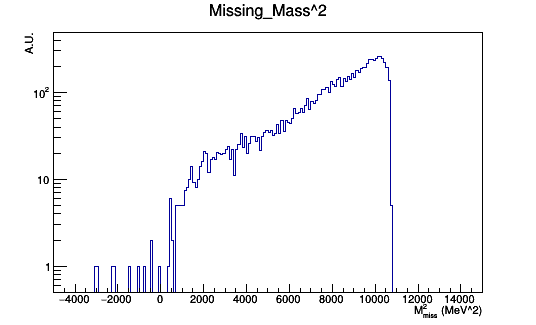}
\caption{Calculated missing mass spectrum of 3 photons events.}
\label{fig:tri}
\end{center}
\end{figure}
Considering the bremsstrahlung background, a full GEANT4~\cite{Ago03} simulation of the positron beam interacting with the target was  performed. The missing mass was computed for all bremsstrahlung photons reaching the calorimeter volume, accounting for the detector angular and momentum resolution. Figure~\ref{fig:bre} shows the obtained spectrum. The total rate of expected bremsstrahlung events for positron on target was scaled accounting for the effect of the veto system. 

The annihiliation into 2 or 3 photons is much less frequent than bremsstrahlung and was therefore studied differently: events were generated directly using CALCHEP~\cite{Puk04} which provided also the total cross sections for the processes. Photons generated in the annihilation were propagated to the calorimeter volume using a custom code and, as in the case of bremsstrahlung, missing mass spectrum was computed for events with a single $\gamma$-hit in the calorimeter. This study proved that, if an energy cut of 600~MeV is applied, the 2$\gamma$-background becomes negligible. This is due to the closed kinematics of the $e^+e^- \rightarrow \gamma\gamma$ process: asking for only one photon to hit the detector translates in a strong constraint on its energy. This argument is not valid for the 3$\gamma$-events: the number of background events from this process is in fact not negligible (see Fig.~\ref{fig:tri} for the missing mass spectrum).

\begin{figure}[t!]
\begin{center}
\includegraphics[width=0.7\textwidth]{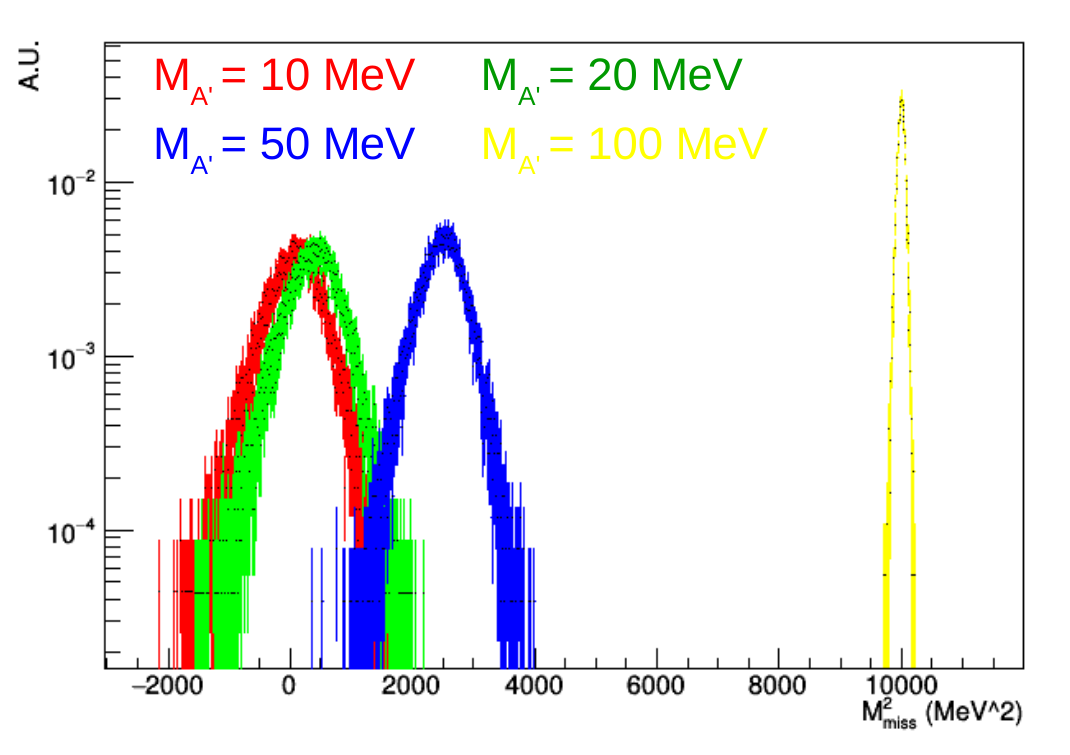}
\caption{Calculated missing mass spectrum of signal events at 4 different $m_{\Apr}$ values.}
\label{fig:sig}
\end{center}
\end{figure}

Signal events were simulated using CALCHEP. The widths $\sigma(m_{\Apr})$ of the missing mass distributions of the measured recoil photon from the $e^+e^- \rightarrow  \gamma \Apr$ process were computed for six different values of the $\Apr$ mass in the 1-103 MeV range. Figure~\ref{fig:sig} shows the corresponding spectra: the missing mass resoluton of the signal is maximum for at high $\Apr$ masses and degrades at low masses ($m_{\Apr} <50$~MeV). As for the annihilation background, CALCHEP provides the total cross section of the process for a full coupling strength ($\varepsilon = 1$). It is then necessary to multiply it with $\epsilon^2$ to obtain the cross section for different coupling values.

\begin{figure}[t!]
\begin{center}
\includegraphics[width=0.80\textwidth]{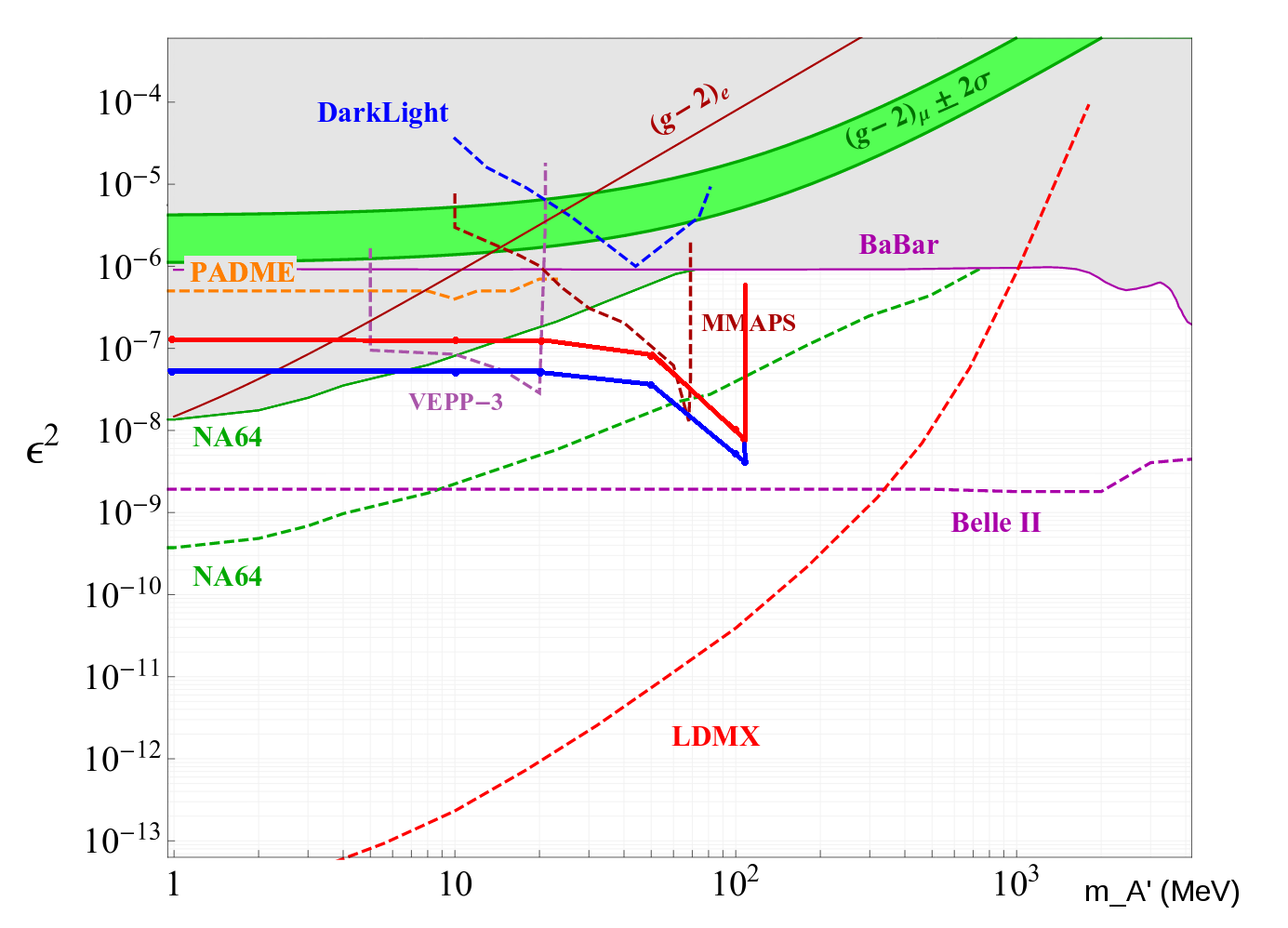}
\caption{Projected exclusion limits in the $\Apr$ invisible decay parameter space for a 180 days experiment with a 10~nA (red curve) and 100~nA (blue curve) 11~GeV positron beam at Jefferson Lab.}
\label{fig:reach}
\end{center}
\end{figure}

The reach of the proposed experiment is obtained from the comparison of the signal and background spectra. A period of 180 days at 10(100)~nA positron beam current is considered. $N_s(m_{\Apr})$ representing the number of expected signal events for a given mass $m_{\Apr}$ at full coupling, $N_B(m_{\Apr})$ representing the number of expected total background events within the missing mass in the interval $[m^2_{\Apr}-2\sigma(m^2_{\Apr});m^2_{\Apr}+2\sigma(m^2_{\Apr})]$, the minimum measurable $\epsilon^2$ coupling writes 
\begin{equation}
\epsilon^2_{min}(m_{\Apr}) = 2\frac{\sqrt{N_B(m_{\Apr})}}{N_S(m_{\Apr})} \, .
\end{equation}
Corresponding in the $(m_{\Apr},\epsilon^2)$ phase space are shown in Fig.~\ref{fig:reach}. Even at low positron beam current (10~nA), an $\Apr$-search experiment at Jefferson Lab will exceed the sensitivity of other current experiments, probing a significant region of the unexplored parameter space. It should be noticed that the sensitivity is maximum for $m_{\Apr}$ values approaching the total energy in the center of mass frame $\sqrt{s} \simeq \sqrt{2m_e E_{beam}} \simeq 100 \,MeV$, due to the $\Apr$ production enhancement near the resonant regime. We considered here a setup with at a fixed energy beam, but this observation suggests that such an experiment would benefit from a postron beam with variable energy, since the value of $\sqrt{s}$ could be optimized to search for given  $m_{\Apr}$ values.

\subsection{Summary}

Making use of the future JLab high energy positron beam with a current in the range of tens of nAs, a PADME-like experiment at JLab running over 180 days will extend the $\Apr$ mass reach up to 100~MeV and will lower the exclusion limit for invisible $\Apr$ decay by up to a factor of 10 in $\epsilon^2$.

%
%

%
%
\end{document}